\documentclass[british]{aa}
\usepackage[T1]{fontenc}
\usepackage{url}
\usepackage{amsmath}
\usepackage{amssymb}
\usepackage{graphicx}
\usepackage[authoryear]{natbib}

\makeatletter

\newcommand{\lyxdot}{.}

\usepackage{txfonts}
\bibpunct{(}{)}{;}{a}{}{,} 

\makeatother

\usepackage{babel}
\begin{document}

\title{The planet search programme at the ESO CES and HARPS\thanks{Based on observations collected at the European Southern Observatory, La Silla Chile, ESO programmes 50.7-0095, 51.7-0054, 52.7-0002, 53.7-0064, 54.E-0424, 55.E-0361, 56.E-0490, 57.E-0142, 58.E-0134, 59.E-0597, 60.E-0386, 61.E-0589, 62.L-0490, 64.L-0568, 66.C-0482, 67.C-0296, 69.C-0723, 70.C-0047, 71.C-0599, 072.C-0513, 073.C-0784, 074.C-0012, 076.C-0878, 077.C-0530, 078.C-0833, 079.C-0681. Also based on data obtained from the ESO Science Archive Facility.}\fnmsep\thanks{Radial velocity data are available in electronic form at the CDS via anonymous ftp to cdsarc.u-strasbg.fr (130.79.128.5) or via http://cdsweb.u-strasbg.fr/cgi-bin/qcat?J/A+A/} }

\subtitle{IV. The search for Jupiter analogues around solar-like stars}

\author{M. Zechmeister\inst{1,2}\and M. K\"{u}rster\inst{2}\and M. Endl\inst{3}\and
G. Lo Curto\inst{4}\and H. Hartman\inst{5,6}\and H. Nilsson\inst{6}\and
T. Henning\inst{2}\and A. P. Hatzes\inst{7}\and W.~D. Cochran\inst{3} }

\institute{Institut f\"{u}r Astrophysik, Georg-August-Universit\"{a}t, Friedrich-Hund-Platz
1, 37077 G\"{o}ttingen, Germany\\ \email{zechmeister@astro.physik.uni-goettingen.de}\and
Max-Planck-Institut f\"{u}r Astronomie, K\"{o}nigstuhl 17, 69117
Heidelberg, Germany\and McDonald Observatory, University of Texas,
Austin, TX78712, USA\and European Southern Observatory, Karl-Schwarzschild-Str.
2, 85748 Garching, Germany\and Group for Materials Science and Applied
Mathematics, School of Technology, Malm\"{o} University, SE-20506
Malm\"{o}, Sweden\and Lund Observatory, Lund University, Box 43,
22100 Lund, Sweden\and Th\"{u}ringer Landessternwarte Tautenburg
(TLS), Sternwarte 5, 07778 Tautenburg, Germany}

\date{Received / Accepted}

\abstract{In 1992 we began a precision radial velocity survey for planets around
solar-like stars with the Coud\'{e} Echelle Spectrograph and the
Long Camera (CES LC) at the 1.4\,m telescope in La Silla (Chile)
resulting in the discovery of the planet $\iota$~Hor~b. We have
continued the survey with the upgraded CES Very Long Camera (VLC)
and the HARPS spectrographs, both at the 3.6\,m telescope, until
2007.}{In this paper we present additional radial velocities for
31 stars of the original sample with higher precision. The observations
cover a time span of up to 15 years and permit a search for Jupiter
analogues.}{The survey was carried out with three different instruments/instrument
configurations using the iodine absorption cell and the ThAr methods
for wavelength calibration. We combine the data sets and perform a
joint analysis for variability, trends, and periodicities. We compute
Keplerian orbits for companions and detection limits in case of non-detections.
Moreover, the HARPS radial velocities are analysed for correlations
with activity indicators (CaII~H\&K and cross-correlation function
shape).}{We achieve a long-term RV precision of 15\,m/s (CES+LC,
1992--1998), 9\,m/s (CES+VLC, 1999--2006), and 2.8\,m/s (HARPS,
2003--2009, including archive data), respectively. This enables us
to confirm the known planetary signals in $\iota$~Hor and HR~506
as well as the three known planets around HR~3259. A steady RV trend
for $\epsilon$~Ind~A can be explained by a planetary companion
and calls for direct imaging campaigns. On the other hand, we find
previously reported trends to be smaller for $\beta$~Hyi and not
present for $\alpha$~Men. The candidate planet $\epsilon$~Eri~b
was not detected despite our better precision. Also the planet announced
for HR~4523 cannot be confirmed. Long-term trends in several of our
stars are compatible with known stellar companions. We provide a spectroscopic
orbital solution for the binary HR~2400 and refined solutions for
the planets around HR~506 and $\iota$~Hor. For some other stars
the variations could be attributed to stellar activity, as e.g. the
magnetic cycle in the case of HR~8323.}{The occurrence of two Jupiter-mass
planets in our sample is in line with the estimate of 10\% for the
frequency of giant planets with periods smaller than 10\,yr around
solar-like stars. We have not detected a Jupiter analogue, while the
detections limits for circular orbits indicate at 5\,AU a sensitivity
for minimum mass of at least 1$M_{\mathrm{Jup}}$ (2$M_{\mathrm{Jup}}$)
for 13\% (61\%) of the stars.}


\authorrunning{Zechmeister et al.}

\maketitle

\section{Introduction}

The search for extra-solar planets has so far revealed approximately
850 exoplanets%
\footnote{\url{http://exoplanet.eu}%
}, most of them discovered by the radial velocity (RV) technique. Interestingly,
many hot Jupiters have been found, a consequence related to the fact
that the RV method as well as the transit method is more sensitive
to short period planets. Out of 850 planets discovered so far, 65\%
have a period shorter than 1 year. Before the discovery of the first
extrasolar planet around a solar-like star, the hot Jupiter 51~Peg~b
\citep{Mayor1995}, it was widely expected that planetary systems
are in general similar to the solar-system and this was also predicted
by most theoretical models as noted by \citet{Marcy2008}. In the
solar system, Jupiter is the dominant planet amongst all other planets
and causes the largest RV amplitude. Therefore surveys were set up
to search for planets with masses of 1\,$M_{\mathrm{Jup}}$ and at
distances of 5\,AU from solar-like stars \citep[e.g.][]{Walker1995}.
The regime of Jupiter analogues is still sparsely explored because
observations with long time-baselines and precise RV measurements
are required; e.g. Jupiter orbits the Sun in 12 years and induces
an RV semi-amplitude of 12\,m/s.

There are many exoplanet search projects e.g. at Lick, AAT \citep{OToole2009,Wittenmyer2011},
Keck \citep{Cumming08}, ELODIE/SOPHIE \citep{Naef2005,Bouchy2009a},
CORALIE \citep{Segransan2010}, and HARPS (\citealp{Naef2010}; \citealp{Mayor2011}).
These high precision RV projects have discovered a large fraction
of the currently known planets and are continuously extending their
time baselines. Examples for discovered Jupiter analogues are GJ~777Ab
\citep{Naef2003}, a 1.33\,$M_{\mathrm{Jup}}$ planet at 4.8\,AU%
\footnote{The planet was confirmed by \citet{Vogt2005} who revised the semi-major
axis to 3.9\,AU and discovered a second inner planet (17.1\,d, 0.057\,$M_{\mathrm{Jup}}$).%
} around a G6IV star, or HD~154345b \citep{Wright2008}, a 0.95\,$M_{\mathrm{Jup}}$
planet at 4.5\,AU around a G8V dwarf (all masses are $M\sin i$ minimum
masses). Two more Jupiter-analogues were also recently reported by
\citet{Boisse2012}: HD150706b (2.7\,$M_{\mathrm{Jup}}$, 7\,AU)
and HD222155b (1.9\,$M_{\mathrm{Jup}}$, 5.1\,AU).

The survey described in this paper was begun in 1992 \citep{Endl02}
with the Coud\'{e} Echelle Spectrograph (CES) Long Camera (LC). With
the advent of the CES Very Long Camera (VLC) in 1999, it was transferred
to this instrument combination, and was continued in 2003 with the
HARPS spectrograph. The last observations for this programme were
taken in September of 2007, although we have also made use of archival
data acquired up to 2009. The survey covers a time span of up to 15
years with RV precisions ranging from 15\,m/s down to 2\,m/s. A
comparable survey was analysed by \citet{Wittenmyer06} and carried
out in the northern hemisphere with the 2.7\,m telescope at the McDonald
Observatory. It started in 1988 with 24 solar-like stars%
\footnote{There are three targets ($\delta$~Eri, $\alpha$~For, and $\tau$~Cet)
common to both samples (we do not combine the measurement of both
samples).%
} and 7 subgiants. When combined with CFHT data \citep{Walker1995},
it gave an even longer temporal coverage of up to 25 years, albeit
with a somewhat lower precision (10--20\,m/s).

\section{The sample}

The original sample of 37 solar-like stars was introduced in detail
in \citet{Endl02}. Of these, the monitoring of six stars was stopped:
HR~448, HR~753, HR~7373, Barnard's star, Proxima Centauri, and
GJ~433. The first three had been observed tempo\-ra\-rily in 1996/97
as once promising metal rich targets, but were soon left out due to
limited observing time. The latter three are M~dwarfs which were
included in a dedicated M dwarf survey with VLT+UVES. For these stars
recent and more precise results are published in \citet{Zechmeister09b}.
So we are left with the 31 stars listed in Table~\ref{Tab:SampleInfo}
along with some of their properties (spectral type, visual magnitude,
distance, and stellar mass).

\begin{table}
\caption{\label{Tab:SampleInfo}Targets with their spectral type SpT \citep{Hoffleit1991},
visual magnitude~$V$ \citep{Perryman1997}, distance $d$ \citep{vanLeeuwen07},
secular acceleration $\dot{v}_{\mathrm{r}}$, and stellar mass $M$.}

\setlength{\tabcolsep}{3pt}

\begin{centering}
\begin{tabular}{lll@{}rrrl@{\,}l}
\hline 
\hline Star & alias & \multicolumn{1}{l}{SpT} & \multicolumn{1}{l}{$V$ {[}mag{]}} & $d$ {[}pc{]} & $\dot{v}_{\mathrm{r}}$ {[}m/s/yr{]} & \multicolumn{2}{c}{$M$ {[}$M_{\odot}${]}}\\
\hline 
\object{HR77} & $\zeta$~Tuc & F9V & 4.23 & 8.59 & 0.84 & 1.06 & [PM]\\
\object{HR98} & $\beta$~Hyi & G2IV & 2.82 & 7.46 & 0.86 & 1.1 & [D]\\
\object{HR209} & HR~209 & G1V & 5.80 & 15.16 & 0.01 & 1.10 & [G]\\
\object{HR370} & $\nu$~Phe & F8V & 4.97 & 15.11 & 0.16 & 1.20 & [G]\\
\object{HR506} & HR~506 & F9V & 5.52 & 17.43 & 0.02 & 1.17 & [G]\\
\object{HR509} & $\tau$~Cet & G8V & 3.49 & 3.65 & 0.31 & 0.78 & [T]\\
\object{HR695} & $\kappa$~For & G0V & 5.19 & 21.96 & 0.02 & 1.12 & [G]\\
\object{HR810} & $\iota$~Hor & G0V & 5.40 & 17.17 & 0.06 & 1.25 & [V]\\
\object{HR963} & $\alpha$~For & F8V & 3.80 & 14.24 & 0.17 & 1.20 & [G]\\
\object{HR1006} & $\zeta^1$~Ret & G2.5V & 5.53 & 12.01 & 0.61 & 1.05 & [G]\\
\object{HR1010} & $\zeta^2$~Ret & G1V & 5.24 & 12.03 & 0.61 & 1.10 & [G]\\
\object{HR1084} & $\epsilon$~Eri & K2V & 3.72 & 3.22 & 0.07 & 0.85 & [DS]\\
\object{HR1136} & $\delta$~Eri & K0IV & 3.52 & 9.04 & 0.12 & 1.23 & [PM]\\
\object{HR2261} & $\alpha$~Men & G6V & 5.08 & 10.20 & 0.01 & 0.95 & [G]\\
\object{HR2400} & HR~2400 & F8V & 5.58 & 36.91 & 0.02 & 1.20 & [G]\\
\object{HR2667} & HR~2667 & G3V & 5.56 & 16.52 & 0.06 & 1.04 & [G]\\
\object{HR3259} & HR~3259 & G7.5V & 5.95 & 12.49 & 0.30 & 0.90 & [G]\\
\object{HR3677} & HR~3677 & G0III & 5.85 & 196.85 & 0.00 & 2.1 & [G]\\
\object{HR4523} & HR~4523 & G3V & 4.89 & 9.22 & 0.53 & 1.04 & [G]\\
\object{HR4979} & HR~4979 & G3V & 4.85 & 20.67 & 0.07 & 1.04 & [G]\\
\object{HR5459} & $\alpha$~Cen~A & G2V & -0.01 & 1.25 & 0.42 & 1.10 & [P]\\
\object{HR5460} & $\alpha$~Cen~B & K1V & 1.35 & 1.32 & 0.40 & 0.93 & [P]\\
\object{HR5568} & GJ~570~A & K4V & 5.72 & 5.84 & 0.54 & 0.71 & [G]\\
\object{HR6416} & HR~6416 & G8V & 5.47 & 8.80 & 0.22 & 0.89 & [G]\\
\object{HR6998} & HR~6998 & G4V & 5.85 & 13.08 & 0.01 & 1.00 & [G]\\
\object{HR7703} & HR~7703 & K3V & 5.32 & 6.02 & 0.37 & 0.74 & [G]\\
\object{HR7875} & $\phi^2$~Pav & F8V & 5.11 & 24.66 & 0.24 & 1.1 & [PM]\\
\object{HR8323} & HR~8323 & G0V & 5.57 & 15.99 & 0.04 & 1.12 & [G]\\
\object{HR8387} & $\epsilon$~Ind~A & K4.5V & 4.69 & 3.62 & 1.84 & 0.70 & [G]\\
\object{HR8501} & HR~8501 & G3V & 5.36 & 13.79 & 0.19 & 1.04 & [G]\\
\object{HR8883} & HR~8883 & G4III & 5.65 & 101.32 & 0.00 & 2.1 & [G]\\

\hline 
\end{tabular}
\par\end{centering}

References for mass estimates: {[}D{]} \citet{Dravins1998}, {[}DS{]}
\citet{Drake1993}, {[}G{]} \citet{Gray1988}, {[}PM{]} Porto de Mello,
priv. comm., {[}P{]} \citet{Pourbaix2002}, {[}T{]} \citet{Teixeira09},
{[}V{]} \citet{Vauclair08}.
\end{table}

All stars have a brightness of $V<6$\,mag and spectral types ranging
from late F to K. There are two subgiant stars ($\beta$~Hyi and
$\delta$~Eri) and two giant stars (HR~3677 and HR~8883) in the
sample%
\footnote{HR~3677 and HR~8883 were indicated in the Bright Star Catalogue
as dwarf stars \citep{Hoffleit1991}. Therefore they entered our sample,
however they are giants as indicated by their distances.%
}.

The sample includes six stars for which planet detections have been
claimed. These are HR~506 (Mayor et al.)%
\footnote{\label{fn:Mayor03}HR~506b was announced by Mayor et al. at the XIX~th
IAP Colloquium in Paris (2003). We found no refereed publication.
Information is available on \url{http://obswww.unige.ch/~udry/planet/hd10647.html}.%
}, $\iota$~Hor \citep{Kuerster00}, $\epsilon$~Eri \citep{Hatzes00},
HR~3259 \citep{Lovis06}, HR~4523 \citep{Tinney2011}, and recently
$\alpha$~Cen~B \citep{Dumusque2012}. In Sect.~5 we provide more
detailed information on individual objects and we will stress those
planet hypotheses.

\section{Instruments and data reduction}

We used three high resolution spectrographs that are briefly described
below with more detail provided for the less known VLC+CES. Table~\ref{Tab:Instruments}
gives an overview of some basic properties of the three instruments.

\begin{table}
\caption{\label{Tab:Instruments}The three used instruments/configurations
with their wavelength reference, chosen spectral coverage and resolving
power, and telescope diameter.}

\centering{}
\begin{tabular}{llrcc}
\hline 
\hline Spectrograph & Ref. & \multicolumn{1}{c}{$\lambda$ {[}\AA{}{]}} & $R$ & Tel.\\
\hline 
CES + LC & I$_{2}$ & 5367 -- 5412 & 100\,000 & 1.4\,m\\
CES + VLC & I$_{2}$ & 5376 -- 5412 & 220\,000 & 3.6\,m\\
HARPS & ThAr & 3800 -- 6900 & 115\,000 & 3.6\,m\\
\hline 
\end{tabular}
\end{table}

\subsection{CES + Long Camera}

In 1992 the survey started (1992-11-03 to 1998-04-04) with the Coud\'{e}
Echelle Spectrograph (CES; \citealp{Enard1982}) and its Long Camera
(LC) fed by the 1.4\,m Coud\'{e} Auxiliary Telescope (CAT) at La
Silla (Chile). The CES+LC had a chosen wavelength coverage of 45\,\AA{}
and a resolution of 100\,000 (Table~\ref{Tab:Instruments}). A 2\,k
$\times$ 2\,k CCD gathered part of one spectral Echelle order. An
iodine gas absorption cell provided the wavelength calibration. More
details about the instrument, data analysis, as well as the obtained
results can be found in \citet{Endl02}. Table~\ref{Tab:RVprecision}
lists the radial velocity results. The median rms is 15.2\,m/s when
excluding the giants and targets with companions and trends as commented
in Table~\ref{Tab:RVprecision} and reflects the typical precision.

\begin{table*}
\caption{\label{Tab:RVprecision}Radial velocity results for all targets. For
each instrument configuration the rms is calculated independently.
RV data are not binned. Comments are on multiplicity (see also Table~\ref{Tab:Binaries}
for more information) and also indicate giants.}

\begin{centering}
\begin{tabular}{l|rrrr|rrrr|rrrr|l}
\hline 
\hline Star & \multicolumn{4}{c|}{CES + LC} & \multicolumn{4}{c|}{CES + VLC} & \multicolumn{4}{c|}{HARPS} & Comment\\
 & \multicolumn{1}{c}{$N$} & \multicolumn{1}{c}{$T$} & \multicolumn{1}{c}{rms} & \multicolumn{1}{c|}{$\overline{\Delta RV}$} & \multicolumn{1}{c}{$N$} & \multicolumn{1}{c}{$T$} & \multicolumn{1}{c}{rms} & \multicolumn{1}{c|}{$\overline{\Delta RV}$} & \multicolumn{1}{c}{$N$} & \multicolumn{1}{c}{$T$} & \multicolumn{1}{c}{rms} & \multicolumn{1}{c|}{$\overline{\Delta RV}$} & \\
 &  & {[}d{]} & {[}m/s{]} & {[}m/s{]} &  & {[}d{]} & {[}m/s{]} & {[}m/s{]} &  & {[}d{]} &  {[}m/s{]} & {[}m/s{]} & \\
\hline 
$\zeta$~Tuc & 51 & 1888 & 19.8 & 14.5 & 48 & 2104 & 9.2 & 7.9 & 1019 & 2206 & 1.9 & 0.4\\
$\beta$~Hyi & 157 & 1887 & 22.9 & 18.6 & 46 & 1920 & 7.7 & 12.5 & 2860 & 1837 & 2.6 & 0.4\\
HR~209 & 35 & 1572 & 17.2 & 17.6 & 36 & 1941 & 11.1 & 9.8 & 48 & 1401 & 8.6 & 0.5\\
$\nu$~Phe & 58 & 1926 & 15.6 & 14.6 & 35 & 1910 & 10.1 & 8.2 & 63 & 1415 & 2.7 & 0.6\\
HR~506 & 23 & 1573 & 28.0 & 20.2 & 42 & 1910 & 18.7 & 17.7 & 119 & 1401 & 10.7 & 1.0 & planet\\
$\tau$~Cet & 116 & 1888 & 11.4 & 13.3 & 61 & 1920 & 8.1 & 8.8 & 5373 & 2244 & 1.4 & 0.3\\
$\kappa$~For & 40 & 1890 & 722.9 & 12.8 & 45 & 2094 & 1134.0 & 10.2 & 74 & 1401 & 657.6 & 0.5 & SB1\\
$\iota$~Hor & 95 & 1976 & 51.3 & 16.9 & 122 & 2186 & 37.1 & 12.6 & 1861 & 1401 & 13.1 & 0.9 & planet\\
$\alpha$~For & 65 & 1889 & 42.2 & 31.6 & 39 & 1856 & 16.7 & 14.7 & 191 & 1401 & 8.5 & 0.7 & trend\\
$\zeta^1$~Ret & 14 & 184 & 17.0 & 14.2 & 42 & 1857 & 15.3 & 10.3 & 63 & 1401 & 8.1 & 0.5\\
$\zeta^2$~Ret & 58 & 1976 & 18.9 & 14.6 & 43 & 1857 & 10.1 & 9.3 & 82 & 1411 & 2.8 & 0.5\\
$\epsilon$~Eri & 66 & 1889 & 12.2 & 9.0 & 69 & 2186 & 10.0 & 8.1 & 521 & 1390 & 5.5 & 0.3 & planet(?)\\
$\delta$~Eri & 48 & 1888 & 12.5 & 11.7 & 42 & 1856 & 7.5 & 7.0 & 152 & 1410 & 2.8 & 0.2\\
$\alpha$~Men & 46 & 1852 & 9.8 & 10.7 & 77 & 2368 & 8.4 & 9.8 & 188 & 1308 & 2.6 & 0.3\\
HR~2400 & 53 & 1924 & 275.1 & 23.0 & 54 & 2039 & 523.8 & 14.4 & 63 & 1296 & 190.3 & 0.8 & SB1\\
HR~2667 & 66 & 1934 & 15.1 & 18.5 & 64 & 2329 & 7.5 & 11.4 & 68 & 1296 & 1.6 & 0.4\\
HR~3259 & 35 & 1851 & 16.5 & 11.3 & 61 & 2367 & 9.2 & 7.7 & 435 & 1294 & 3.7 & 0.3 & three~planets\\
HR~3677 & 34 & 1924 & 492.4 & 15.1 & 38 & 2044 & 1253.9 & 8.2 & 66 & 1287 & 870.4 & 0.5 & SB1,~giant\\
HR~4523 & 27 & 1925 & 14.9 & 12.2 & 57 & 2276 & 6.7 & 8.9 & 253 & 1608 & 3.4 & 0.3\\
HR~4979 & 52 & 1933 & 11.7 & 10.8 & 58 & 2329 & 9.2 & 10.2 & 460 & 1286 & 3.7 & 0.3\\
$\alpha$~Cen~A & 205 & 1852 & 166.5 & 10.7 & 1074 & 1929 & 97.7 & 10.2 & 5029 & 1206 & 21.0 & 0.2 & SB1\\
$\alpha$~Cen~B & 291 & 1852 & 203.5 & 9.3 & 54 & 1770 & 247.6 & 7.7 & 255 & 1206 & 191.9 & 0.2 & SB1\\
GJ~570~A & 40 & 384 & 6.9 & 11.4 & 87 & 2284 & 10.2 & 6.5 & 47 & 1853 & 2.7 & 0.3\\
HR~6416 & 57 & 1845 & 23.8 & 12.6 & 59 & 2278 & 23.4 & 9.4 & 76 & 1310 & 7.5 & 0.4 & trend\\
HR~6998 & 51 & 1789 & 15.3 & 20.8 & 23 & 2062 & 9.6 & 9.8 & 70 & 1044 & 1.2 & 0.4\\
HR~7703 & 30 & 1042 & 10.3 & 11.6 & 31 & 2039 & 7.6 & 8.0 & 79 & 1735 & 4.8 & 0.4 & trend\\
$\phi^2$~Pav & 90 & 1969 & 32.1 & 25.6 & 200 & 2062 & 17.1 & 23.9 & 63 & 2158 & 4.0 & 1.2\\
HR~8323 & 20 & 1067 & 14.4 & 14.6 & 31 & 2124 & 11.4 & 10.1 & 318 & 1413 & 3.7 & 0.5\\
$\epsilon$~Ind~A & 73 & 1888 & 11.9 & 9.1 & 54 & 2124 & 7.1 & 7.9 & 457 & 2170 & 6.3 & 0.3 & trend(?)\\
HR~8501 & 66 & 1889 & 36.8 & 24.7 & 45 & 2125 & 36.6 & 16.1 & 58 & 1413 & 15.2 & 0.6 & trend\\
HR~8883 & 31 & 1258 & 63.1 & 31.3 & 30 & 2125 & 66.9 & 23.6 & 45 & 1401 & 61.7 & 1.6 & giant\\

\hline 
\end{tabular}
\par\end{centering}

Listed are the number of observations $N$, the time baseline $T$,
the weighted rms of the time series and the effective mean internal
radial velocity error $\overline{\Delta RV}$.
\end{table*}

\subsection{CES + Very Long Camera}

The Very Long Camera (VLC; \citealp{Piskunov1997}) of the Coud\'{e}
Echelle Spectrograph (CES) was commissioned at the ESO 3.6\,m telescope
in La Silla (Chile) in April 1998 and decommissioned in 2007. The
VLC was an upgrade of the CES that doubled the resolving power to
$R=220\,000-235\,000$ as well as the CCD length so that 80\% of the
spectral coverage compared to the LC was retained (cf. Table~\ref{Tab:Instruments}).
This upgrade together with improved internal stability, and also the
larger telescope aperture promised an improvement of the RV precision.
For our sample we collected VLC spectra from 1999-11-21 to 2006-05-24.

\begin{figure}[tbh]
\centering

\includegraphics[bb=36bp 255bp 515bp 510bp,clip,width=1\linewidth]{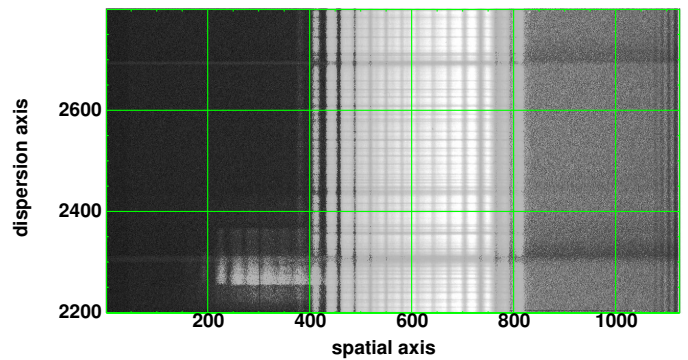}

\caption{\label{Fig:VLCspectrum}A small section of a VLC raw spectrum (star
seen through the iodine cell; broad absorption features are stellar
lines whereas narrow lines were introduced by the molecular iodine
gas). The 14 slices span 400 pixels. The bright feature to the left
near pixel row 2300 is a grating ghost. The area on the left side
has a lower bias. The readout register is in the lower right corner
of the chip (parallel clocking down, serial clocking to the right).
Deeper stellar lines near row 2300 and 2700 have tails to the left
and right. (The intensity scale is non-linear to bring out the discussed
effects.)}
\end{figure}

\begin{figure}[tbh]
\centering

\includegraphics[clip,width=1\linewidth]{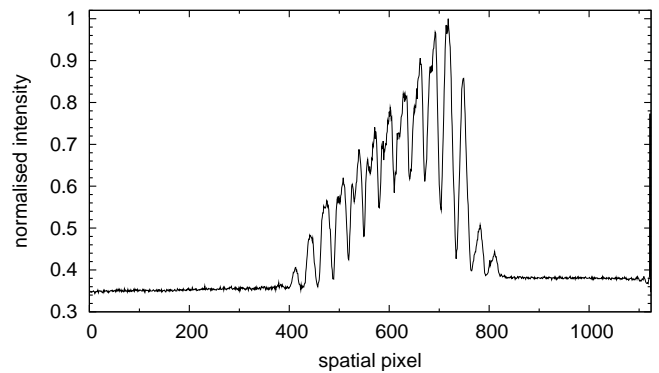}

\caption{\label{Fig:VLC_crosssection}Spatial profile of a VLC spectrum (cross-section
for the rows 2400--2410 of the raw spectrum in Fig.~\ref{Fig:VLCspectrum};
linear intensity scale).}
\end{figure}

The VLC was fed by a fibre link from the Cassegrain focus of the 3.6\,m
telescope. A modified Bowen-Walraven image slicer provided an efficient
light throughput at the high resolving power. It redistributed the
light from the fibre with a 2\arcsec{} aperture via 14 slices to
an effective slit width of 0.16\arcsec{} and resulted in a complex
illumination profile in the spatial direction, i.e. perpendicular
to the dispersion axis (Figs.~\ref{Fig:VLCspectrum} and \ref{Fig:VLC_crosssection}).
The right half of a 4\,k $\times$ 2\,k EEV CCD recorded part of
one spectral order with the wavelength range of 5376\,--\,5412\,\AA .
In 2000-06-15 CCD\#59 was replaced by CCD\#61 and in 2001-11-23 the
CES fibre was exchanged.

The CES+VLC employed the same iodine cell as the CES+LC for wavelength
calibration. This cell was controlled at a temperature of 50$^{\circ}$C.
The RV modelling (Sect.~\ref{Sub:VLC_RVs}) requires a high resolution
and high signal-to-noise iodine spectrum to reconstruct confidently
the instrumental line profile (IP) of the spectrograph. In November
2008 we obtained a laboratory spectrum for our iodine cell with $R=925\,000$
and S/N$\sim$1000 using a Bruker IFS125HR high-resolution Fourier
Transform Spectrometer (FTS) at Lund Observatory. A filament lamp
was used as a background source. To limit the light from adjacent
wavelength regions a set of filters were applied: a coloured glass
filter type VG11, a notch-filter to reduce the internal laser light,
and a hot mirror to suppress the red spectrum. While the iodine cell
laboratory spectrum previously used by \citet{Endl02} in their analysis
of the CES LC data had only $R=400\,000$, the new scan ensures an
iodine spectrum with a resolution almost 5 times higher than the resolution
of the CES+VLC.

The following properties of the CES+VLC spectra must be considered
in the data analysis: The VLC spectra are conta\-mi\-na\-ted by
a grating ghost located in the middle of the CCD (Fig.~\ref{Fig:VLCspectrum})
and suffered also from stray light produced by the image slicer. Ripples
are visible in the continuum of high S/N ($\sim$1000) spectra caused
by interference in the fibre. This can be seen, for instance, in flatfield
exposures. Also visible in flats are less efficient rows on the chip
every 512 pixels, due to a smaller pixel size resulting from the manufacturing
process, which affect the wavelength solution. Moreover, as a peculiarity
of the CES CCD electronics, a lower bias level is observed to the
left of the spectra, caused by an electronic offset that occurs after
processing a strong signal. This effect is attri\-bu\-ted to the
video amplifier electronics and requires the readout of several CCD
rows to properly discharge (P. Sinclair, ESO, 2011, priv. comm.).
Hence subsequent CCD rows are affected which may cause systematic
spectral line asymmetries and RV shifts depending on the spectral
line depth. Moreover, since the iodine lines are weaker, they may
not receive the same shift as the stronger stellar lines and cannot
correct completely for this effect.%
\footnote{This effect looks similar to charge transfer inefficiency (CTI), which
can also cause RV shifts of several m/s \citep{Bouchy2009}. However,
CTI is caused by local defects on the CCD itself.%
}

The spatial profile has a width spanning more than 400 pixels offering
a large cross-section for cosmic ray hits (so-called cosmics). For
this reason the observing strategy aimed at three consecutive spectra
in one night to be able to identify cosmics as outliers. However,
we did not use this cosmics detection method because cosmics could
also be efficiently identified as deviations from the spatial profile
in the optimum extraction.

The VLC spectra were reduced with standard IRAF-tasks including subtraction
of the overscan and a nightly master-bias, 2D flat-fielding, scattered
light subtraction, and optimum extraction \citep{Horne1986} which
also removes cosmics. The scattered light was defined left and right
of the aperture with a low-degree polynomial used to interpolate across
the aperture. This was done row-by-row and afterwards ``smoothed''
in the dispersion direction with a high-order spline to account for
the above mentioned features and then as scattered light subtracted
from the spectra. Finally, the science spectra were roughly calibrated
with a nightly ThAr spectrum to provide an initial guess for the wavelength
solution which is later refined with the iodine spectrum in the subsequent
modelling process. The whole data reduction process largely removed
the artefacts described above, however residual deviations are likely
to still exist in the RVs of the VLC data. The typical precision is
9.4\,m/s calculated as the median rms in Table~\ref{Tab:RVprecision}
for the stars without comments.

\subsection{\label{Sub:HARPS}HARPS}

With HARPS we monitored our targets from 2003-11-06 to 2007-09-21
(2009-12-19 including archive data). The HARPS spectrograph is described
in the literature \citep[e.g.][]{Mayor03,Pepe04}. It is fibre fed
from the Cassegrain focus of the 3.6\,m telescope and located in
a pressure and temperature stabilised environment. An optical fibre
sends light from a ThAr lamp to the Cassegrain adapter for wavelength
calibration. For the RV computation via cross-crorelation with a binay
mask 72 Echelle orders ranging from 3800\,\AA{} to 6900\,\AA{}
are available, a region much larger than for the CES.

We made use of the ESO advanced data products (ADP) to complement
our time series which sometimes also extended the timebase. This archive
provides fully reduced HARPS spectra including the final radial velocities
processed by the pipeline DRS%
\footnote{\url{http://www.eso.org/sci/facilities/lasilla/instruments/harps/doc/index.html}%
} 3.5 (data reduction software). The radial velocities are corrected
for the wavelength drift of the spectrograph (if measured by the simultaneous
calibration fibre) and the RV uncertainty estimated assuming photon
noise%
\footnote{The pertinent information can be found in the {*}CCF\_A.fits-file
header (keywords RVC and DVRMS).%
}. The mean RV uncertainties range from 0.2 to 0.8\,m/s and do not
include calibration errors, guiding errors, and residual instrumental
errors. For data analysis a stellar jitter term ($\ge$1.6\,m/s)
will be added in quadrature (see Sect.~\ref{Sec:RV_Analysis}).

We recomputed with the HARPS DRS some of these archival RVs that suffered
in the cross-correlation process from a misadjusted initial RV guess
(off by more than 2\,km/s) or from an inappropriate binary correlation
mask. A different mask, e.g. K5 instead of G2, can produce RV shifts
up to 20\,m/s. The publicly available archive data originate from
other programs such as short-term asteroseismology campaigns or the
HARPS GTO (guaranteed time observations). The latter complemented
our data with additional measurements, and in some cases provided
a data set that outnumbers our own in terms of number of measurements
and time base. We use only data taken in HARPS high accuracy mode
(HAM), while we leave out data taken with iodine absorption cell or
in high efficiency mode (EGGS, ``Extra Good General Spectroscopy'')
which uses a different fibre, a different injection method, and no
scrambler and has a lower stability and a different zero point. Furthermore,
spectra with a signal to noise of $\mathrm{S/N}<50$ are also left
out.

The HARPS data provide an absolute RV scale which is shown in Figs.~\ref{Fig:RV-1}--\ref{Fig:RV-5}
and serves as our references frame into which the other instruments
are transferred. Note that relative RV measurements are more precise
than the absolute RVs (i.e. more precise than accurate)%
\footnote{Systematics in absolute (spectroscopic) RVs arise e.g. from the stellar
mask, gravitational red shift and convective blue shift of the star
(e.g. \citealp{Pourbaix2002}).%
}. The median rms in Table~\ref{Tab:RVprecision} is 2.8\,m/s for
stars without comments. Note that several of our stars are active,
so that this value is higher than the precision of 1\,m/s usually
quoted for HARPS.

To improve the combination of the HARPS and VLC data, spectra were
taken in a few nights with both spectrographs immediately after each
other making use of an easy switch possible with the common fibre
adapter installed in May 2004.

\subsection{\label{Sub:VLC_RVs}Details of the RV computation for the CES+VLC
data}

\begin{figure}
\centering

\includegraphics[width=1\linewidth]{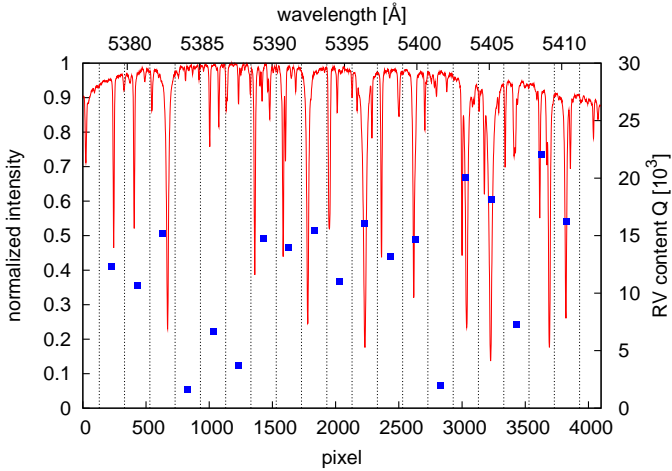}

\caption{\label{Fig:taucet.vlc}A VLC spectrum of $\tau$~Cet (without iodine
cell) and the arrangement of the 200~pixel chunks with their individual
$Q$-factors (blue squares). The intensity maximum of the spectrum
is set to unity. The intensity declines to the edges due to instrumental
effects (Blaze function).}
\end{figure}

To compute the RVs of the VLC spectra we used the AUSTRAL code described
in \citet{Endl00} which is based on the mo\-del\-ling technique
outlined in \citet{Butler1996}. The spectral order was divided into
19 spectral segments (chunks) with a size of 200 pixels (1.8\,\AA )
which we empirically found to yield the optimal RV precision. The
reasons could be that for smaller chunks the stellar RV information
content becomes too small. For a larger chunk size, on the other hand,
probably the assumption that the instrumental profile (IP) is constant
over the chunk breaks down, or the discontinuities of the wavelength
solution by the mentioned smaller and less efficient pixel rows are
more problematic.

The stellar spectrum can be shifted across the CCD by several pixels
due to the barycentric velocity of the Earth (calculated with the
JPL ephemerides DE200, e.g. \citealt{Standish1990}) and offsets in
the instrument setup. To ensure that the same stellar lines fell in
the same chunk, we shifted the chunks to the proper spectral location
according to the barycentric correction (1 pixel is $\sim$ 500\,m/s).
So instead of having fixed chunk positions with respect to the CCD
as originally implemented in the AUSTRAL code, this modification ensures
always the same weighting factor for each chunk. The final wavelength
solution in each chunk is provided by the iodine lines which record
the instrumental drifts and offsets.

Figure~\ref{Fig:taucet.vlc} illustrates the alignment of the chunks
with respect to the stellar spectrum. This placement of the chunks
tries to avoid splitting up stellar lines between adjacent chunks.
As one can see, the chunks contain only a few deep stellar lines or
sometimes none. To quantify this, we calculated the quality factor
$Q$ \citep{Connes85,Butler1996,Bouchy01} for each chunk in a stellar
template%
\footnote{This template was used in the modelling and obtained via deconvolution
from a stellar spectrum taken without iodine cell as described in
\citet{Endl02}.%
}. This factor sums in a flux-weighted way the squared gradients in
a spectrum%
\footnote{${\displaystyle Q=\sqrt{\frac{\sum A(i)\left(\frac{\partial\ln A(i)}{\partial\ln\lambda(i)}\right)^{2}}{\sum A(i)}}}$,
where $A(i)$ is the flux in the $i$-th pixel.%
}, hence measuring its RV information content. For photon noise, the
estimated RV uncertainty is inversely proportional to $Q$, i.e. $\Delta RV\sim\frac{1}{Q}$.
Hence, we weight each chunk RV with $Q^{2}$ when computing the RV
mean. Chunks with $Q<7\,000$ were discarded (cf. Fig.~\ref{Fig:taucet.vlc},
right axis). For comparison, the quality factor is $Q$=12857 for
the whole spectral range in Fig.~\ref{Fig:taucet.vlc} and $Q=67\,000$
for an iodine spectrum (e.g. the spectrum of a featureless B-star
taken through the iodine cell).

For $\tau$~Cet (GJ~71, HIP~8102, HR~509, HD~10700) which is
known as an RV constant star (the HARPS data have an rms scatter of
$\sim$1.1\,m/s HARPS data, \citealt{Pepe2011}, this work), we achieve
with the CES+VLC a long-term precision of 8.1\,m/s (Fig.~\ref{Fig:RV-1},
Table~\ref{Tab:RVprecision}). The internal RV errors of the individual
spectra ($\sim$8.8\,m/s), calculated as the errors of the mean RV
of the chunks ($\mathrm{rms}/\sqrt{N_{\mathrm{chunk}}}$), are of
the same order as the rms of the time series implying a fair error
estimation.

\subsection{Combining the LC and VLC data\label{Sub:LC_VLC}}

The problem of instrumental offsets, i.e. different radial velocity
zero points, occurs when data sets originate from different instruments
\citep[e.g.][]{Wittenmyer06} or after instrumental changes/upgrades.
For instance, an offset of -1.8\,m/s was reported by \citet{Rivera2010}
after upgrading the Keck/HIRES spectrograph with a new CCD. An offset
of only 0.9\,m/s was mentioned by \citet{Vogt2010} when combining
Keck and AAT data.

As described above we have used three different instruments/instrument
configurations and we are also faced with the problem of the instrumental
offset. There are basically two different methods for combining the
data sets: (1) Simply fitting the offset, i.e. the data sets are considered
to be completely independent and the zero points are free parameters
in the model fitting. (2) If possible, measuring the offset physically
by making use of some known relation between the data sets/instruments
to keep the offset fixed.

In fact, we can measure the offset for the LC and VLC data albeit
with a limited precision. The LC and VLC spectra were taken through
the same iodine cell, i.e. the same wavelength calibrator. Because
\citet{Endl02} calculated the LC RVs with different stellar templates
and an iodine spectrum of lower resolution than used in this work,
we re-calculated the RVs for all LC spectra with the same VLC stellar
template (which is shorter than the LC spectra) and the new iodine
cell scan to have the same reference for the LC and VLC. The re-calculated
RVs are verified to have a precision similar to the published LC data.

\begin{figure}
\centering

\includegraphics[width=1\linewidth]{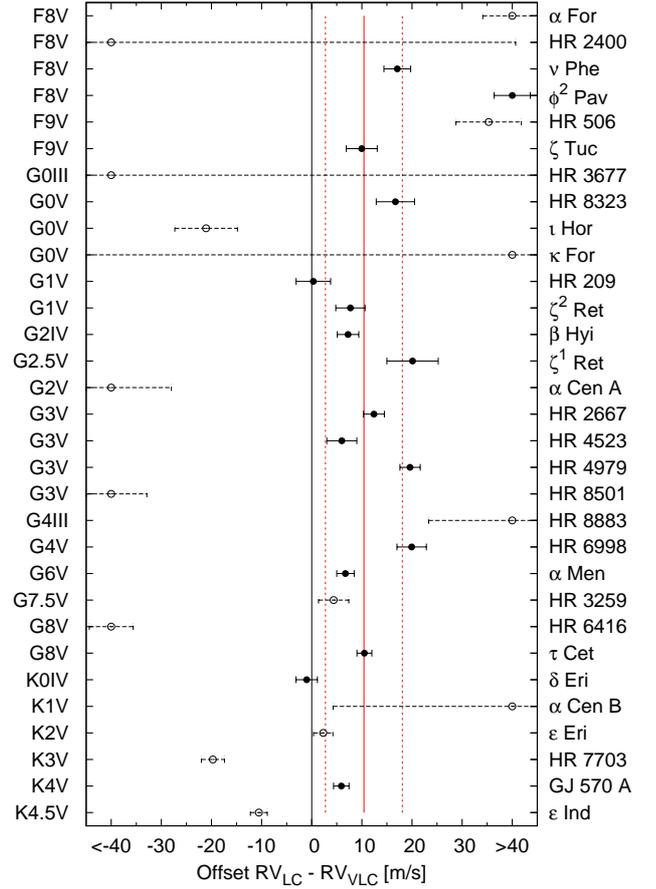}

\caption{\label{Fig:Offset_LC_VLC}Difference between the means of the VLC
and re-computed LC time series for all our stars ordered by spectral
type. For RV constant stars (black filled circles) there occurs a
systematic offset of 10.4$\pm$7.7\,m/s (red solid line and red dashed
lines). Stars shown with open circles were not included in the offset
analysis (for reasons see comments in Table~\ref{Tab:RVprecision}).
The shown error bars correspond to the uncertainty in the means, i.e.
parameter $s$ from Eq.~(\ref{Eq:ttest}).}

\end{figure}

Then we computed the mean of the re-calculated LC and VLC time series.
If a star has a constant RV, one would expect that the means of both
time series are the same, i.e. the offset $\mathrm{\overline{RV}_{VLC}-\overline{RV}_{LC}}=0$
within the uncertainties of the means ($\sigma_{\mathrm{LC}}$ and
$\sigma_{V\mathrm{LC}}$). This can be tested with the $t$-statistics,
in particular Welch's $t$-test (for two independent samples with
unequal sizes and variances). We suggest that keeping the offset fixed
is valid, if the quantity
\begin{equation}
t=\frac{\overline{RV}_{\mathrm{LC}}-\overline{RV}_{\mathrm{VLC}}}{s}\quad\text{with}\quad s=\sqrt{\frac{\sigma_{\mathrm{LC}}^{2}}{N_{\mathrm{LC}}}+\frac{\sigma_{V\mathrm{LC}}^{2}}{N_{\mathrm{VLC}}}}\label{Eq:ttest}
\end{equation}
is not rejected by the Null-hypothesis. The parameter $s$ is an estimate
for the standard error of the difference in the means and is calculated
from sample variances $\sigma_{i}^{2}$ and sample sizes $N_{i}$.
The variable $t$ follows a $t$-distribution with $\nu$ degrees
of freedom%
\footnote{The effective degree of freedom is ${\displaystyle \nu=\frac{(s_{1}^{2}/N_{1}+s_{2}^{2}/N_{2})^{2}}{(s_{1}^{2}/N_{1})^{2}/(N_{1}-1)+(s_{2}^{2}/N_{2})^{2}/(N_{2}-1)}}$
where $s_{1}^{2}$ and $s_{2}^{2}$ are the sample variances.%
}. For instance, for $|t|>1.7$ and $\nu>30$ the difference in the
means is significant with a false alarm probability of FAP$<$10\%.
For some of our stars the FAP for the offset difference is not significant:
$\delta$~Eri (64\%), $\epsilon$~Eri (23\%), HR~209 (92\%), and
HR~3259 (15\%). However, from Fig.~\ref{Fig:Offset_LC_VLC} it can
be seen that there are also stars having significant offsets leaving
doubts whether the offset can be kept fixed in general.

Figure~\ref{Fig:Offset_LC_VLC} shows that for our sample an average
offset of 10.4\,m/s ($\pm$7.7\,m/s) remains for the RV constant
stars when comparing the RV means of the VLC and LC data. This offset
might be due to systematics in the deconvolution process of the stellar
template or in the modelling. For example, due to the different resolution,
the LC data have to be modelled with a different chunk size (154 pixels
to cover two VLC chunks). We corrected all re-calculated LC RVs for
this systematic offset. Finally, we adjusted the RV mean of the published
LC time series \citep{Endl02} to fit the RV mean of the re-calculated
time series. In Figs.~\ref{Fig:RV-1}--\ref{Fig:RV-5} the LC \citep{Endl02}
and VLC data are always shown relative to each other with the measured
and corrected offset (and \emph{not} with a fitted RV offsets that
could have been taken from our fit results presented below in Sect.~\ref{Sec:RV_Analysis})
to conserve the true measurements.

The uncertainty of the offset found in the sample is rather large
for it to be considered a fixed value. On the other hand the approximately
known offset can hold important information, in particular in the
case of HR~2400 or $\epsilon$~Ind~A. Therefore we choose a compromise
between a fixed offset and a free offset when fitting a function.
Because one expects the difference of the zero point parameters to
be zero ($c_{\mathrm{VLC}}-c_{\mathrm{LC}}\approx0$), we introduce
in the $\chi^{2}$-fitting a counteracting potential term $\eta^{2}$
(also called penalty function, e.g. \citealt{Shporer2010}), that
increases when the zero point difference becomes larger
\begin{equation}
\hat{\chi}^{2}=\chi^{2}+\eta^{2}\quad\text{with}\quad\eta=\frac{c_{\mathrm{VLC}}-c_{\mathrm{LC}}}{s}.
\end{equation}

The resulting $\chi^{2}$ (when minimising $\hat{\chi}^{2}$) will
be higher compared to that obtained when fitting with free offsets
but lower than for fixed offsets. The parameter $s$ determines the
coupling between the offsets. After performing the fit it can be checked,
if the fit has spread the zero points too much (if $\eta\gg1$ or
if there are large jumps in the model curves in Figs.~\ref{Fig:RV-1}--\ref{Fig:RV-5}).
For $s$ we attributed the uncertainty of the offset correction of
7.7\,m/s leading to a weak coupling.

It is worth mentioning, that in Bayesian analysis $\hat{\chi}^{2}$
can be identified with the likelihood when assuming a Gaussian distribution
for the prior information that the expected zero point difference
is zero.

\subsection{Combining the CES and HARPS data}

In principle, VLC and HARPS data could be combined in a simi\-lar
way. They have different wavelength calibrators, but, since there
are some nights with almost simultaneous observations (within minutes),
they are closely related in time. The difference between these consecutive
measurements should be zero so that it is tempting to bind directly
the time series by means of those nights. However, this does not account
for fluctuations due to the individual uncertainties. Again a coupling
term%
\footnote{For one simultaneous measurement taken at the time $t_{s}$, this
term could be written as ${\displaystyle \eta=\frac{c_{\mathrm{H}}-c_{\mathrm{VLC}}}{\sqrt{\Delta RV_{\mathrm{H}}(t_{s})^{2}+\Delta RV_{\mathrm{VLC}}(t_{s})^{2}}}}$
where $c$ is the zero point parameter, $\Delta RV(t_{s})$ are the
individual errors of the simultaneous measurements, and the indices
H (HARPS) and VLC indicate the instruments. The VLC time series must
be a priori adjusted by a zero point such that $RV_{\mathrm{VLC}}(t_{s}):=RV_{\mathrm{H}}(t_{s})$.%
} would be a more secure approach.

However, for reasons of simplicity we choose a fully free offset between
the HARPS and the CES data. Because the VLC and the HARPS time series
overlap well this is less critical, in contrast to the LC and VLC
time series which are separated by a 2-year gap. The relative offsets
between the CES and HARPS data as illustrated in Figs.~\ref{Fig:RV-1}--\ref{Fig:RV-5}
correspond to the common best fitting model (constant, slope, sinusoid,
or Keplerian; cf. Sect.~\ref{Sec:RV_Analysis}).

\section{\label{Sec:RV_Analysis}Analysis of the radial velocities}

In this section we describe our data analysis and the general re\-sults
of the survey, while some individual objects are discussed in detail
in Sect.~\ref{Sec:Discussion}. The tests which we perform hereafter
were repeated on the residuals of the binaries and planet hosting
stars to search for additional companions and are indicated as objects
with index r in Table~\ref{Tab:TestsCombinedSets}.

\subsection{\label{Sub:Jitter}Preparation of RV data and jitter consideration}

Before the data analysis we binned the data into 2-hr intervals by
calculating weighted means for the temporal midpoint, RV, and RV error.
The 2-hr interval will especially down-weight nights from asteroseismology
campaigns (see Sect.~\ref{Sub:HARPS}) and reduce the impact of red
noise \citep{Baluev2012}, while resulting in nightly averages for
most other nights and still permitting to search for planets with
periods as short as one day. Such intervals are also employed for
solar-like stars (e.g. \citealp{Rivera2010}) to average out the stellar
jitter, i.e. intrinsic stellar RV variation caused by, e.g., oscillation
or granulation in the atmospheres of the stars. While the Sun has
an oscillation timescale of $\sim$5\,min, its granulation%
\footnote{There is also meso- and supergranulation (life times up to $\sim30$\,hours)
which take place on different size scales \citet{Dumusque2011}.%
} has lifetime $\lesssim$25\,min (see \citealt{Dumusque2011} for
adequate observing strategies). However note, that in our own survey
we have usually taken three consecutive spectra in one night covering
in total only 5--10\,min, which is not sufficient to average out
all those intrinsic stellar RV variations. To investigate the short-period
jitter, we calculated the weighted scatter%
\footnote{\label{fn:binweight}Weighting of the $i$-th measurement with its
internal error $w_{i}\sim1/\sigma_{\mathrm{int,}i}^{2}$.%
} in each 2-hr bin with at least 2 measurements and then the weighted
mean of these scatters%
\footnote{\label{fn:bintime}Weighting of the $j$-th bin with the number of
measurements $n_{j}$ and the mean internal error in that bin: $w_{j}\sim n_{j}/\sigma_{\mathrm{int,}j}^{2}$.
Note that bins with more measurements usually cover larger time intervals
and get more weight.%
}. Table~\ref{Tab:Jitter} lists the jitter estimate $\sigma_{\mathrm{jit},\tau}$
from the HARPS data for each star and the mean time scale $\tau$
accessible for this estimate within the 2\,h bins. Note that these
time scales may not sufficiently cover the real jitter time scale
in all cases. Therefore these estimated jitter values were not used
in a further analysis.

There can be also a long-term jitter with time scales of few days
to weeks related to the rotation period (due to the appearance and
disappearance of spots) or up to few years due to the magnetic cycle
of a star. \citet{Isaacson2010} provide jitter estimates as a function
of $B-V$ colour and chromospheric activity index $S_{\mathrm{HK}}$
based on Keck observations for more than 2600 main sequence stars
and subgiants. Using these relations and the median $S_{\mathrm{HK}}$
values in the HARPS data (which in most cases agree well with other
literature values; see Table~\ref{Tab:Jitter}), we estimate the
jitter $\sigma_{\mathrm{jit,long}}$ for our stars (Table~\ref{Tab:Jitter}).
The jitter terms are usally $\gtrsim2$\,m/s. For GJ~570~A and
$\epsilon$~Ind~A the expected jitter is only 1.6\,m/s. Both are
K dwarfs with $1.0<B-V<1.3$ and, according to \citet{Isaacson2010},
those stars have the lowest level of velocity jitter decoupled from
their chromospheric activity. The jitter terms $\sigma_{\mathrm{jit,long}}$
were added in quadrature to the internal errors for all stars and
lead to a more balanced fit with the CES data. Morevover, to cross-check
whether detected RV signals might be caused by those kinds of stellar
activity we will also analyse in the HARPS data correlations between
the RV data and activity indicators such as Ca~II~H\&K emission
and variations of the bisector (BIS) and the FWHM of the cross-correlation
profile (Sect. \ref{sub:Correlations-with-Ca}). All RVs and HARPS
activity indicators are online available.

In fitting the data we accounted for the secular acceleration of the
RVs (as given in Table~\ref{Tab:SampleInfo}). This perspective effect
can become a measurable po\-si\-tive trend in some high proper motion
stars \citep{Schlesinger1917,Kuerster03,Zechmeister09b}. In our sample,
$\epsilon$~Ind~A has the highest secular acceleration with 1.8\,m/s/yr.
Its contribution is depicted in Fig.~\ref{Fig:RV-5} by a dashed
line.

\subsection{Excess variability}

To investigate objects for excess variability it is common to compare
the observed scatter with a noise estimate. A significantly larger
scatter indicates variability. Because internal errors $\Delta RV_{i}$
and jitter estimations $\sigma_{\mathrm{jit}}$ are available, the
quality of each mea\-su\-re\-ment is assessed and allows us to
weight the measurements in the $\chi^{2}$-statistics, $w_{i}=\frac{1}{\sigma_{i}^{2}}=\frac{1}{\Delta RV_{i}^{2}+\sigma_{\mathrm{jit}}^{2}}$.
As the scatter we calculate the weighted rms which is here defined
as
\begin{equation}
\mathrm{rms}=\sqrt{\frac{N}{N-\nu}\frac{1}{W}\sum_{i=1}^{N}\frac{(RV_{i}-f(t_{i}))^{2}}{\sigma_{i}^{2}}}=\sqrt{\frac{N}{N-\nu}\frac{\chi^{2}}{W}}\label{Eq:rms}
\end{equation}
where $W=\sum w_{i}$ is the sum of the weights and $\nu$ the number
of model parameters. Outliers with a large uncertainty will con\-tri\-bute
less to the rms. The factor $\frac{N}{N-\nu}$ is a correction that
converts the uncorrected and biased variance into an unbiased variance,
i.e. to have an unbiased estimator for the population variance%
\footnote{Note however, that the square root of this variance, $\mathrm{rms}=\sqrt{\mathrm{rms}^{2}}$
is not a unbiased estimate of the population standard deviation \citep{Deakin1999}.%
}. In the unweighted case ($w_{i}=1$, $W=N$) we obtain the well known
formula for the unbiased rms: $\mathrm{rms}=\sqrt{\frac{1}{N-\nu}\sum(RV_{i}-f(t_{i}))^{2}}$.

Furthermore, we define the weighted mean noise term $\sigma$ via
the mean of the weights%
\footnote{Another point of view leads to the same result: Gaussian errors are
added in quadrature. Hence the trivial weighted mean is $\sqrt{\frac{1}{W}\sum w_{i}\sigma_{i}^{2}}=\sqrt{\frac{1}{W}\sum1}=\sqrt{\frac{N}{W}}$.%
}
\begin{equation}
\sigma=\frac{1}{\sqrt{\langle w_{i}\rangle}}=\sqrt{\frac{N}{W}}=\sqrt{\frac{N}{\sum_{i=1}^{N}\frac{1}{\sigma_{i}^{2}}}}.\label{Eq:meanerror}
\end{equation}
Again lower-quality measurements will contribute less to the mean
noise level.

With these definitions the reduced $\chi^{2}$ can be easily expressed
as the ratio of weighted rms to weighted mean noise level
\begin{equation}
\chi_{\mathrm{red}}^{2}=\frac{\chi^{2}}{N-\nu}=\frac{\mathrm{rms}^{2}}{\sigma^{2}}.\label{Eq:Chi2red}
\end{equation}

To test for excess variability we have to fit a constant and to calculate
the scatter around the fit. For the joint analysis we account for
the zero point parameter of each data set when fitting a constant
as outlined in Sect.~\ref{Sub:LC_VLC}. Note that the probability
for the excess variability $\mathrm{Prob}(\chi_{\mathrm{red}}^{2})$
is directly reliant on a proper estimate for the noise level $\sigma$.
Also note that the tests in the next sections employ model comparisons
and the jitter estimate enters only indirectly through fitting with
modified weights. Table~\ref{Tab:TestsCombinedSets} summarises for
the combined data set the weighted noise term $\sigma$, the weighted
rms, and the $\chi^{2}$-probability for this test. Table~\ref{Tab:TestsCombinedSetsSingle}
lists additionally the individual rms (columns 5-7 labelled $\mathrm{rms}_{\mathrm{constant}}$)
for each instrument. These values can differ from Table~\ref{Tab:RVprecision},
because in Table~\ref{Tab:TestsCombinedSets} secular acceleration
is accounted for, jitter has been added, the data are binned, and
the LC and VLC offsets are coupled. Because the HARPS data have a
much higher precision, they dominate the statistics.

\begin{table*}
\caption{\label{Tab:TestsCombinedSets}Summary of the tests for excess variability,
slope, and periodicities (sinusoidal and Keplerian) for the combined
data set. Significant trends and sinusoidal periods are printed in
bold face (FAP$<1$\textperthousand ). An index ``r'' to the name
of the star (first column) indicates tests on the residuals as derived
from the most significant model to the original data.}

\setlength{\tabcolsep}{3pt}

\begin{centering}
\begin{tabular}{l|rrrrll|rrl|rrl|rrl}
\hline 
\hline Star & $N_{\mathrm{bin}}$ & \multicolumn{1}{c}{$T$} & \multicolumn{1}{c}{$\sigma$} & \multicolumn{1}{c}{rms} & S & $\mathrm{Prob}(\chi_{\mathrm{red}}^{2})$ & rms$_{\mathrm{slope}}$ & \multicolumn{1}{c}{slope} & \multicolumn{1}{c|}{FAP} & rms$_{\mathrm{sin}}$ & \multicolumn{1}{c}{$P_{\mathrm{sin}}$} & \multicolumn{1}{c|}{FAP} & rms$_{\mathrm{Kep}}$ & \multicolumn{1}{c}{$P_{\mathrm{Kep}}$} & \multicolumn{1}{c}{FAP}\\
 &  & {[}yr{]} & {[}m/s{]} & {[}m/s{]} &  &  & {[}m/s{]} & {[}m/s/yr{]} &  & {[}m/s{]} & {[}d{]} &  & {[}m/s{]} & {[}d{]} & \\
\hline 
$\zeta$~Tuc & 197 & 17.0 & 3.08 & 2.24 & ~ & 1 & 2.24 & 0.04 & 0.57 & 2.18 & 14.8~~ & 1 & 1.99 & 551.~~~~ & $3\!\cdot\!10^{-6}$\\
$\beta$~Hyi & 109 & 16.0 & 3.34 & 3.84 & ~ & 0.014 & 3.29 & 1.79 & $\bf{8.9\!\cdot\!10^{-9}}$ & 3.07 & 4340.~~~~ & $\bf{1.8\!\cdot\!10^{-7}}$ & 2.96 & 1.03 & $5\!\cdot\!10^{-8}$\\
HR~209 & 53 & 14.0 & 6.57 & 9.44 & A & $1.4\!\cdot\!10^{-5}$ & 9.28 & 1.81 & 0.1 & 7.26 & 1.24 & 0.0064 & 5.92 & 2.52 & $5\!\cdot\!10^{-6}$\\
$\nu$~Phe & 71 & 14.9 & 4.72 & 5.23 & ~ & 0.1 & 5.17 & -0.96 & 0.13 & 4.59 & 1.98 & 0.3 & 4.37 & 1.09 & 0.037\\
HR~506 & 60 & 14.0 & 4.68 & 12.2~~ & P & 0 & 12.3~~ & -0.69 & 0.73 & 6.69 & 963.~~~~ & $\bf{8.7\!\cdot\!10^{-12}}$ & 6.44 & 963.~~~~ & $3\!\cdot\!10^{-11}$\\
HR~506$_{\mathrm{r}}$ & ~ & ~ & ~ & 6.53 & A & $2.5\!\cdot\!10^{-5}$ & 5.89 & -2.41 & \bf{0.00041} & 5.31 & 3.60 & 0.021 & 4.68 & 3.61 & 0.00023\\
$\tau$~Cet & 339 & 17.1 & 2.41 & 1.37 & ~ & 1 & 1.33 & -0.19 & $\bf{5.2\!\cdot\!10^{-6}}$ & 1.28 & 380.~~~~ & $\bf{5.5\!\cdot\!10^{-7}}$ & 1.25 & 382.~~~~ & $7\!\cdot\!10^{-9}$\\
$\kappa$~For & 78 & 14.8 & 3.69 & 717.~~~~ & B & 0 & 49.4~~ & -701.~~~~ & \bf{0} & 8.13 & $>$30\,yr~~ & $\bf{2\!\cdot\!10^{-139}}$ & 4.03 & 10700.~~~~ & $5\!\cdot\!10^{-156}$\\
$\kappa$~For$_{\mathrm{r}}$ & ~ & ~ & ~ & 3.87 & ~ & 0.18 & 3.88 & 0.29 & 0.38 & 3.50 & 4010.~~~~ & 0.67 & 3.11 & 3720.~~~~ & 0.0018\\
$\iota$~Hor & 154 & 14.8 & 5.79 & 31.7~~ & P & 0 & 31.8~~ & -0.11 & 0.92 & 15.4~~ & 307.~~~~ & $\bf{2.6\!\cdot\!10^{-44}}$ & 12.9~~ & 307.~~~~ & $2\!\cdot\!10^{-53}$\\
$\iota$~Hor$_{\mathrm{r}}$ & ~ & ~ & ~ & 12.5~~ & A & 0 & 12.3~~ & -2.14 & 0.029 & 10.9~~ & 5.72 & $\bf{5.6\!\cdot\!10^{-6}}$ & 9.83 & 5.81 & $6\!\cdot\!10^{-13}$\\
$\alpha$~For & 75 & 14.8 & 4.28 & 13.2~~ & B & 0 & 4.25 & -11.5~~ & $\bf{7.7\!\cdot\!10^{-37}}$ & 4.25 & $>$30\,yr~~ & $\bf{7.2\!\cdot\!10^{-32}}$ & 4.24 & 8230.~~~~ & $7\!\cdot\!10^{-30}$\\
$\alpha$~For$_{\mathrm{r}}$ & ~ & ~ & ~ & 4.19 & ~ & 0.58 & 4.22 & 0.03 & 0.93 & 3.77 & 350.~~~~ & 0.73 & 3.62 & 72.6~~ & 0.37\\
$\zeta^1$~Ret & 48 & 9.9 & 5.09 & 12.4~~ & A & $1\!\cdot\!10^{-33}$ & 11.5~~ & -4.14 & 0.0059 & 9.01 & 21.5~~ & 0.0013 & 7.51 & 21.4~~ & $9\!\cdot\!10^{-6}$\\
$\zeta^2$~Ret & 89 & 14.8 & 4.69 & 5.23 & ~ & 0.064 & 5.13 & 0.85 & 0.037 & 4.95 & 2.48 & 1 & 4.41 & 3.90 & 0.01\\
$\epsilon$~Eri & 79 & 14.8 & 5.41 & 8.22 & A & $8.1\!\cdot\!10^{-10}$ & 8.25 & 0.54 & 0.49 & 6.85 & 3.11 & 0.0029 & 6.02 & 1.44 & $4\!\cdot\!10^{-6}$\\
$\delta$~Eri & 75 & 14.8 & 2.78 & 3.23 & ~ & 0.027 & 3.24 & -0.09 & 0.61 & 2.85 & 3.07 & 0.29 & 2.69 & 2.66 & 0.045\\
$\alpha$~Men & 102 & 14.2 & 3.23 & 3.12 & ~ & 0.66 & 3.12 & -0.26 & 0.27 & 2.75 & 880.~~~~ & 0.0095 & 2.61 & 837.~~~~ & 0.00051\\
HR~2400 & 77 & 14.4 & 4.53 & 266.~~~~ & B & 0 & 215.~~~~ & -103.~~~~ & $\bf{1.4\!\cdot\!10^{-8}}$ & 107.~~~~ & 5860.~~~~ & $\bf{6.1\!\cdot\!10^{-26}}$ & 5.89 & 9490.~~~~ & $3\!\cdot\!10^{-112}$\\
HR~2400$_{\mathrm{r}}$ & ~ & ~ & ~ & 5.56 & ~ & 0.0044 & 5.38 & -1.46 & 0.016 & 5.23 & 1.25 & 1 & 4.48 & 1.88 & 0.0028\\
HR~2667 & 88 & 14.4 & 4.82 & 3.09 & ~ & 1 & 3.08 & -0.31 & 0.26 & 2.91 & 3.49 & 1 & 2.63 & 1.63 & 0.03\\
HR~3259 & 191 & 14.2 & 2.49 & 4.12 & P & $4.4\!\cdot\!10^{-32}$ & 4.10 & 0.58 & 0.086 & 3.21 & 8.67 & $\bf{1.2\!\cdot\!10^{-17}}$ & 3.22 & 8.67 & $4\!\cdot\!10^{-16}$\\
HR~3677 & 61 & 14.4 & 3.97 & 1000.~~~~ & B & 0 & 257.~~~~ & 808.~~~~ & $\bf{1.6\!\cdot\!10^{-35}}$ & 35.2~~ & $>$30\,yr~~ & $\bf{7.8\!\cdot\!10^{-79}}$ & 8.68 & $>$30\,yr~~ & $1\!\cdot\!10^{-107}$\\
HR~3677$_{\mathrm{r}}$ & ~ & ~ & ~ & 8.39 & ~ & $6.6\!\cdot\!10^{-27}$ & 8.46 & 0.16 & 0.82 & 6.16 & 8.79 & $\bf{6.2\!\cdot\!10^{-5}}$ & 5.84 & 31.2~~ & $4\!\cdot\!10^{-5}$\\
HR~4523 & 100 & 15.3 & 3.35 & 3.47 & ~ & 0.29 & 3.49 & -0.01 & 0.97 & 3.32 & 2.92 & 1 & 2.91 & 1.04 & 0.00082\\
HR~4979 & 151 & 14.6 & 2.78 & 3.96 & A & $2\!\cdot\!10^{-12}$ & 3.91 & -0.75 & 0.026 & 3.45 & 1.00 & $\bf{3.7\!\cdot\!10^{-6}}$ & 3.25 & 1.00 & $6\!\cdot\!10^{-9}$\\
$\alpha$~Cen~A & 121 & 14.2 & 3.31 & 101.~~~~ & B & 0 & 10.2~~ & 131.~~~~ & \bf{0} & 3.03 & $>$30\,yr~~ & $\bf{5.1\!\cdot\!10^{-174}}$ & 2.84 & $>$30\,yr~~ & $8\!\cdot\!10^{-173}$\\
$\alpha$~Cen~A$_{\mathrm{r}}$ & ~ & ~ & ~ & 2.79 & ~ & 0.99 & 2.81 & 0.00 & 1 & 2.51 & 89.1~~ & 0.0066 & 2.29 & 179.~~~~ & $2\!\cdot\!10^{-6}$\\
$\alpha$~Cen~B & 82 & 14.2 & 3.96 & 179.~~~~ & B & 0 & 19.1~~ & -154.~~~~ & \bf{0} & 5.06 & $>$30\,yr~~ & $\bf{1.2\!\cdot\!10^{-116}}$ & 4.92 & 9370.~~~~ & $1\!\cdot\!10^{-113}$\\
$\alpha$~Cen~B$_{\mathrm{r}}$ & ~ & ~ & ~ & 4.79 & ~ & 0.055 & 4.81 & 0.18 & 0.61 & 4.08 & 390.~~~~ & 0.0091 & 3.51 & 387.~~~~ & $1\!\cdot\!10^{-6}$\\
GJ~570~A & 64 & 12.0 & 2.89 & 5.12 & B & $2\!\cdot\!10^{-15}$ & 3.71 & -2.72 & $\bf{3.5\!\cdot\!10^{-10}}$ & 3.74 & $>$30\,yr~~ & $\bf{1.6\!\cdot\!10^{-5}}$ & 3.46 & 3390.~~~~ & $2\!\cdot\!10^{-6}$\\
HR~6416 & 77 & 14.5 & 3.99 & 11.0~~ & B & 0 & 3.83 & 9.40 & $\bf{1.7\!\cdot\!10^{-35}}$ & 3.72 & $>$30\,yr~~ & $\bf{2.1\!\cdot\!10^{-31}}$ & 3.78 & $>$30\,yr~~ & $6\!\cdot\!10^{-29}$\\
HR~6416$_{\mathrm{r}}$ & ~ & ~ & ~ & 3.67 & ~ & 0.82 & 3.70 & 0.06 & 0.85 & 3.51 & 3.41 & 1 & 3.16 & 1.05 & 0.19\\
HR~6998 & 64 & 14.3 & 4.91 & 3.19 & ~ & 1 & 3.19 & -0.60 & 0.34 & 3.02 & 1.25 & 1 & 2.61 & 1.18 & 0.076\\
HR~7703 & 57 & 13.2 & 3.21 & 5.08 & B & $6.5\!\cdot\!10^{-9}$ & 2.00 & 3.64 & $\bf{2.3\!\cdot\!10^{-23}}$ & 2.02 & $>$30\,yr~~ & $\bf{2.4\!\cdot\!10^{-18}}$ & 2.00 & 5690.~~~~ & $8\!\cdot\!10^{-17}$\\
HR~7703$_{\mathrm{r}}$ & ~ & ~ & ~ & 1.98 & ~ & 1 & 2.00 & -0.01 & 0.94 & 1.82 & 1.23 & 1 & 1.65 & 2.87 & 0.47\\
$\phi^2$~Pav & 88 & 16.9 & 5.72 & 6.75 & ~ & 0.0099 & 6.70 & -1.01 & 0.15 & 6.01 & 10.2~~ & 0.14 & 5.95 & 1.64 & 0.47\\
HR~8323 & 104 & 12.8 & 3.12 & 3.94 & A & 0.00015 & 3.71 & -1.17 & \bf{0.0003} & 2.89 & 1370.~~~~ & $\bf{7.8\!\cdot\!10^{-11}}$ & 2.89 & 1300.~~~~ & $1\!\cdot\!10^{-9}$\\
$\epsilon$~Ind~A & 132 & 16.9 & 2.25 & 4.49 & P? & 0 & 3.05 & 2.36 & $\bf{1.6\!\cdot\!10^{-23}}$ & 3.02 & $>$30\,yr~~ & $\bf{2.7\!\cdot\!10^{-19}}$ & 2.99 & 6350.~~~~ & $3\!\cdot\!10^{-18}$\\
$\epsilon$~Ind~A$_{\mathrm{r}}$ & ~ & ~ & ~ & 2.99 & ~ & $1.8\!\cdot\!10^{-7}$ & 3.01 & 0.01 & 0.93 & 2.70 & 35.7~~ & 0.0049 & 2.63 & 35.7~~ & 0.0013\\
HR~8501 & 78 & 14.8 & 5.47 & 22.5~~ & B & 0 & 4.52 & 17.3~~ & \bf{0} & 4.54 & $>$30\,yr~~ & $\bf{4\!\cdot\!10^{-48}}$ & 4.47 & 7650.~~~~ & $4\!\cdot\!10^{-46}$\\
HR~8501$_{\mathrm{r}}$ & ~ & ~ & ~ & 4.46 & ~ & 0.99 & 4.49 & 0.11 & 0.69 & 4.12 & 64.1~~ & 1 & 3.77 & 1.61 & 0.053\\
HR~8883 & 45 & 13.3 & 7.23 & 63.7~~ & A & 0 & 64.0~~ & -6.90 & 0.45 & 32.6~~ & 7.60 & $\bf{4.3\!\cdot\!10^{-9}}$ & 28.2~~ & 13.1~~ & $4\!\cdot\!10^{-10}$\\
HR~8883$_{\mathrm{r}}$ & ~ & ~ & ~ & 31.7~~ & ~ & 0 & 32.1~~ & -0.76 & 1 & 19.9~~ & 1.58 & $\bf{1.6\!\cdot\!10^{-5}}$ & 18.1~~ & 1.22 & $5\!\cdot\!10^{-6}$\\

\hline 
\end{tabular}
\par\end{centering}

Listed are the number of binned observations $N_{\mathrm{bin}}$,
the combined time baseline $T$, the mean combined noise term $\sigma$
(including jitter), the combined weighted rms of the time series,
the flag S for the probable main source of the RV variations as concluded
in this work in Sect.~5 (A - activity, B - binary/wide stellar companion,
P - planet), the $\chi^{2}$-probability for fitting a constant, and
the false alarm probabilities (FAP) for the other tests. Also listed
are the weighted scatter of the residuals (rms$_{\mathrm{slope}}$,
rms$_{\mathrm{sin}}$, rms$_{\mathrm{Kep}}$) and some best-fitting
parameters (slope and the periods $P_{\mathrm{sin}}$ and $P_{\mathrm{Kep}}$).
\end{table*}

The small $\chi^{2}$-probabilities for most of the stars indicate
that they are variable with respect to our noise estimate $\sigma$.
However, 10 stars (and the RV orbit residuals of 8 stars) have $\mathrm{Prob}(\chi^{2})>1\%$,
i.e. they show only low or no excess variability. In five cases the
scatter is smaller than the noise level, i.e. $\chi_{\mathrm{red}}^{2}<1$,
implying an overestimation of the noise level. Indeed, for four stars
($\zeta$~Tuc, $\tau$~Cet, HR~2667, and HR~6998) the jitter estimate
$\sigma_{\mathrm{jit,long}}$ in Table~\ref{Tab:Jitter} is higher
than the scatter of the HARPS mea\-sure\-ments (Table~\ref{Tab:TestsCombinedSetsSingle}).
The reason for jitter overestimation might be a somewhat lower precision
of the Keck sample from which the jitter relation was derived \citep{Isaacson2010}.

\subsection{\label{Sub:Long-term-trend}Long-term trends}

Because potential planets or companions can have orbital periods much
longer than our observations, these objects may betray themselves
by a trend in the RVs. We searched for trends by fitting a slope to
the data and derived its significance via the fit improvement with
respect to the constant model (previous Sect.) via
\begin{equation}
F_{\mathrm{slope}}=(N-4)\frac{\chi_{\mathrm{constant}}^{2}-\chi_{\mathrm{slope}}^{2}}{\chi_{\mathrm{slope}}^{2}}
\end{equation}
or when expressed with unbiased weighted variances
\begin{equation}
F_{\mathrm{slope}}=\frac{(N-3)\,\mathrm{rms}_{\mathrm{constant}}^{2}-(N-4)\,\mathrm{rms}_{\mathrm{slope}}^{2}}{\mathrm{rms}_{\mathrm{slope}}^{2}}.
\end{equation}

The associated probability for this $F$-value follows a $F_{1,N-4}$-distribution
(4 parameters: 1 slope, 3 zero points). Again Table~\ref{Tab:TestsCombinedSets}
summarises the test for long-term trends. When adopting a false alarm
probability threshold of $<10^{-3}$ fitting a slope improves significantly
the rms of all binaries as well as that of $\beta$~Hyi, $\tau$~Cet,
GJ~570~A, $\epsilon$~Ind~A and the residuals of HR~506. We note
that for $\beta$~Hyi, $\alpha$~For, GJ~570~A, HR~6416, HR~7703,
$\epsilon$~Ind~A, and HR~8501 the trend is a sufficient model
(regarding sinusoid and Keplerian fit, see next Section), because
of the smaller FAP or weighted rms (i.e. smaller $\chi_{\mathrm{red}}^{2})$.
For these stars the trend is depicted in Figs.~\ref{Fig:RV-1}--\ref{Fig:RV-5}.

Some of our stars have known wide visual companions with a known separation
$\rho$ listed in Table~\ref{Tab:Binaries}. Whether these objects
are able to cause the observed trend, can be verified by the estimate
($|\ddot{\vec{r}}_{\mathrm{A}}|=G\frac{m_{\mathrm{B}}}{r^{2}}$) 
\begin{equation}
m_{_{\mathrm{B}}}\sin i\ge\frac{|\ddot{z}_{\mathrm{A}}|r^{2}}{G}\ge\frac{|\ddot{z}_{\mathrm{A}}|\rho^{2}}{G}=5.6\cdot10^{-3}M_{\mathrm{Jup}}\frac{|\ddot{z}_{\mathrm{A}}|}{\mathrm{m/s/yr}}\left(\frac{\rho}{\mathrm{AU}}\right)^{2}\label{Eq:TrendMass}
\end{equation}
with the radial acceleration $|\ddot{z}_{\mathrm{A}}|\le|\ddot{\vec{r}}_{\mathrm{A}}|\sin i$
of the observed component A and the projected separation $\rho\le r$
between both components. For comparison, Jupiter at 5.2\,AU can accelerate
the Sun by 6.6\,m/s/yr. Table~\ref{Tab:Binaries} summaries the
information about known wide companions and shows that the minimum
companion masses derived from the measured slopes are below 0.5\,$M_{\odot}$
for $\alpha$~For, GJ~570~A, HR~6416, HR~7703, and HR~8501.
These masses are in agreement with the masses as expected from the
spectral type of their companions. However, $\epsilon$~Ind~B cannot
explain the trend seen for $\epsilon$~Ind~A (see Sect.~\ref{Sec:Discussion}
for details).

The other possibility for a trend is an unknown and unseen companion.
Whether the strength and the long duration of a trend is still compatible
with a planetary companion, can be estimated more conveniently, when
Eq.~(\ref{Eq:TrendMass}) is expressed in terms of the orbital period
$P$ which is also unknown but has to be (for circular orbits) at
least twice as large as the time span of observations $T$. With Kepler's
3$^{\mathrm{rd}}$ law $\frac{a^{3}}{P^{2}}=G\frac{M+m}{4\pi^{2}}$,
Eq.~(\ref{Eq:TrendMass}) can be written as 
\begin{align}
m_{\mathrm{B}} & \sin i\ge G^{-1/3}|\ddot{z}|\left(\frac{P}{2\pi}\right)^{4/3}\left(M_{\mathrm{A}}+m_{\mathrm{B}}\right)^{2/3}\nonumber \\
 & =5.6\cdot10^{-3}M_{\mathrm{Jup}}\frac{|\ddot{z}|}{\mathrm{m/s/yr}}\left(\frac{P}{\mathrm{yr}}\right)^{4/3}\left(\frac{M_{\mathrm{A}}+m_{\mathrm{B}}}{\mathrm{M}_{\odot}}\right)^{2/3}\label{Eq:TrendMass2}
\end{align}

For $\epsilon$~Ind~A we find its companion to have $m\sin i\gtrsim0.97\, M_{\mathrm{Jup}}$
for $P>30$\,yr.

\begin{table*}
\caption{\label{Tab:Binaries}Information about wide companions. Listed are
separation $\rho$, period estimate $P$, references, and our estimates
for minimum masses $M_{\min}$ derived via Eq.~(\ref{Eq:TrendMass})
from the slope of the linear fit to the data (Table~\ref{Tab:TestsCombinedSets})
and $\rho$. For minimum masses in brackets the trends were not significant.}

\begin{centering}
\begin{tabular}{llcrrlrl}
\hline 
\hline Star & Companion & $\rho$ {[}\arcsec{]} & \multicolumn{1}{c}{$\rho$ {[}AU{]}} & \multicolumn{1}{c}{$P$ {[}yr{]}} & Ref. & $M_{\min}$ {[}$M_{\mathrm{Jup}}${]} & further companions\\
\hline 
$\alpha$~For & GJ~127~B (G7V) & ~~~~4.4~~ & 62 & 314 & {[}BP, H, P{]} & 248 & \\
$\alpha$~Men & HD~43834~B (M3.5) & ~~~~3.05 & 31 &  & {[}E{]} & (1) & \\
HR~2667 & GJ~9223~B (K0V) & ~~20.5~~ & 332 &  & {[}F, WD{]} & (191) & \\
HR~4523 & GJ~442~B (M4V) & ~~25.4~~ & 234 &  & {[}P{]} & (3) & \\
GJ~570~A & GJ~570~BC (M1.5V+M3V) & ~~24.7~~ & 146 &  & {[}B{]} & 325 & GJ~570~D (T, 258\farcs3)\\
HR~6416 & GJ~666~B (M0V) & ~~10.4~~ & 92 & 550 & {[}LH, P{]} & 446 & GJ~666~C (M6.5V, 41\farcs8)\\
 &  &  &  &  &  &  & GJ~666~D (M7V, 40\farcs7)\\
HR~7703 & GJ~783~B (M3.5)  & ~~~~7.1~~ & 43 &  & {[}P{]} & 38 & \\
$\epsilon$~Ind~A & GJ~845~Bab (T1+T6) & 402.3~~ & 1459 &  & {[}S{]} & 28130 & \\
HR~8501 & GJ~853~B ($V<10$\,mag) & 2.5--3.4 & 41 &  & {[}M, WD{]} & 163 & \\
\hline 
\end{tabular}
\par\end{centering}

References: {[}BP{]} \citet{Baize1989}, {[}B{]} \citet{Burgasser2000},
{[}E{]} \citet{Eggenberger07}, {[}F{]} \citet{Favata1997}, {[}H{]}
\citet{Heintz1978}, {[}LH{]} \citet{Luyten1980}, {[}M{]} \citet{Mason2001},
{[}P{]} \citet{Poveda1994}, {[}S{]} \citet{Scholz2003}, {[}WD{]}
\citet{Worley1997}.
\end{table*}

\subsection{Search for Periodicities and Keplerian orbits}

To search for the best-fitting sinusoidal and Keplerian orbits, we
employed the generalised Lomb-Scargle (GLS) algorithm described in
\citet{Zechmeister09}. It was adapted to treat all three data sets
with different offsets and also incorporates the weak offset coupling
described before (Sect.~\ref{Sub:LC_VLC}). Searching for sine waves
is a robust method to find periodicities and orbits with low eccentricities,
while for highly eccentric orbits the Keplerian model should be applied.

\begin{figure}
\centering

\includegraphics[width=1\linewidth]{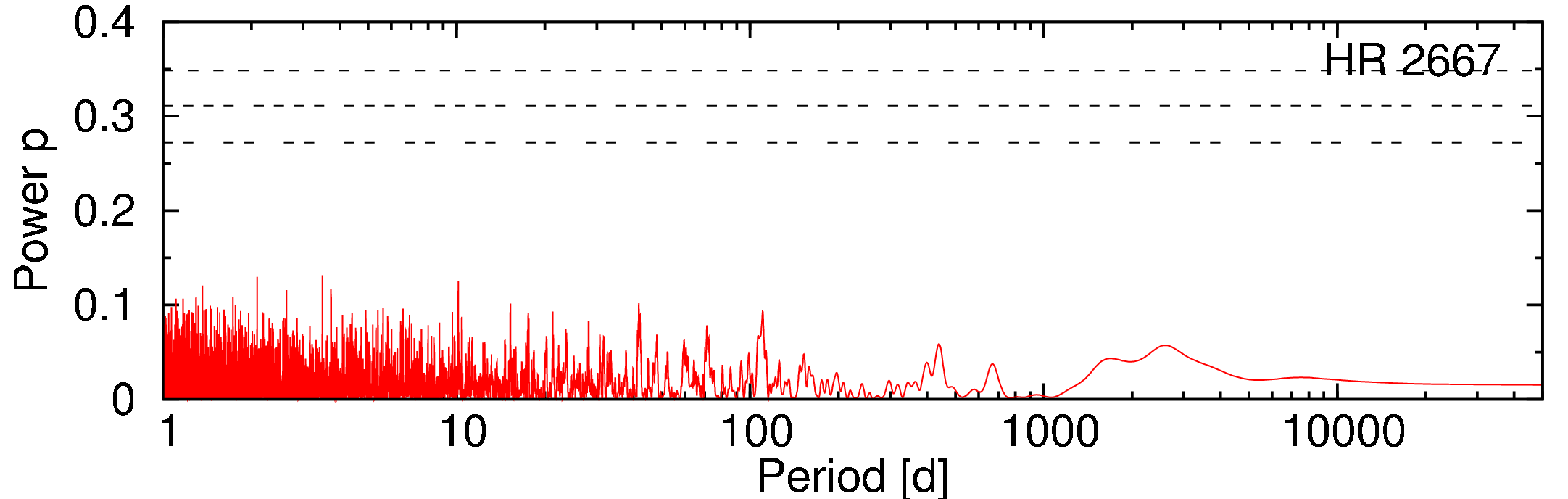}

\includegraphics[width=1\linewidth]{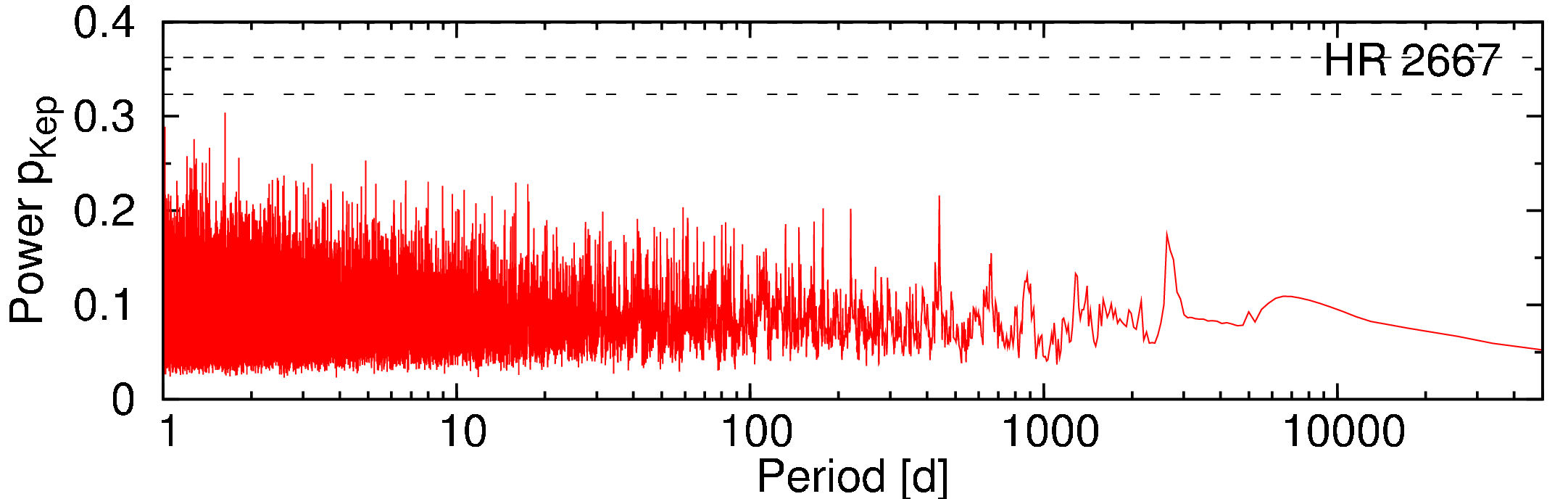}

\caption{\label{Fig:GLS_hr2667}GLS (top) and Keplerian (bottom) periodogram
for HR~2667 which is constant to 3.1\,m/s. The horizontal lines
mark the $10^{-2}$, $10^{-3}$, and $10^{-4}$ false alarm probabilities
(FAP).}
\end{figure}

Figures~\ref{Fig:GLS_hr2667}, \ref{Fig:GLS_iotahor}, \ref{Fig:GLS_zet1ret},
\ref{Fig:GLS_epseri}, and \ref{Fig:GLS_hr3259} show the periodograms
for some of the stars discussed here. The periodograms are normalised
as
\begin{equation}
p=\frac{\chi_{\mathrm{constant}}^{2}-\chi_{\mathrm{sin}}^{2}}{\chi_{\mathrm{constant}}^{2}}\quad\mathrm{and}\quad p_{\mathrm{Kep}}=\frac{\chi_{\mathrm{constant}}^{2}-\chi_{\mathrm{Kep}}^{2}}{\chi_{\mathrm{constant}}^{2}}.
\end{equation}
involving the $\chi^{2}$ of the constant, sinusoidal, and Keplerian
model, respectively. Analogous to \citealt{Cumming08} and \citet{Zechmeister09},
we calculated the probabilities of the power values for the best-fitting
sinusoid and Keplerian orbit ($p_{\mathrm{best}}$ and $p_{\mathrm{Kep,best}}$)
via 
\begin{align}
\mathrm{Prob}(p>p_{\mathrm{best}}) & =(1-p_{\mathrm{best}})^{\frac{N-5}{2}}\quad\mathrm{and}\\
\mathrm{Prob}(p_{\mathrm{Kep}}>p_{\mathrm{Kep,best}}) & =(1+\tfrac{N-7}{2}p_{\mathrm{Kep,best}})(1-p_{\mathrm{Kep,best}})^{\frac{N-7}{2}},
\end{align}
respectively. Compared to the probability functions given by these
authors which account for one offset, here are slight modifications
in the equations (numerator in the fractional terms decreased by 2)
arising from the three zero points, i.e. two more free parameters%
\footnote{The corresponding normalisation as ${\displaystyle z=\frac{(\chi_{\mathrm{constant}}^{2}-\chi^{2})/(\nu-\nu_{\mathrm{constant}})}{\chi_{\mathrm{best}}^{2}/(N-\nu)}}$
follows a $F_{2,N-5}$ and $F_{4,N-7}$-distribution, respectively.%
}.

The final false alarm probability (FAP) for the period search accounts
for the number of independent frequencies $M$ with the simple estimate
$M\approx fT$ \citep{Cumming04}, i.e. the frequency range $f$ and
the time baseline $T$, and is given by
\begin{equation}
\mathrm{FAP}=1-[1-\mathrm{Prob}(p>p_{\mathrm{best}})]^{M}
\end{equation}
and can be approximated by $\mathrm{FAP}\approx M\cdot\mathrm{Prob}(p>p_{\mathrm{best}})$
for $\mathrm{FAP}\ll1$. Since our frequency search interval ranges
from 0 to 1\,d$^{-1}$, we have typically $M\sim5500$ for a 15 year
time baseline.

Table~\ref{Tab:TestsCombinedSets} summarises the formal best-fitting
sinusoidal and Keplerian periods ($P_{\mathrm{sin}}$ and $P_{\mathrm{Kep}}$)
found by the periodograms along with their residual weighted rms and
FAP.%
\footnote{We list the formal, analytic FAP for the Keplerian orbits, but we
do not highlight them in the table. As remarked in \citet{Zechmeister09},
this FAP is likely to be underestimated due to an underestimated number
of independent frequencies. Also Keplerian solutions tend to fit outliers
(likely orginating from non-Gaussian noise) making them less robust
for period search.%
} Our approach recovers all stars that exhibit long-term trends emulated
by long periods and generally decreases the rms down to a few m/s.

We identify for HR~506, $\iota$~Hor, and HR~3259 the same periods
that were previously announced as planetary signals (\citealt{Kuerster00},
Mayor et al.\textsuperscript{\ref{fn:Mayor03}}, \citealt{Lovis06}).
In Sect.~\ref{Sec:Discussion} we derive for HR~506 and $\iota$~Hor
refined orbital solutions (see also Table~\ref{Tab:OrbitParamHR506}
and \ref{Tab:OrbitParamIotaHor}) and investigate the correlation
between RV and activity indicators.

For the refinement of the orbital parameters and the error estimation
we used the program GaussFit \citep{Jefferys1988} which can solve
general nonlinear fit-problems by weighted least squares and robust
estimation. As initial guess we provided the parameters found with
the Keplerian periodogram in the previous section. All offsets were
free parameters.

We provide also a first orbital solution for the spectroscopic binary
HR~2400 (Table~\ref{Tab:OrbitParamHR2400}). However, for $\kappa$~For,
$\alpha$~Cen~A+B and the giant HR~3677 it is not possible to give
a reliable orbital solution since our measurements cover only a small
piece of their orbits. Companion masses are estimated in Sect.~5.

\subsection{\label{sub:Correlations-with-Ca}Correlations with Ca II H\&K, BIS,
and FWHM}

Radial velocity variations can be caused by stellar variability such
as oscillations, granulations, spots, and magnetic activity cycles.
They can affect the stellar line profile and, in case of spots and
magnetic cycles, also the amount of Ca~II H\&K emission. To test
this, one can analyse the $R'_{\mathrm{HK}}$-index and the shape
of the cross-correlation function (CCF) function in particular its
full width at half maximum (FWHM) and bisector span (BIS, \citealp{Queloz2001})
which are measures for the averaged stellar line width and asymmetry,
respectively. The activity indicators are products of the HARPS pipeline
and the computation of ($S_{\mathrm{HK}}$ and) $R'_{\mathrm{HK}}$
is described in \citet{Lovis2011}. Note that these indicators do
not cover the whole survey, because they cannot be derived from the
CES spectra (which do not include the Ca~II region and are contaminated
by the iodine lines). The $S_{\mathrm{HK}}$ errors are derived from
photon noise \citep{Lovis2011}, the BIS span errors are taken as
twice%
\footnote{The precision of measuring the bisector velocity in the upper and
lower part of the CCF (i.e. each uses only a half of the gradients
in the CCF) is $\approx\sqrt{2}\sigma_{RV}$ and when taking their
difference adding both errors in quadrature yields another factor
of $\sqrt{2}$, i.e. a factor of two in total for the BIS span.%
} the internal RV errors, and the FWHM errors are 2.35 times%
\footnote{For a Gaussian fit the mean parameter errors for center ($RV$) and
width ($s$) are the same $\sigma_{RV}=\sigma_{s}$ (e.g. Eq.~(5.8)
in \citealp{Kaper1966}). Moreover, since for a Gaussian function
$\mathrm{FWHM}=2\sqrt{2\ln2}\cdot s$, we have $\sigma_{\mathrm{FWHM}}=2.35\sigma_{RV}$.%
} the internal RV errors.

The time series of the activity indicators and their corre\-lations
with RVs are shown for each star in Figs.~\ref{onlFig:zettuc}--\ref{onlFig:hr8883}.
Tables~\ref{Tab:rhk} and \ref{Tab:fwhm} summarise for each star
and activity indi\-ca\-tor their mean values, scatter, and the correlation
coefficients. Statistically significant linear correlations with $\mathrm{FAP}<0.01$
are highlighted in bold font in the Tables and depicted by a solid
line in the Figures.

Note that a high statistical correlation does not necessarily mean
a physical correlation, in particular when both quantities exhibit
just trends which could coincide just by chance and temporarily. However,
if the correlation is present during more complex temporal variations,
e.g., both quantities have the same period, a planetary hypothesis
should be excluded. But note also, that the Sun hosts a Jupiter in
a 12\,yr orbit and shows a comparably long magnetic cycle (11\,yr).

In the sample, $\tau$~Cet has the smallest variations in $\log R'_{\mathrm{HK}}$
(0.005\,dex), while $\zeta^{1}$~Ret has the largest variations
(0.048\,dex). We find significant correlations between RV and $\log R'_{\mathrm{HK}}$
for the stars $\beta$~Hyi, HR~209, $\zeta^{1}$~Ret, $\alpha$~Men,
HR~4979, HR~8323, and for the two stars with planet candidates $\iota$~Hor
and $\epsilon$~Ind~A. However, the correlation seen for $\alpha$~Cen~B
(probably also HR~7703 and $\epsilon$~Ind~A) is artificial since
the RV trend is largely caused by a wide companion, respectively,
instead of a magnetic cycle (Sect.~5).

\citet{Lovis2011} provide also a relation to estimate the slope of
the RV and $R'_{\mathrm{HK}}$ correlations based on the stellar temperature
$T_{\mathrm{eff}}$ and metallicity {[}Fe/H{]}. After conversion%
\footnote{$\frac{\Delta RV}{\Delta\log RHK}=\ln10\frac{\Delta RV}{\Delta\ln RHK}=\ln10\cdot RHK\frac{\Delta RV}{\Delta RHK}=\ln10\cdot RHK\cdot C_{RV}$
where the sensitivity $C_{RV}=C_{RV}(T_{\mathrm{eff}},[\mathrm{Fe}/\mathrm{H}])$
is given by Eq.~(9) in \citet{Lovis2011}.%
} to a slope w.r.t. $\log R'_{\mathrm{HK}}$ by multiplication with
a factor of $\ln10\cdot R'_{\mathrm{HK}}/10^{5}$, these estimates
can be compared to the derived slopes given in Table~\ref{Tab:rhk}.
As an additional cross-check, we will do this comparison occasionally
in Sect.~\ref{Sec:Discussion} when we conclude for a magnetic cycle
hypothesis. We note that eight stars%
\footnote{$\zeta$~Tuc (HD\,1581), $\tau$~Cet (HD\,10700), $\zeta^{2}$~Ret
(HD\,20807), $\delta$~Eri (HD\,23249), HR~3259 (HD\,69830),
HR~4979 (HD\,114613), HR~8323 (HD\,207129), and $\epsilon$~Ind~A
(HD\,209100)%
} were included also in a sample analysed by \citet{Lovis2011} for
magnetic cycles via $R'_{\mathrm{HK}}$. With the exception of the
RV standard star $\tau$~Cet and the subgiant $\delta$~Eri, these
authors reported $R'_{\mathrm{HK}}$ cycles/trends for these stars.

Significant correlations between RV and BIS are found for 7 stars
($\tau$~Cet, $\iota$~Hor, $\epsilon$~Eri, $\alpha$~Men, HR~4979,
HR~8323, and HR~8883). The correlation for three more stars ($\alpha$~Cen~B,
GJ~570~A, and HR~8501) should be artificial, since the RV trends
can be attributed to a wide stellar companion.

Finally, RV-FWHM correlations are found for 7 stars ($\beta$~Hyi,
HR~209, $\tau$~Cet, $\zeta^{1}$~Ret, $\zeta^{2}$~Ret, $\alpha$~Men,
and HR~8323). For 6 other stars (HR 2400, HR~3677, $\alpha$~Cen~A,
$\alpha$~Cen~B, GJ~570~A, and probably also $\epsilon$~Ind~A)
the correlations are artificial due the RV trends caused by their
wide companions.

RV variations caused by the magnetic cycle should result in positive
correlations with all three indicators \citep{Lovis2011}. The stars
$\alpha$~Men and HR~8323 are nice showcase examples for this effect.
On the other hand, an RV-BIS anti-correlation (cf. the active stars
$\iota$~Hor, $\epsilon$~Eri, and also the giant HR~8883) is expected
for rotating spots \citep{Boisse2011} and should be therefore related
to the stellar rotation period.

\subsection{Detection limits}

\begin{figure}
\centering

\includegraphics[width=1\linewidth]{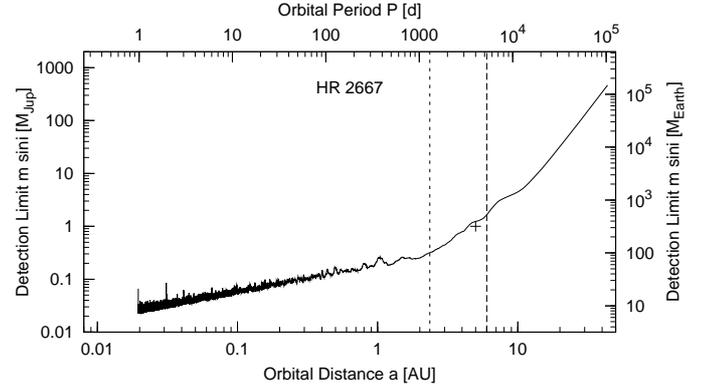}

\caption{\label{Fig:Limits}Detection limit for HR~2667 considering circular
orbits. The cross marks the distance and $m\sin i$ of a Jupiter analogue
for $i=90^{\circ}$. The vertical lines indi\-cate the time baseline
of the HARPS and all combined measurements, respectively.}
\end{figure}

To demonstrate the sensitivity of our survey, we have calculated for
each star conservative 99.9\% detection limits for circular orbits
following the method outlined in \citet{Zechmeister09b}. As an example,
Fig.~\ref{Fig:Limits} illustrates the upper mass limit for HR~2667
(one of our most constant stars) showing that we are sensitive approximately
to Jupiter analogues. Because the more precise HARPS data typically
cover only 1500\,d, there is a loss of sensitivity for longer periods
indicated by a steep increase of the upper mass limit. The longer
time baseline gained with the CES data pushes a bit down the limit
at longer periods.

\begin{figure}
\centering

\includegraphics[width=1\linewidth]{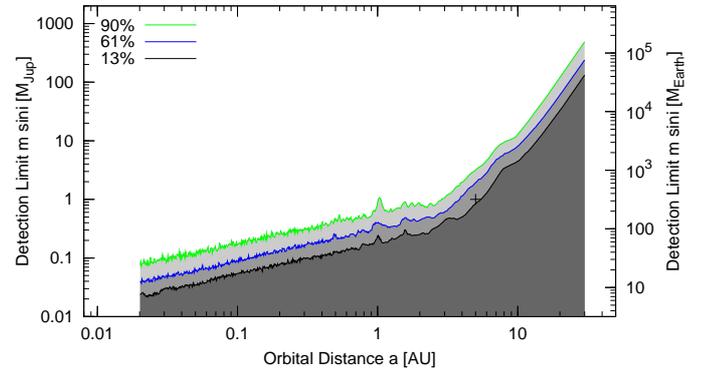}

\caption{\label{Fig:Limits_Survey}Sensitivity of the survey. For each orbital
distance the mi\-ni\-mum mass is indicated that could have been
detected for 4, 19, and 28 stars of the 31 stars (corresponding to
a fraction of 13\%, 61\%, and 90\%) with 99.9\% significance. The
cross marks again a Jupiter analogue.}
\end{figure}

The detection limits of the other stars have a qualitatively similar
shape to that shown in Fig.~\ref{Fig:Limits}. For four stars ($\zeta$~Tuc,
$\tau$~Cet, and the residuals of HR~7703 as well as $\epsilon$~Ind~A)
the upper mass limit is lower than 1\,$M_{\mathrm{Jup}}$ at 5\,AU
(due to the lower stellar mass of $0.7\, M_{\odot}$). For 19 stars
the limit is still lower than 2\,$M_{\mathrm{Jup}}$ and for 28 stars
lower than 4\,$M_{\mathrm{Jup}}$ at 5\,AU (see Figure~\ref{Fig:Limits_Survey}).

\section{\label{Sec:Discussion}Discussion on individual targets}

\begin{figure*}
\centering

\includegraphics[width=0.5\linewidth]{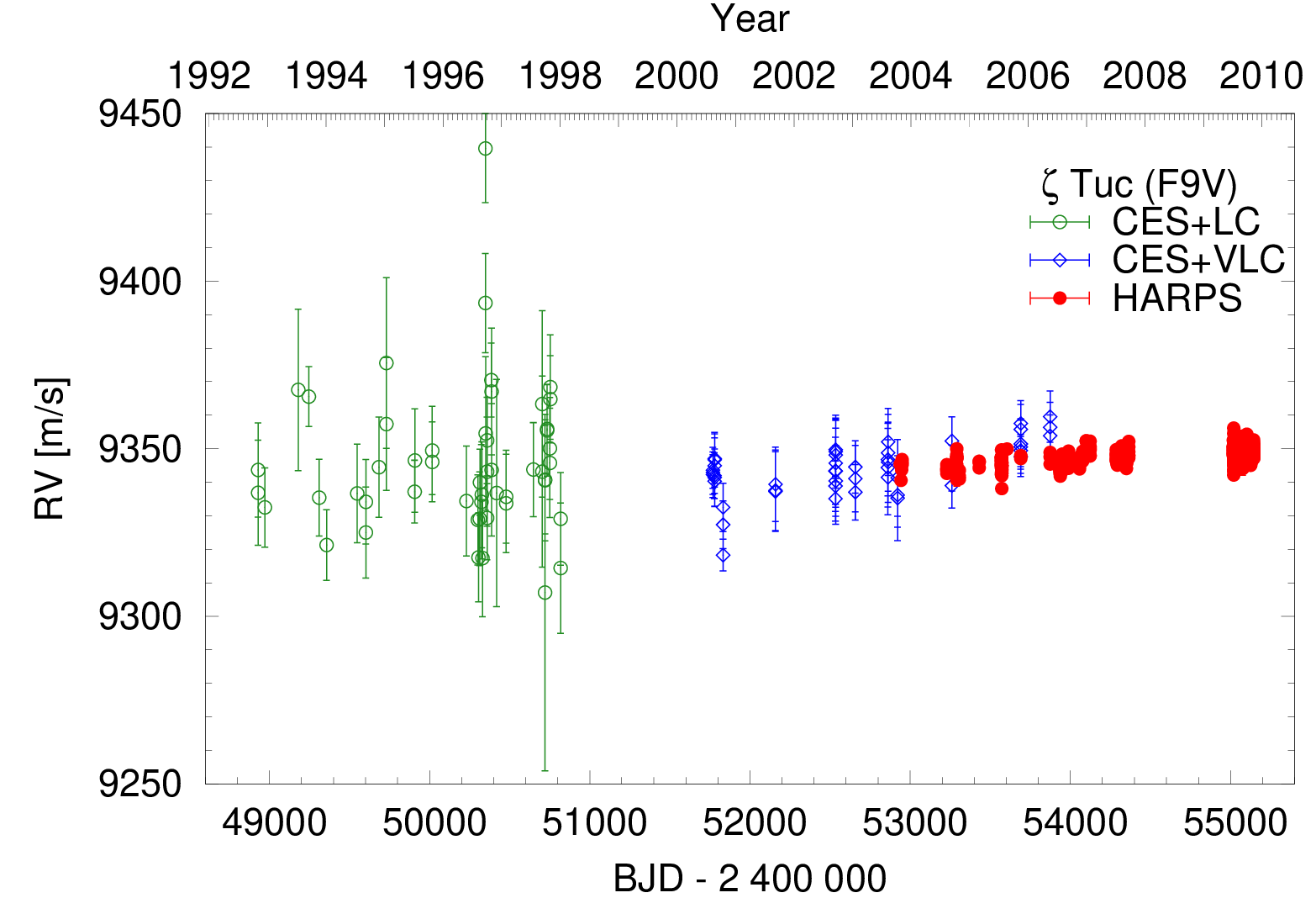}\includegraphics[width=0.5\linewidth]{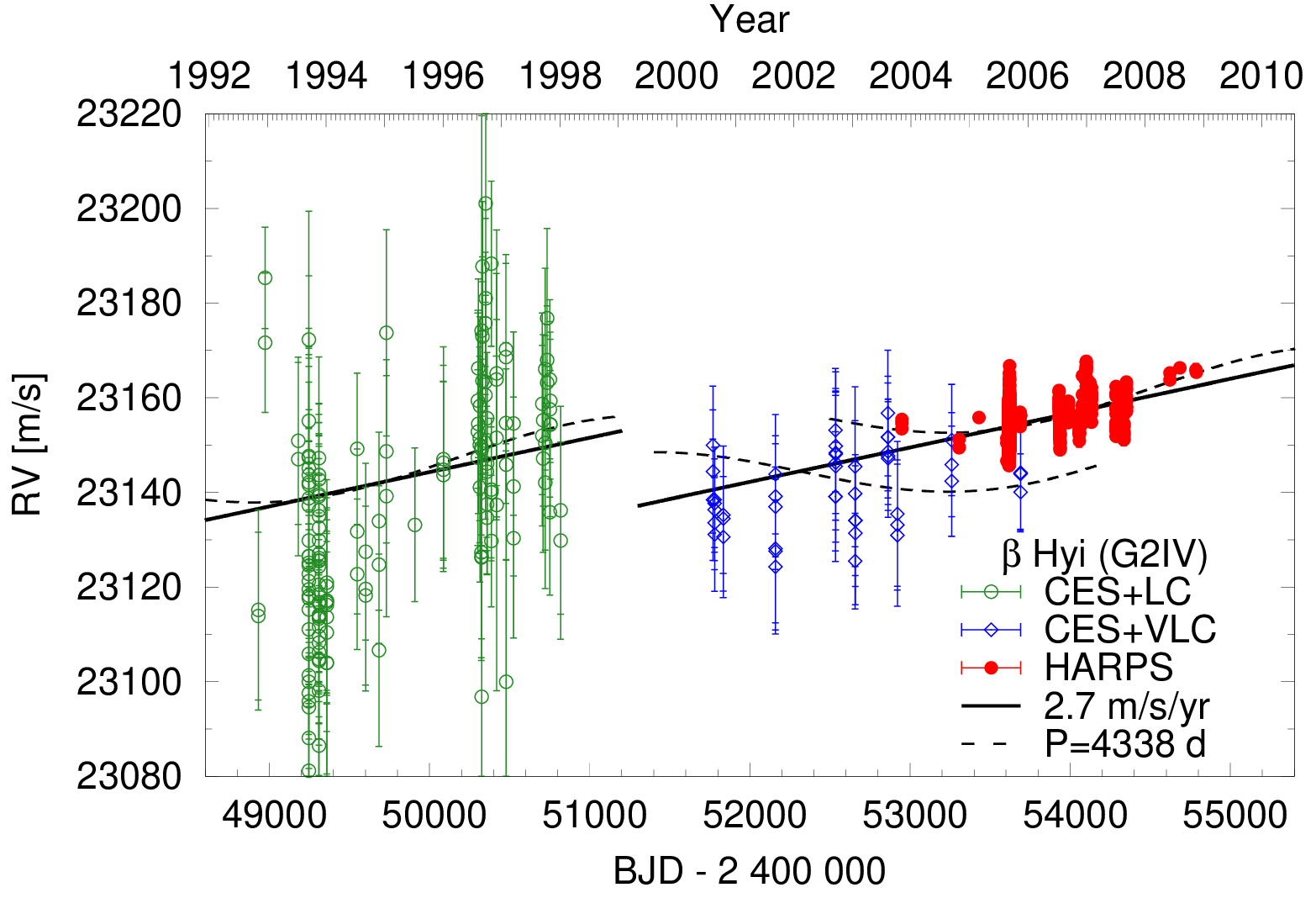}

\includegraphics[width=0.5\linewidth]{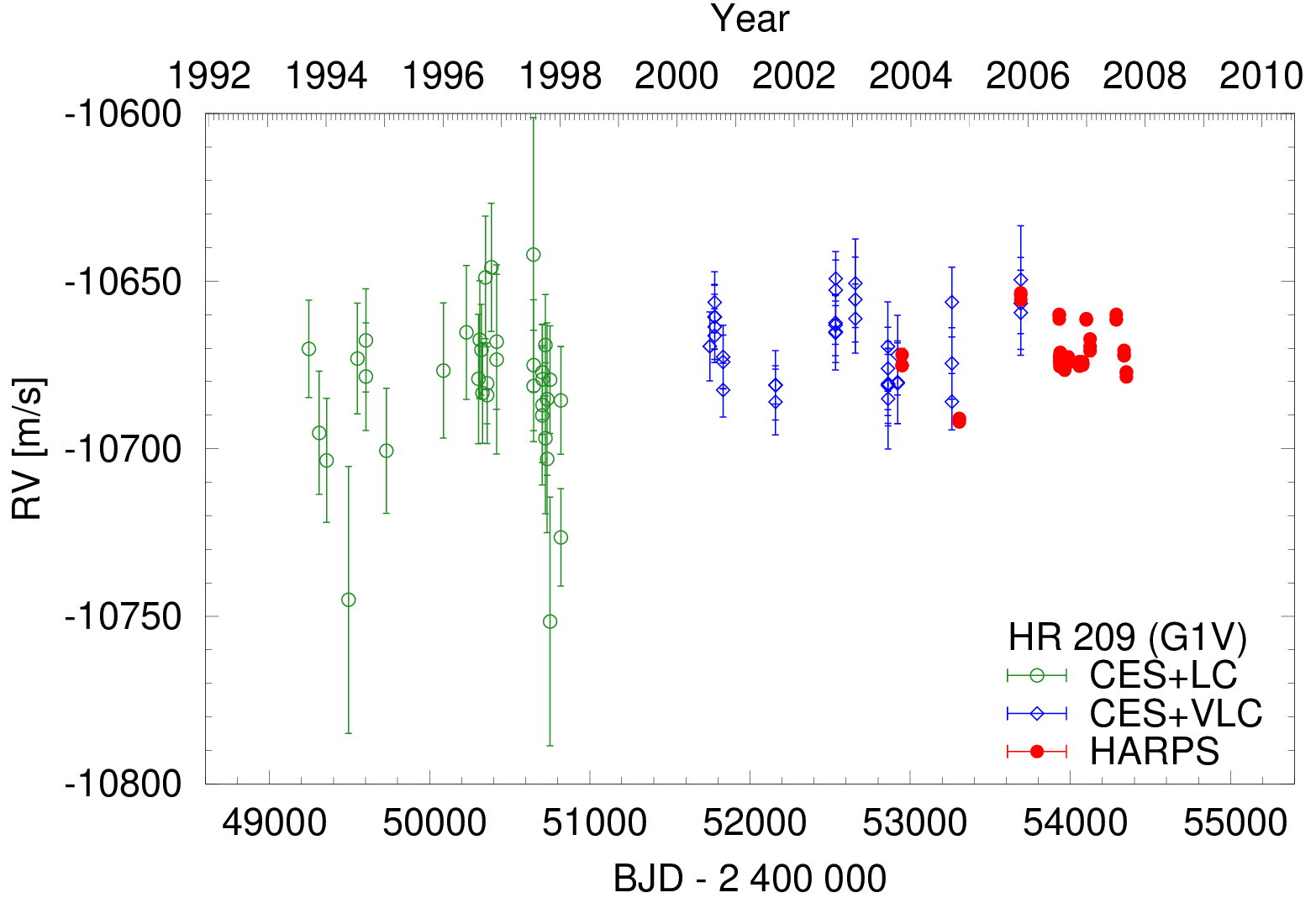}\includegraphics[width=0.5\linewidth]{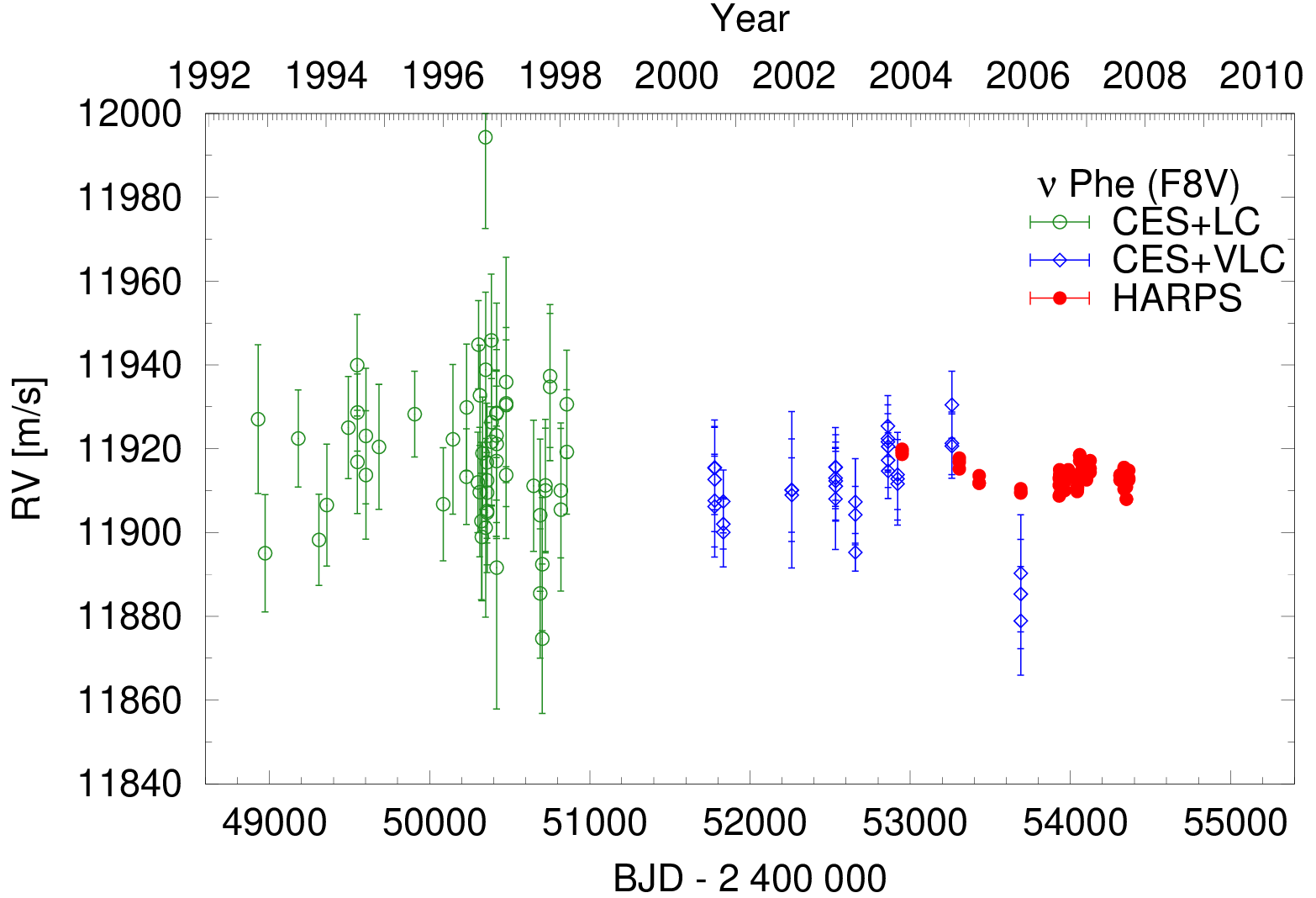}

\includegraphics[width=0.5\linewidth]{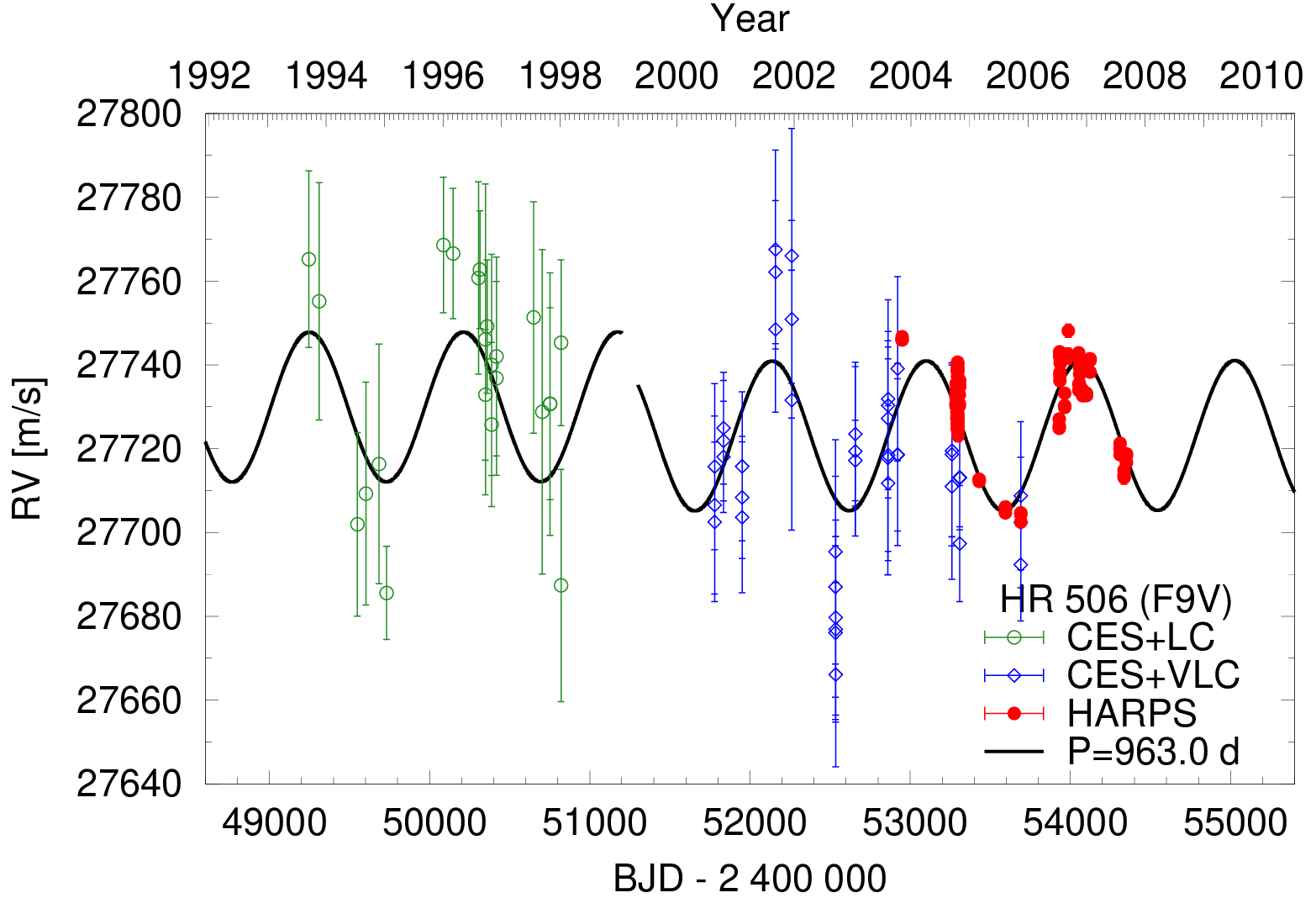}\includegraphics[width=0.5\linewidth]{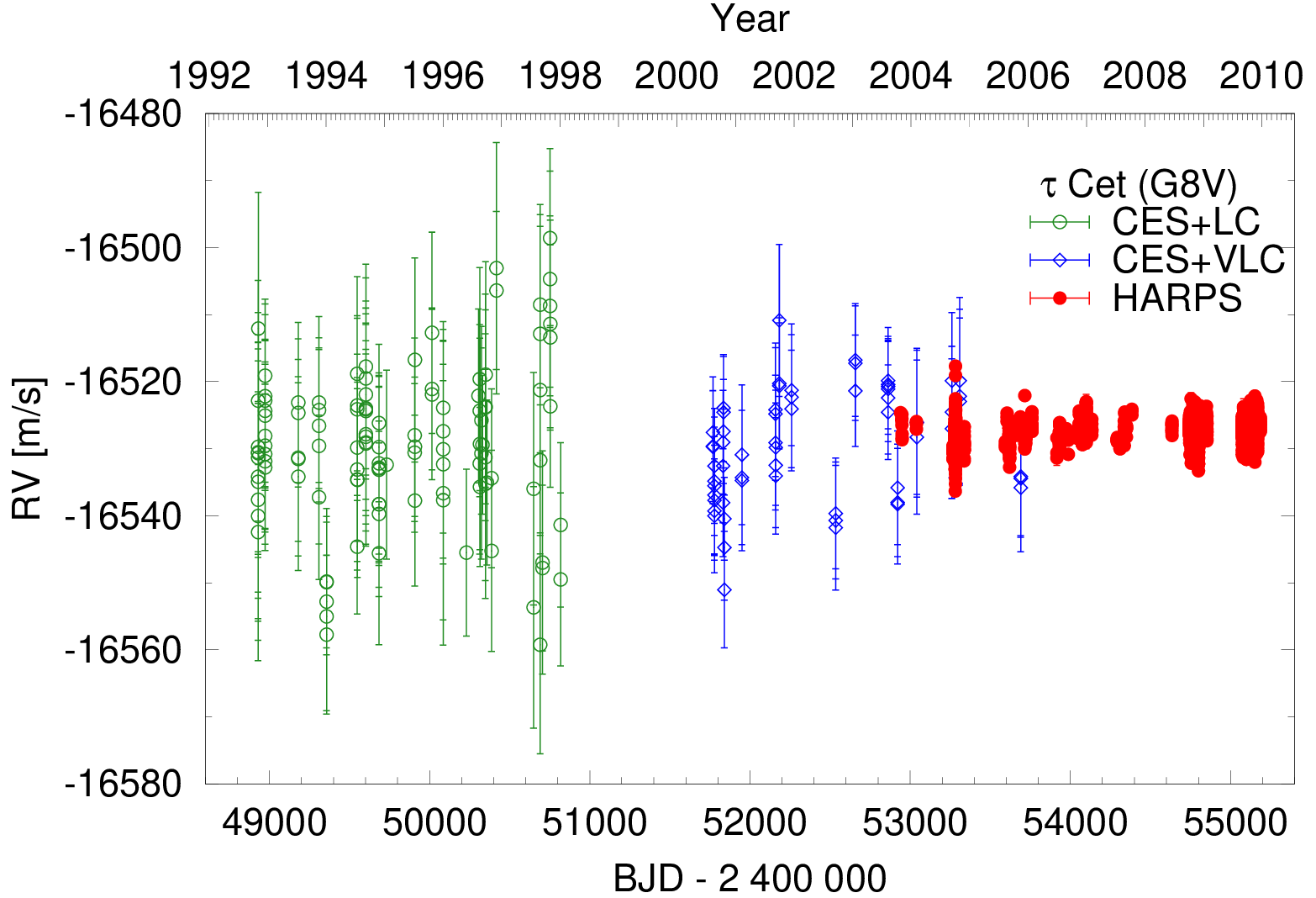}

\caption{\label{Fig:RV-1}Radial velocity time series (unbinned data). The
error bars depict the internal measurement errors $\Delta RV_{i}$,
i.e. not including jitter. LC (green open circles) and VLC (blue open
diamonds) data are displayed with their measured offsets. Jumps in
the curves indicates the difference between the measured (Sect.~\ref{Sub:LC_VLC})
and fitted (Sect.~\ref{Sec:RV_Analysis}) LC-VLC offset. HARPS data
are in red filled circles. The solid lines indicate significant models,
while dashed lines illustrate less or non-significant alternative
models. All models include secular acceleration. Model curves are
shown for $\beta$~Hyi (trend and long-period sinusoid, see text
for discussion), and HR~506 (sinusoid).}
\end{figure*}

\begin{figure*}
\centering

\includegraphics[width=0.5\linewidth]{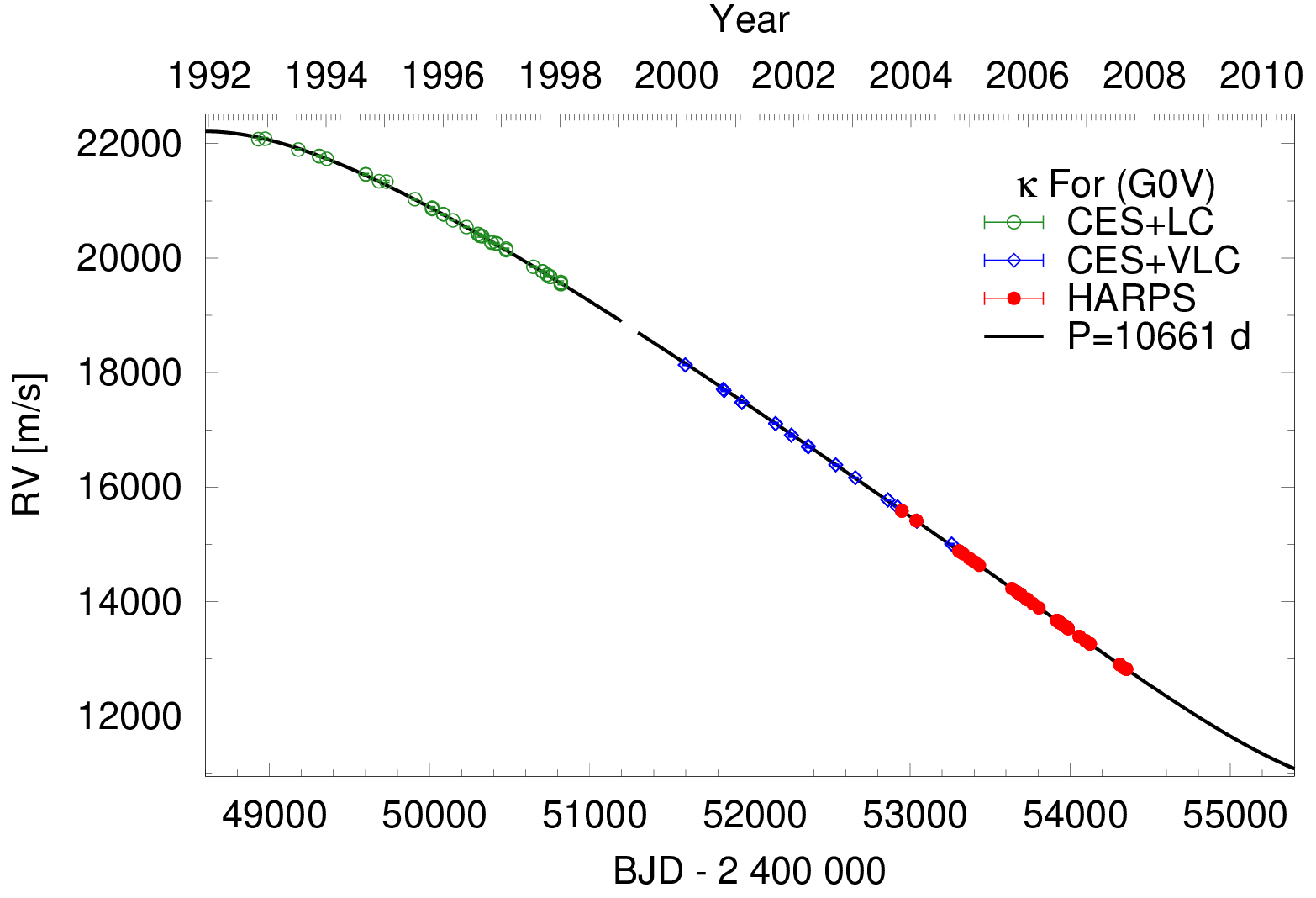}\includegraphics[width=0.5\linewidth]{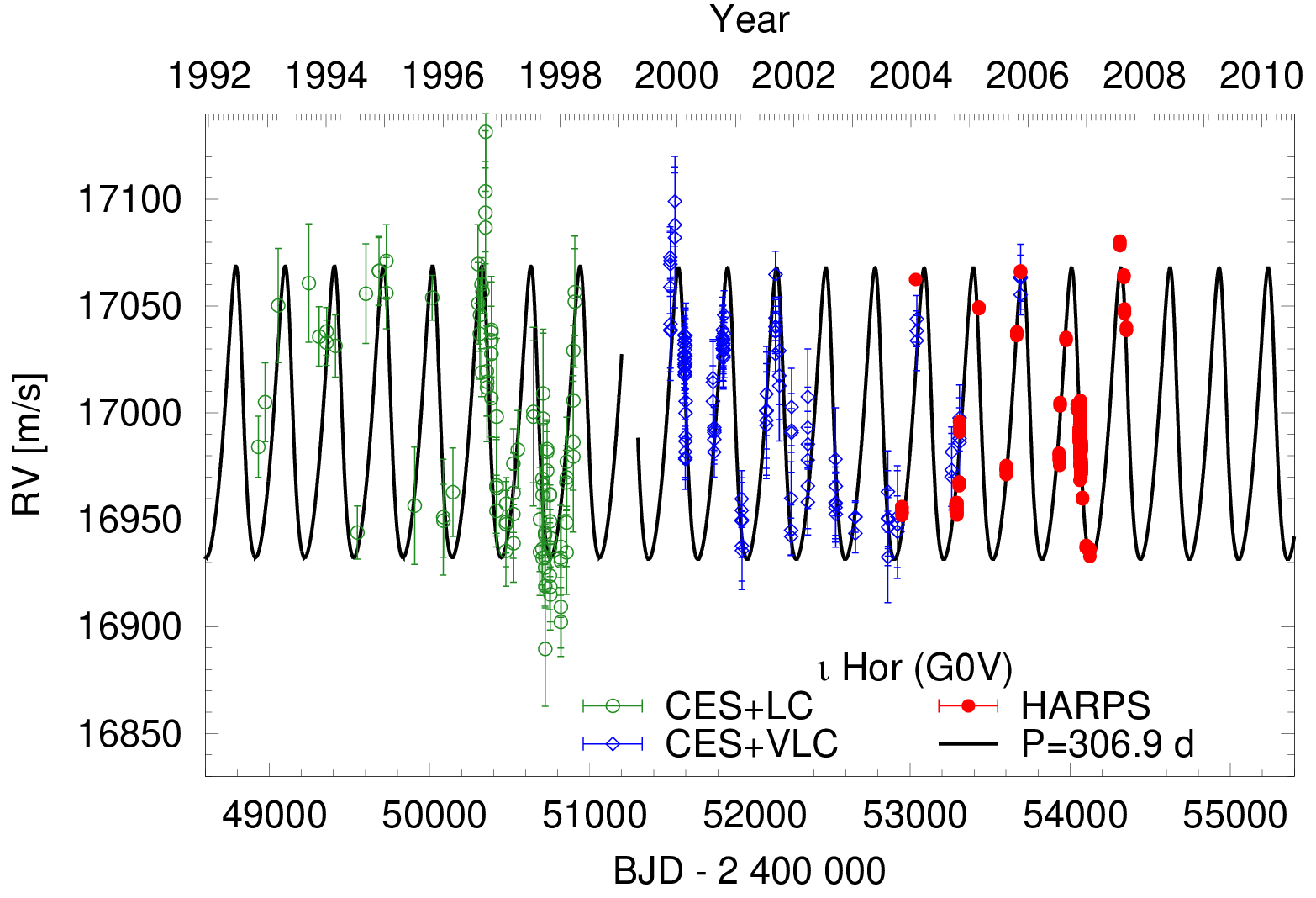}

\includegraphics[width=0.5\linewidth]{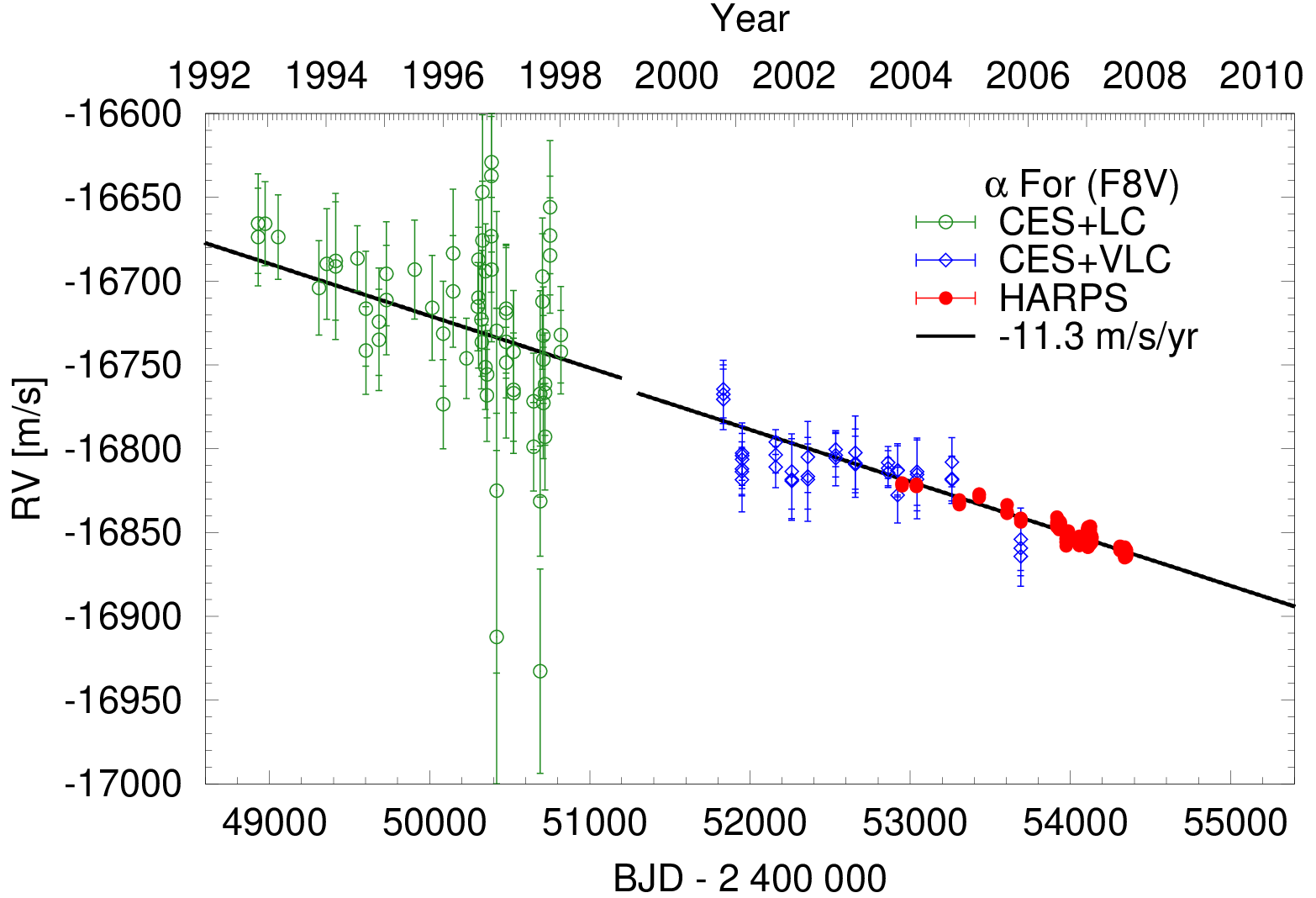}\includegraphics[width=0.5\linewidth]{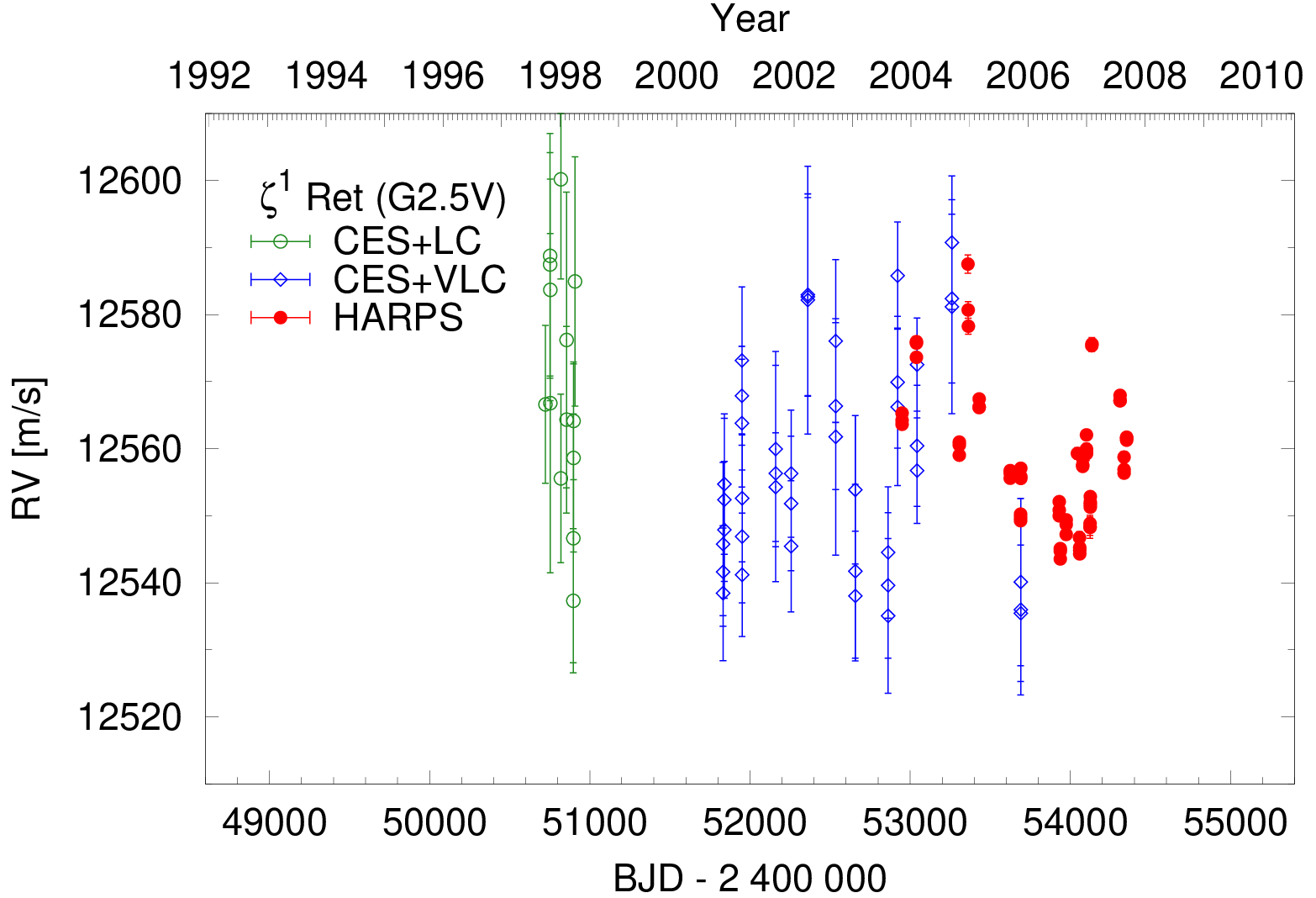}

\includegraphics[width=0.5\linewidth]{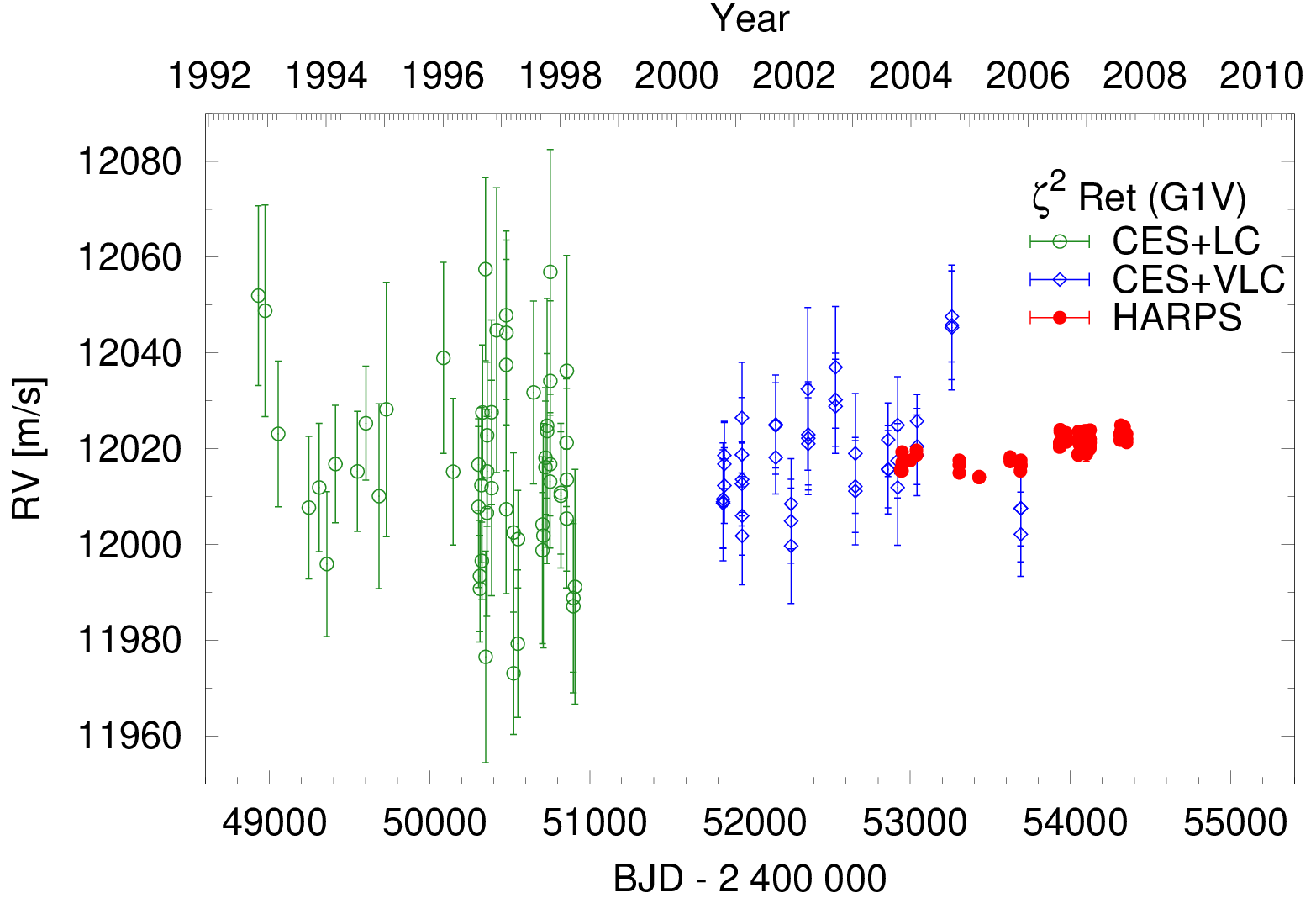}\includegraphics[width=0.5\linewidth]{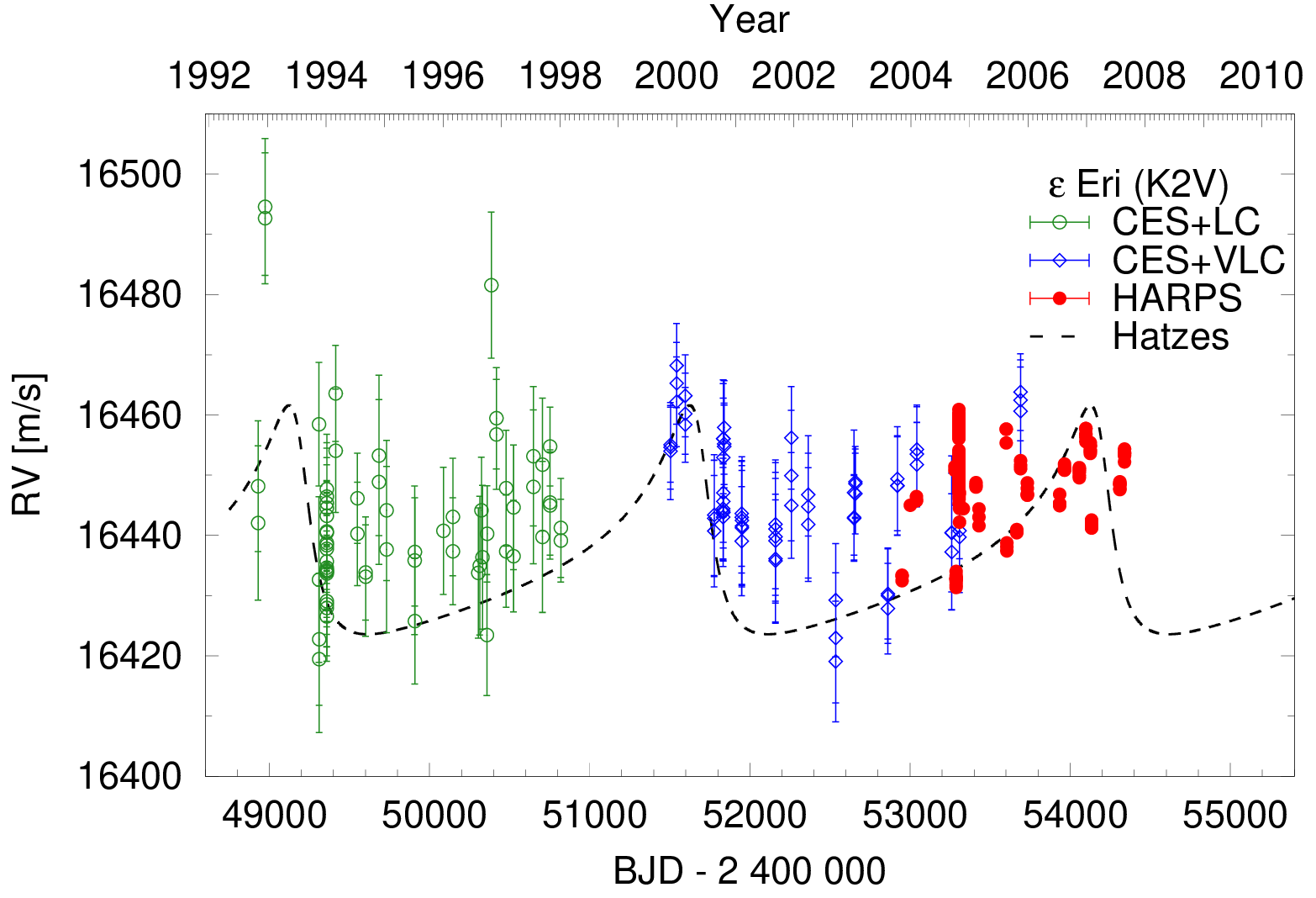}

\caption{\label{Fig:RV-2}Radial velocities. Continuation of Fig.~\ref{Fig:RV-1}.
Model curves are shown for $\kappa$~For (Keplerian), $\iota$~Hor
(Keplerian), and $\alpha$~For (trend). The Keplerian orbit for $\epsilon$~Eri
taken from \citet{Hatzes00} is not significant in this work.}
\end{figure*}

\begin{figure*}
\centering

\includegraphics[width=0.5\linewidth]{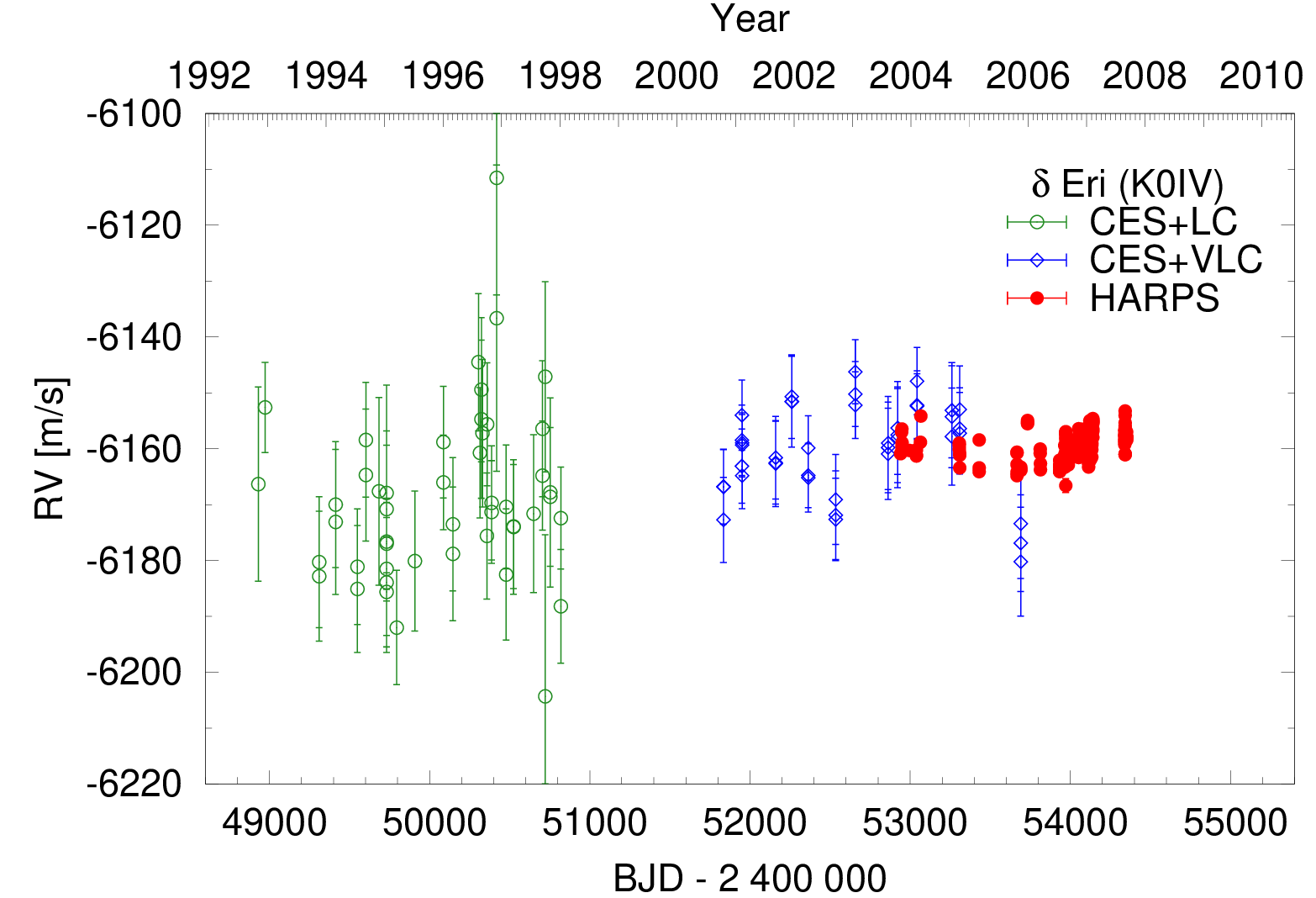}\includegraphics[width=0.5\linewidth]{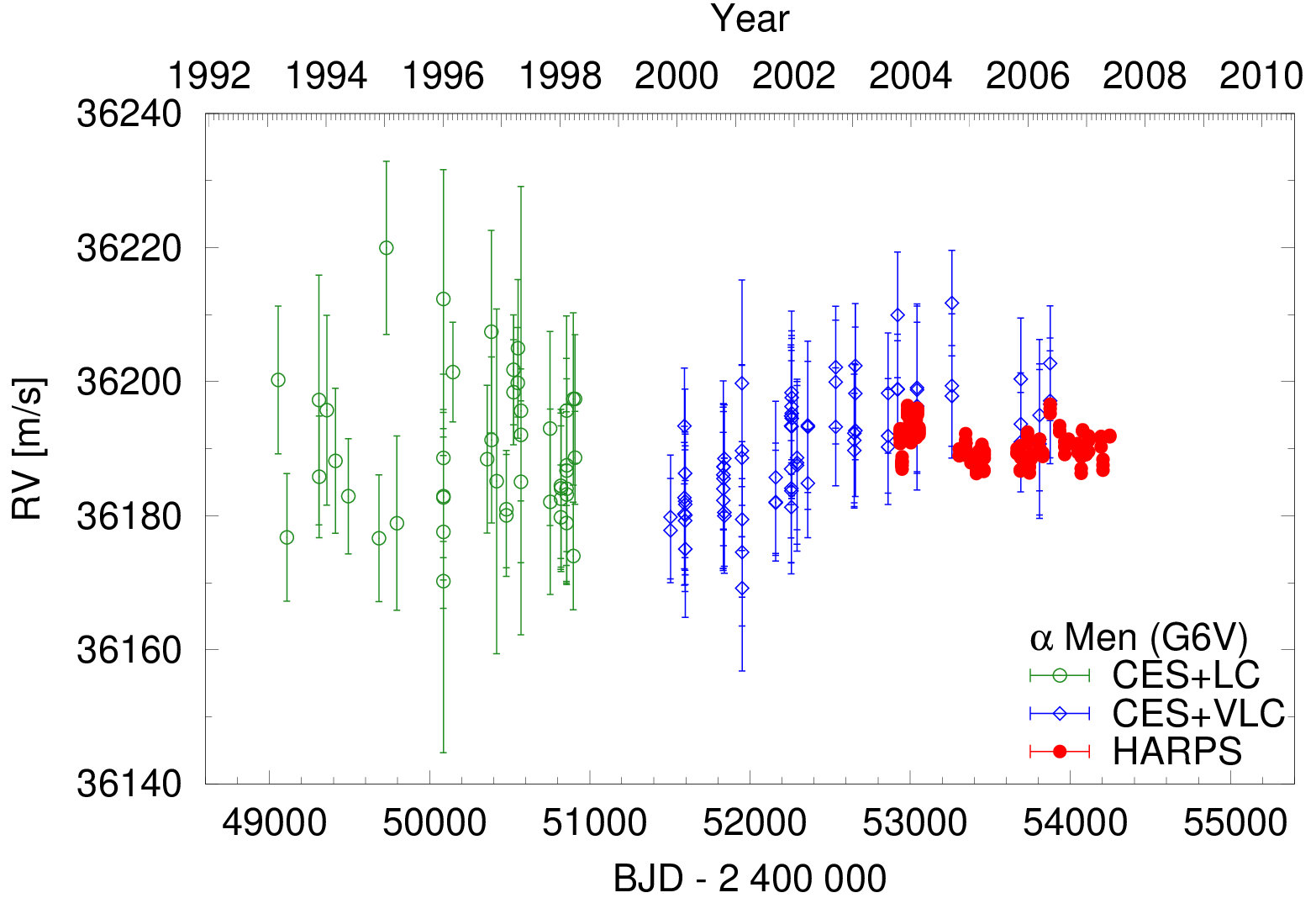}

\includegraphics[width=0.5\linewidth]{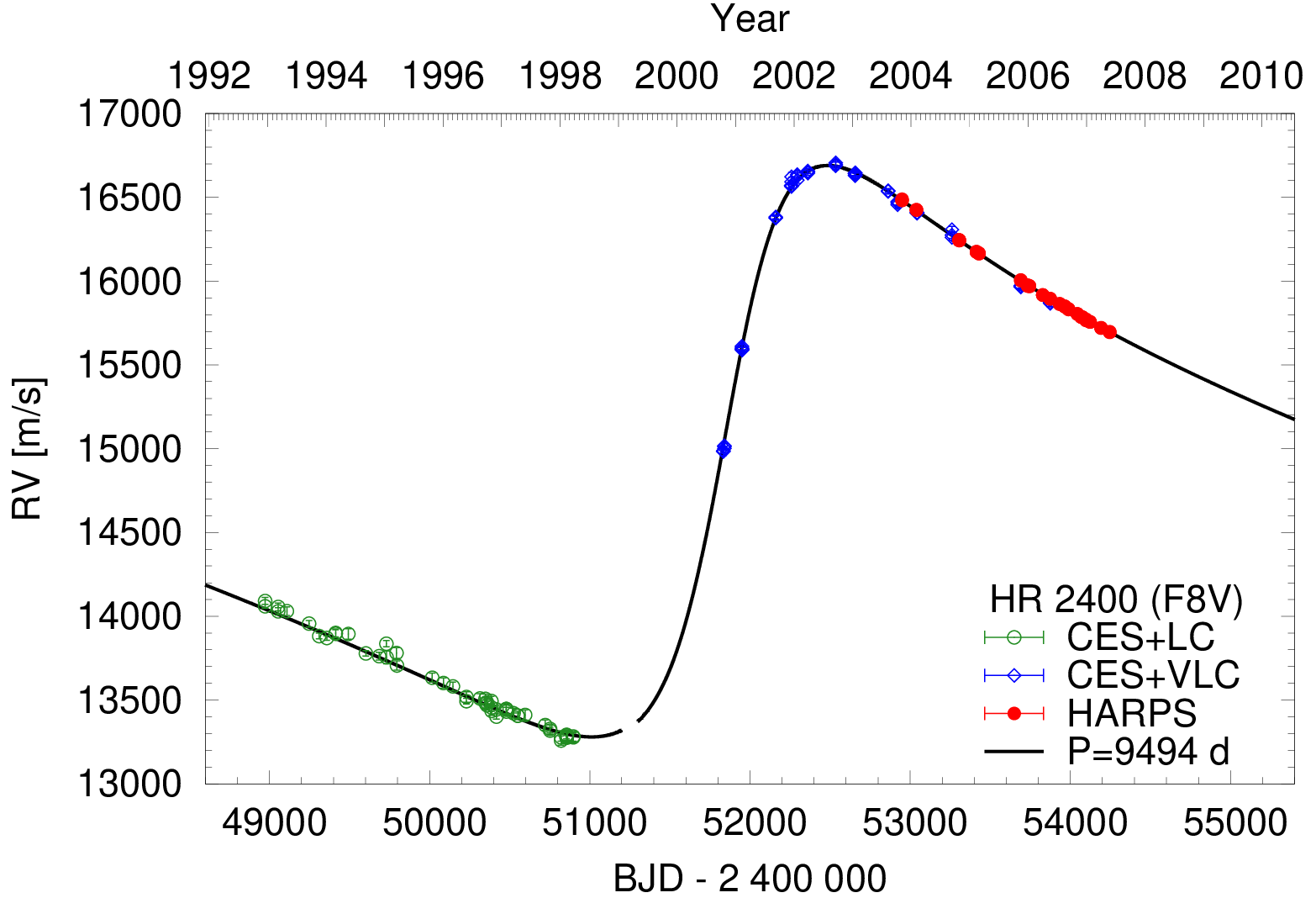}\includegraphics[width=0.5\linewidth]{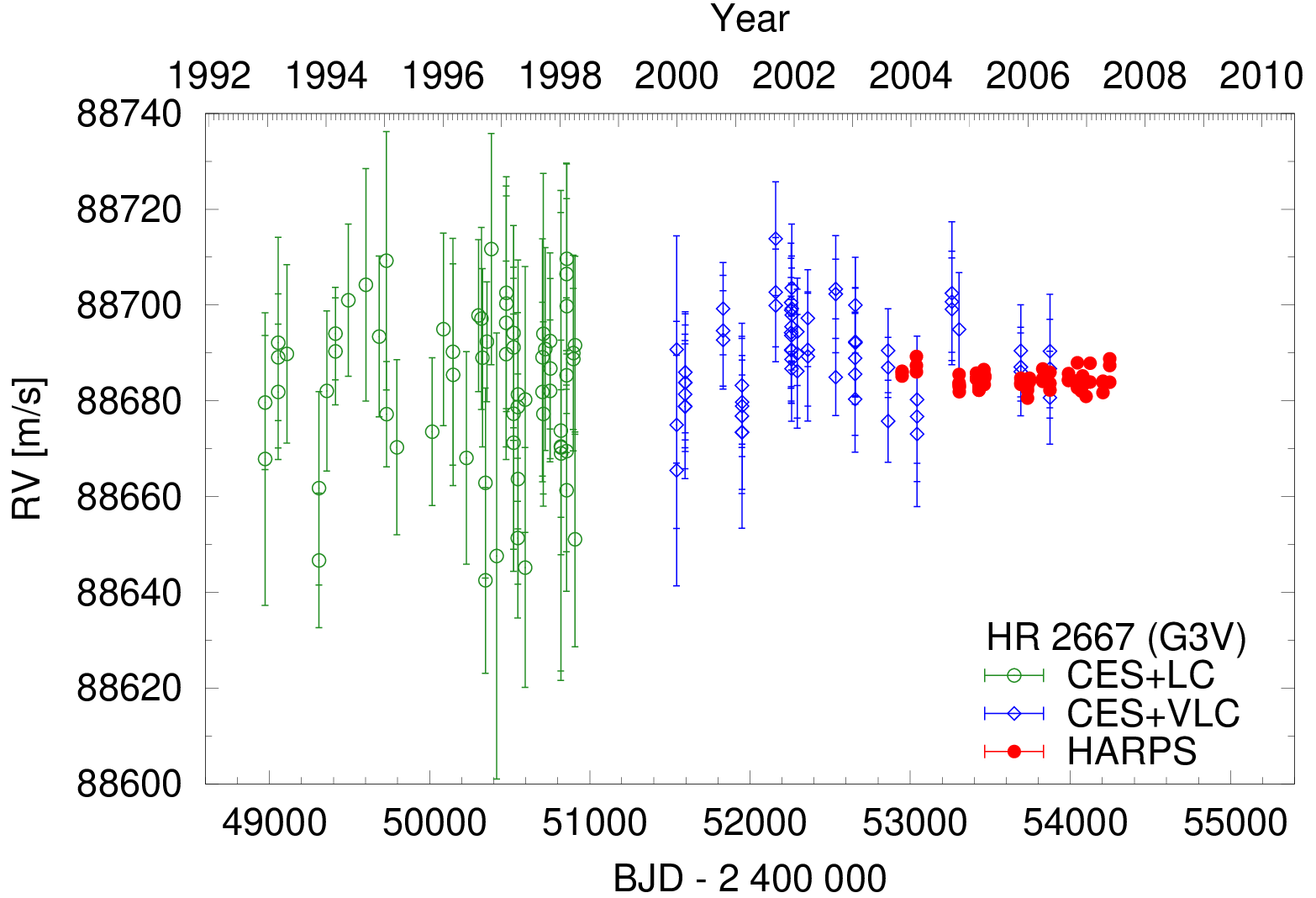}

\includegraphics[width=0.5\linewidth]{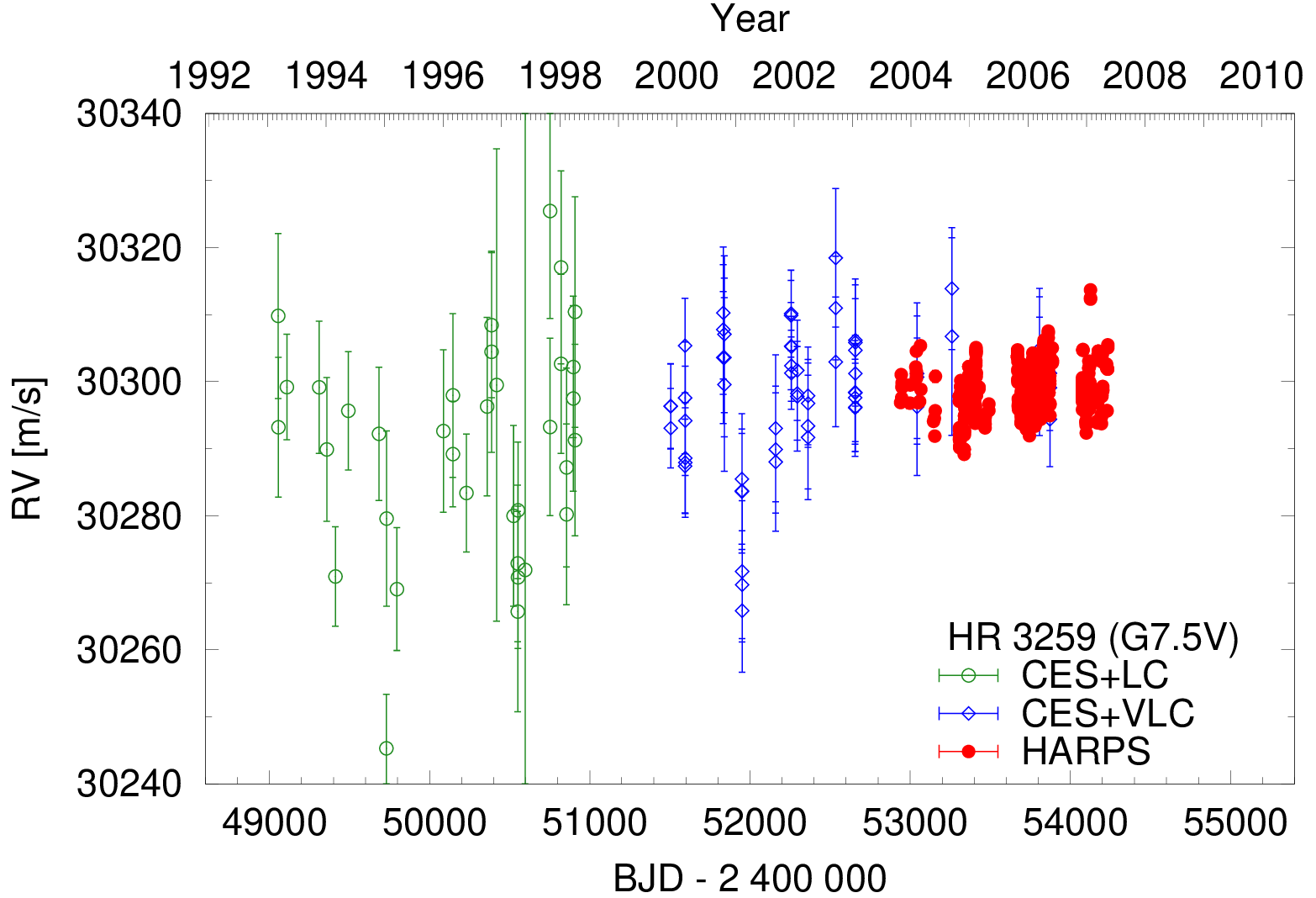}\includegraphics[width=0.5\linewidth]{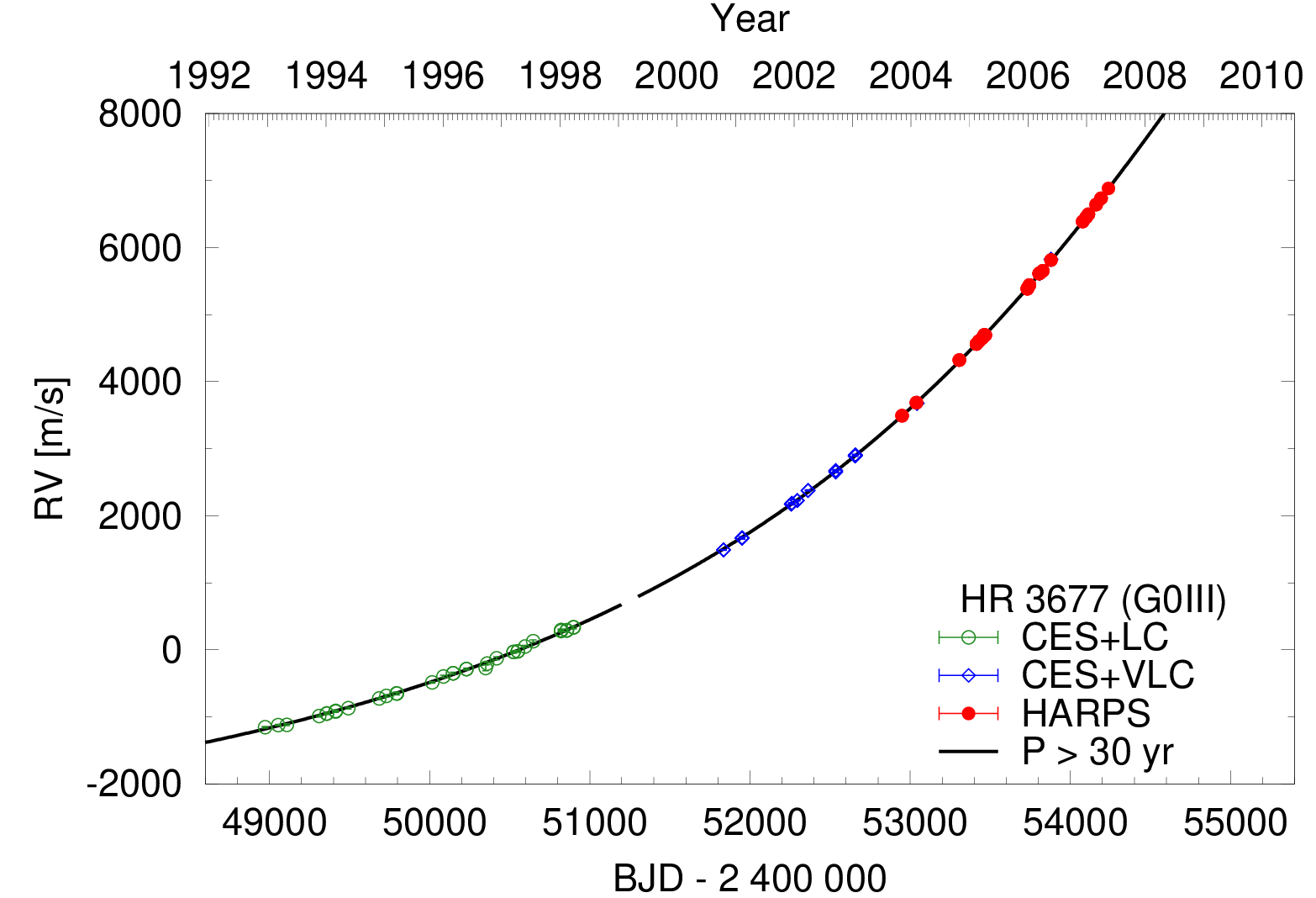}

\caption{\label{Fig:RV-3}Radial velocities. Continuation of Figs.~\ref{Fig:RV-1}
and \ref{Fig:RV-2}. Model curves are shown for HR~2400 (Keplerian)
and HR~3677 (Keplerian).}
\end{figure*}

\begin{figure*}
\centering

\includegraphics[width=0.5\linewidth]{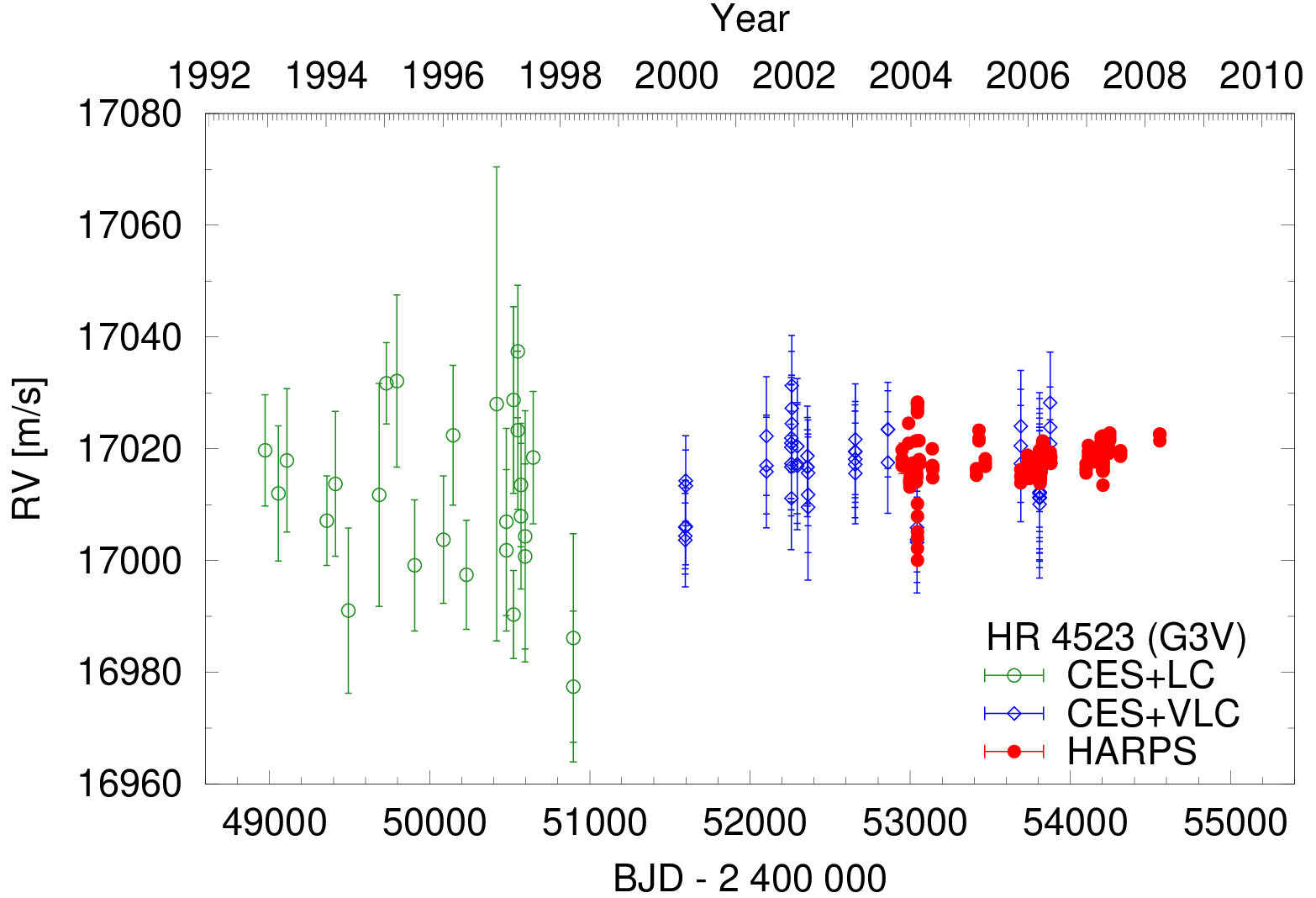}\includegraphics[width=0.5\linewidth]{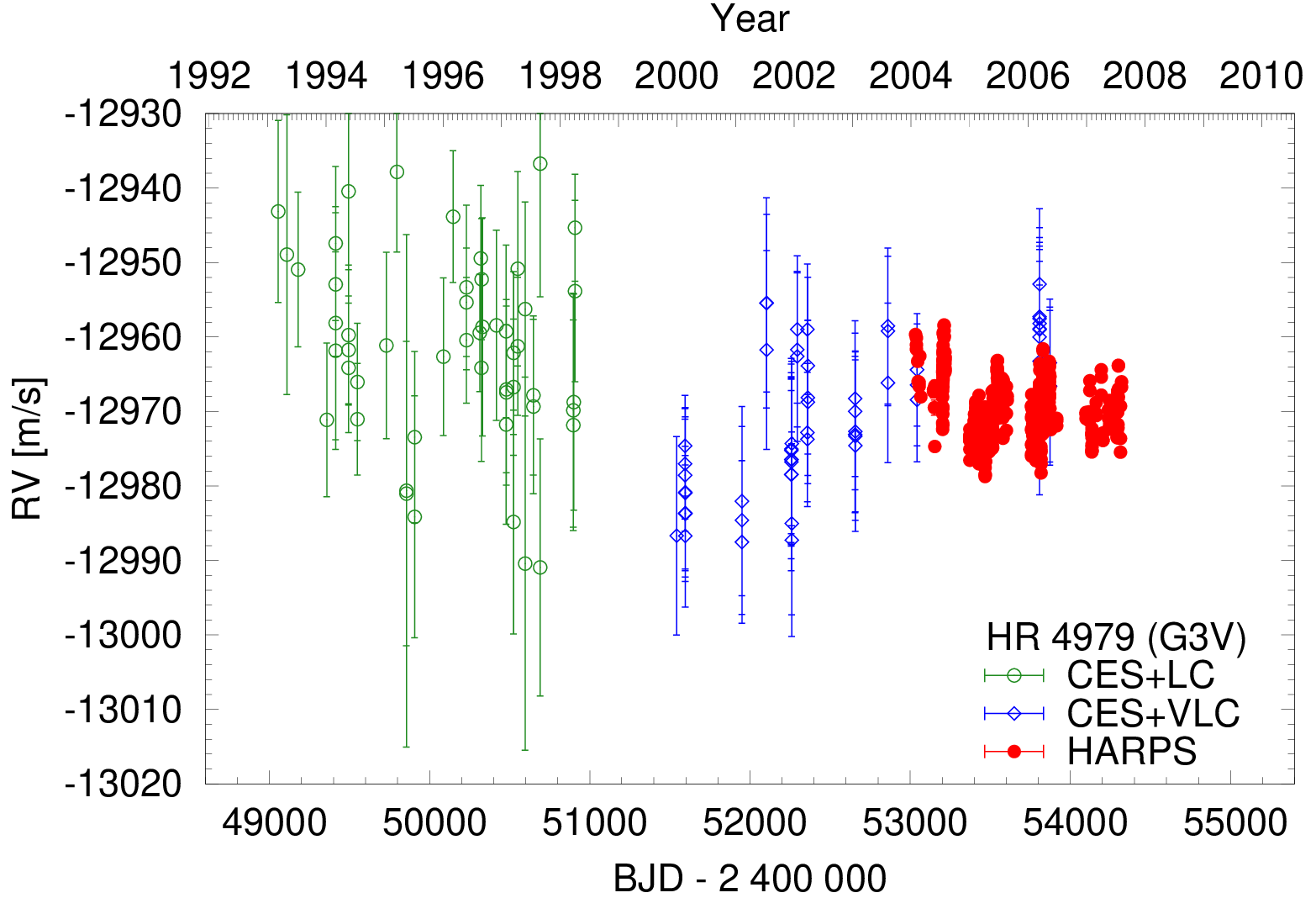}

\includegraphics[width=0.5\linewidth]{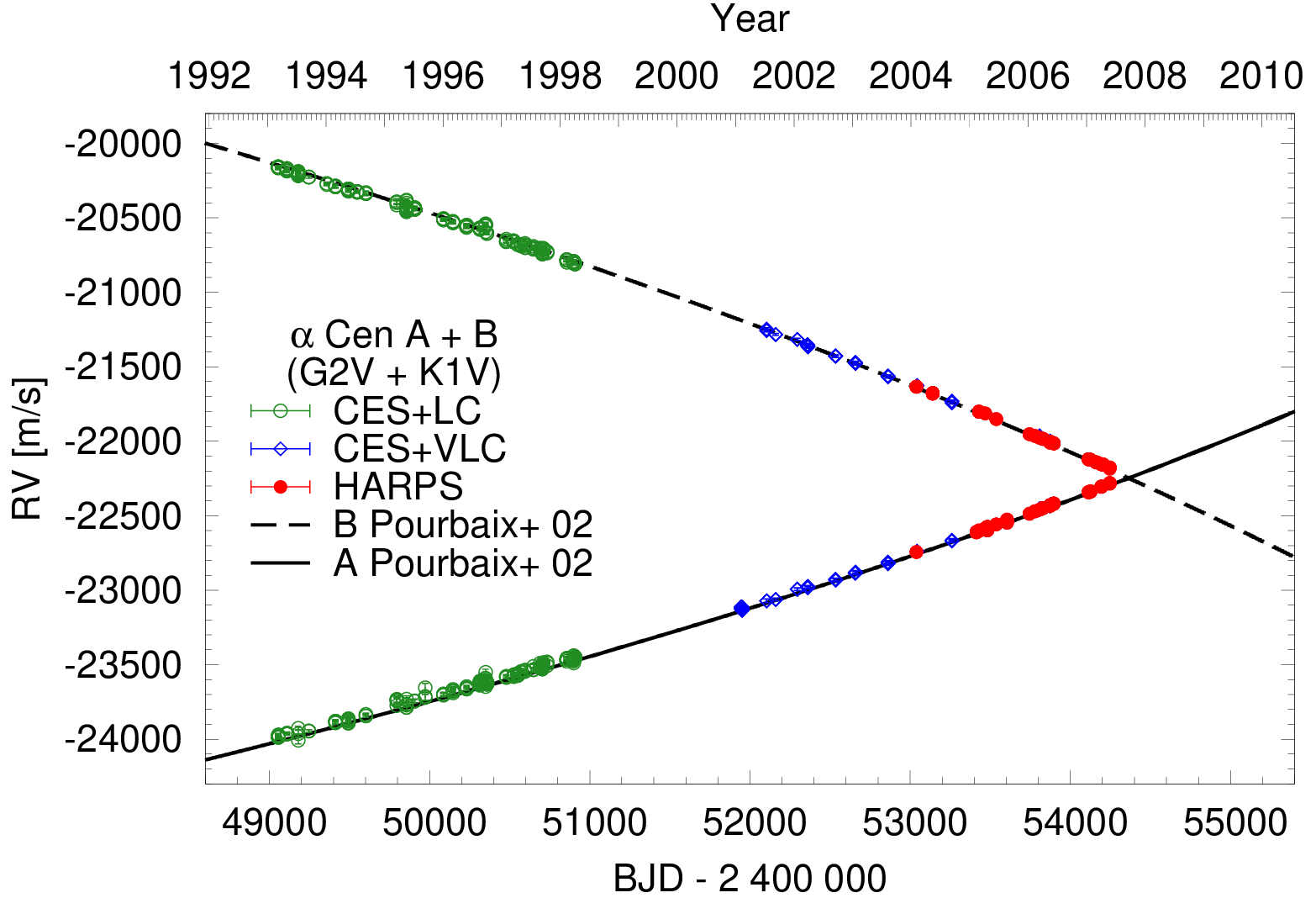}\includegraphics[width=0.5\linewidth]{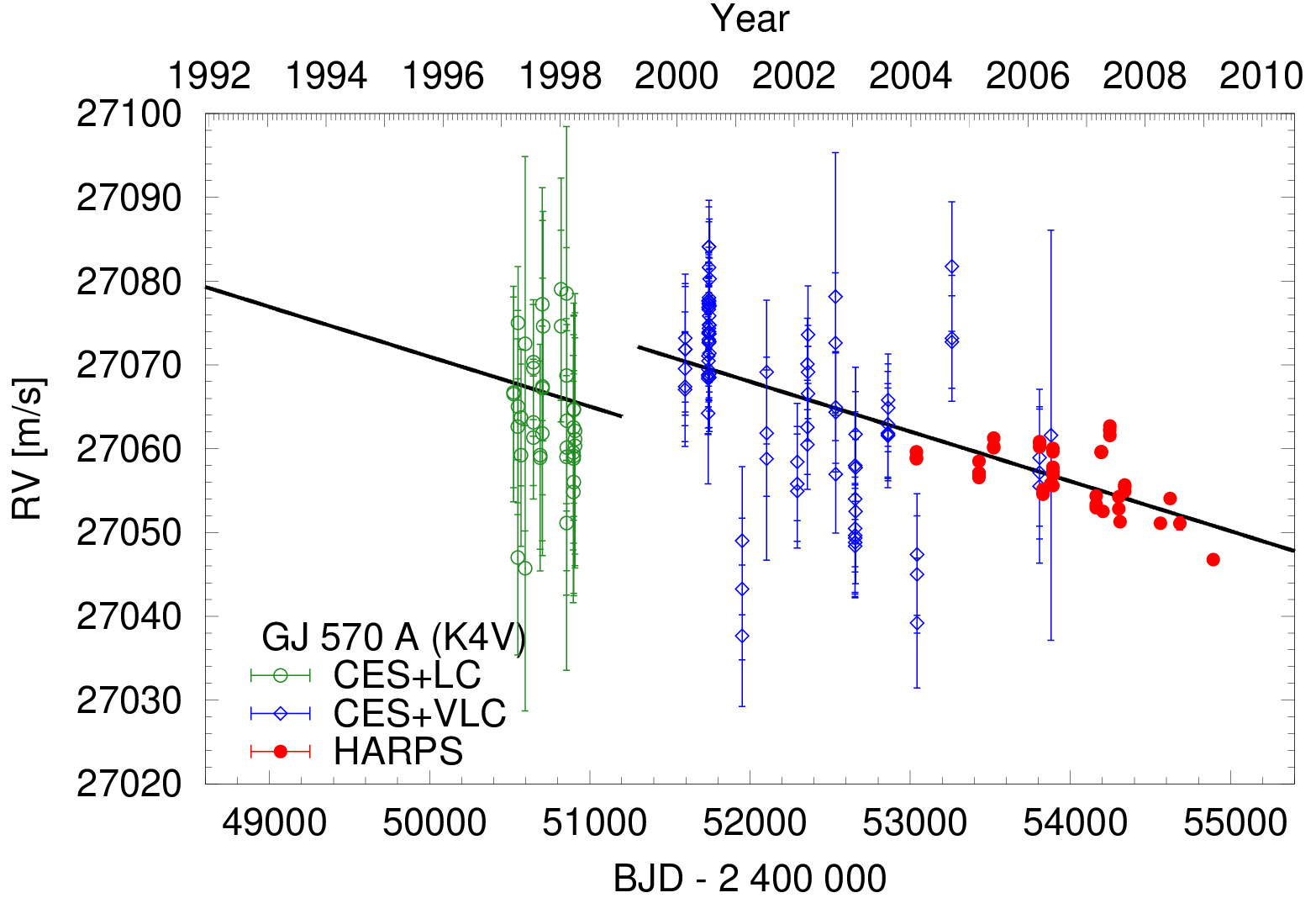}

\includegraphics[width=0.5\linewidth]{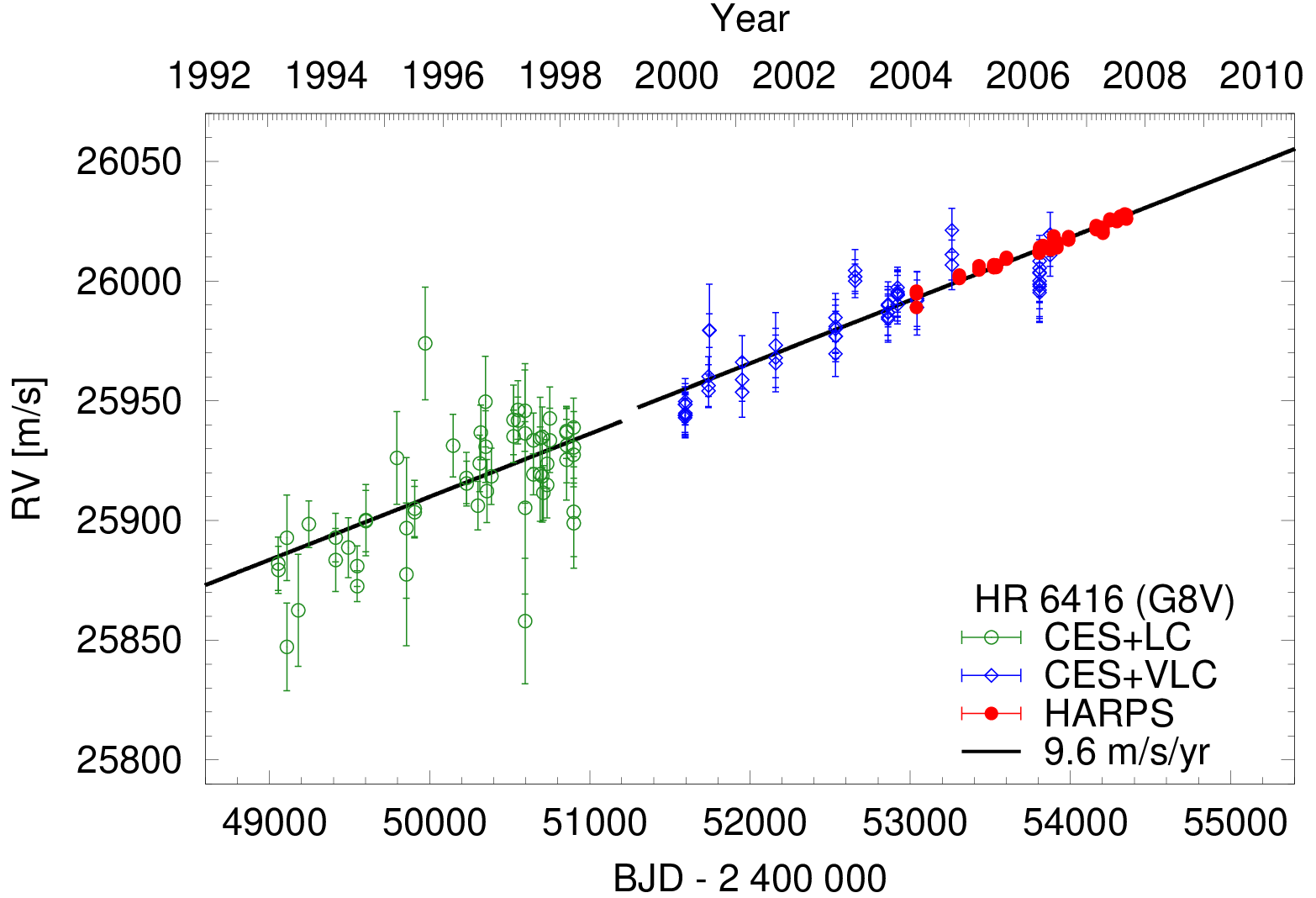}\includegraphics[width=0.5\linewidth]{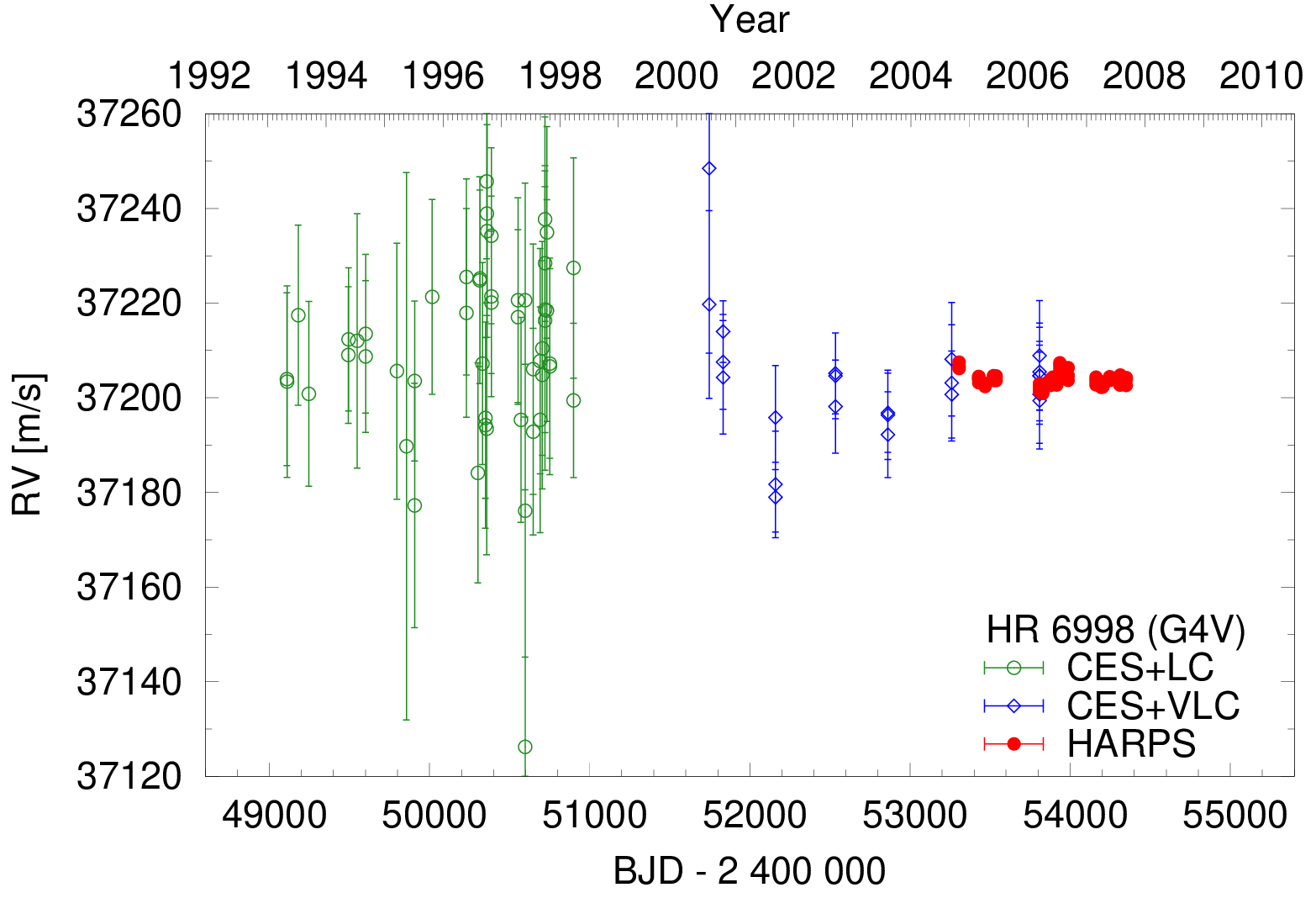}

\caption{\label{Fig:RV-4}Radial velocities. Continuation of Figs.~\ref{Fig:RV-1}--\ref{Fig:RV-3}.
Model curves are shown for $\alpha$~Cen~A and $\alpha$~Cen~B
(Keplerian orbit from \citealt{Pourbaix2002}), GJ~570~A (trend),
and HR~6416 (trend).}
\end{figure*}

\begin{figure*}
\centering

\includegraphics[width=0.5\linewidth]{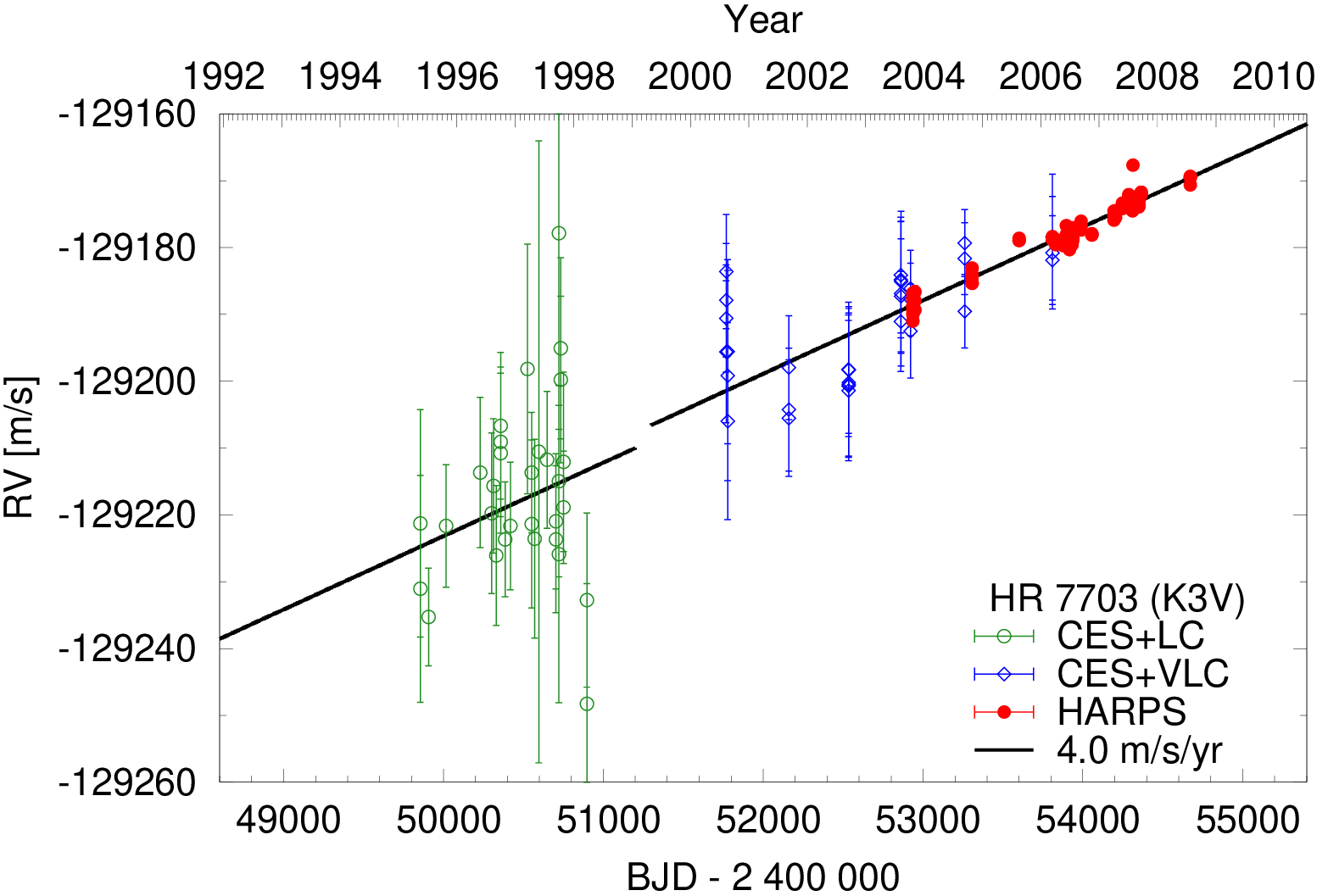}\includegraphics[width=0.5\linewidth]{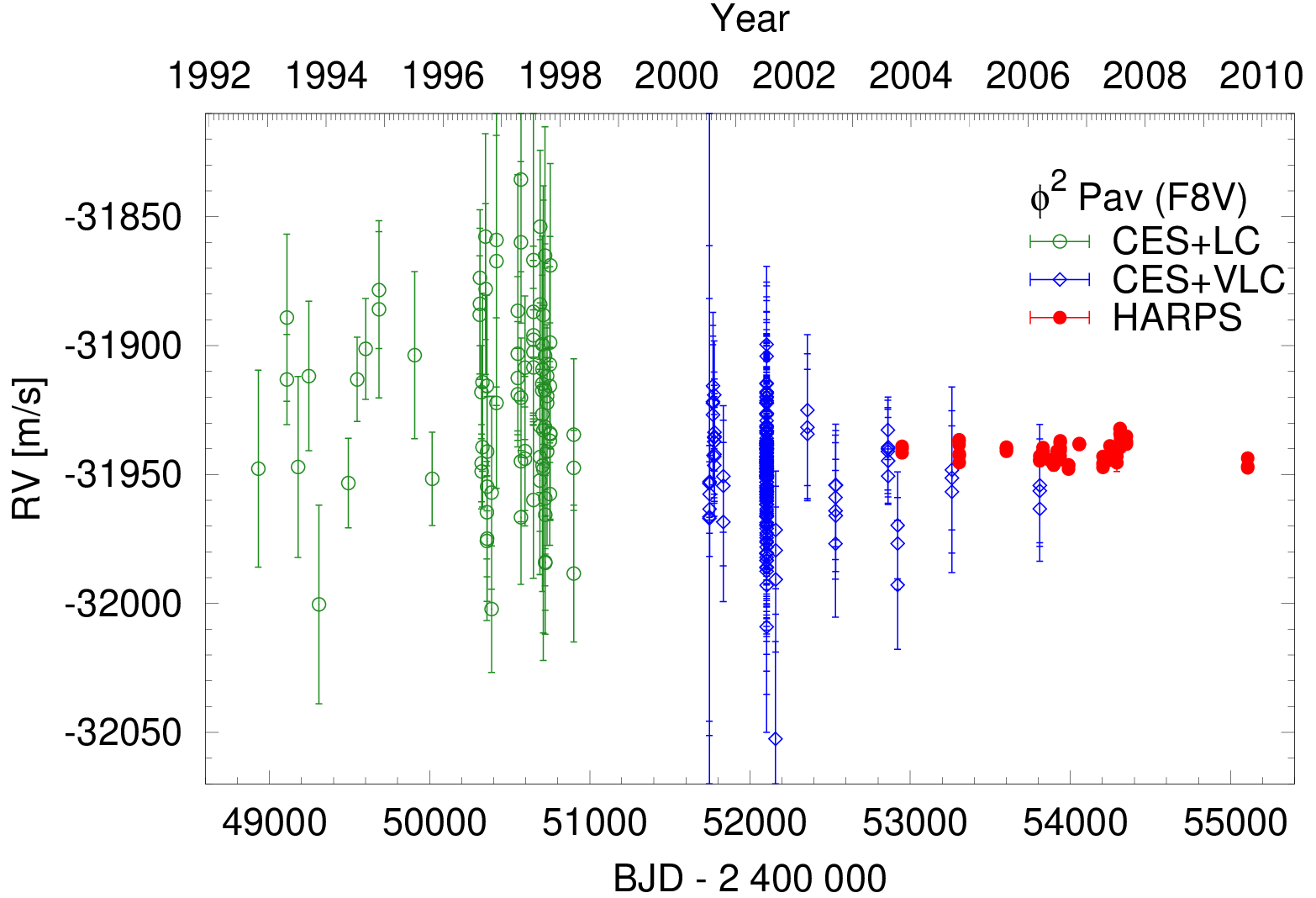}

\includegraphics[width=0.5\linewidth]{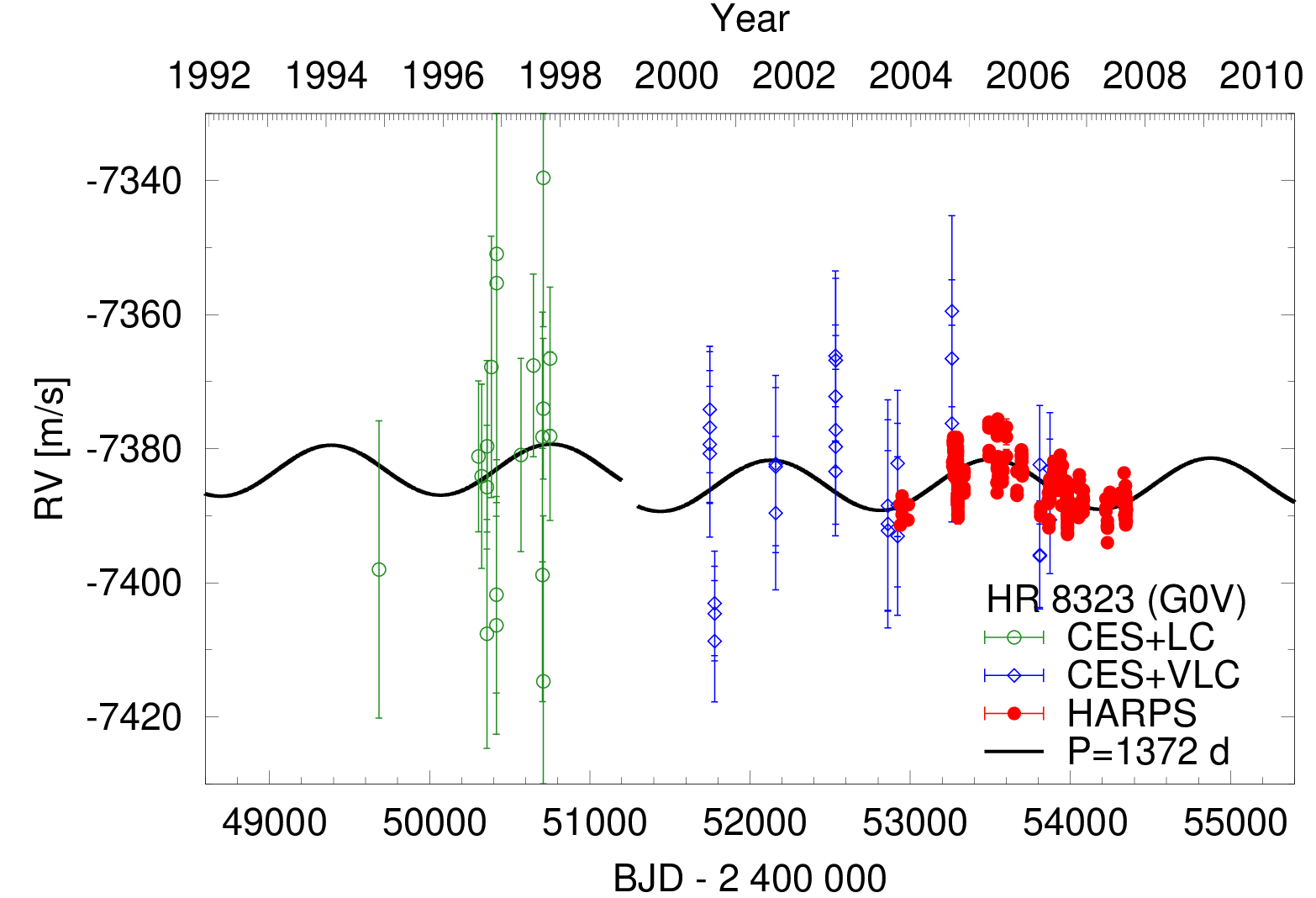}\includegraphics[width=0.5\linewidth]{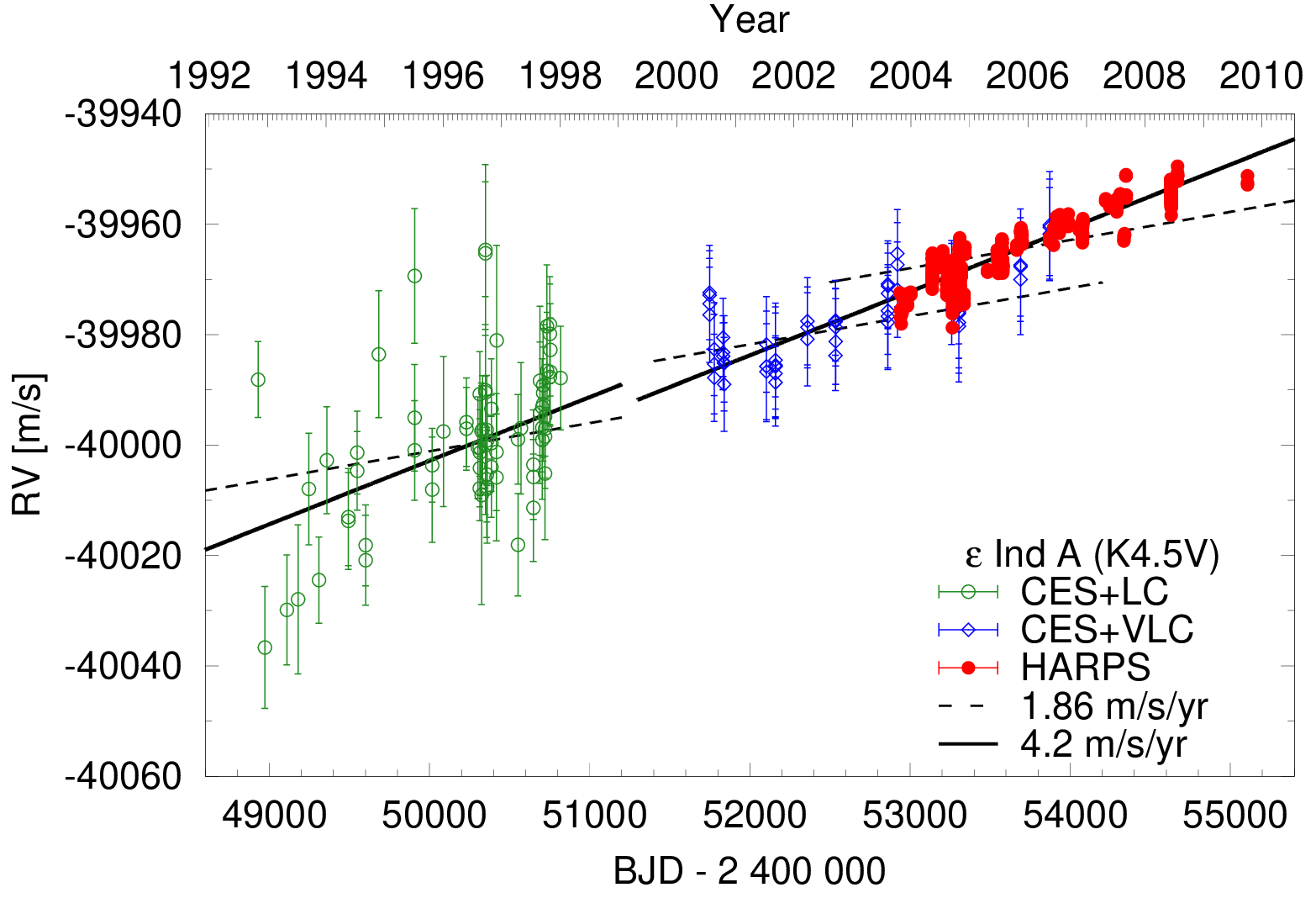}

\includegraphics[width=0.5\linewidth]{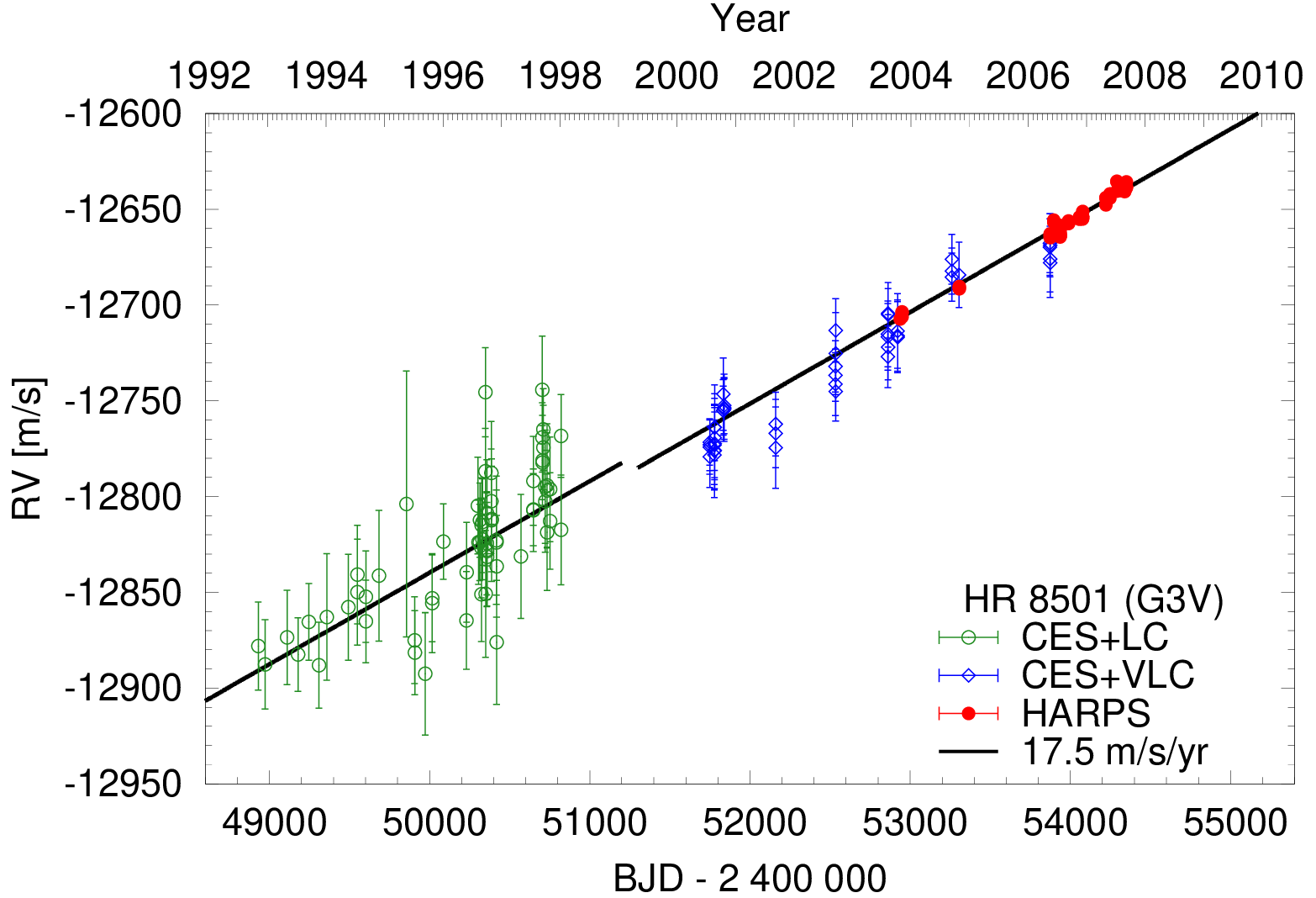}\includegraphics[width=0.5\linewidth]{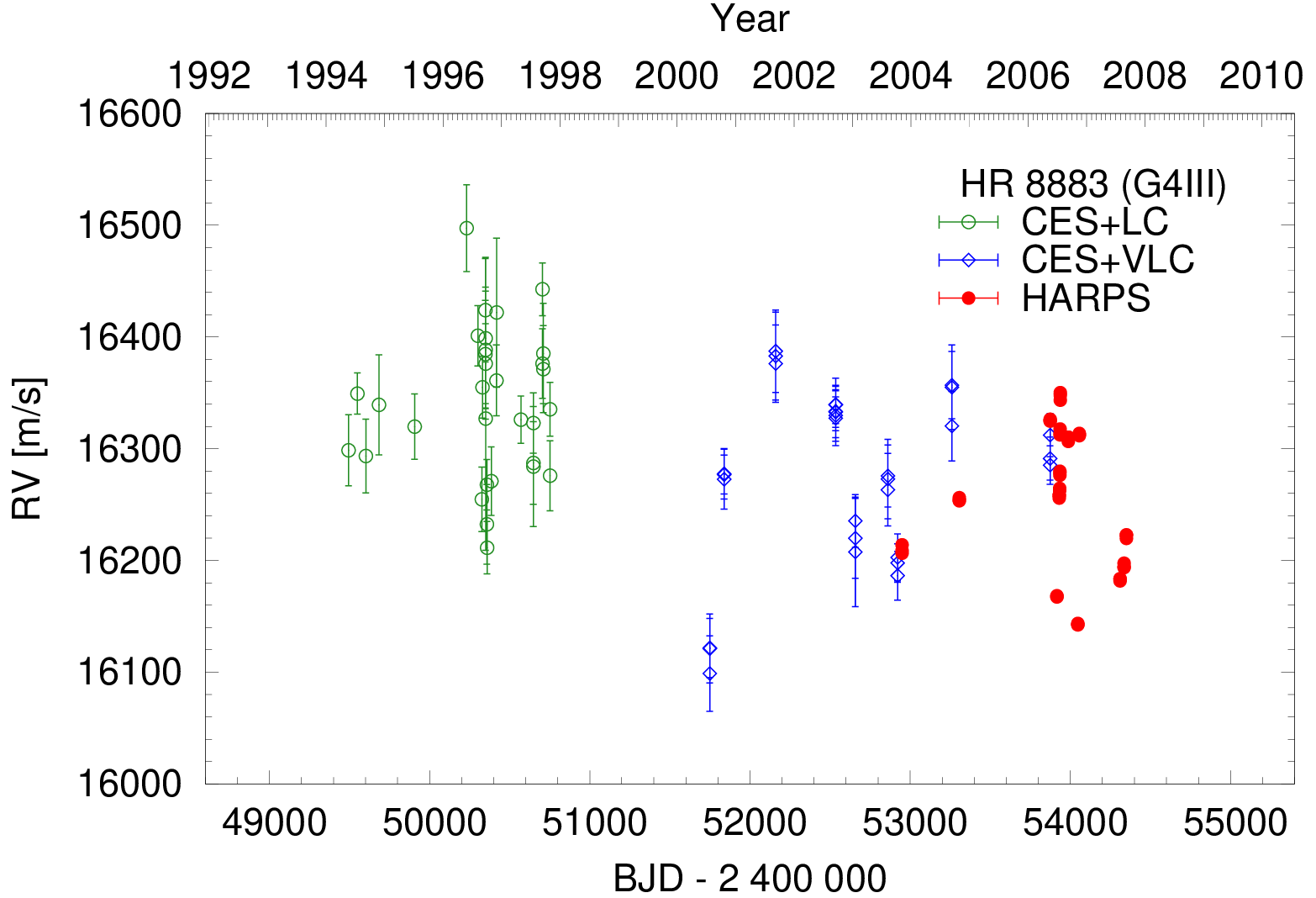}

\caption{\label{Fig:RV-5}Radial velocities. Continuation of Figs.~\ref{Fig:RV-1}--\ref{Fig:RV-4}.
Model curves are shown for HR~7703 (trend), HR~8323 (sinusoid),
$\epsilon$~Ind~A (constant, i.e. only secular acceleration of 1.86\,m/s/yr,
and trend), and HR~8501 (trend).}
\end{figure*}

In this section we discuss individually those stars that exhibit variability.
\begin{description}
\item [{$\beta$~Hyi:}] For $\beta$~Hyi \citet{Endl02} announced a
trend of 7\,m/s/yr with a remaining scatter of 19\,m/s. Here the
best common trend is only 1.79\,m/s/yr depicted with a black solid
line in Fig.~\ref{Fig:RV-1} (plus the secular acceleration of 0.86\,m/s).
The scatter around the HARPS data decreases to only 2.3\,m/s (see
Table~\ref{Tab:TestsCombinedSetsSingle}). However, the trend increases
the VLC scatter from 7.4\,m/s to 9.0\,m/s and the fitted LC-VLC
offset departs by 2.2$\sigma$ (-17\,m/s, Table~\ref{Tab:eta})
from the measured offset. Thus, it is unclear whether the trend is
steady.\\
Additionally, we plot the 4300\,d period tabulated for $\beta$~Hyi
(Table~\ref{Tab:TestsCombinedSets}) with a black dashed line in
Fig.~\ref{Fig:RV-1}. This period matches that of a Jupiter analogue,
while the amplitude of 6.5\,m/s would result in a formal minimum
mass of 0.56\,$M_{\mathrm{Jup}}$. Compared to the trend in the previous
section the fitted LC-VLC offset is less discrepant (-8.0\,m/s),
but the sine fit is less significant than trend and still not supported
by the VLC data, because their scatter increases from 7.4\,m/s to
8.9\,m/s. Moreover, $\log R'_{\mathrm{HK}}$ and FWHM correlate with
the RVs. Hence the long-term variations might be related to the magnetic
cycle.
\item [{HR~209:}] $\log R'_{\mathrm{HK}}$ and FWHM correlate with the
RVs. Hence the RV variations are related to stellar activity probably
induced by spots.
\item [{HR~506~(HD~10647):}] A planet candidate was presented by Mayor
et al.$^{\text{\ref{fn:Mayor03}}}$ based on CORALIE measurements.
\citet{Jones04} found also weak evidence for a similar signal with
the AAT, but did not exclude stellar activity as the cause. \citet{Butler06}
listed AAT RV data and derived orbital parameters.\\
For HR~506 we clearly recovered the long RV period in the period
analysis. Hence, we combine our observations with AAT data and CORALIE
data to fit the orbit. Because more cycles have been covered, our
combined solution gives a more precise period compared to the solutions
given by the other authors ($P=(1003\,\pm\,56)$\,d, $e=0.16\pm0.22$
and $P=(1040\,\pm\,37)$\,d, $e=0.18\,\pm\,0.08$, respectively).
For our three combined data sets an eccentric orbit does not fit much
better (Table~\ref{Tab:TestsCombinedSets}) and also in the solution
for the five combined data sets the eccentricity vanished. Therefore
a circular obit was fitted ($e$ and $\omega$ fixed to zero, Table~\ref{Tab:OrbitParamHR506},
Fig.~\ref{Fig:hr506}). The semi-axis and the companion minimum-mass
were derived by assuming a stellar mass of $1.17\pm0.1\, M_{\oplus}$
(Table~\ref{Tab:SampleInfo}).
\begin{figure*}
\centering

\includegraphics[width=1\linewidth]{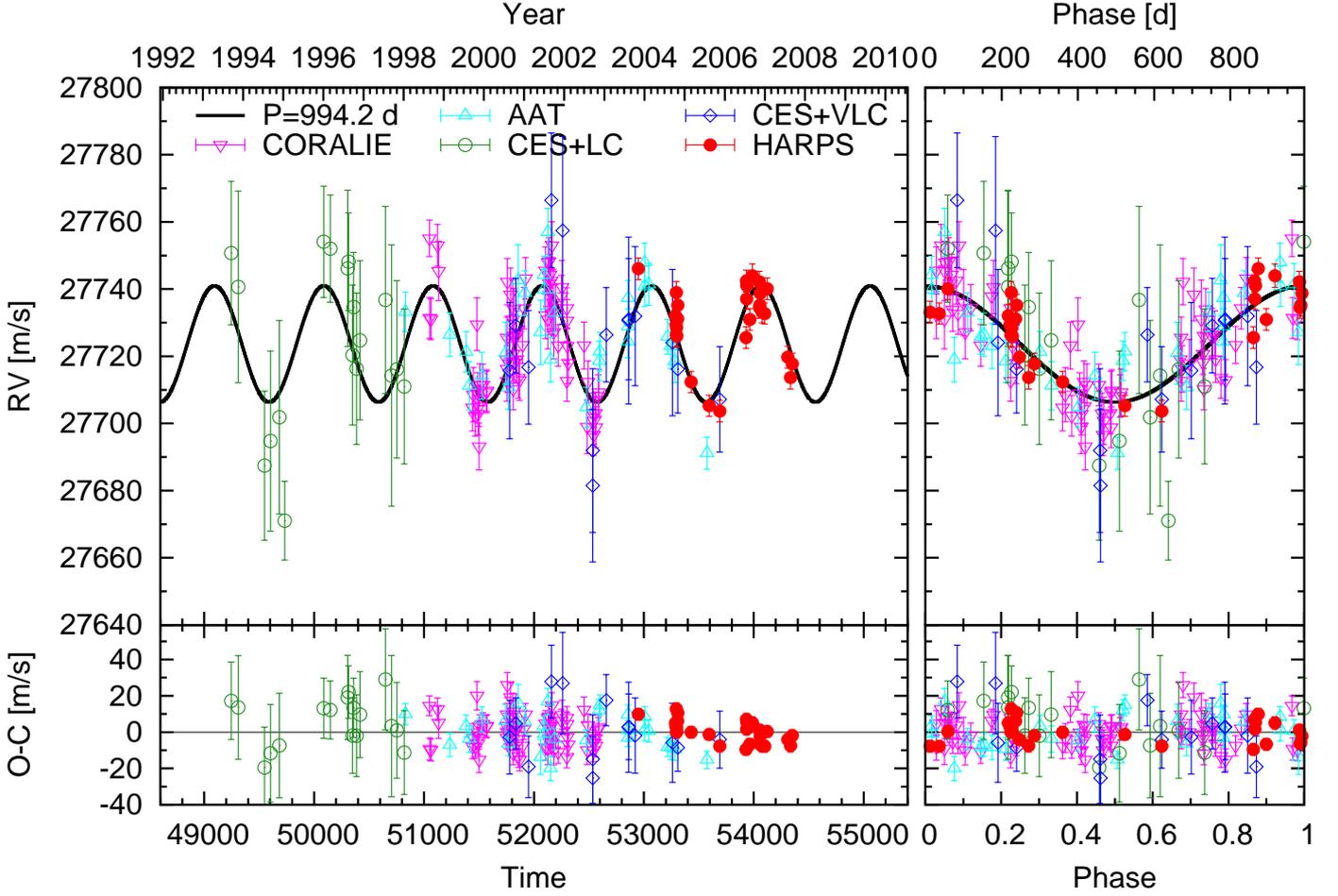}

\caption{\label{Fig:hr506}Left: RV time series for HR~506 combined with AAT
and CORALIE data. Right: RVs phase folded to the period of $P=995$\,d
and the residuals (bottom).}
\end{figure*}
\begin{table}
\caption{\label{Tab:OrbitParamHR506}Orbital parameters for the planetary companion
to HR~506.}

\centering{}
\begin{tabular}{llr@{}l@{}l}
\hline 
\hline Parameter &  & \multicolumn{3}{c}{Value}\\
\hline 
$P$ & {[}d{]} & 994.2 & \,$\pm$\, & 8.6\\
$K$ & {[}m/s{]} & 17.3 & \,$\pm$\, & 1.0\\
$T_{0}$ & {[}JD{]} & 2\,450\,088 & \,$\pm$\, & 25\\
$\omega$  & {[}$^{\circ}${]} & 0 & \multicolumn{2}{l}{(fixed)}\\
$e$ &  & 0 & \multicolumn{2}{l}{(fixed)}\\
$a$ & {[}AU{]} & 2.05 & \,$\pm$\, & 0.24\\
$M\sin i$ & {[}$M_{\mathrm{Jup}}${]} & 0.94 & \,$\pm$\, & 0.05\\
$N$ &  & 158 &  & \\
rms & {[}m/s{]} & 7.8 &  & \\
\hline 
\end{tabular}
\end{table}
\\
There is no clear RV-$\log R'_{\mathrm{HK}}$ correlation ($r=0.45$,
$\mathrm{FAP}=1.7\%$). However, when subtracting the 994\,d RV period,
there is a significant correlation ($r=0.63$, $\mathrm{FAP}=0.033\%$),
implying that the residuals of this active star are affected by stellar
activity. The RV-BIS and RV-FWHM correlations are not significant,
also not for the residual RVs. The RV residuals of HR~506 also have
some excess variability and a marginally significant trend of -2.41\,m/s/yr
which however increases the rms of the LC data (from 18.8\,m/s to
20.0\,m/s, Table~\ref{Tab:TestsCombinedSetsSingle}).
\item [{$\tau$~Cet:}] The small trend reduces the scatter from 1.37\,m/s
to just 1.33\,m/s and is just significant due to the large number
of observations. However, at this level an instrumental cause is likely
for the small trend and the RV-BIS and RV-FWHM correlation. In this
respect we also like to point out that the FWHM of $\tau$~Cet (but
also some others stars, e.g. $\zeta$~Tuc) exhibits a noticeable
positive long-term trend (Fig.~\ref{onlFig:taucet}) which might
be due to a drifting focus of HARPS. In this case, assuming a constant
line equivalent width, a negative trend is expected and indeed seen
in the contrast (depth) of the CCF (recently also noted by \citealp{GomesdaSilva2012}).
$\tau$~Cet has the smallest variations in $\log R'_{\mathrm{HK}}$
in the sample.
\item [{$\kappa$~For:}] The RVs of $\kappa$~For decline over the whole
time baseline of 14\,yr which indicates an orbital period longer
than the estimate of 21\,yr given in \citet{Endl02}. The Keplerian
period of 10700\,d listed in Table~\ref{Tab:TestsCombinedSets}
(29.3\,yr) is not well constrained. However, again with the slope
and Eq.~(\ref{Eq:TrendMass2}) this period might be used to assess
a minimum mass of 0.36\,$M_{\odot}$ for the companion. The RV residuals
do not exhibit significant variability.
\item [{$\iota$~Hor:}] For this active star \citet{Kuerster00} discovered
a planet. The signal was also seen by \citet{Naef01} using the CORALIE
spectrograph and by \citet{Butler2001} with the AAT. Using the HARPS
data of \citet{Vauclair08} taken for an asteroseismology campaign,
\citealt{Boisse2011} searched for short period companions ($P<7$\,d),
but detected no further planet.\\
In our analysis we recovered the planetary signal seen for $\iota$~Hor
by \citet{Kuerster00}. We refined the orbital parameters with our
data sets and included also the AAT \citep{Butler06} and CORALIE
data \citep{Naef01} as shown in Fig.~\ref{Fig:IotaHor}. For these
two data sets the jitter estimate (from Table~\ref{Tab:Jitter})
was added to the measurements error. Due to the activity of the star
the residuals have a high scatter. The orbital parameters are listed
in Table~\ref{Tab:OrbitParamIotaHor}. The semi-axis and the companion
minimum-mass were derived by assuming a stellar mass of $1.25\pm0.02\, M_{\oplus}$
(Table~\ref{Tab:SampleInfo}). An astrometric upper mass limit of
18.4\,$M_{\mathrm{Jup}}$ was placed by \citet{Reffert2011} in combination
with the orbital solution of \citet{Butler06}. The orbital period
of 307\,d is different from the rotation period of $\sim$8\,d (see
below), while a relation with the magnetic cycle has to be discussed.\\
\citet{Metcalfe2010} reported that a magnetic activity cycle of $\sim$1.6\,yr
bears no obvious relation to the orbital period, but we find that
there is a mode\-rately significant RV-$\log R'_{\mathrm{HK}}$ correlation
(Fig.~\ref{onlFig:iotahor}) and that the refined orbital period
(307\,d) is half of the formal best-fitting period (630\,d) for
the combined HARPS and SMARTS $S_{\mathrm{HK}}$-index mea\-sure\-ments
(Fig.~\ref{Fig:rhk_iotahor}; we use the $S_{\mathrm{HK}}$-index
instead of $\log R'_{\mathrm{HK}}$ for the comparison, since \citet{Metcalfe2010}
published $S_{\mathrm{HK}}$-index measurements). However, if the
307\,d RV period were be caused by a magnetic cycle, we would expect
the same period for the $S_{\mathrm{HK}}$-index and a positive correlation
with the BIS \citep{Lovis2011}, while there is a formal RV-BIS anti-correlation
probably artificially induced by the short-term variations (Fig.~\ref{Fig:bis_rv_iotahor}).
Therefore we tend to prefer the planet hypothesis as a more simple
and likely explanation. Moreover, after subtracting the 307\,d period,
the correlation with $\log R'_{\mathrm{HK}}$ remains and the anti-correlation
with BIS increases and becomes more significant.
\begin{figure*}
\centering

\includegraphics[width=1\linewidth]{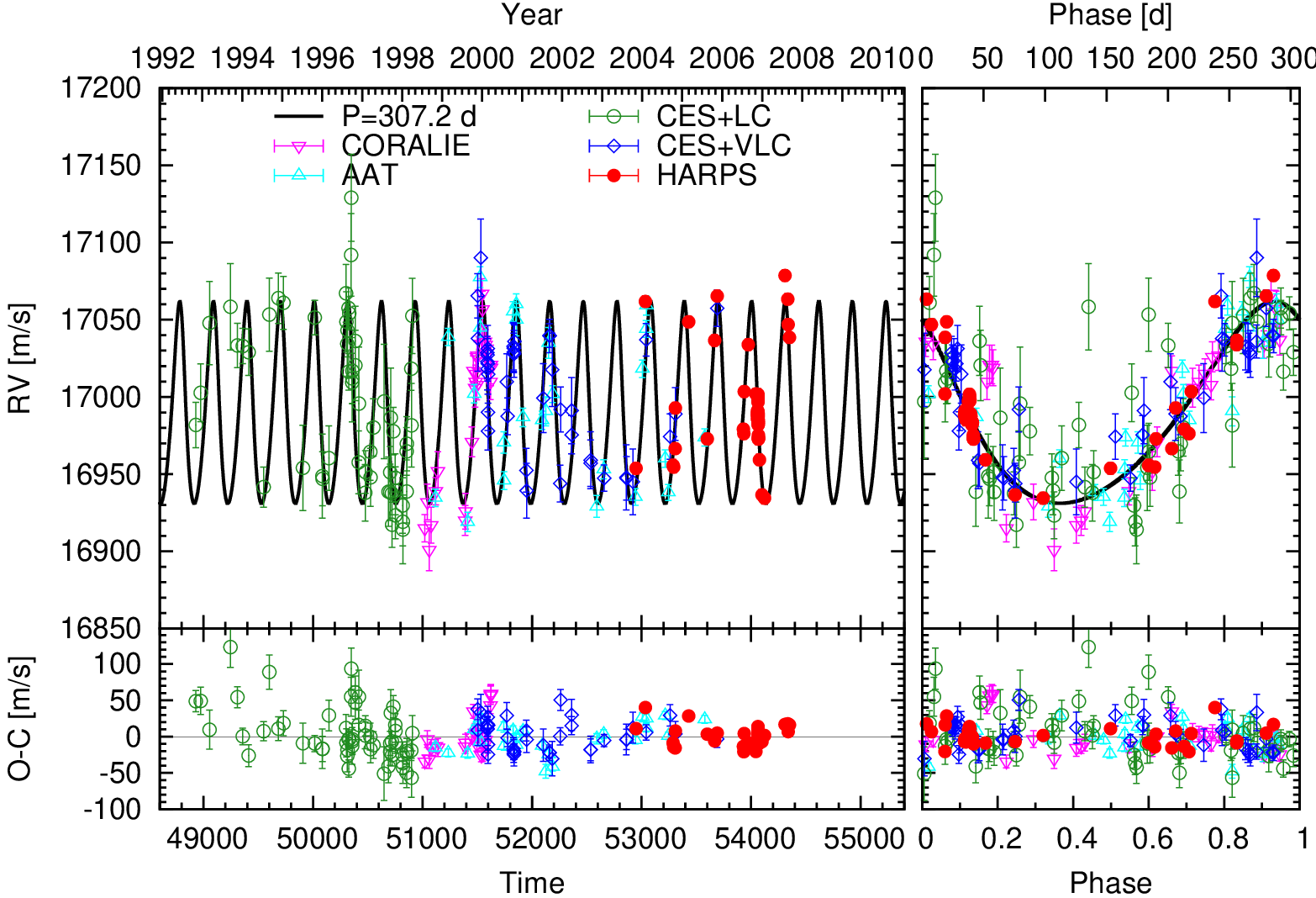}

\caption{\label{Fig:IotaHor}(Left) RV time series for $\iota$~Hor combined
with AAT and CORALIE data. (Right) RVs phase folded to the period
of $P=307$\,d and the residuals (bottom).}
\end{figure*}
\begin{table}
\caption{\label{Tab:OrbitParamIotaHor}Orbital parameters for the planetary
companion to $\iota$~Hor.}

\centering{}
\begin{tabular}{llr@{}l@{}l}
\hline 
\hline Parameter &  & \multicolumn{3}{c}{Value}\\
\hline 
$P$ & {[}d{]} & 307.2 & \,$\pm$\, & 0.3\\
$K$ & {[}m/s{]} & 65.3 & \,$\pm$\, & 2.2\\
$T_{0}$ & {[}JD{]} & 2\,449\,110 & \,$\pm$\, & 9\\
$\omega$  & {[}$^{\circ}${]} & 35 & \,$\pm$\, & 10\\
$e$ &  & 0.18 & \,$\pm$\, & 0.03\\
$a$ & {[}AU{]} & 0.96 & \,$\pm$\, & 0.05\\
$M\sin i$ & {[}$M_{\mathrm{Jup}}${]} & 2.48 & \,$\pm$\, & 0.08\\
$N$ &  & 205 &  & \multicolumn{1}{l}{}\\
rms & {[}m/s{]} & 14.5 & \\
\hline 
\end{tabular}
\end{table}
\\
The periodogram of the residuals for $\iota$~Hor (Fig.~\ref{Fig:GLS_iotahor})
shows power at periods of 5.7\,d as well as 7.94\,d and 8.45\,d
(also found by \citealt{Boisse2011} based on HARPS RVs). The latter
peak coincides with periodic variations (8.5\,d) found in the $S_{\mathrm{HK}}$
index by \citet{Metcalfe2010}, which therefore probably indicates
the rotation of this star. The residual RVs anti-correlate with BIS
which is indeed expected for rotating spots \citep{Boisse2011}.
\begin{figure}
\centering

\includegraphics[width=1\linewidth]{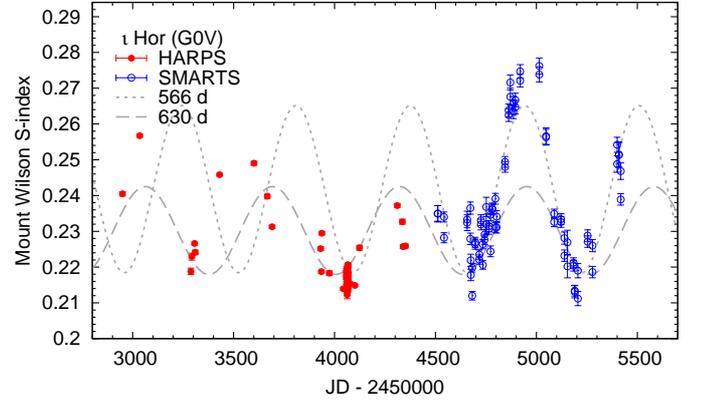}

\caption{\label{Fig:rhk_iotahor}S-index measurements for $\iota$~Hor. \citet{Metcalfe2010}
published the SMARTS data (blue open circles) and the 566\,d period
(dotted line). The formal best fit for the combined data set is 630\,
d (dashed line).}
\end{figure}
\begin{figure}
\centering

\includegraphics[width=1\linewidth]{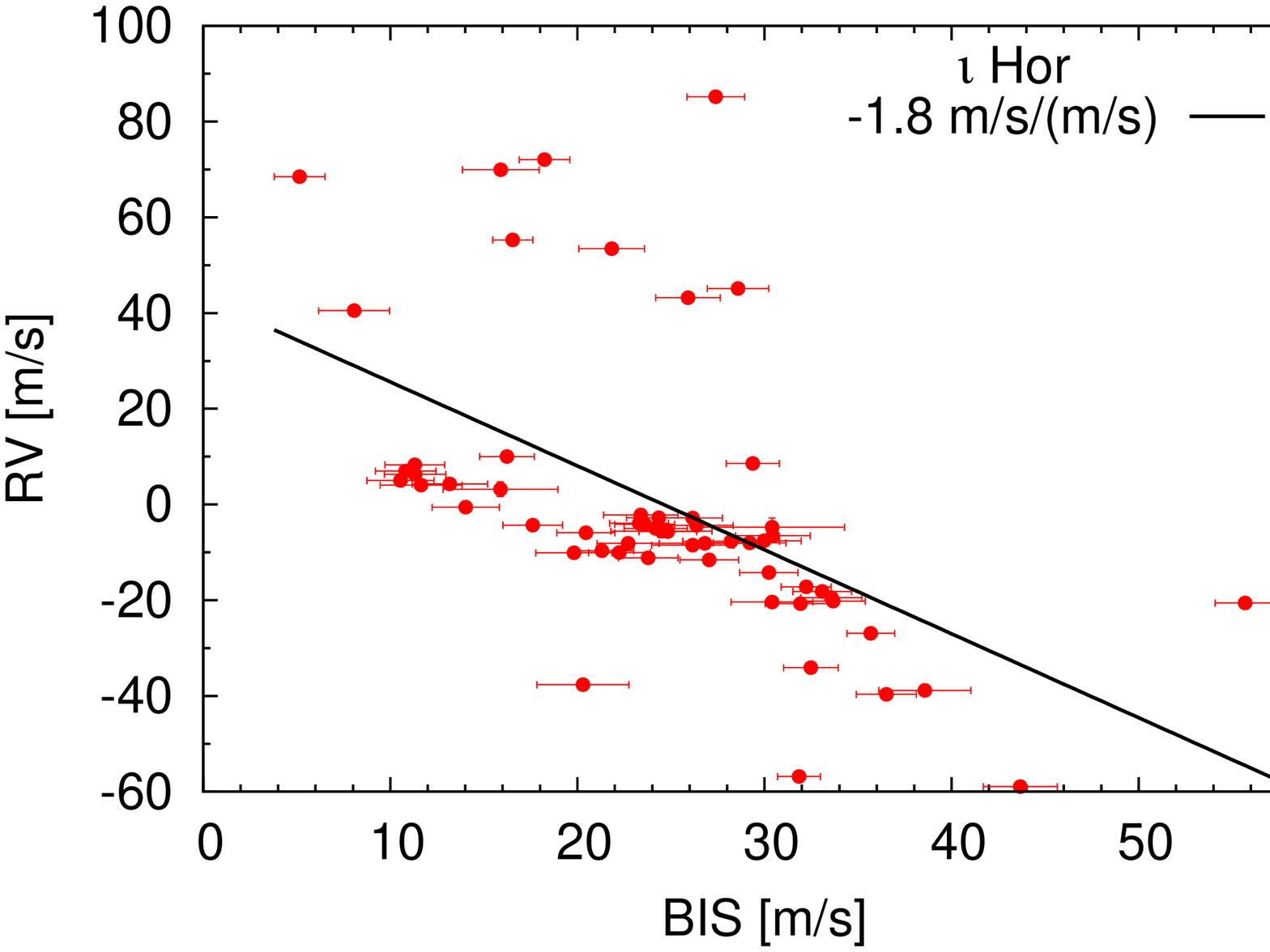}

\includegraphics[width=1\linewidth]{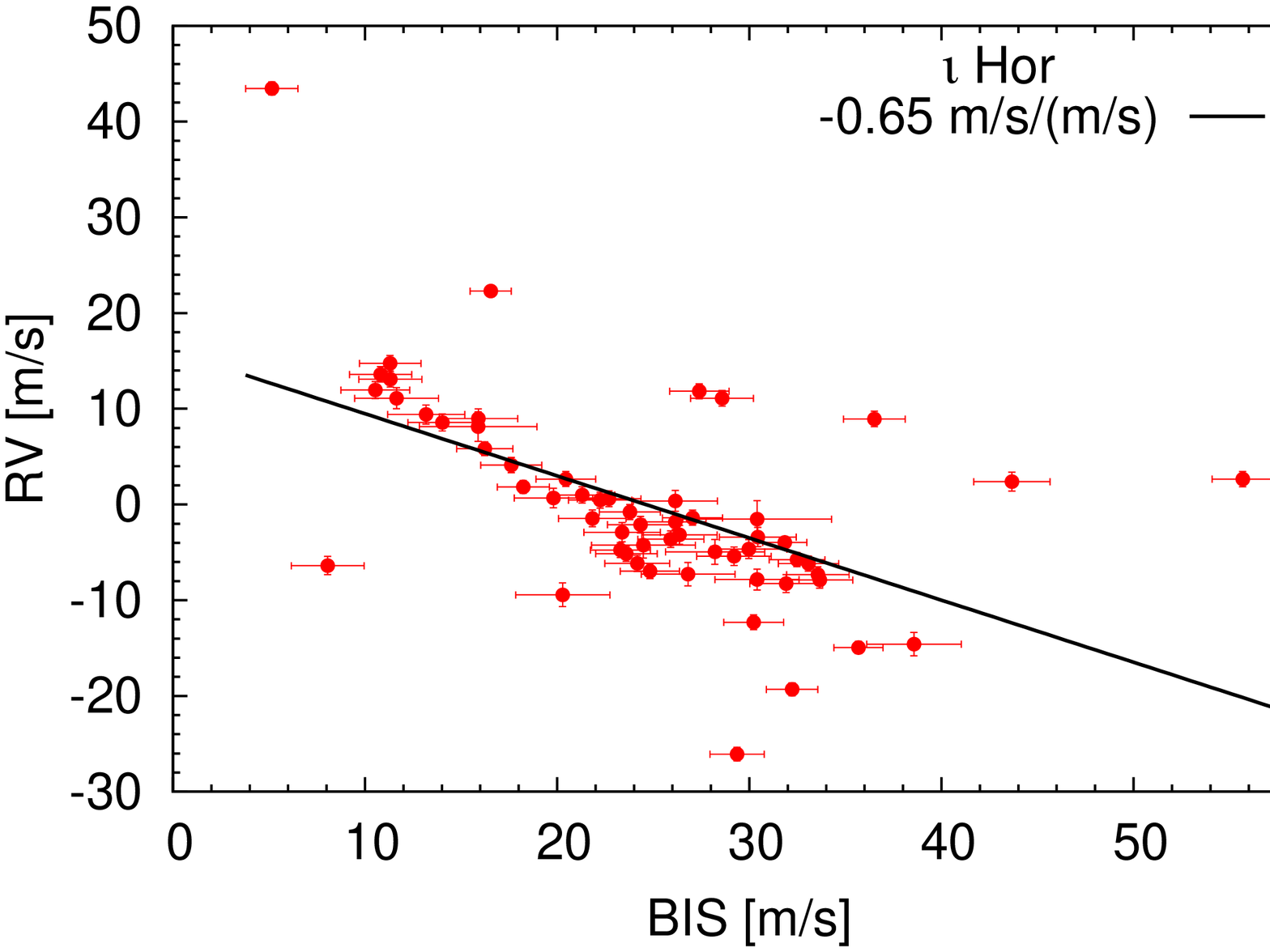}

\caption{\label{Fig:bis_rv_iotahor}RV vs. BIS for $\iota$~Hor. (Top) The
linear correlation coefficient is $r=-0.53$, but the correlation
appears more complex and loop-like. (Bottom) After subtracting the
307\,d period from the RVs, $r$ is $-0.56$.}
\end{figure}
\begin{figure}
\centering

\includegraphics[width=1\linewidth]{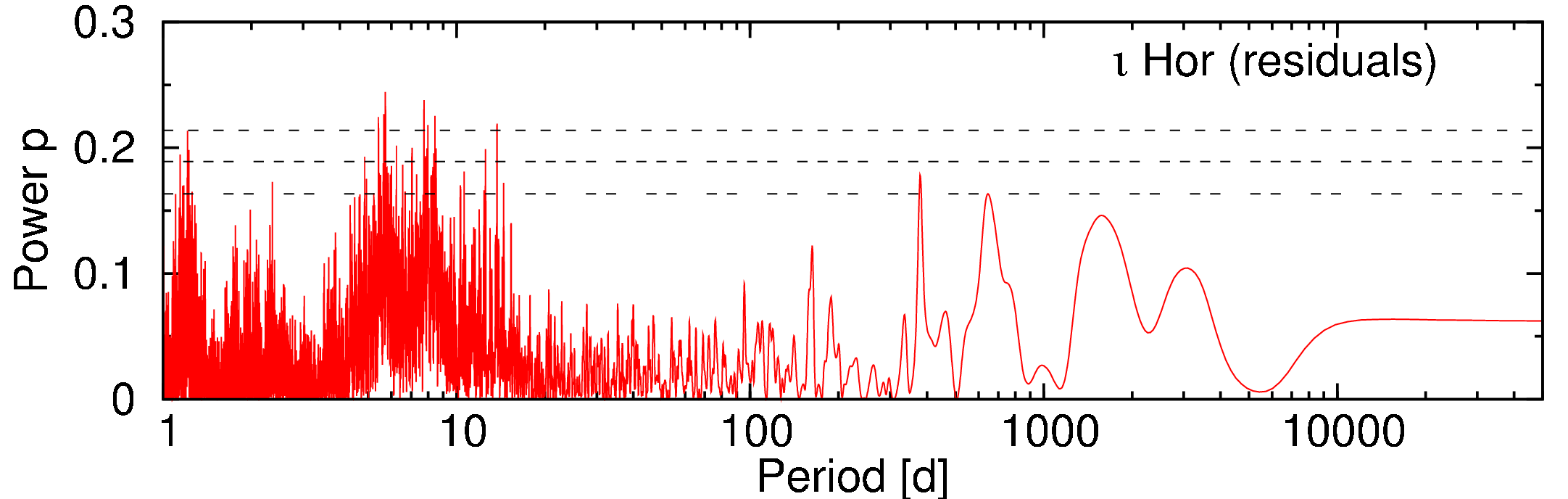}

\caption{\label{Fig:GLS_iotahor}GLS periodogram on the residuals for $\iota$~Hor.
In this and the subsequent figures (Figs. \ref{Fig:GLS_zet1ret}-\ref{Fig:GLS_hr3259})
the horizontal dashed lines depict FAP levels as in Fig. \ref{Fig:GLS_hr2667}.}
\end{figure}

\item [{$\alpha$~For:}] GJ~127~B can explain the trend of -11.5\,m/s
(Sect.~\ref{Sub:Long-term-trend}, Table~\ref{Tab:Binaries}, Fig.~\ref{Fig:RV-residuals}).
So far a trend is a sufficient model. The RV residuals do not exhibit
significant variability.
\item [{$\zeta^{1}$~Ret:}] The relatively large scatter of 12.5\,m/s
is indicative of excess variability and the periodogram of the RVs
shows peaks close to our FAP threshold at 21.5\,d and 426\,d (Fig.~\ref{Fig:GLS_zet1ret})
with the first giving the lowest for the combined rms value (but increasing
the rms of the VLC data). The strong RV-$\log R'_{\mathrm{HK}}$ (also
RV-FWHM) correlation shows that most of the scatter is due to stellar
activity, likely not only the magnetic cycle, since the observed correlation
slope $219\pm24$\,m/s/dex exceeds the predicted value$140\pm7$\,m/s/dex
the for a magnetic cycle (Sect.~\ref{sub:Correlations-with-Ca},
assuming $5746\pm27$\,K and $\mathrm{[Fe/H]}=-0.22\pm0.05$ from
\citealp{Peloso2000}). $r=0.88$ indicates that the subtraction of
this correlation will reduce the scatter by $\sqrt{1-r^{2}}$ to 5.9\,m/s.
$\zeta^{1}$~Ret has the largest $\log R'_{\mathrm{HK}}$-variations
and a quite high jitter estimate of 3.8 m/s within our sample.
\begin{figure}
\centering

\includegraphics[width=1\linewidth]{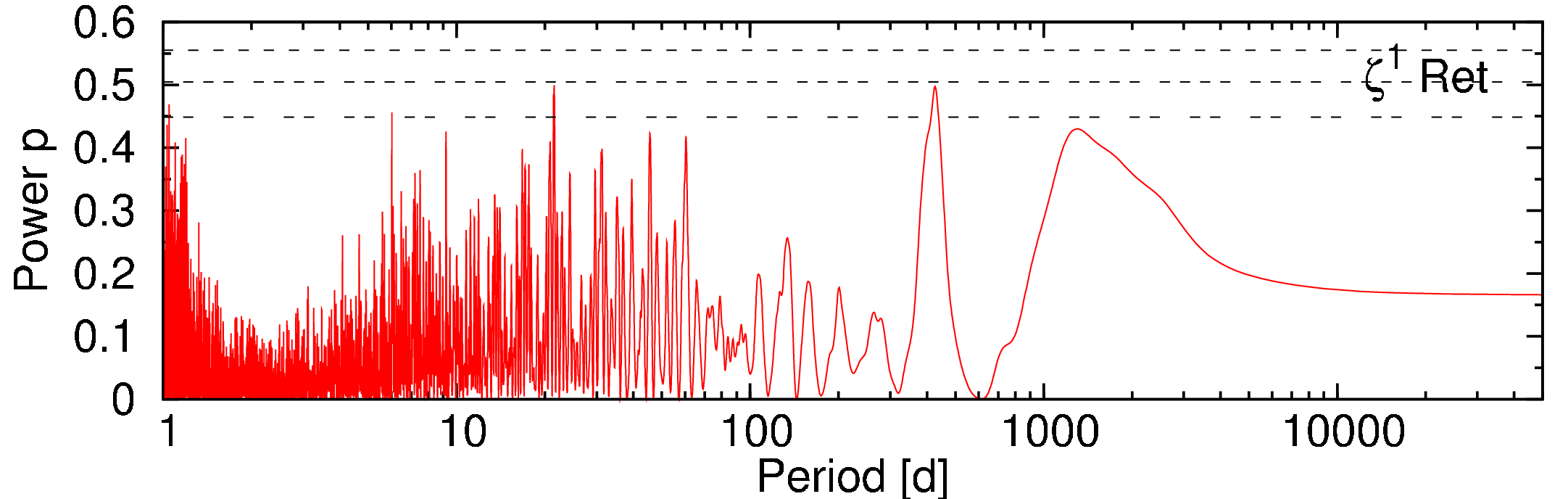}

\caption{\label{Fig:GLS_zet1ret}GLS periodogram for $\zeta^{1}$~Ret.}
\end{figure}

\item [{$\zeta^{2}$~Ret:}] Given the very similar stellar parameters
of the binary pair $\zeta^{1}$~Ret and $\zeta^{2}$~Ret, \citet{Peloso2000}
already pointed out the baffling fact of their very different activity
level. We do not find any significant RV excess variability for $\zeta^{2}$~Ret.
However, there is a significant RV-FHWM correlation and all three
activity indicators seem to exhibit a small trend, which might be
either due to a magnetic cycle or instrumental (cf. $\tau$~Cet). 
\item [{$\epsilon$~Eri:}] \citet{Hatzes00} announced a planet in an
eccentric orbit ($e=0.6$) around this active star. \citet{Benedict2006}
refined the orbital solution and combined the RVs with astrometric
observations with the HST Fine Guidance Sensor indicating an orbital
inclination of $i=30^{\circ}$. Likewise, \citet{Reffert2011} derived
a similar value for the inclination and an upper mass limit of 6.1\,$M_{\mathrm{Jup}}$
using Hipparcos astrometry and the RV orbit solution.\\
However, in the periodogram for $\epsilon$~Eri (Fig.~\ref{Fig:GLS_epseri})
we cannot find any evidence for the long-period planet ($P$=2500\,d)
suggested by \citet{Hatzes00} whose orbital solution is plotted in
Fig.~\ref{Fig:RV-2} for comparison. Due to the higher precision
of our data, the combined rms is 8.2\,m/s for fitting a constant,
while \citet{Hatzes00} list an rms ranging from 11 to 22\,m/s for
different instruments in the residuals of their Keplerian model. $\epsilon$~Eri
is an active star and all three activity indicators are variable,
and the BIS shows an anti-correlation with RV. The stellar rotation
period is 11.2\,d \citep{Donahue1996,Froehlich2007}. On a short
time scale of 85\,min its variability is 0.86\,m/s (Table~\ref{Tab:Jitter}),
while the long term jitter estimate is 3.6\,m/s (Table~\ref{Tab:Jitter}).
Our best-fitting sine function has a period of 3.11\,d, reduces the
scatter from 8.2\,m/s to only 6.9\,m/s, but is not signifi\-cant.
We note that, combining the RVs of HARPS and all the RVs given in
\citet{Benedict2006}, \citet{Anglada-Escude2012} could also not
confirm the planet solution, though, without evaluating the significance,
they suggested another best fitting, similar long-period. 
\begin{figure}
\centering

\includegraphics[width=1\linewidth]{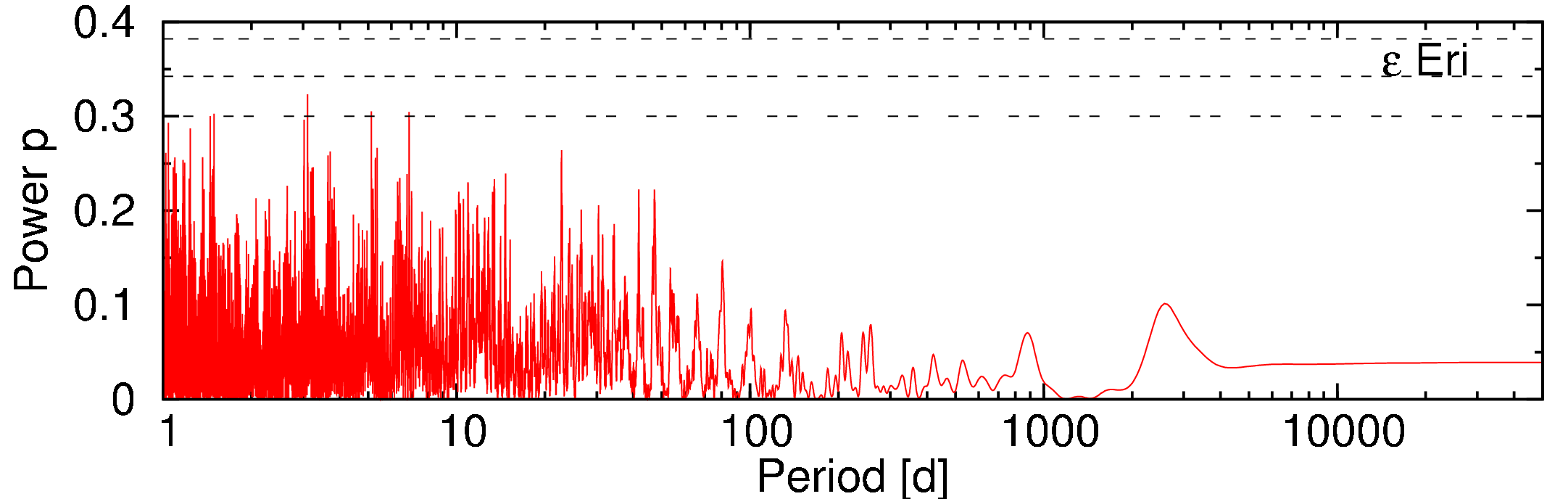}

\includegraphics[width=1\linewidth]{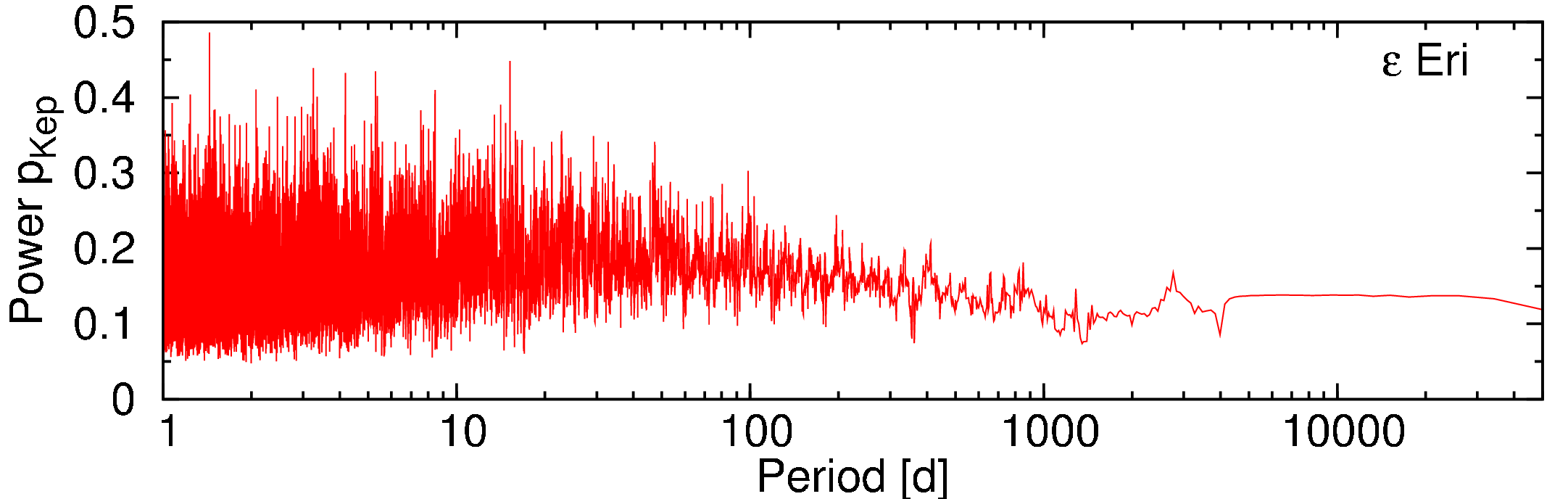}

\caption{\label{Fig:GLS_epseri}GLS (top) and Keplerian (bottom) periodogram
for $\epsilon$~Eri. There is no significant power at the period
of the putative planet $\epsilon$~Eri~b ($P=2500$\,d).}
\end{figure}

\item [{$\alpha$~Men:}] \citet{Eggenberger07} reported a companion for
$\alpha$~Men which is likely an M3.5\,\textendash{}\,M6.5 dwarf
that was seen at a separation of 3.05'' with the VLT+NACO. They also
mentioned (but did not quantify) a radial velocity drift for $\alpha$~Men
seen with the CORALIE spectrograph and suggested that the companion
causes this trend.\\
In our data however, the star has no significant excess variability
($\mathrm{rms}=3.1$\,m/s) or trend. Therefore we cannot confirm
an RV trend for $\alpha$~Men, whose value should be unchanged over
years if induced by an M dwarf with a separation of 3\arcsec. However,
since no value for the trend was given by \citet{Eggenberger07},
a quantitative comparison cannot be made.\\
Moreover, the RVs correlate with all three activity indicators. The
observed RV-$\log R'_{\mathrm{HK}}$ slope of $57\pm10$\,m/s/dex
(Table~\ref{Tab:rhk}) agrees with the predicted sensitivity of $C_{RV}=69\pm4$\,m/s/dex
(assuming $T_{\mathrm{eff}}=5594\pm36$\,K and $\mathrm{[Fe/H]}=0.10\pm0.05$
from \citealp{Santos2004}) indicating a magnetic cycle as cause of
the RV variations.
\item [{HR~2400:}] The orbital solution for HR~2400 is given in Table~\ref{Tab:OrbitParamHR2400}.
Assuming a stellar mass of $1.2\pm0.1\, M_{\oplus}$ (Table~\ref{Tab:SampleInfo})
the companion has a minimum mass of 0.17\,$M_{\mathrm{\odot}}$ indicating
an M~dwarf at a separation of 9.6\,AU (0\farcs26). Hence, HR~2400
is a single-lined spectroscopic binary (SB1). Figure~\ref{Fig:RV-residuals}
displays the RV residuals for HR~2400 which do not have a significant
variability. In the FWHM there seems to be a trend which might be
due to stellar activity or spectral contamination by the companion.
In the residuals the RV-FWHM correlation vanished.\\
\begin{table}
\caption{\label{Tab:OrbitParamHR2400}Orbital parameters for the companion
to HR~2400.}

\centering{}
\begin{tabular}{llr@{}l@{}l}
\hline 
\hline Parameter &  & \multicolumn{3}{c}{Value}\\
\hline 
$P$ & {[}d{]} & 9346 & \,$\pm$\, & 554\\
$K$ & {[}m/s{]} & 1717 & \,$\pm$\, & 83\\
$T_{0}$ & {[}JD{]} & 2\,451\,881 & \,$\pm$\, & 16\\
$\omega$  & {[}$^{\circ}${]} & 279 & \,$\pm$\, & 1\\
$e$ &  & 0.58 & \,$\pm$\, & 0.01\\
$a$ & {[}AU{]} & 9.6 & \,$\pm$\, & 1.1\\
$M\sin i$ & {[}$M_{\odot}${]} & 0.17 & \,$\pm$\, & 0.01\\
$N$ &  & 77 &  & \\
rms & {[}m/s{]} & 4.8 &  & \\
\hline 
\end{tabular}
\end{table}

\item [{HR~3259:}] For HR~3259 \citet{Lovis06} reported a planetary
system with three Neptune-mass planets with periods (amplitudes) of
8.667\,d (3.51\,m/s), 31.56\,d (2.66\,m/s), and 197\,d (2.20\,m/s).
All three periods are seen in the periodogram (Fig.~\ref{Fig:GLS_hr3259})
and the corresponding peak heights are in the same order as suggested
by the given amplitudes. We refer to \citet{Anglada-Escude2012} for
a recent and detailed analysis of the HARPS data and updated orbital
parameters of the short period planets. We only note that with our
own and archive (but without GTO) HARPS data alone we also have seen
excess power at these periods, although these peaks did not reach
our FAP threshold due to too few observations (only 18 binned HARPS
data points) and considerable spectral leakage (aliasing). 
\begin{figure}
\centering

\includegraphics[width=1\linewidth]{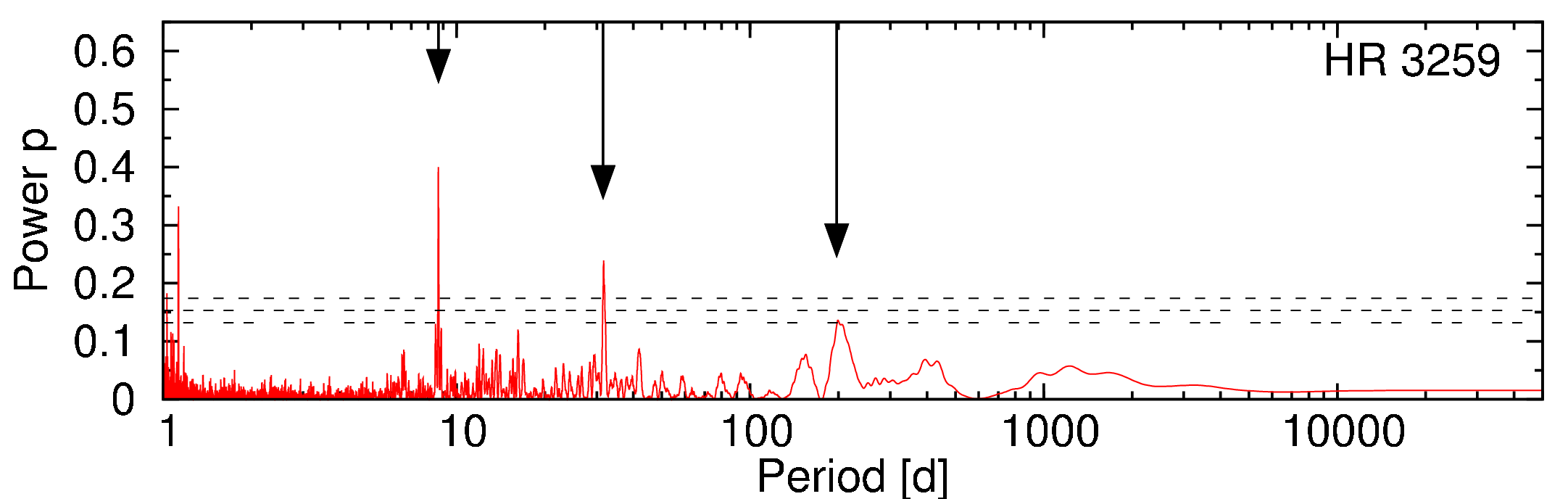}

\caption{\label{Fig:GLS_hr3259}GLS periodogram for HR~3259. The vertical
arrows indicate the periods for the three planets announced by \citet{Lovis06}.}
\end{figure}

\item [{HR~3677:}] The orbit of the companion to the giant HR~3677 is
eccentric indicated by the much lower residuals when fitting a Keplerian
orbit (8.7\,m/s) compared to a circular orbit (35\,m/s). Using the
slope and Eq.~(\ref{Eq:TrendMass2}) a raw estimate for the minimum
companion mass is 0.69\,$M_{\odot}$. The significant excess variability
of the RV residuals (Fig.~\ref{Fig:RV-residuals}) is likely explained
by an underestimated jitter of this giant. The RV-FWHM correlation
disappears in the residuals.
\item [{HR~4523:}] \citet{Tinney2011} reported a Neptune-like planet
with 16\,$M_{\oplus}$ minimum mass in a 122\,d orbit.\\
However, the periodogram of our data shows no power at periods of
122\,d. Therefore we cannot confirm the planet reported by \citet{Tinney2011}.
Their 145 measurements over 7\,yr with UCLES have a scatter of $\mathrm{rms}=3.0$\,m/s.
For comparison, the 62 HARPS RVs over 4.4\,yr have also $\mathrm{rms}=2.9$\,m/s.
Figure~\ref{Fig:UCLES_hr4523} shows the RV data phase folded to
the period of the proposed planet. When subtracting the proposed orbital
solution, the scatter decreases to 2.56\,m/s for UCLES, but increases
to 3.29\,m/s for HARPS. 
\begin{figure}
\centering

\includegraphics[width=1\linewidth]{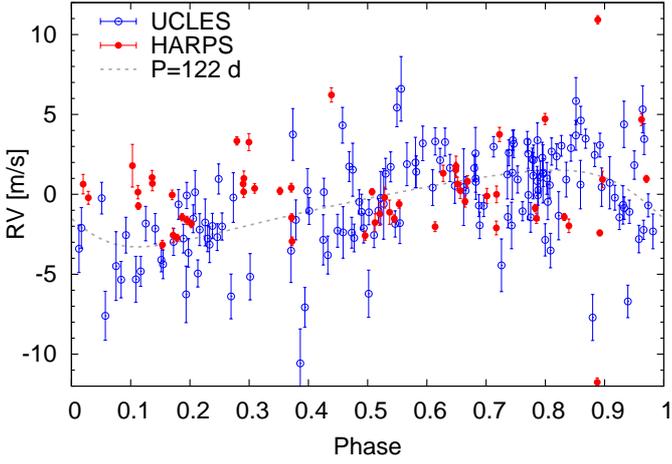}

\caption{\label{Fig:UCLES_hr4523}RV data for HR~4523 phase folded to 122.1\,d.
The dashed curve is an orbital solution proposed by \citet{Tinney2011}
based on the UCLES data. HARPS RVs are shown for comparison.}
\end{figure}

\item [{HR~4979:}] Most of the RV variations is likely caused by a magnetic
cycle, because the RV correlates with $\log R'_{\mathrm{HK}}$ and
BIS. Moreover, the observed correlation slope is $81\pm12$\,m/s/dex
which is of the order of the value expected for magnetic cycles ($65\pm2$\,m/s/dex;
Sect. \ref{sub:Correlations-with-Ca}, assuming $T_{\mathrm{eff}}=5729\pm17$\,K
and $\mathrm{[Fe/H]}=0.19\pm0.01$ from \citealp{Sousa2008}).
\item [{$\alpha$~Cen~A+B:}] We refer to \citet{Pourbaix2002} for the
orbital solution for $\alpha$~Cen~A+B with a period of 80\,yr.
We would just like to point out that the fitted trends in Sect.~\ref{Sub:Long-term-trend},
although obviously not a sufficient model, can be interpreted as a
mean acceleration and that the ratio of these slopes is a measure
of the mass ratio $\frac{M_{\mathrm{A}}}{M_{\mathrm{B}}}=-\frac{\dot{RV}_{\mathrm{B}}}{\dot{RV}_{\mathrm{A}}}=\frac{154.4\,\mathrm{m/s/yr}}{130.9\,\mathrm{m/s/yr}}=1.180$
which agrees with the value of $\frac{M_{\mathrm{A}}}{M_{\mathrm{B}}}=\frac{1.105\, M_{\odot}}{0.934\, M_{\odot}}=1.183$
derived from \citet{Pourbaix2002}.\\
Neither the residuals of $\alpha$~Cen~A nor $\alpha$~Cen~B exhibit
a significant variability. However, using HARPS measurements and a
very complex analysis, \citet{Dumusque2012} have recently announced
a planet candidate for $\alpha$~Cen~B with a very small amplitude
(0.51\,m/s, 3.236\,d). Due to the lower number of the (complementary)
HARPS measurements in this work (21 vs. 459) we are not sensitive
to this planet.\\
The RV-$\log R'_{\mathrm{HK}}$ and RV-BIS correlations listed for
$\alpha$~Cen~B are only formally significant, since obviously the
RV trend is caused by $\alpha$~Cen~A and not by a stellar activity,
and vanishes after subtracting the orbit. The trend seen in $\log R'_{\mathrm{HK}}$
is part of a longer magnetic cycle which be seen more clearly when
adding recently published HARPS data by \citet{Dumusque2012}. We
estimate a period of $P\sim2420$\,d (Fig.~\ref{Fig:alpcenb_rhk})
for the magnetic cycle, but a true periodicity is not secured, since
only one cycle is covered and e.g. \citet{Buccino2008} reported a
period of 3061\,d with a FAP of 24\%.\\
\begin{figure}
\centering

\includegraphics[width=1\linewidth]{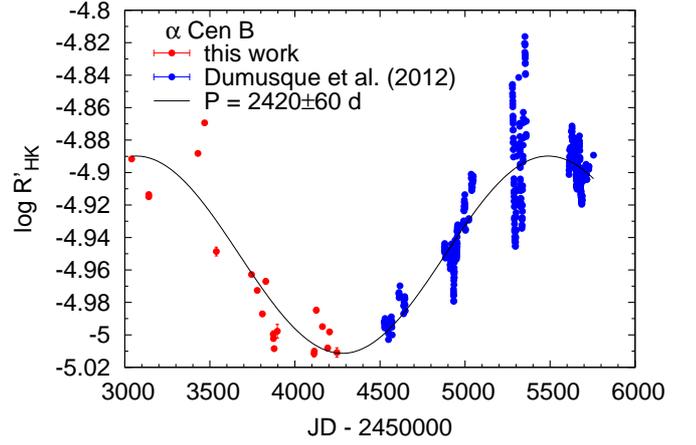}

\caption{\label{Fig:alpcenb_rhk}The time behaviour of $\log R'_{\mathrm{HK}}$
that shows the magnetic cycle of $\alpha$~Cen~B.}
\end{figure}

\item [{GJ~570~A:}] The trend has a smaller FAP compared to sinusoidal
and Keplerian model and is therefore a sufficient model for the RV.
GJ~570~BC can explain the trend of -2.7\,m/s (Sect.~\ref{Sub:Long-term-trend},
Table~\ref{Tab:Binaries}). Also, there is a marginally significant
li\-near correlation of the RVs with BIS and FWHM and therefore stellar
activity could contribute to the trend as well.
\item [{HR~6416:}] GJ~666~B can explain the trend of 9.4\,m/s (Sect.~\ref{Sub:Long-term-trend},
Table~\ref{Tab:Binaries}). A trend is a sufficient model. The residuals
do not exhibit significant excess variability.
\item [{HR~7703:}] A trend is a sufficient model and the value of 3.6\,m/s
(Sect.~\ref{Sub:Long-term-trend}, Table~\ref{Tab:Binaries}) can
be explained by GJ~783~B. Therefore the formally significant RV-$\log R'_{\mathrm{HK}}$correlation
should be artificial, however an additional (positive or negative)
contribution to the trend by activity cannot be excluded. After removing
the trend, there is no significant correlation or excess variability.
\item [{HR~8323:}] The listed significant period of 1372\,d is caused
by a stellar magnetic cycle. All three activity indicators show si\-mi\-lar
variations and strong correlations with RVs (Fig.~\ref{onlFig:hr8323},
Table~\ref{Tab:rhk} and \ref{Tab:fwhm}). The observed slope of
$98\pm10$\,m/s/dex (Table~\ref{Tab:rhk}) agrees well with the
predicted RV-$\log R'_{\mathrm{HK}}$ sensitivity of $C_{RV}=108\pm2$\,m/s/dex
for a magnetic cycle (assuming $T_{\mathrm{eff}}=5937\pm13$\,K and
$\mathrm{[Fe/H]}=0.00\pm0.01$ from \citealp{Sousa2008}).
\item [{$\epsilon$~Ind~A:}] The trend for $\epsilon$~Ind~A originally
announced by \citet{Endl02} is probably of highest interest, since
it might be caused by a planet. The wide binary brown dwarf companion
$\epsilon$~Ind~B ($M_{Ba}+M_{Bb}=112\, M_{\mathrm{Jup}}$, \citealt{King2010})
is too far from $\epsilon$~Ind~A (Table~\ref{Tab:Binaries}, Sect~\ref{Sub:Long-term-trend}).
It can induce only an acceleration of 0.009\,m/s/yr and hence cannot
explain the measured trend. The common slope of 2.4\,m/s/yr is predominantly
based on the more precise HARPS data. This trend improves the individual
rms of the LC and VLC only slightly (see Table~\ref{Tab:TestsCombinedSetsSingle}),
however the rms for both data sets is close to or below the stated
precision. The scatter of the VLC data around the trend is even only
4.4\,m/s, making it the most precise star in the VLC data set%
\footnote{$\epsilon$~Ind~A has the spectral type K which has numerous stellar
lines (high quality factor $Q$, \citealt{Bouchy01}) allowing more
precise RV measurements.%
}. In the overall picture the trend seems to be present for the whole
observation. The fit places the offset between LC and VLC at -4.0\,m/s
(Table~\ref{Tab:eta}, visible as the small jump in the fit in Fig.~\ref{Fig:RV-5},
$\eta=0.52$), i.e. is consistent with our physically estimated offset.\\
With in our sample, $\epsilon$~Ind~A exhibits comparably large
variations in $\log R'_{\mathrm{HK}}$ ($\mathrm{rms}_{\mathrm{RHK}}=0.032$\,dex,
Table~\ref{Tab:rhk}). Since both the RVs and $\log R'_{\mathrm{HK}}$
(as well as FWHM) values have a long-term trend during the HARPS observations
(Fig.~\ref{onlFig:epsind}), this results in a significant correlation
($r=0.60$). However, both trends could coincide just by chance temporarily,
so that it is still not clear whether they are associated with the
same long-term period (if any). Note also that the last $\log R'_{\mathrm{HK}}$
observations seem to depart from a steady trend in contrast to the
RVs.
\item [{HR~8501:}] A trend is a sufficient model and GJ~853~B can explain
its value of 17.3\,m/s (Sect.~\ref{Sub:Long-term-trend}, Table~\ref{Tab:Binaries}).
The RV-BIS correlation should be artificial and vanishes in the residuals
which also do not exhibit significant variability.
\item [{HR~8883:}] The period of 7.6\,d found for the giant HR~8883
improves only the HARPS rms, but not the LC and VLC rms (Table~\ref{Tab:TestsCombinedSetsSingle}).
\citet{Endl02} found strong Ca II H\&K emission in a FEROS spectrum
(cf. $S_{\mathrm{HK}}=0.533$ in Table~\ref{Tab:Jitter}) and mentioned
the high X-ray luminosity. Furthermore, the star is indicated in the
Hipparcos catalogue \citep{Perryman1997} as a photometrically variable
star (variability flag 'U', i.e. unresolved variable, e.g. irregular
or semi-irregular variables). Finally, the large variation of the
bisector span and the striking anti-correlation ($r=-0.96$) with
the RVs as seen in the HARPS data (Fig.~\ref{Fig:bis_hr8883}) reveal
that the variations and periods of this star are intrinsic to the
giant (e.g. rotation, activity, or solar-like oscillations%
\footnote{With a mass estimate of 2.1\,$M_{\odot}$ (Table~\ref{Tab:SampleInfo})
and a luminosity estimate of $L=60\, L_{\odot}$ from visual magnitude
V, distance (Table~\ref{Tab:SampleInfo}), and bolometric correction
-0.34\,mag (\citealp{Cox2000}), we obtain with scaling relations
for solar-like oscillations \citep{Kjeldsen1995} an amplitude of
$v_{\mathrm{osc}}=0.234\,\mathrm{m/s}\cdot\frac{L/L_{\odot}}{M/M_{\odot}}\sim7$\,m/s.%
}). The RVs do not correlate with $\log R'_{\mathrm{HK}}$ or FWHM
(Table~\ref{Tab:rhk} and \ref{Tab:fwhm}, Fig.~\ref{onlFig:hr8883}).
\end{description}
\begin{figure}
\centering

\includegraphics[width=1\linewidth]{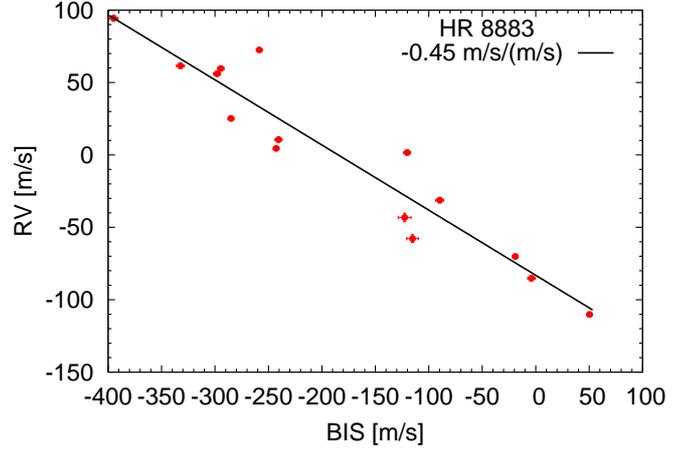}

\caption{\label{Fig:bis_hr8883}Correlation between bisector span (BIS) and
RV for the giant HR~8883.}
\end{figure}

\section{Summary and Conclusion}

Our sample consists of three planet hosting stars. These planets are
already known and we are able to trace their periodic RV signals allowing
confirmation of their reality. The planet hosting stars are $\iota$~Hor,
discovered by this survey \citep{Kuerster00}, as well as HR~506
(Mayor et al.$^{\text{\ref{fn:Mayor03}}}$) and HR~3259 \citep{Lovis06}
which even has a system of three planets. However, we have no clear
additional planet detection in our sample, while some promising or
controversial cases like $\beta$~Hyi, $\epsilon$~Ind~A, or $\epsilon$~Eri
call for follow-up observations.

Compared to our previous results in \citet{Endl02}, we confirm that:
\begin{itemize}
\item $\kappa$~For, HR 2400, and HR 3677 are SB1 spectroscopic binaries
with long periods.
\item $\alpha$~For, HR 6416, and HR 8501 exhibits trends consistent with
their wide stellar companions.
\item $\epsilon$~Ind~A has a steady long-term trend still explained by
a planetary companion.
\end{itemize}
Moreover, we see trends in HR~7703 and GJ~570~A which can also
be explained by the known wide stellar companions listed in Table~\ref{Tab:Binaries}.
However, the RV trend of $7$\,m/s/yr reported for $\beta$~Hyi
\citep{Endl02} is only 1.8\,m/s/yr in the combined data set and
is not seen in the VLC data. We cannot confirm a linear drift reported
by \citet{Eggenberger07} for $\alpha$~Men, neither can we confirm
the planet HR~4523~b announced by \citet{Tinney2011}, nor the planet
$\epsilon$~Eri~b reported by \citet{Hatzes00}. For $\epsilon$~Eri~b,
the astro\-me\-tric results \citep{Benedict2006,Reffert2011} seem
to confirm the planet, but they do not constitute independent detections,
since they rely on the combination with less precise RV measurements.
Also the astrometric phase coverage is not complete. However, the
same is true for our HARPS measurements. Hence, this planet is not
yet fully disproved with our RV non-detection or by imaging non-detections
(e.g. \citealp{Janson2007,Janson2008}). Furthermore, there are also
discussions whether an observed inner disk of warm dust can coexist
with this inner planet \citep{Reidemeister2011}.

It has to be pointed out that stellar magnetic cycles (the solar cycle
length is 2$\times$\,11\,yr) may cause long-term RV variations
which might be mistaken as companions. The CaII~H\&K lines are an
appropriate indicator to check this \citep{Santos2010,Dumusque2010,Lovis2011}.
Indeed, some stars, like e.g. HR~209, $\zeta^{1}$~Ret, and HR~8323
exhibit clear RV-$\log R'_{\mathrm{HK}}$ correlations which originates
from stellar activity and prevented us from claiming planets from
the RV variations/periods. Hence, also in the search for Jupiter analouges,
it is important to monitor activity indicators. Unfortunately, only
the 4--5\,yr of HARPS data contain the CaII~H\&K lines, so that
the whole time baseline is not covered, which would have been especially
desirable for HR~506 and $\iota$~Hor.

We have not detected a Jupiter analogue, while our upper mass limits
demonstrate that we are sensitive to Jupiter-mass planets up to 5\,AU,
i.e. Jupiter analogues (Fig.~\ref{Fig:Limits_Survey}). In our sample
the planet candidate with the longest period is HR~506~b (2.7\,yr).
While it has a minimum mass of one Jupiter mass, its period is still
4 times shorter than that of Jupiter. Although our sample size is
too small to provide a meaningful number for the occurrence rate of
Jupiters and Jupiter analogues, the two Jupiter-mass planets in our
sample imply that our results are in agreement with the planet frequency
of other much larger surveys. For example, from the ELODIE survey
\citet{Naef2005} estimated a fraction of $7.5\pm1.5$\,\% for stars
hosting giant planets with periods less than 10\,yr, while \citet{Cumming08}
derived a frequency of $12\pm1.6$ \% from the Keck survey and \citet{Mayor2011}
$9.7\pm1.3$\,\% (for $>100\, M_{\oplus}$) from the HARPS/CORALIE
survey. For Jupiter analogs, i.e. only in the range of 3--6\,AU,
\citet{Lineweaver2003} estimated a frequency of 5$\pm$2\,\% which
is e.g. consistent with the discovery rate in the AAPS survey (3.3$\pm$1.4\,\%,
\citealt{Wittenmyer2011}). Such low observed frequencies are compatible
with the non-detection of a Jupiter analogue in our sample and a decreasing
frequency of giant planets at larger distances ($\gtrsim5$\,AU)
is also predicted with core accretion theory \citep{Mordasini2012}.

The main reason for the limited VLC precision of $\sim$9\,m/s is
the short wavelength coverage of only 39\,\AA{}. Useful iodine lines
cover a total wavelength range of 1000\,\AA{}. If this 25 time larger
range could have been used, the extrapolated VLC precision would be
5 times higher, i.e. 1.8\,m/s. The actively stabilised HARPS spectrograph
has a larger wavelength coverage and an outperforming precision. The
higher resolution of the CES+VLC is not necessarily an advantage because
the stellar lines are nearly resolved already at the lower resolution
of HARPS, i.e. they become not sharper and therefore do not provide
much more RV information%
\footnote{The thermal line broadening of a line with frequency $f$ is given
by (e.g. \citealp{Unsoeld2002}) 
\begin{eqnarray*}
\frac{\Delta f_{\mathrm{FWHM}}}{f} & = & \sqrt{8\ln2\frac{kT}{mc^{2}}}=\frac{1}{1.4\cdot10^{6}}\sqrt{\frac{T/\mathrm{K}}{m/\mathrm{amu}}}
\end{eqnarray*}
and depends on the temperature $T$ and the atomic/molecular mass~$m$.
Therefore resolving hydrogen lines ($m_{\mathrm{H}}=1$\,amu) in
a solar-like star ($T=6000$\,K) requires a resolution of $R=\frac{\lambda}{\Delta\lambda}\approx18\,000$
and for iron lines ($m_{\mathrm{Fe}}=56$\,amu) $R=135\,000$. A
higher resolution as provided by the CES leads to an oversampling
or stretching of the lines but does not resolve new lines or sharper
features which would improve the RV measurements. Contrariwise, for
iodine gas ($T=323$\,K, $m_{\mathrm{I_{2}}}=154$\,amu) the thermal
line broadening effect corresponds to $R=2\,400\,000$.%
}. Nevertheless the CES data are valuable for extending the time baseline
to 15 years.

The problem of combining long-term precision RV data from different
instruments complicates the analysis. We could derive the zero point
offset between LC and VLC only with a limited precision which leads
to a loss of sensitivity for trends and long periods. However, this
problem can occur in long-term surveys quite frequently as spectrographs
receive upgrades or survey projects are transferred to new instruments.
Long-term access to the same instrument/instrument configuration is
therefore important.

While the RV method probes the inner region of the stellar environment,
the outer regions can be explored for planets with direct imaging.
Since both methods complement each other, their combination leads
to a more complete picture about the existence and nature of planets
around stars. An example is $\epsilon$~Ind~A, where despite the
high imaging sensitivity with HST/NICMOS and VLT/NACO \citep{Geissler2007,Janson09}
the non-detection puts constraints to the companion which induces
the observed RV trend of 2.4\,m/s/yr. Hence the results and detection
limits from our survey can be valuable for other campaigns which target
these bright stars.

Some of our nearby and bright stars are also the subject of projects
searching and studying surrounding debris disks. Structures in these
disks permit conclusions about the presence of outer planets. For
example, the \emph{Herschel} satellite has resolved exo-Kuiper belts
around $\zeta^{2}$~Ret \citep{Eiroa2010} and around HR~506 \citep[q$^1$~Eri,][]{Liseau2010}.
In the case of the RV planet-hosting star HR~506, the structure of
a ring at 35-40\,AU provides a hint for another planet. Similarly,
the structure of the known debris disk of $\epsilon$~Eri suggests
an outer planet \citep{Quillen2002}. But then for HR~8501 that has
an RV trend likely due to its wide stellar companion, \citet{Eiroa2010}
exclude a cold debris disk.

\begin{figure*}
\includegraphics[width=0.5\linewidth]{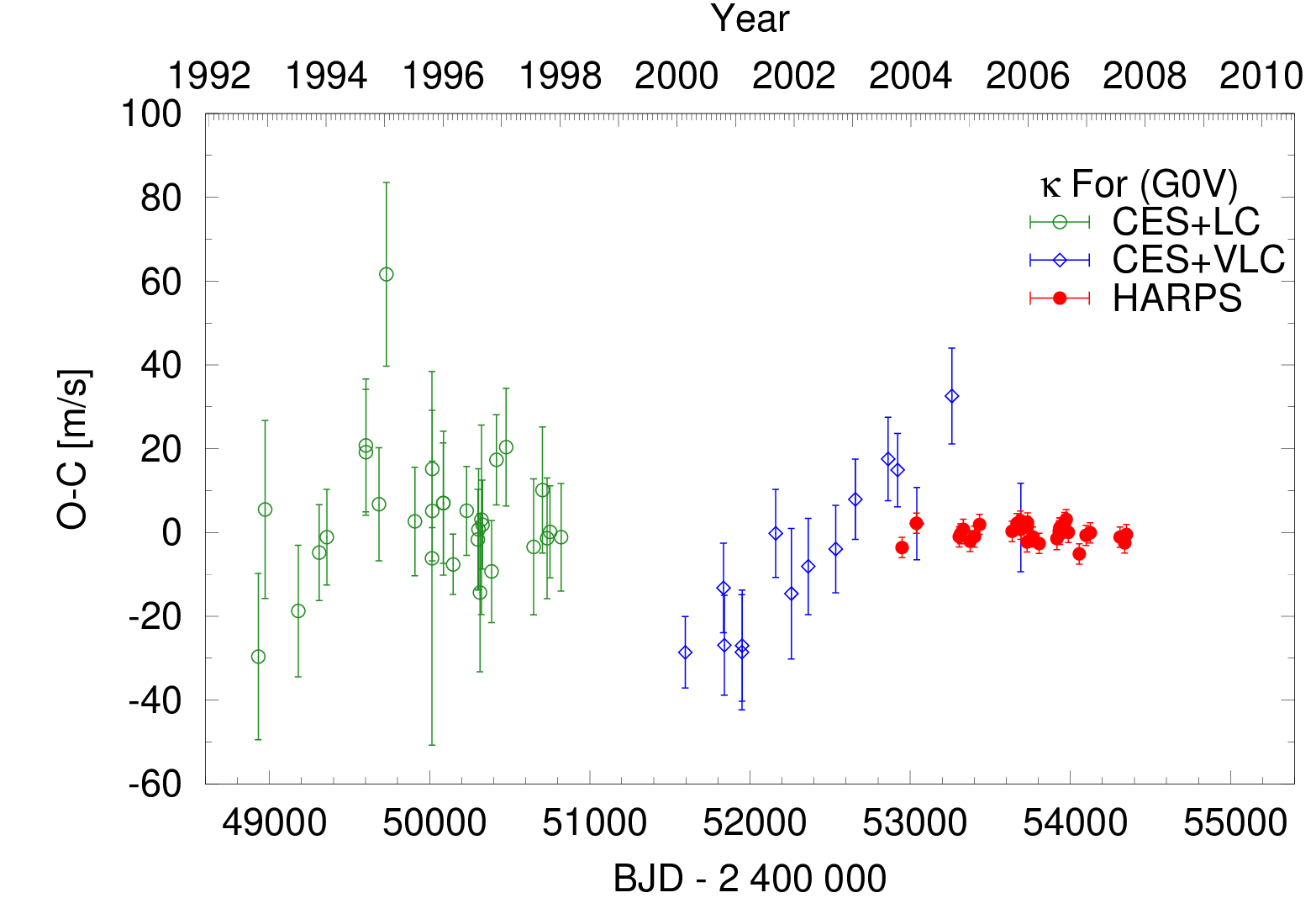}\includegraphics[width=0.5\linewidth]{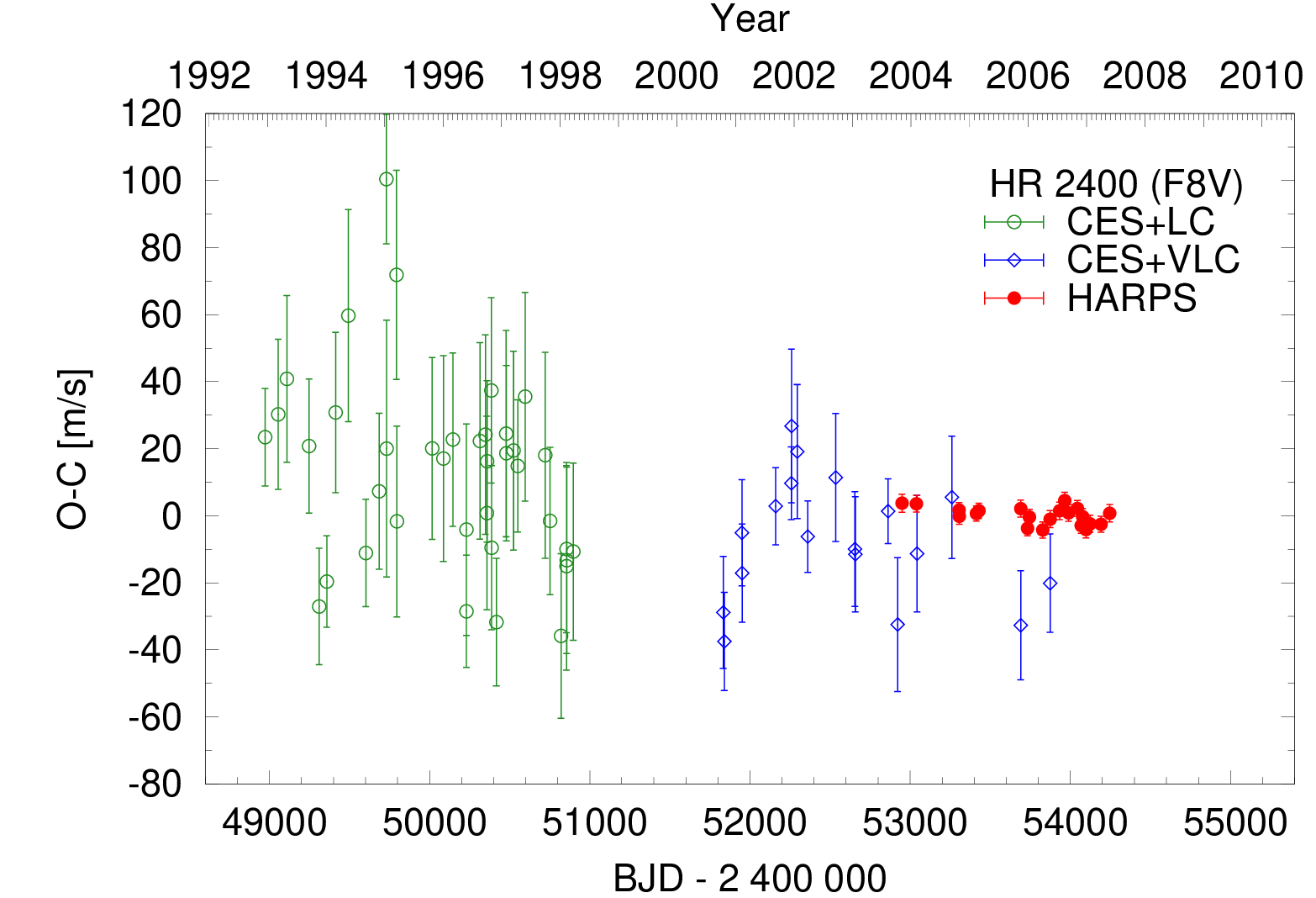}

\includegraphics[width=0.5\linewidth]{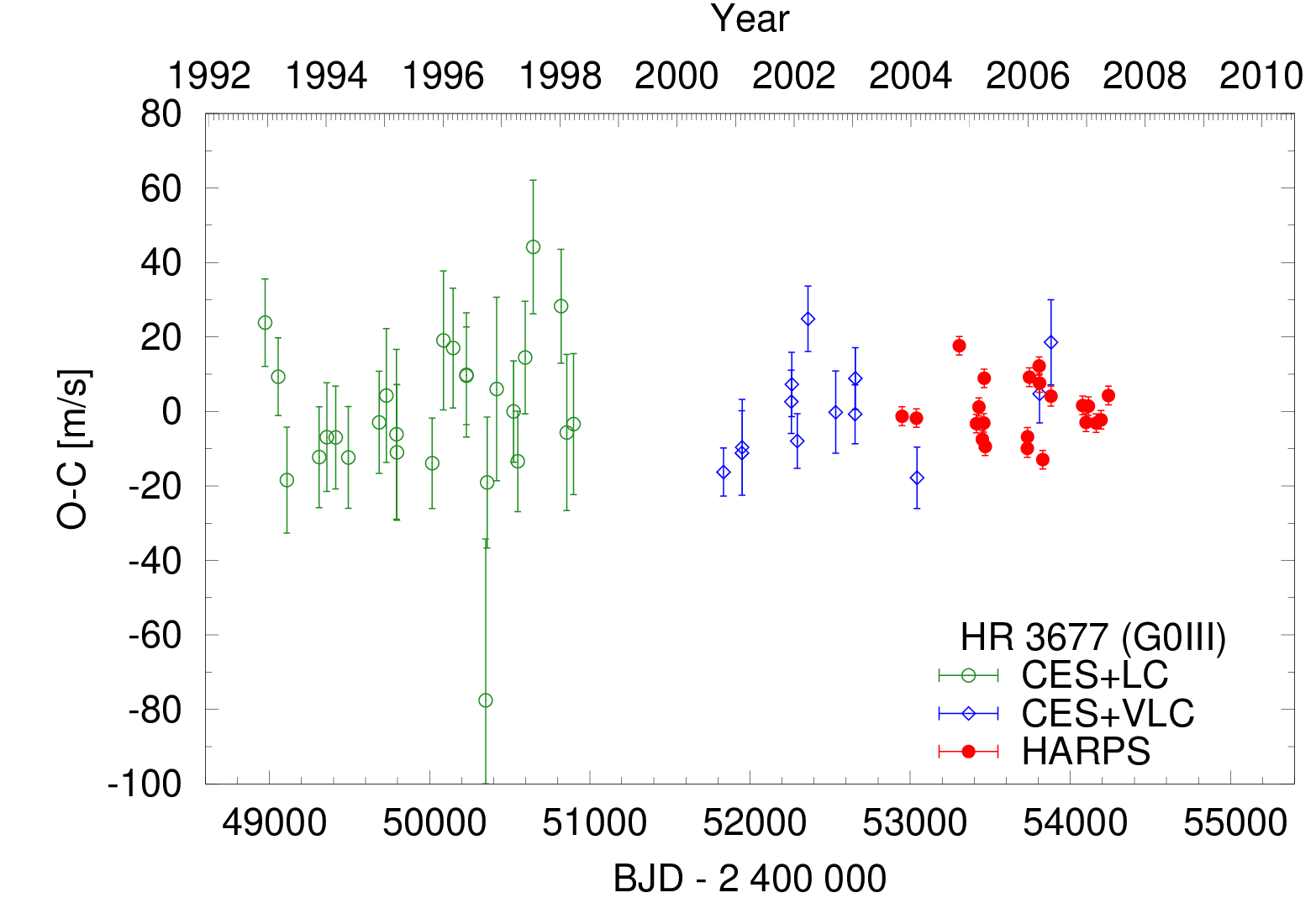}\includegraphics[width=0.5\linewidth]{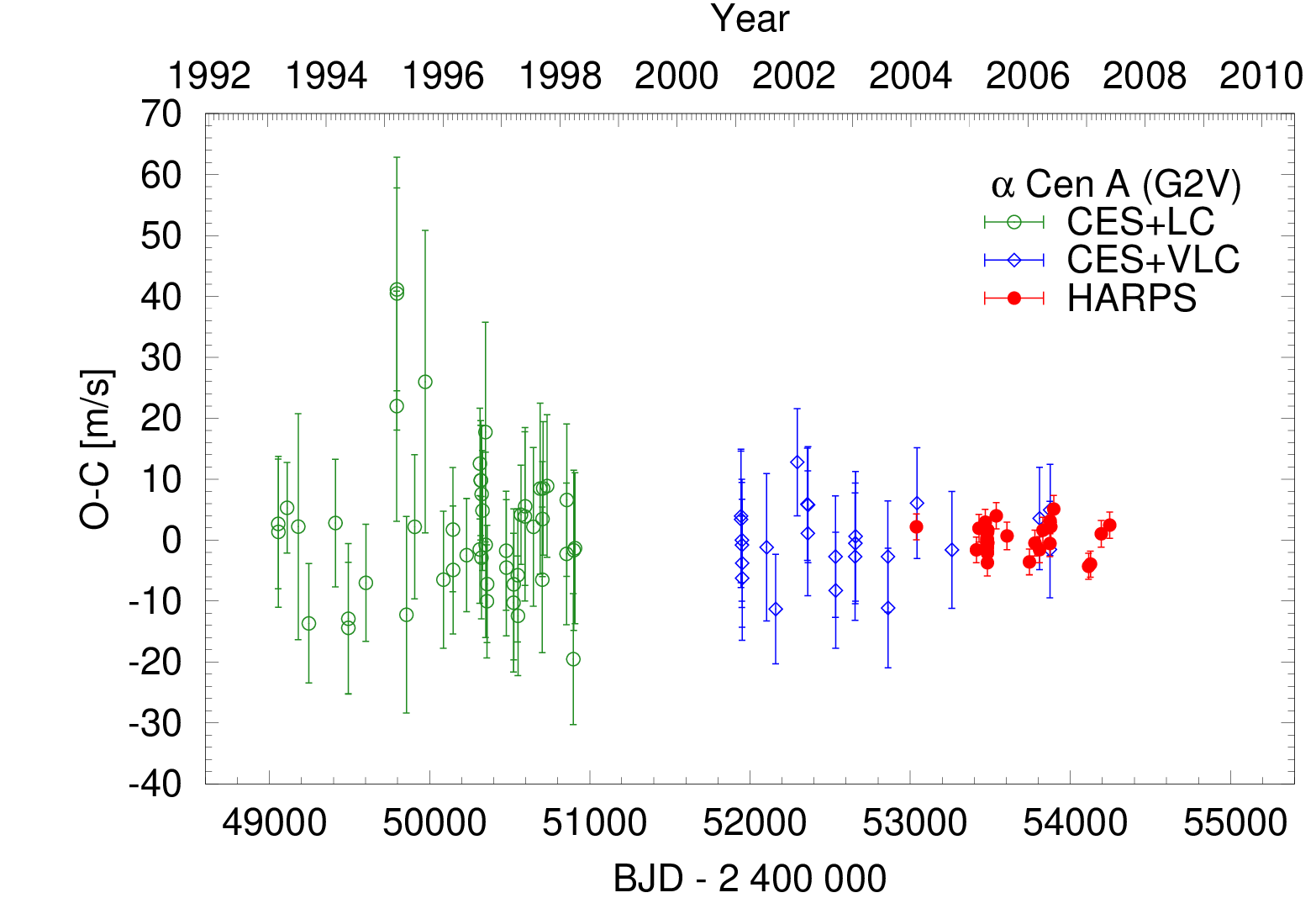}

\includegraphics[width=0.5\linewidth]{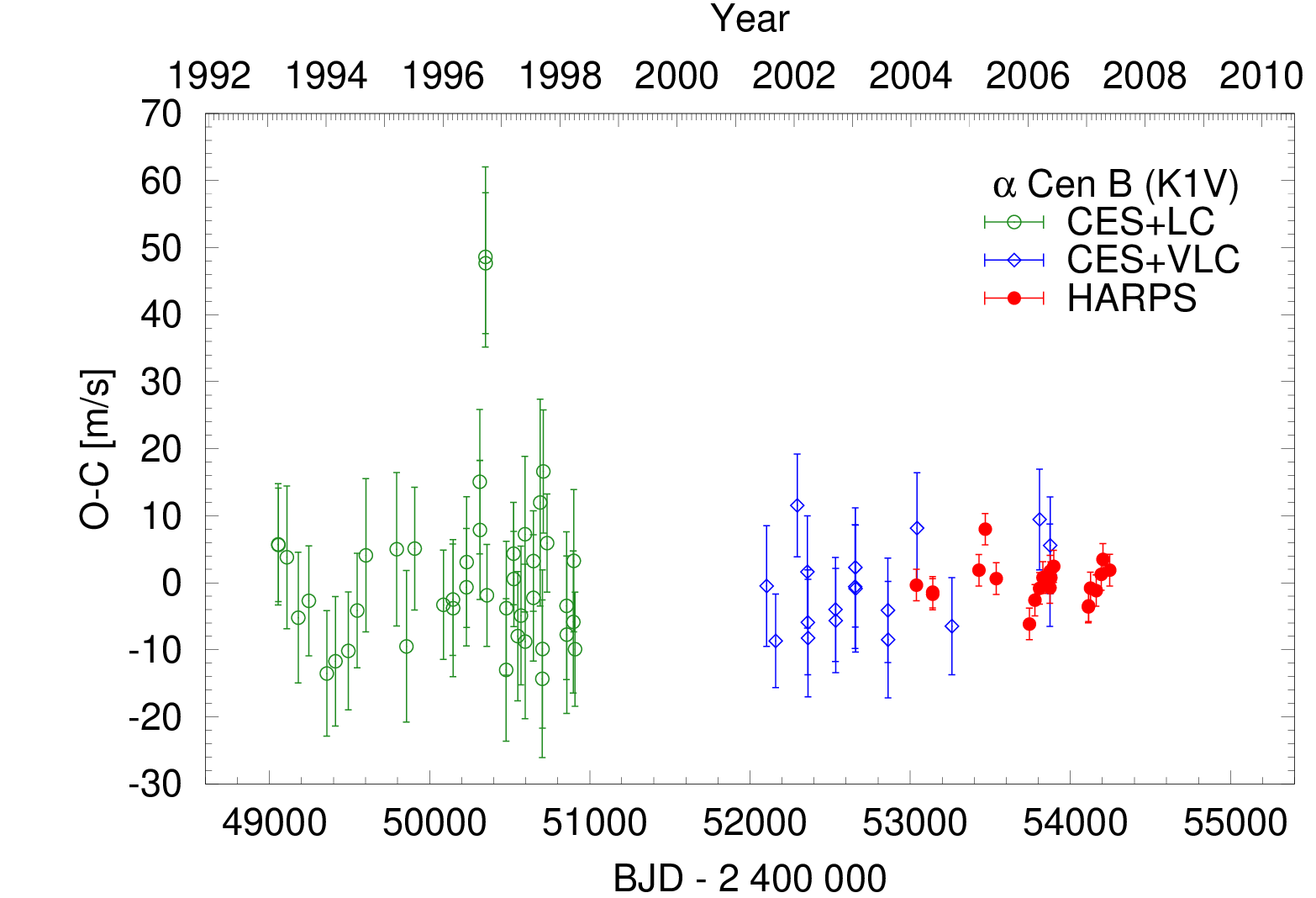}

\caption{\label{Fig:RV-residuals}Residual RVs (binned as described in Sect.~\ref{Sec:RV_Analysis})
for the spectroscopic binaries.}
\end{figure*}

\begin{acknowledgements}
We are grateful to K. Dennerl and S. D{\"o}bereiner for their help
with obtaining CES-LC data in the early phase of the survey as well
as F. Rodler for obtaining CES-VLC and HARPS data. S. Els helped with
the CES-VLC observation and the data reduction. We thank G. Anglada-Escud\'{e}
for his useful hints and discussions and for providing the HARPS-TERRA
software. Discussions with P. Bristow, D. Baade, and P. Sinclair on
VLC-CCD effects are gratefully acknowledged. We are thankful to the
ESO OPC for generous allocation of observing time to the CES/HARPS
planet search programme. MZ acknowledges fi{}nancial support from
the Deutsche Forschungsgemeinschaft (DFG) under RE 1664/4-1. The Lund
Observatory obtained the IRFTS with support from Knut and Alice Wallenberg
Foundation. HH and HN acknowledge support from the Swedish Research
Council (VR) through contract 621-2006-3085 and Linnaeus Grant through
Lund Laser Centre.
\end{acknowledgements}
\bibliographystyle{aa}
\bibliography{CES}

\begin{thebibliography}{104}
\expandafter\ifx\csname natexlab\endcsname\relax\def\natexlab#1{#1}\fi

\bibitem[{{Anglada-Escud{\'e}} \& {Butler}(2012)}]{Anglada-Escude2012}
{Anglada-Escud{\'e}}, G. \& {Butler}, R.~P. 2012, \apjs, 200, 15

\bibitem[{{Baize} \& {Petit}(1989)}]{Baize1989}
{Baize}, P. \& {Petit}, M. 1989, \aaps, 77, 497

\bibitem[{{Baluev}(2012)}]{Baluev2012}
{Baluev}, R.~V. 2012, ArXiv e-prints

\bibitem[{{Benedict} {et~al.}(2006){Benedict}, {McArthur}, {Gatewood}, {Nelan},
  {Cochran}, {Hatzes}, {Endl}, {Wittenmyer}, {Baliunas}, {Walker}, {Yang},
  {K{\"u}rster}, {Els}, \& {Paulson}}]{Benedict2006}
{Benedict}, G.~F., {McArthur}, B.~E., {Gatewood}, G., {et~al.} 2006, \aj, 132,
  2206

\bibitem[{{Boisse} {et~al.}(2011){Boisse}, {Bouchy}, {H{\'e}brard}, {Bonfils},
  {Santos}, \& {Vauclair}}]{Boisse2011}
{Boisse}, I., {Bouchy}, F., {H{\'e}brard}, G., {et~al.} 2011, \aap, 528, A4

\bibitem[{{Boisse} {et~al.}(2012){Boisse}, {Pepe}, {Perrier}, {Queloz},
  {Bonfils}, {Bouchy}, {Santos}, {Arnold}, {Beuzit}, {D{\'{\i}}az}, {Delfosse},
  {Eggenberger}, {Ehrenreich}, {Forveille}, {H{\'e}brard}, {Lagrange}, {Lovis},
  {Mayor}, {Moutou}, {Naef}, {Santerne}, {S{\'e}gransan}, {Sivan}, \&
  {Udry}}]{Boisse2012}
{Boisse}, I., {Pepe}, F., {Perrier}, C., {et~al.} 2012, \aap, 545, A55

\bibitem[{{Bouchy} {et~al.}(2009{\natexlab{a}}){Bouchy}, {H{\'e}brard}, {Udry},
  {Delfosse}, {Boisse}, {Desort}, {Bonfils}, {Eggenberger}, {Ehrenreich},
  {Forveille}, {Lagrange}, {Le Coroller}, {Lovis}, {Moutou}, {Pepe}, {Perrier},
  {Pont}, {Queloz}, {Santos}, {S{\'e}gransan}, \& {Vidal-Madjar}}]{Bouchy2009a}
{Bouchy}, F., {H{\'e}brard}, G., {Udry}, S., {et~al.} 2009{\natexlab{a}}, \aap,
  505, 853

\bibitem[{{Bouchy} {et~al.}(2009{\natexlab{b}}){Bouchy}, {Isambert}, {Lovis},
  {Boisse}, {Figueira}, {H{\'e}brard}, \& {Pepe}}]{Bouchy2009}
{Bouchy}, F., {Isambert}, J., {Lovis}, C., {et~al.} 2009{\natexlab{b}}, in EAS
  Publications Series, Vol.~37, EAS Publications Series, ed. {P.~Kern},
  247--253

\bibitem[{{Bouchy} {et~al.}(2001){Bouchy}, {Pepe}, \& {Queloz}}]{Bouchy01}
{Bouchy}, F., {Pepe}, F., \& {Queloz}, D. 2001, \aap, 374, 733

\bibitem[{{Buccino} \& {Mauas}(2008)}]{Buccino2008}
{Buccino}, A.~P. \& {Mauas}, P.~J.~D. 2008, \aap, 483, 903

\bibitem[{{Burgasser} {et~al.}(2000){Burgasser}, {Kirkpatrick}, {Cutri},
  {McCallon}, {Kopan}, {Gizis}, {Liebert}, {Reid}, {Brown}, {Monet}, {Dahn},
  {Beichman}, \& {Skrutskie}}]{Burgasser2000}
{Burgasser}, A.~J., {Kirkpatrick}, J.~D., {Cutri}, R.~M., {et~al.} 2000, \apjl,
  531, L57

\bibitem[{{Butler} {et~al.}(1996){Butler}, {Marcy}, {Williams}, {McCarthy},
  {Dosanjh}, \& {Vogt}}]{Butler1996}
{Butler}, R.~P., {Marcy}, G.~W., {Williams}, E., {et~al.} 1996, \pasp, 108, 500

\bibitem[{{Butler} {et~al.}(2001){Butler}, {Tinney}, {Marcy}, {Jones}, {Penny},
  \& {Apps}}]{Butler2001}
{Butler}, R.~P., {Tinney}, C.~G., {Marcy}, G.~W., {et~al.} 2001, \apj, 555, 410

\bibitem[{{Butler} {et~al.}(2006){Butler}, {Wright}, {Marcy}, {Fischer},
  {Vogt}, {Tinney}, {Jones}, {Carter}, {Johnson}, {McCarthy}, \&
  {Penny}}]{Butler06}
{Butler}, R.~P., {Wright}, J.~T., {Marcy}, G.~W., {et~al.} 2006, \apj, 646, 505

\bibitem[{{Connes}(1985)}]{Connes85}
{Connes}, P. 1985, \apss, 110, 211

\bibitem[{{Cox}(2000)}]{Cox2000}
{Cox}, A.~N. 2000, \skytel, 100, 72

\bibitem[{{Cumming}(2004)}]{Cumming04}
{Cumming}, A. 2004, \mnras, 354, 1165

\bibitem[{{Cumming} {et~al.}(2008){Cumming}, {Butler}, {Marcy}, {Vogt},
  {Wright}, \& {Fischer}}]{Cumming08}
{Cumming}, A., {Butler}, R.~P., {Marcy}, G.~W., {et~al.} 2008, \pasp, 120, 531

\bibitem[{{Deakin} \& {Kildea}(1999)}]{Deakin1999}
{Deakin}, R.~E. \& {Kildea}, D.~G. 1999, The Australian Surveyor, Vol.44, No.1,
  74

\bibitem[{{del Peloso} {et~al.}(2000){del Peloso}, {da Silva}, \& {Porto de
  Mello}}]{Peloso2000}
{del Peloso}, E.~F., {da Silva}, L., \& {Porto de Mello}, G.~F. 2000, \aap,
  358, 233

\bibitem[{{Donahue} {et~al.}(1996){Donahue}, {Saar}, \&
  {Baliunas}}]{Donahue1996}
{Donahue}, R.~A., {Saar}, S.~H., \& {Baliunas}, S.~L. 1996, \apj, 466, 384

\bibitem[{{Drake} \& {Smith}(1993)}]{Drake1993}
{Drake}, J.~J. \& {Smith}, G. 1993, \apj, 412, 797

\bibitem[{{Dravins} {et~al.}(1998){Dravins}, {Lindegren}, \&
  {Vandenberg}}]{Dravins1998}
{Dravins}, D., {Lindegren}, L., \& {Vandenberg}, D.~A. 1998, \aap, 330, 1077

\bibitem[{{Dumusque} {et~al.}(2011{\natexlab{a}}){Dumusque}, {Lovis}, {Udry},
  \& {Santos}}]{Dumusque2010}
{Dumusque}, X., {Lovis}, C., {Udry}, S., \& {Santos}, N.~C. 2011{\natexlab{a}},
  in IAU Symposium, Vol. 276, IAU Symposium, ed. A.~{Sozzetti}, M.~G.
  {Lattanzi}, \& A.~P. {Boss}, 530--532

\bibitem[{{Dumusque} {et~al.}(2012){Dumusque}, {Pepe}, {Lovis}, {Segransan},
  {Sahlmann}, {Benz}, {Bouchy}, {Mayor}, {Queloz}, {Santos}, \&
  {Udry}}]{Dumusque2012}
{Dumusque}, X., {Pepe}, F., {Lovis}, C., {et~al.} 2012, \nat, 491, 207

\bibitem[{{Dumusque} {et~al.}(2011{\natexlab{b}}){Dumusque}, {Udry}, {Lovis},
  {Santos}, \& {Monteiro}}]{Dumusque2011}
{Dumusque}, X., {Udry}, S., {Lovis}, C., {Santos}, N.~C., \& {Monteiro},
  M.~J.~P.~F.~G. 2011{\natexlab{b}}, \aap, 525, A140

\bibitem[{{Eggenberger} {et~al.}(2007){Eggenberger}, {Udry}, {Chauvin},
  {Beuzit}, {Lagrange}, {S{\'e}gransan}, \& {Mayor}}]{Eggenberger07}
{Eggenberger}, A., {Udry}, S., {Chauvin}, G., {et~al.} 2007, \aap, 474, 273

\bibitem[{{Eiroa} {et~al.}(2010){Eiroa}, {Fedele}, {Maldonado},
  {Gonz{\'a}lez-Garc{\'{\i}}a}, {Rodmann}, {Heras}, {Pilbratt}, {Augereau},
  {Mora}, {Montesinos}, {Ardila}, {Bryden}, {Liseau}, {Stapelfeldt},
  {Launhardt}, {Solano}, {Bayo}, {Absil}, {Ar{\'e}valo}, {Barrado},
  {Beichmann}, {Danchi}, {Del Burgo}, {Ertel}, {Fridlund}, {Fukagawa},
  {Guti{\'e}rrez}, {Gr{\"u}n}, {Kamp}, {Krivov}, {Lebreton}, {L{\"o}hne},
  {Lorente}, {Marshall}, {Mart{\'{\i}}nez-Arn{\'a}iz}, {Meeus}, {Montes},
  {Morbidelli}, {M{\"u}ller}, {Mutschke}, {Nakagawa}, {Olofsson}, {Ribas},
  {Roberge}, {Sanz-Forcada}, {Th{\'e}bault}, {Walker}, {White}, \&
  {Wolf}}]{Eiroa2010}
{Eiroa}, C., {Fedele}, D., {Maldonado}, J., {et~al.} 2010, \aap, 518, L131+

\bibitem[{{Enard}(1982)}]{Enard1982}
{Enard}, D. 1982, in Presented at the Society of Photo-Optical Instrumentation
  Engineers (SPIE) Conference, Vol. 331, Society of Photo-Optical
  Instrumentation Engineers (SPIE) Conference Series, 232--242

\bibitem[{{Endl} {et~al.}(2000){Endl}, {K{\"u}rster}, \& {Els}}]{Endl00}
{Endl}, M., {K{\"u}rster}, M., \& {Els}, S. 2000, \aap, 362, 585

\bibitem[{{Endl} {et~al.}(2002){Endl}, {K{\"u}rster}, {Els}, {Hatzes},
  {Cochran}, {Dennerl}, \& {D{\"o}bereiner}}]{Endl02}
{Endl}, M., {K{\"u}rster}, M., {Els}, S., {et~al.} 2002, \aap, 392, 671

\bibitem[{{Favata} {et~al.}(1997){Favata}, {Micela}, \&
  {Sciortino}}]{Favata1997}
{Favata}, F., {Micela}, G., \& {Sciortino}, S. 1997, \aap, 323, 809

\bibitem[{{Fr{\"o}hlich}(2007)}]{Froehlich2007}
{Fr{\"o}hlich}, H. 2007, Astronomische Nachrichten, 328, 1037

\bibitem[{{Gei{\ss}ler} {et~al.}(2007){Gei{\ss}ler}, {Kellner}, {Brandner},
  {Masciadri}, {Hartung}, {Henning}, {Lenzen}, {Close}, {Endl}, \&
  {K{\"u}rster}}]{Geissler2007}
{Gei{\ss}ler}, K., {Kellner}, S., {Brandner}, W., {et~al.} 2007, \aap, 461, 665

\bibitem[{{Gomes da Silva} {et~al.}(2012){Gomes da Silva}, {Santos}, {Bonfils},
  {Delfosse}, {Forveille}, {Udry}, {Dumusque}, \& {Lovis}}]{GomesdaSilva2012}
{Gomes da Silva}, J., {Santos}, N.~C., {Bonfils}, X., {et~al.} 2012, \aap, 541,
  A9

\bibitem[{{Gray}(1988)}]{Gray1988}
{Gray}, D.~F. 1988, {Lectures on spectral-line analysis: F,G, and K stars}

\bibitem[{{Gray} {et~al.}(2006){Gray}, {Corbally}, {Garrison}, {McFadden},
  {Bubar}, {McGahee}, {O'Donoghue}, \& {Knox}}]{Gray2006}
{Gray}, R.~O., {Corbally}, C.~J., {Garrison}, R.~F., {et~al.} 2006, \aj, 132,
  161

\bibitem[{{Hatzes} {et~al.}(2000){Hatzes}, {Cochran}, {McArthur}, {Baliunas},
  {Walker}, {Campbell}, {Irwin}, {Yang}, {K{\"u}rster}, {Endl}, {Els},
  {Butler}, \& {Marcy}}]{Hatzes00}
{Hatzes}, A.~P., {Cochran}, W.~D., {McArthur}, B., {et~al.} 2000, \apjl, 544,
  L145

\bibitem[{{Heintz}(1978)}]{Heintz1978}
{Heintz}, W.~D. 1978, \apjs, 37, 515

\bibitem[{{Henry} {et~al.}(1996){Henry}, {Soderblom}, {Donahue}, \&
  {Baliunas}}]{Henry1996}
{Henry}, T.~J., {Soderblom}, D.~R., {Donahue}, R.~A., \& {Baliunas}, S.~L.
  1996, \aj, 111, 439

\bibitem[{{Hoffleit} \& {Jaschek}(1991)}]{Hoffleit1991}
{Hoffleit}, D. \& {Jaschek}, C. 1991, {The Bright star catalogue}

\bibitem[{{Horne}(1986)}]{Horne1986}
{Horne}, K. 1986, \pasp, 98, 609

\bibitem[{{Isaacson} \& {Fischer}(2010)}]{Isaacson2010}
{Isaacson}, H. \& {Fischer}, D. 2010, \apj, 725, 875

\bibitem[{{Janson} {et~al.}(2009){Janson}, {Apai}, {Zechmeister}, {Brandner},
  {K{\"u}rster}, {Kasper}, {Reffert}, {Endl}, {Lafreni{\`e}re}, {Gei{\ss}ler},
  {Hippler}, \& {Henning}}]{Janson09}
{Janson}, M., {Apai}, D., {Zechmeister}, M., {et~al.} 2009, \mnras, 399, 377

\bibitem[{{Janson} {et~al.}(2007){Janson}, {Brandner}, {Henning}, {Lenzen},
  {McArthur}, {Benedict}, {Reffert}, {Nielsen}, {Close}, {Biller}, {Kellner},
  {G{\"u}nther}, {Hatzes}, {Masciadri}, {Geissler}, \& {Hartung}}]{Janson2007}
{Janson}, M., {Brandner}, W., {Henning}, T., {et~al.} 2007, \aj, 133, 2442

\bibitem[{{Janson} {et~al.}(2008){Janson}, {Reffert}, {Brandner}, {Henning},
  {Lenzen}, \& {Hippler}}]{Janson2008}
{Janson}, M., {Reffert}, S., {Brandner}, W., {et~al.} 2008, \aap, 488, 771

\bibitem[{{Jefferys} {et~al.}(1988){Jefferys}, {Fitzpatrick}, \&
  {McArthur}}]{Jefferys1988}
{Jefferys}, W.~H., {Fitzpatrick}, M.~J., \& {McArthur}, B.~E. 1988, Celestial
  Mechanics, 41, 39

\bibitem[{{Jones} {et~al.}(2004){Jones}, {Butler}, {Tinney}, {Marcy},
  {McCarthy}, {Penny}, \& {Carter}}]{Jones04}
{Jones}, H.~R.~A., {Butler}, R.~P., {Tinney}, C.~G., {et~al.} 2004, in
  Astronomical Society of the Pacific Conference Series, Vol. 321, Extrasolar
  Planets: Today and Tomorrow, ed. {J.~Beaulieu, A.~Lecavelier Des Etangs, \&
  C.~Terquem}, 298--+

\bibitem[{{Kaper} {et~al.}(1966){Kaper}, {Smits}, {Schwarz}, {Takakubo}, \&
  {van Woerden}}]{Kaper1966}
{Kaper}, H.~G., {Smits}, D.~W., {Schwarz}, U., {Takakubo}, K., \& {van
  Woerden}, H. 1966, \bain, 18, 465

\bibitem[{{King} {et~al.}(2010){King}, {McCaughrean}, {Homeier}, {Allard},
  {Scholz}, \& {Lodieu}}]{King2010}
{King}, R.~R., {McCaughrean}, M.~J., {Homeier}, D., {et~al.} 2010, \aap, 510,
  A99+

\bibitem[{{Kjeldsen} \& {Bedding}(1995)}]{Kjeldsen1995}
{Kjeldsen}, H. \& {Bedding}, T.~R. 1995, \aap, 293, 87

\bibitem[{{K{\"u}rster} {et~al.}(2000){K{\"u}rster}, {Endl}, {Els}, {Hatzes},
  {Cochran}, {D{\"o}bereiner}, \& {Dennerl}}]{Kuerster00}
{K{\"u}rster}, M., {Endl}, M., {Els}, S., {et~al.} 2000, \aap, 353, L33

\bibitem[{{K{\"u}rster} {et~al.}(2003){K{\"u}rster}, {Endl}, {Rouesnel}, {Els},
  {Kaufer}, {Brillant}, {Hatzes}, {Saar}, \& {Cochran}}]{Kuerster03}
{K{\"u}rster}, M., {Endl}, M., {Rouesnel}, F., {et~al.} 2003, \aap, 403, 1077

\bibitem[{{Lineweaver} \& {Grether}(2003)}]{Lineweaver2003}
{Lineweaver}, C.~H. \& {Grether}, D. 2003, \apj, 598, 1350

\bibitem[{{Liseau} {et~al.}(2010){Liseau}, {Eiroa}, {Fedele}, {Augereau},
  {Olofsson}, {Gonz{\'a}lez}, {Maldonado}, {Montesinos}, {Mora}, {Absil},
  {Ardila}, {Barrado}, {Bayo}, {Beichman}, {Bryden}, {Danchi}, {Del Burgo},
  {Ertel}, {Fridlund}, {Heras}, {Krivov}, {Launhardt}, {Lebreton}, {L{\"o}hne},
  {Marshall}, {Meeus}, {M{\"u}ller}, {Pilbratt}, {Roberge}, {Rodmann},
  {Solano}, {Stapelfeldt}, {Th{\'e}bault}, {White}, \& {Wolf}}]{Liseau2010}
{Liseau}, R., {Eiroa}, C., {Fedele}, D., {et~al.} 2010, \aap, 518, L132+

\bibitem[{{Lovis} {et~al.}(2011){Lovis}, {Dumusque}, {Santos}, {Bouchy},
  {Mayor}, {Pepe}, {Queloz}, {S{\'e}gransan}, \& {Udry}}]{Lovis2011}
{Lovis}, C., {Dumusque}, X., {Santos}, N.~C., {et~al.} 2011, ArXiv e-prints

\bibitem[{{Lovis} {et~al.}(2006){Lovis}, {Mayor}, {Pepe}, {Alibert}, {Benz},
  {Bouchy}, {Correia}, {Laskar}, {Mordasini}, {Queloz}, {Santos}, {Udry},
  {Bertaux}, \& {Sivan}}]{Lovis06}
{Lovis}, C., {Mayor}, M., {Pepe}, F., {et~al.} 2006, \nat, 441, 305

\bibitem[{{Luyten} \& {Hughes}(1980)}]{Luyten1980}
{Luyten}, W.~J. \& {Hughes}, H.~S. 1980, Proper motion survey with the
  forty-eight inch Schmidt telescope.~LV.~First supplement to the NLTT
  catalogue., by Luyten, W.~J.; Hughes, H.~S..~ Sep.~print University of
  Minnesota, Minneapolis, MN (USA), 16 p., 55, 1

\bibitem[{{Marcy} {et~al.}(2008){Marcy}, {Butler}, {Vogt}, {Fischer}, {Wright},
  {Johnson}, {Tinney}, {Jones}, {Carter}, {Bailey}, {O'Toole}, \&
  {Upadhyay}}]{Marcy2008}
{Marcy}, G.~W., {Butler}, R.~P., {Vogt}, S.~S., {et~al.} 2008, Physica Scripta
  Volume T, 130, 014001

\bibitem[{{Mason} {et~al.}(2001){Mason}, {Wycoff}, {Hartkopf}, {Douglass}, \&
  {Worley}}]{Mason2001}
{Mason}, B.~D., {Wycoff}, G.~L., {Hartkopf}, W.~I., {Douglass}, G.~G., \&
  {Worley}, C.~E. 2001, \aj, 122, 3466

\bibitem[{{Mayor} {et~al.}(2011){Mayor}, {Marmier}, {Lovis}, {Udry},
  {S{\'e}gransan}, {Pepe}, {Benz}, {Bertaux}, {Bouchy}, {Dumusque}, {Lo Curto},
  {Mordasini}, {Queloz}, \& {Santos}}]{Mayor2011}
{Mayor}, M., {Marmier}, M., {Lovis}, C., {et~al.} 2011, ArXiv e-prints

\bibitem[{{Mayor} {et~al.}(2003){Mayor}, {Pepe}, {Queloz}, {Bouchy},
  {Rupprecht}, {Lo Curto}, {Avila}, {Benz}, {Bertaux}, {Bonfils}, {Dall},
  {Dekker}, {Delabre}, {Eckert}, {Fleury}, {Gilliotte}, {Gojak}, {Guzman},
  {Kohler}, {Lizon}, {Longinotti}, {Lovis}, {Megevand}, {Pasquini}, {Reyes},
  {Sivan}, {Sosnowska}, {Soto}, {Udry}, {van Kesteren}, {Weber}, \&
  {Weilenmann}}]{Mayor03}
{Mayor}, M., {Pepe}, F., {Queloz}, D., {et~al.} 2003, The Messenger, 114, 20

\bibitem[{{Mayor} \& {Queloz}(1995)}]{Mayor1995}
{Mayor}, M. \& {Queloz}, D. 1995, \nat, 378, 355

\bibitem[{{Metcalfe} {et~al.}(2010){Metcalfe}, {Basu}, {Henry}, {Soderblom},
  {Judge}, {Kn{\"o}lker}, {Mathur}, \& {Rempel}}]{Metcalfe2010}
{Metcalfe}, T.~S., {Basu}, S., {Henry}, T.~J., {et~al.} 2010, \apjl, 723, L213

\bibitem[{{Mordasini} {et~al.}(2012){Mordasini}, {Alibert}, {Benz}, {Klahr}, \&
  {Henning}}]{Mordasini2012}
{Mordasini}, C., {Alibert}, Y., {Benz}, W., {Klahr}, H., \& {Henning}, T. 2012,
  \aap, 541, A97

\bibitem[{{Naef} {et~al.}(2005){Naef}, {Mayor}, {Beuzit}, {Perrier}, {Queloz},
  {Sivan}, \& {Udry}}]{Naef2005}
{Naef}, D., {Mayor}, M., {Beuzit}, J., {et~al.} 2005, in ESA Special
  Publication, Vol. 560, 13th Cambridge Workshop on Cool Stars, Stellar Systems
  and the Sun, ed. {F.~Favata, G.~A.~J.~Hussain, \& B.~Battrick}, 833--+

\bibitem[{{Naef} {et~al.}(2003){Naef}, {Mayor}, {Korzennik}, {Queloz}, {Udry},
  {Nisenson}, {Noyes}, {Brown}, {Beuzit}, {Perrier}, \& {Sivan}}]{Naef2003}
{Naef}, D., {Mayor}, M., {Korzennik}, S.~G., {et~al.} 2003, \aap, 410, 1051

\bibitem[{{Naef} {et~al.}(2010){Naef}, {Mayor}, {Lo Curto}, {Bouchy}, {Lovis},
  {Moutou}, {Benz}, {Pepe}, {Queloz}, {Santos}, {S{\'e}gransan}, {Udry},
  {Bonfils}, {Delfosse}, {Forveille}, {H{\'e}brard}, {Mordasini}, {Perrier},
  {Boisse}, \& {Sosnowska}}]{Naef2010}
{Naef}, D., {Mayor}, M., {Lo Curto}, G., {et~al.} 2010, \aap, 523, A15+

\bibitem[{{Naef} {et~al.}(2001){Naef}, {Mayor}, {Pepe}, {Queloz}, {Santos},
  {Udry}, \& {Burnet}}]{Naef01}
{Naef}, D., {Mayor}, M., {Pepe}, F., {et~al.} 2001, \aap, 375, 205

\bibitem[{{O'Toole} {et~al.}(2009){O'Toole}, {Jones}, {Tinney}, {Butler},
  {Marcy}, {Carter}, {Bailey}, \& {Wittenmyer}}]{OToole2009}
{O'Toole}, S.~J., {Jones}, H.~R.~A., {Tinney}, C.~G., {et~al.} 2009, \apj, 701,
  1732

\bibitem[{{Pepe} {et~al.}(2011){Pepe}, {Lovis}, {S{\'e}gransan}, {Benz},
  {Bouchy}, {Dumusque}, {Mayor}, {Queloz}, {Santos}, \& {Udry}}]{Pepe2011}
{Pepe}, F., {Lovis}, C., {S{\'e}gransan}, D., {et~al.} 2011, \aap, 534, A58

\bibitem[{{Pepe} {et~al.}(2004){Pepe}, {Mayor}, {Queloz}, {Benz}, {Bonfils},
  {Bouchy}, {Lo Curto}, {Lovis}, {M{\'e}gevand}, {Moutou}, {Naef}, {Rupprecht},
  {Santos}, {Sivan}, {Sosnowska}, \& {Udry}}]{Pepe04}
{Pepe}, F., {Mayor}, M., {Queloz}, D., {et~al.} 2004, \aap, 423, 385

\bibitem[{{Perryman} {et~al.}(1997){Perryman}, {Lindegren}, {Kovalevsky},
  {Hoeg}, {Bastian}, {Bernacca}, {Cr{\'e}z{\'e}}, {Donati}, {Grenon}, {van
  Leeuwen}, {van der Marel}, {Mignard}, {Murray}, {Le Poole}, {Schrijver},
  {Turon}, {Arenou}, {Froeschl{\'e}}, \& {Petersen}}]{Perryman1997}
{Perryman}, M.~A.~C., {Lindegren}, L., {Kovalevsky}, J., {et~al.} 1997, \aap,
  323, L49

\bibitem[{{Piskunov} {et~al.}(1997){Piskunov}, {Edvardsson}, {Magain}, \&
  {Swings}}]{Piskunov1997}
{Piskunov}, N., {Edvardsson}, B., {Magain}, P., \& {Swings}, J.~P. 1997,
  \url{http://www.astro.uu.se/obs_ast/ces/}

\bibitem[{{Pourbaix} {et~al.}(2002){Pourbaix}, {Nidever}, {McCarthy}, {Butler},
  {Tinney}, {Marcy}, {Jones}, {Penny}, {Carter}, {Bouchy}, {Pepe}, {Hearnshaw},
  {Skuljan}, {Ramm}, \& {Kent}}]{Pourbaix2002}
{Pourbaix}, D., {Nidever}, D., {McCarthy}, C., {et~al.} 2002, \aap, 386, 280

\bibitem[{{Poveda} {et~al.}(1994){Poveda}, {Herrera}, {Allen}, {Cordero}, \&
  {Lavalley}}]{Poveda1994}
{Poveda}, A., {Herrera}, M.~A., {Allen}, C., {Cordero}, G., \& {Lavalley}, C.
  1994, Revista Mexicana de Astronomia y Astrofisica, 28, 43

\bibitem[{{Queloz} {et~al.}(2001){Queloz}, {Henry}, {Sivan}, {Baliunas},
  {Beuzit}, {Donahue}, {Mayor}, {Naef}, {Perrier}, \& {Udry}}]{Queloz2001}
{Queloz}, D., {Henry}, G.~W., {Sivan}, J.~P., {et~al.} 2001, \aap, 379, 279

\bibitem[{{Quillen} \& {Thorndike}(2002)}]{Quillen2002}
{Quillen}, A.~C. \& {Thorndike}, S. 2002, \apjl, 578, L149

\bibitem[{{Reffert} \& {Quirrenbach}(2011)}]{Reffert2011}
{Reffert}, S. \& {Quirrenbach}, A. 2011, \aap, 527, A140

\bibitem[{{Reidemeister} {et~al.}(2011){Reidemeister}, {Krivov}, {Stark},
  {Augereau}, {L{\"o}hne}, \& {M{\"u}ller}}]{Reidemeister2011}
{Reidemeister}, M., {Krivov}, A.~V., {Stark}, C.~C., {et~al.} 2011, \aap, 527,
  A57

\bibitem[{{Rivera} {et~al.}(2010){Rivera}, {Butler}, {Vogt}, {Laughlin},
  {Henry}, \& {Meschiari}}]{Rivera2010}
{Rivera}, E.~J., {Butler}, R.~P., {Vogt}, S.~S., {et~al.} 2010, \apj, 708, 1492

\bibitem[{{Santos} {et~al.}(2010){Santos}, {Gomes da Silva}, {Lovis}, \&
  {Melo}}]{Santos2010}
{Santos}, N.~C., {Gomes da Silva}, J., {Lovis}, C., \& {Melo}, C. 2010, \aap,
  511, A54+

\bibitem[{{Santos} {et~al.}(2004){Santos}, {Israelian}, \&
  {Mayor}}]{Santos2004}
{Santos}, N.~C., {Israelian}, G., \& {Mayor}, M. 2004, \aap, 415, 1153

\bibitem[{{Schlesinger}(1917)}]{Schlesinger1917}
{Schlesinger}, F. 1917, \aj, 30, 137

\bibitem[{{Scholz} {et~al.}(2003){Scholz}, {McCaughrean}, {Lodieu}, \&
  {Kuhlbrodt}}]{Scholz2003}
{Scholz}, R., {McCaughrean}, M.~J., {Lodieu}, N., \& {Kuhlbrodt}, B. 2003,
  \aap, 398, L29

\bibitem[{{S{\'e}gransan} {et~al.}(2010){S{\'e}gransan}, {Udry}, {Mayor},
  {Naef}, {Pepe}, {Queloz}, {Santos}, {Demory}, {Figueira}, {Gillon},
  {Marmier}, {M{\'e}gevand}, {Sosnowska}, {Tamuz}, \& {Triaud}}]{Segransan2010}
{S{\'e}gransan}, D., {Udry}, S., {Mayor}, M., {et~al.} 2010, \aap, 511, A45+

\bibitem[{{Shporer} {et~al.}(2010){Shporer}, {Winn}, {Dreizler}, {Col{\'o}n},
  {Wood-Vasey}, {Choi}, {Morley}, {Moutou}, {Welsh}, {Pollaco}, {Starkey},
  {Adams}, {Barros}, {Bouchy}, {Cabrera-Lavers}, {Cerutti}, {Coban},
  {Costello}, {Deeg}, {D{\'{\i}}az}, {Esquerdo}, {Fernandez}, {Fleming},
  {Ford}, {Fulton}, {Good}, {H{\'e}brard}, {Holman}, {Hunt}, {Kadakia},
  {Lander}, {Lockhart}, {Mazeh}, {Morehead}, {Nelson}, {Nortmann}, {Reyes},
  {Roebuck}, {Rudy}, {Ruth}, {Simpson}, {Vincent}, {Weaver}, \&
  {Xie}}]{Shporer2010}
{Shporer}, A., {Winn}, J.~N., {Dreizler}, S., {et~al.} 2010, \apj, 722, 880

\bibitem[{{Sousa} {et~al.}(2008){Sousa}, {Santos}, {Mayor}, {Udry},
  {Casagrande}, {Israelian}, {Pepe}, {Queloz}, \& {Monteiro}}]{Sousa2008}
{Sousa}, S.~G., {Santos}, N.~C., {Mayor}, M., {et~al.} 2008, \aap, 487, 373

\bibitem[{{Standish}(1990)}]{Standish1990}
{Standish}, Jr., E.~M. 1990, \aap, 233, 252

\bibitem[{{Teixeira} {et~al.}(2009){Teixeira}, {Kjeldsen}, {Bedding}, {Bouchy},
  {Christensen-Dalsgaard}, {Cunha}, {Dall}, {Frandsen}, {Karoff}, {Monteiro},
  \& {Pijpers}}]{Teixeira09}
{Teixeira}, T.~C., {Kjeldsen}, H., {Bedding}, T.~R., {et~al.} 2009, \aap, 494,
  237

\bibitem[{{Tinney} {et~al.}(2011){Tinney}, {Butler}, {Jones}, {Wittenmyer},
  {O'Toole}, {Bailey}, \& {Carter}}]{Tinney2011}
{Tinney}, C.~G., {Butler}, R.~P., {Jones}, H.~R.~A., {et~al.} 2011, \apj, 727,
  103

\bibitem[{{Uns{\"o}ld} \& {Baschek}(2002)}]{Unsoeld2002}
{Uns{\"o}ld}, A. \& {Baschek}, B. 2002, {Der neue Kosmos. Einf{\"u}hrung in die
  Astronomie und Astrophysik}

\bibitem[{{van Leeuwen}(2007)}]{vanLeeuwen07}
{van Leeuwen}, F. 2007, \aap, 474, 653

\bibitem[{{Vauclair} {et~al.}(2008){Vauclair}, {Laymand}, {Bouchy}, {Vauclair},
  {Hui Bon Hoa}, {Charpinet}, \& {Bazot}}]{Vauclair08}
{Vauclair}, S., {Laymand}, M., {Bouchy}, F., {et~al.} 2008, \aap, 482, L5

\bibitem[{{Vogt} {et~al.}(2005){Vogt}, {Butler}, {Marcy}, {Fischer}, {Henry},
  {Laughlin}, {Wright}, \& {Johnson}}]{Vogt2005}
{Vogt}, S.~S., {Butler}, R.~P., {Marcy}, G.~W., {et~al.} 2005, \apj, 632, 638

\bibitem[{{Vogt} {et~al.}(2010){Vogt}, {Wittenmyer}, {Butler}, {O'Toole},
  {Henry}, {Rivera}, {Meschiari}, {Laughlin}, {Tinney}, {Jones}, {Bailey},
  {Carter}, \& {Batygin}}]{Vogt2010}
{Vogt}, S.~S., {Wittenmyer}, R.~A., {Butler}, R.~P., {et~al.} 2010, \apj, 708,
  1366

\bibitem[{{Walker} {et~al.}(1995){Walker}, {Walker}, {Irwin}, {Larson}, {Yang},
  \& {Richardson}}]{Walker1995}
{Walker}, G.~A.~H., {Walker}, A.~R., {Irwin}, A.~W., {et~al.} 1995, Icarus,
  116, 359

\bibitem[{{Wittenmyer} {et~al.}(2006){Wittenmyer}, {Endl}, {Cochran}, {Hatzes},
  {Walker}, {Yang}, \& {Paulson}}]{Wittenmyer06}
{Wittenmyer}, R.~A., {Endl}, M., {Cochran}, W.~D., {et~al.} 2006, \aj, 132, 177

\bibitem[{{Wittenmyer} {et~al.}(2011){Wittenmyer}, {Tinney}, {O'Toole},
  {Jones}, {Butler}, {Carter}, \& {Bailey}}]{Wittenmyer2011}
{Wittenmyer}, R.~A., {Tinney}, C.~G., {O'Toole}, S.~J., {et~al.} 2011, \apj,
  727, 102

\bibitem[{{Worley} \& {Douglass}(1997)}]{Worley1997}
{Worley}, C.~E. \& {Douglass}, G.~G. 1997, \aaps, 125, 523

\bibitem[{{Wright} {et~al.}(2004){Wright}, {Marcy}, {Butler}, \&
  {Vogt}}]{Wright2004}
{Wright}, J.~T., {Marcy}, G.~W., {Butler}, R.~P., \& {Vogt}, S.~S. 2004, \apjs,
  152, 261

\bibitem[{{Wright} {et~al.}(2008){Wright}, {Marcy}, {Butler}, {Vogt}, {Henry},
  {Isaacson}, \& {Howard}}]{Wright2008}
{Wright}, J.~T., {Marcy}, G.~W., {Butler}, R.~P., {et~al.} 2008, \apjl, 683,
  L63

\bibitem[{{Zechmeister} \& {K{\"u}rster}(2009)}]{Zechmeister09}
{Zechmeister}, M. \& {K{\"u}rster}, M. 2009, \aap, 496, 577

\bibitem[{{Zechmeister} {et~al.}(2009){Zechmeister}, {K{\"u}rster}, \&
  {Endl}}]{Zechmeister09b}
{Zechmeister}, M., {K{\"u}rster}, M., \& {Endl}, M. 2009, \aap, 505, 859

\end{thebibliography}

\begin{appendix}

\section{Accompanying Tables}

\begin{table*}
\caption{\label{Tab:TestsCombinedSetsSingle}Residual rms detailed for each
instrument resulting from the joint analysis for fitting a constant,
slope, sinusoid, and Keplerian orbit.}

\setlength{\tabcolsep}{3.5pt}

\centering{}
\begin{tabular}{l|rrr|rrr|rrr|rrr|rrr}
\hline 
\hline Star & \multicolumn{3}{c|}{$N_{\mathrm{bin}}$} & \multicolumn{3}{c|}{rms$_{\mathrm{constant}}$ {[}m/s{]}} & \multicolumn{3}{c|}{rms$_{\mathrm{slope}}$ {[}m/s{]}} & \multicolumn{3}{c|}{rms$_{\mathrm{sine}}$ {[}m/s{]}} & \multicolumn{3}{c}{rms$_{\mathrm{Kep}}$ {[}m/s{]}}\\
 & LC & VLC & HARPS & LC & VLC & HARPS & LC & VLC & HARPS & LC & VLC & HARPS  & LC & VLC & HARPS \\
\hline 
$\zeta$~Tuc & 36 & 14 & 147 & 17.88 & 7.34 & 1.33 & 17.88 & 7.30 & 1.33 & 17.80 & 7.03 & 1.23 & 13.89 & 7.26 & 1.31\\
$\beta$~Hyi & 40 & 14 & 55 & 19.01 & 7.36 & 3.08 & 19.05 & 8.97 & 2.30 & 18.45 & 8.93 & 2.02 & 17.15 & 7.09 & 2.03\\
HR~209 & 25 & 12 & 16 & 15.80 & 9.81 & 8.44 & 15.73 & 9.78 & 8.14 & 14.08 & 8.33 & 5.91 & 13.71 & 9.64 & 3.60\\
$\nu$~Phe & 41 & 11 & 19 & 15.78 & 9.72 & 2.67 & 15.85 & 9.96 & 2.39 & 14.71 & 8.40 & 1.81 & 12.04 & 9.34 & 2.02\\
HR~506 & 18 & 14 & 28 & 29.88 & 18.90 & 11.19 & 29.87 & 18.19 & 11.20 & 19.18 & 14.99 & 5.71 & 18.60 & 15.95 & 5.28\\
HR~506$_{\mathrm{r}}$ & ~ & ~ & ~ & 18.77 & 14.43 & 5.71 & 20.02 & 13.52 & 4.87 & 17.90 & 15.49 & 4.17 & 16.49 & 14.98 & 3.43\\
$\tau$~Cet & 32 & 19 & 288 & 10.93 & 7.71 & 1.13 & 10.94 & 7.78 & 1.08 & 10.72 & 7.84 & 1.02 & 9.48 & 7.73 & 1.02\\
$\kappa$~For & 30 & 15 & 33 & 868.31 & 1327.23 & 674.28 & 248.61 & 62.67 & 20.79 & 30.33 & 33.00 & 3.13 & 12.20 & 17.82 & 1.90\\
$\kappa$~For$_{\mathrm{r}}$ & ~ & ~ & ~ & 11.97 & 17.57 & 1.90 & 11.96 & 17.22 & 1.94 & 11.94 & 13.17 & 1.92 & 11.66 & 9.19 & 1.79\\
$\iota$~Hor & 57 & 39 & 58 & 50.50 & 36.88 & 29.94 & 50.43 & 36.84 & 29.94 & 33.00 & 18.80 & 13.53 & 30.79 & 19.27 & 10.61\\
$\iota$~Hor$_{\mathrm{r}}$ & ~ & ~ & ~ & 30.60 & 21.61 & 9.96 & 29.46 & 21.42 & 9.87 & 30.31 & 22.82 & 7.50 & 29.27 & 21.02 & 6.23\\
$\alpha$~For & 37 & 13 & 25 & 48.79 & 26.94 & 11.69 & 31.50 & 12.13 & 2.32 & 30.84 & 12.09 & 2.35 & 31.07 & 11.43 & 2.25\\
$\alpha$~For$_{\mathrm{r}}$ & ~ & ~ & ~ & 30.83 & 12.07 & 2.35 & 30.83 & 12.06 & 2.35 & 30.92 & 10.93 & 1.48 & 25.80 & 11.45 & 1.89\\
$\zeta^1$~Ret & 8 & 14 & 26 & 14.81 & 14.46 & 11.81 & 13.74 & 16.44 & 10.51 & 14.30 & 18.47 & 7.19 & 13.87 & 15.09 & 5.90\\
$\zeta^2$~Ret & 44 & 15 & 30 & 19.78 & 9.60 & 2.19 & 20.11 & 9.54 & 1.65 & 19.23 & 8.69 & 1.73 & 15.46 & 8.09 & 2.13\\
$\epsilon$~Eri & 28 & 23 & 28 & 13.61 & 9.88 & 6.83 & 13.74 & 10.12 & 6.71 & 12.83 & 9.47 & 4.88 & 10.85 & 8.76 & 4.26\\
$\delta$~Eri & 27 & 14 & 34 & 12.44 & 7.24 & 2.41 & 12.45 & 7.26 & 2.39 & 12.03 & 6.80 & 1.91 & 11.83 & 6.28 & 1.70\\
$\alpha$~Men & 30 & 26 & 46 & 9.84 & 7.68 & 2.39 & 9.86 & 8.06 & 2.33 & 9.55 & 7.56 & 1.88 & 8.21 & 7.51 & 1.82\\
HR~2400 & 38 & 18 & 21 & 673.45 & 743.59 & 222.11 & 804.57 & 895.41 & 119.56 & 310.82 & 408.97 & 73.39 & 31.27 & 17.20 & 2.55\\
HR~2400$_{\mathrm{r}}$ & ~ & ~ & ~ & 30.40 & 16.06 & 2.55 & 29.77 & 15.65 & 2.28 & 30.01 & 15.30 & 1.80 & 22.76 & 14.97 & 1.85\\
HR~2667 & 43 & 22 & 23 & 13.28 & 6.38 & 1.39 & 13.28 & 6.33 & 1.35 & 12.68 & 6.35 & 1.09 & 10.85 & 5.81 & 1.13\\
HR~3259 & 25 & 20 & 146 & 16.23 & 8.30 & 3.82 & 16.18 & 8.11 & 3.79 & 15.81 & 7.17 & 2.84 & 15.76 & 7.18 & 2.83\\
HR~3677 & 26 & 13 & 22 & 794.73 & 1356.64 & 957.62 & 999.46 & 343.32 & 172.92 & 136.41 & 78.25 & 18.72 & 16.27 & 12.38 & 7.50\\
HR~3677$_{\mathrm{r}}$ & ~ & ~ & ~ & 16.21 & 12.35 & 7.52 & 16.19 & 12.27 & 7.53 & 16.37 & 13.72 & 4.48 & 15.17 & 12.02 & 4.28\\
HR~4523 & 21 & 17 & 62 & 13.56 & 6.07 & 2.88 & 13.56 & 6.08 & 2.88 & 13.18 & 6.08 & 2.68 & 11.44 & 5.49 & 2.30\\
HR~4979 & 33 & 18 & 100 & 11.50 & 8.86 & 3.64 & 11.15 & 9.76 & 3.56 & 10.83 & 9.07 & 3.07 & 10.69 & 9.34 & 2.82\\
$\alpha$~Cen~A & 50 & 24 & 47 & 159.66 & 247.40 & 89.60 & 39.54 & 13.23 & 6.48 & 10.21 & 6.83 & 2.03 & 9.46 & 5.75 & 1.94\\
$\alpha$~Cen~A$_{\mathrm{r}}$ & ~ & ~ & ~ & 9.46 & 5.75 & 1.94 & 9.46 & 5.75 & 1.94 & 8.93 & 5.27 & 1.61 & 7.65 & 4.98 & 1.53\\
$\alpha$~Cen~B & 43 & 18 & 21 & 204.72 & 264.67 & 164.98 & 47.19 & 12.60 & 12.67 & 11.09 & 8.91 & 3.19 & 11.05 & 6.38 & 2.88\\
$\alpha$~Cen~B$_{\mathrm{r}}$ & ~ & ~ & ~ & 11.04 & 6.34 & 2.88 & 11.02 & 6.23 & 2.89 & 10.21 & 4.52 & 2.05 & 7.40 & 5.09 & 2.17\\
GJ~570~A & 19 & 26 & 19 & 5.51 & 9.94 & 4.38 & 5.69 & 8.17 & 2.88 & 5.54 & 8.48 & 2.83 & 5.77 & 7.37 & 2.62\\
HR~6416 & 38 & 17 & 22 & 23.10 & 24.35 & 8.65 & 13.45 & 9.69 & 1.31 & 13.12 & 9.34 & 1.19 & 13.14 & 9.31 & 1.19\\
HR~6416$_{\mathrm{r}}$ & ~ & ~ & ~ & 13.12 & 9.34 & 1.19 & 13.12 & 9.32 & 1.19 & 12.48 & 8.92 & 1.00 & 11.72 & 4.81 & 1.30\\
HR~6998 & 36 & 7 & 21 & 13.84 & 9.91 & 1.23 & 13.88 & 9.65 & 1.20 & 13.39 & 9.23 & 0.95 & 11.88 & 4.56 & 1.11\\
HR~7703 & 21 & 10 & 26 & 10.03 & 7.88 & 4.60 & 8.56 & 4.88 & 1.02 & 8.51 & 4.94 & 1.03 & 8.29 & 4.78 & 0.99\\
HR~7703$_{\mathrm{r}}$ & ~ & ~ & ~ & 8.51 & 4.94 & 1.03 & 8.51 & 4.94 & 1.03 & 8.26 & 4.50 & 0.76 & 7.28 & 4.08 & 0.66\\
$\phi^2$~Pav & 47 & 20 & 21 & 30.76 & 21.05 & 3.59 & 30.57 & 20.11 & 3.56 & 29.62 & 19.97 & 2.27 & 28.54 & 17.01 & 2.75\\
HR~8323 & 13 & 10 & 81 & 10.76 & 10.64 & 3.70 & 11.02 & 11.26 & 3.41 & 10.05 & 10.69 & 2.55 & 10.15 & 10.06 & 2.53\\
$\epsilon$~Ind~A & 47 & 17 & 68 & 10.42 & 4.97 & 4.21 & 10.02 & 4.38 & 2.63 & 10.00 & 4.88 & 2.57 & 9.10 & 4.25 & 2.60\\
$\epsilon$~Ind~A$_{\mathrm{r}}$ & ~ & ~ & ~ & 10.00 & 4.84 & 2.57 & 10.00 & 4.86 & 2.57 & 10.06 & 4.45 & 2.20 & 10.06 & 4.46 & 2.08\\
HR~8501 & 44 & 15 & 19 & 42.82 & 46.85 & 20.21 & 21.17 & 9.75 & 2.29 & 21.56 & 9.29 & 2.23 & 20.70 & 10.16 & 2.10\\
HR~8501$_{\mathrm{r}}$ & ~ & ~ & ~ & 21.49 & 9.05 & 2.23 & 21.45 & 9.05 & 2.23 & 20.71 & 8.68 & 1.62 & 17.55 & 8.22 & 1.77\\
HR~8883 & 21 & 9 & 15 & 64.91 & 73.45 & 61.22 & 63.39 & 72.40 & 60.82 & 66.92 & 69.29 & 27.87 & 58.97 & 71.77 & 22.32\\
HR~8883$_{\mathrm{r}}$ & ~ & ~ & ~ & 66.05 & 67.13 & 27.87 & 66.11 & 66.15 & 27.91 & 67.24 & 63.03 & 13.04 & 59.98 & 56.04 & 10.86\\

\hline 
\end{tabular}
\end{table*}

\begin{table*}
\caption{\label{Tab:Jitter}Jitter estimation. $\sigma_{\mathrm{jit},\tau}$
is the weighted$^{\text{\ref{fn:binweight}}}$ scatter of the HARPS
data in the 2\,h bins and $\tau$ is the weighted$^{\ref{fn:bintime}}$
mean of the time coverage in these bins (which may not cover the real
jitter time scales in all cases). $\log R'_{\mathrm{HK}}$ and activity
index $S_{\mathrm{HK}}$ are median values from the HARPS data of
this work. The jitter estimation on long time scales $\sigma_{\mathrm{jit,long}}$
was derived as described in \citet{Isaacson2010} using B-V \citep{Perryman1997}
and $S_{\mathrm{HK}}$ (this work). For comparison some literature
$S_{\mathrm{HK}}$ values are given ({[}H{]} \citealt{Henry1996},
{[}G{]} \citealt{Gray2006}, {[}W{]} \citealt{Wright2004}) and the
jitter $\sigma_{\mathrm{jit,IF10}}$ (``quadrature difference of
velocity rms minus the formal internal errors'') as given by \citet{Isaacson2010}
for stars monitored with Keck.}

\centering{}
\begin{tabular}{l|cr|ccclcc}
\hline 
\hline Star & $\sigma_{\mathrm{jit},\tau}$ {[}m/s{]} & $\tau$ {[}min{]} & B-V {[}mag{]} & $\log R'_{\mathrm{HK}}$ (this work) & $S_{\mathrm{HK}}$ (this work) & \multicolumn{1}{c}{$S_{\mathrm{HK}}$} & $\sigma_{\mathrm{jit,long}}$ {[}m/s{]}  & $\sigma_{\mathrm{jit,IF10}}$ {[}m/s{]}\\
\hline 
$\zeta$~Tuc & 1.12 & 56.4 & 0.576 & -4.952 & 0.164 & 0.179~[H] & 2.63 & -~\\
$\beta$~Hyi & 2.31 & 112.7 & 0.618 & -5.065 & 0.148 & 0.158~[H] & 2.31 & -~\\
HR~209 & 0.71 & 9.9 & 0.635 & -4.666 & 0.237 & 0.279~[H] & 3.84 & -~\\
$\nu$~Phe & 1.10 & 8.8 & 0.571 & -4.992 & 0.155 & 0.161~[H] & 2.48 & -~\\
HR~506 & 1.07 & 10.3 & 0.551 & -4.773 & 0.189 & 0.217~[G] & 3.09 & -~\\
$\tau$~Cet & 0.97 & 72.7 & 0.727 & -4.952 & 0.172 & 0.171~[H] & 2.20 & 2.27\\
$\kappa$~For & 1.42 & 19.7 & 0.608 & -5.023 & 0.153 & 0.158~[G] & 2.40 & -~\\
$\iota$~Hor & 1.44 & 112.7 & 0.561 & -4.680 & 0.218 & 0.225~[H] & 3.59 & -~\\
$\alpha$~For & 2.23 & 17.2 & 0.543 & -5.009 & 0.147 & 0.162~[G] & 2.36 & -~\\
$\zeta^1$~Ret & 0.67 & 14.4 & 0.641 & -4.667 & 0.237 & 0.245~[H] & 3.83 & -~\\
$\zeta^2$~Ret & 0.81 & 12.0 & 0.600 & -4.888 & 0.174 & 0.196~[H] & 2.77 & -~\\
$\epsilon$~Eri & 0.86 & 84.8 & 0.881 & -4.468 & 0.482 & 0.483~[H] & 3.59 & 3.87\\
$\delta$~Eri & 1.20 & 9.2 & 0.915 & -5.198 & 0.134 & 0.129~[H] & 1.90 & 4.24\\
$\alpha$~Men & 0.57 & 43.4 & 0.714 & -4.959 & 0.171 & 0.175~[H] & 2.19 & -~\\
HR~2400 & 1.68 & 13.5 & 0.534 & -5.083 & 0.137 & 0.146~[G] & 2.20 & -~\\
HR~2667 & 0.93 & 12.3 & 0.624 & -4.990 & 0.160 & 0.169~[H] & 2.51 & -~\\
HR~3259 & 0.44 & 18.6 & 0.754 & -5.005 & 0.167 & 0.167~[G] & 2.17 & 2.29\\
HR~3677 & 0.69 & 13.1 & 0.827 & -4.812 & 0.222 & -~ & 2.41 & -~\\
HR~4523 & 0.79 & 36.0 & 0.664 & -4.942 & 0.170 & 0.168~[H] & 2.65 & 2.97\\
HR~4979 & 1.72 & 20.7 & 0.693 & -5.086 & 0.148 & 0.153~[H] & 2.26 & 4.24\\
$\alpha$~Cen~A & 2.19 & 115.4 & 0.710 & -5.062 & 0.152 & 0.162~[H] & 2.11 & -~\\
$\alpha$~Cen~B & 1.02 & 25.6 & 0.900 & -4.987 & 0.182 & 0.209~[H] & 2.15 & -~\\
GJ~570~A & 0.55 & 21.1 & 1.024 & -4.813 & 0.456 & 0.709~[H] & 1.60 & 1.57\\
HR~6416 & 0.63 & 37.9 & 0.764 & -4.997 & 0.170 & 0.179~[H] & 2.19 & -~\\
HR~6998 & 0.54 & 20.4 & 0.673 & -4.866 & 0.182 & 0.181~[H] & 2.86 & 2.93\\
HR~7703 & 0.64 & 13.4 & 0.868 & -4.996 & 0.180 & 0.180~[H] & 2.18 & 2.32\\
$\phi^2$~Pav & 1.54 & 8.1 & 0.544 & -4.971 & 0.154 & 0.177~[G] & 2.48 & -~\\
HR~8323 & 0.71 & 14.6 & 0.601 & -4.904 & 0.171 & 0.192~[H] & 2.72 & -~\\
$\epsilon$~Ind~A & 0.61 & 24.4 & 1.056 & -4.777 & 0.416 & 0.668~[H] & 1.60 & -~\\
HR~8501 & 0.96 & 9.2 & 0.614 & -4.937 & 0.169 & 0.181~[H] & 2.68 & -~\\
HR~8883 & 1.50 & 7.6 & 0.817 & -4.319 & 0.533 & 0.492~[W] & 3.88 & -~\\

\hline 
\end{tabular}
\end{table*}

\begin{table*}
\caption{\label{Tab:eta}Offset differences $c_{\mathrm{VLC}}-c_{\mathrm{LC}}$
{[}m/s{]} in the fits. For a reasonable model the expected value for
the offset difference is zero with an uncertainty of $\sim8$ m/s
(Sect.~\ref{Sub:LC_VLC}).}

\centering{}
\begin{tabular}{lrrrr}
\hline 
\hline Star & const & slope & sine & Kep\\
\hline 
$\zeta$~Tuc & -4.98 & -5.21 & -5.18 & -1.84\\
$\beta$~Hyi & -9.18 & -16.59 & -7.96 & -8.40\\
HR~209 & 5.87 & -0.68 & 4.87 & 5.53\\
$\nu$~Phe & -6.52 & -1.54 & -6.29 & -4.40\\
HR~506 & -9.02 & -7.02 & -6.99 & -7.71\\
$\tau$~Cet & -0.73 & 0.22 & -0.63 & 0.04\\
$\kappa$~For & -3010.58 & 252.22 & -26.30 & -13.50\\
$\iota$~Hor & 8.67 & 9.09 & -3.99 & -0.86\\
$\alpha$~For & -40.37 & -5.83 & -1.54 & -3.16\\
$\zeta^1$~Ret & -6.12 & 2.80 & -7.80 & -6.67\\
$\zeta^2$~Ret & 1.02 & -2.89 & 0.65 & 0.47\\
$\epsilon$~Eri & 3.06 & 0.34 & 3.59 & 1.37\\
$\delta$~Eri & 6.41 & 6.90 & 5.87 & 6.84\\
$\alpha$~Men & 0.49 & 1.80 & 0.28 & 1.31\\
HR~2400 & 1426.64 & 1830.13 & 529.31 & -18.71\\
HR~2667 & -0.37 & 1.04 & -0.28 & -1.53\\
HR~3259 & 8.42 & 5.30 & 8.18 & 8.02\\
HR~3677 & 2312.74 & -1760.01 & -109.17 & -5.35\\
HR~4523 & 0.26 & 0.30 & 0.15 & 2.05\\
HR~4979 & -11.45 & -7.81 & -11.88 & -11.90\\
$\alpha$~Cen~A & 661.11 & -83.26 & -10.67 & -0.04\\
$\alpha$~Cen~B & -860.86 & 90.80 & 12.11 & -6.74\\
GJ~570~A & 0.01 & 8.91 & 5.48 & -9.99\\
HR~6416 & 49.10 & 3.08 & -2.41 & -2.66\\
HR~6998 & -6.47 & -4.22 & -5.99 & -4.36\\
HR~7703 & 19.57 & 2.32 & 0.82 & 1.96\\
$\phi^2$~Pav & -12.30 & -9.55 & -11.92 & -12.10\\
HR~8323 & -3.18 & 0.86 & -2.56 & -3.35\\
$\epsilon$~Ind~A & 9.17 & -3.94 & -5.37 & -4.87\\
HR~8501 & 53.87 & -7.32 & -4.38 & -4.34\\
HR~8883 & -13.17 & -2.41 & -12.89 & -17.12\\

\hline 
\end{tabular}
\end{table*}

\begin{table*}
\caption{\label{Tab:rhk}Correlations of RV with $\log R'_{\mathrm{HK}}$ and
BIS for the HARPS data. Listed for both are the mean, the rms, the
slope, the correlation coefficient $r$ and its false alarm probability.
Significant correlation coefficients with $\mathrm{FAP}<0.01$ are
in bold font.}

\centering{}
\begin{tabular}{l|rrr@{$\pm$}rrl|rrr@{$\pm$}rrl}
\hline 
\hline Star & $\langle\log R'_{\mathrm{HK}}\rangle$ & $\mathrm{rms}_{\log\mathrm{RHK}}$ & \multicolumn{2}{c}{slope} & $r_{\mathrm{RV-\log RHK}}$ & FAP & $\langle\mathrm{BIS}\rangle$ & $\mathrm{rms}_{\mathrm{BIS}}$ & \multicolumn{2}{c}{slope} & $r_{\mathrm{RV-BIS}}$ & FAP\\
 & {[}dex{]} & {[}dex{]} & \multicolumn{2}{c}{ {[}m/s/dex{]}} &  &  & {[}m/s{]} & {[}m/s{]} & \multicolumn{2}{c}{{[}m/s/(m/s){]}} &  & \\
\hline 
$\zeta$~Tuc & -4.954 & 0.008 & -20 & 14 & -0.12 & 0.16 & -2.3 & 0.9 & -0.15 & 0.12 & -0.11 & 0.2\\
$\beta$~Hyi & -5.061 & 0.019 & 105 & 17 & \bf{0.64} & $1.5\cdot10^{-7}$ & 27.5 & 1.8 & 0.49 & 0.22 & 0.29 & 0.032\\
HR~209 & -4.660 & 0.029 & 241 & 44 & \bf{0.82} & $8.6\cdot10^{-5}$ & -0.5 & 5.5 & -0.08 & 0.41 & -0.05 & 0.85\\
$\nu$~Phe & -4.986 & 0.014 & 89 & 42 & 0.46 & 0.05 & 21.3 & 1.3 & 0.60 & 0.47 & 0.30 & 0.22\\
HR~506 & -4.775 & 0.017 & 286 & 112 & 0.45 & 0.017 & 27.9 & 3.2 & 1.36 & 0.63 & 0.39 & 0.042\\
HR~506$_\mathrm{r}$ & ~ & ~ & 206 & 50 & \bf{0.63} & 0.00033 & ~ & ~ & -0.56 & 0.33 & -0.31 & 0.1\\
$\tau$~Cet & -4.953 & 0.005 & -24 & 15 & -0.10 & 0.099 & -46.3 & 0.5 & 0.48 & 0.13 & \bf{0.22} & 0.00018\\
$\kappa$~For & -5.018 & 0.032 & 6126 & 3609 & 0.29 & 0.1 & 27.8 & 1.8 & 88.05 & 63.57 & 0.24 & 0.18\\
$\kappa$~For$_\mathrm{r}$ & ~ & ~ & -15 & 10 & -0.25 & 0.17 & ~ & ~ & 0.36 & 0.17 & 0.35 & 0.048\\
$\iota$~Hor & -4.670 & 0.030 & 475 & 119 & \bf{0.47} & 0.00019 & 24.6 & 9.1 & -1.75 & 0.37 & \bf{-0.53} & $1.6\cdot10^{-5}$\\
$\iota$~Hor$_\mathrm{r}$ & ~ & ~ & 168 & 42 & \bf{0.47} & 0.00019 & ~ & ~ & -0.65 & 0.13 & \bf{-0.56} & $5.4\cdot10^{-6}$\\
$\alpha$~For & -5.009 & 0.007 & -164 & 350 & -0.10 & 0.64 & 67.1 & 2.3 & -1.50 & 1.03 & -0.29 & 0.16\\
$\alpha$~For$_\mathrm{r}$ & ~ & ~ & -37 & 70 & -0.11 & 0.6 & ~ & ~ & 0.09 & 0.21 & 0.09 & 0.68\\
$\zeta^1$~Ret & -4.662 & 0.048 & 219 & 24 & \bf{0.88} & $2.2\cdot10^{-9}$ & -7.1 & 6.9 & 0.59 & 0.33 & 0.35 & 0.084\\
$\zeta^2$~Ret & -4.892 & 0.015 & 66 & 24 & 0.46 & 0.011 & -14.0 & 1.4 & 0.34 & 0.29 & 0.22 & 0.25\\
$\epsilon$~Eri & -4.474 & 0.024 & 19 & 55 & 0.07 & 0.74 & 27.0 & 6.8 & -0.55 & 0.17 & \bf{-0.54} & 0.0029\\
$\delta$~Eri & -5.201 & 0.008 & -1 & 52 & -0.00 & 0.99 & 14.4 & 0.9 & -0.05 & 0.45 & -0.02 & 0.92\\
$\alpha$~Men & -4.959 & 0.026 & 57 & 11 & \bf{0.61} & $5.8\cdot10^{-6}$ & -35.1 & 2.5 & 0.51 & 0.12 & \bf{0.53} & 0.00013\\
HR~2400 & -5.086 & 0.017 & 2197 & 3039 & 0.16 & 0.48 & 44.6 & 3.3 & -17.42 & 14.96 & -0.26 & 0.26\\
HR~2400$_\mathrm{r}$ & ~ & ~ & -54 & 33 & -0.35 & 0.12 & ~ & ~ & -0.19 & 0.17 & -0.25 & 0.27\\
HR~2667 & -4.993 & 0.007 & -60 & 38 & -0.32 & 0.13 & -9.0 & 0.9 & 0.56 & 0.32 & 0.35 & 0.099\\
HR~3259 & -5.005 & 0.010 & -3 & 32 & -0.01 & 0.93 & -7.3 & 1.1 & -0.14 & 0.29 & -0.04 & 0.64\\
HR~3677 & -4.812 & 0.007 & -39685 & 28001 & -0.30 & 0.17 & 60.6 & 2.9 & -101.83 & 69.70 & -0.31 & 0.16\\
HR~3677$_\mathrm{r}$ & ~ & ~ & 104 & 230 & 0.10 & 0.66 & ~ & ~ & -0.33 & 0.57 & -0.13 & 0.57\\
HR~4523 & -4.942 & 0.010 & 29 & 39 & 0.10 & 0.46 & -33.0 & 1.2 & 0.59 & 0.31 & 0.24 & 0.059\\
HR~4979 & -5.081 & 0.025 & 81 & 12 & \bf{0.55} & $2.8\cdot10^{-9}$ & 8.8 & 1.7 & 1.26 & 0.17 & \bf{0.60} & $5.2\cdot10^{-11}$\\
$\alpha$~Cen~A & -5.063 & 0.007 & -3542 & 1958 & -0.26 & 0.077 & -16.9 & 0.7 & -38.38 & 19.36 & -0.28 & 0.054\\
$\alpha$~Cen~A$_\mathrm{r}$ & ~ & ~ & -131 & 39 & \bf{-0.44} & 0.0018 & ~ & ~ & 0.53 & 0.43 & 0.18 & 0.22\\
$\alpha$~Cen~B & -4.969 & 0.045 & 3218 & 406 & \bf{0.88} & $1.9\cdot10^{-7}$ & 0.6 & 2.4 & 58.38 & 8.25 & \bf{0.85} & $9.9\cdot10^{-7}$\\
$\alpha$~Cen~B$_\mathrm{r}$ & ~ & ~ & 17 & 14 & 0.27 & 0.24 & ~ & ~ & 0.36 & 0.26 & 0.30 & 0.18\\
GJ~570~A & -4.812 & 0.029 & 79 & 32 & 0.52 & 0.022 & 21.4 & 2.5 & 1.14 & 0.32 & \bf{0.65} & 0.0024\\
HR~6416 & -4.999 & 0.008 & 511 & 217 & 0.47 & 0.029 & -43.5 & 0.6 & -4.83 & 3.20 & -0.32 & 0.15\\
HR~6416$_\mathrm{r}$ & ~ & ~ & -33 & 33 & -0.22 & 0.33 & ~ & ~ & 0.07 & 0.46 & 0.03 & 0.88\\
HR~6998 & -4.874 & 0.020 & -26 & 13 & -0.41 & 0.062 & -42.1 & 1.0 & -0.26 & 0.26 & -0.22 & 0.34\\
HR~7703 & -4.994 & 0.016 & -166 & 49 & \bf{-0.57} & 0.0025 & 1.4 & 1.4 & -0.42 & 0.67 & -0.13 & 0.54\\
HR~7703$_\mathrm{r}$ & ~ & ~ & 10 & 13 & 0.16 & 0.44 & ~ & ~ & 0.08 & 0.15 & 0.11 & 0.59\\
$\phi^2$~Pav & -4.969 & 0.020 & 53 & 39 & 0.30 & 0.19 & 49.0 & 3.2 & -0.45 & 0.24 & -0.40 & 0.076\\
HR~8323 & -4.900 & 0.028 & 98 & 10 & \bf{0.74} & $1.8\cdot10^{-15}$ & -9.3 & 2.1 & 1.31 & 0.13 & \bf{0.76} & $3.5\cdot10^{-16}$\\
$\epsilon$~Ind~A & -4.774 & 0.032 & 79 & 13 & \bf{0.60} & $7.7\cdot10^{-8}$ & 20.9 & 2.5 & 0.10 & 0.20 & 0.06 & 0.62\\
$\epsilon$~Ind~A$_\mathrm{r}$ & ~ & ~ & 19 & 10 & 0.23 & 0.059 & ~ & ~ & -0.23 & 0.12 & -0.22 & 0.067\\
HR~8501 & -4.936 & 0.011 & -764 & 426 & -0.40 & 0.09 & -11.3 & 1.9 & -6.87 & 1.91 & \bf{-0.66} & 0.0022\\
HR~8501$_\mathrm{r}$ & ~ & ~ & 65 & 49 & 0.31 & 0.2 & ~ & ~ & 0.54 & 0.25 & 0.47 & 0.04\\
HR~8883 & -4.313 & 0.036 & 605 & 445 & 0.35 & 0.2 & -184.8 & 130.1 & -0.45 & 0.04 & \bf{-0.96} & $2.8\cdot10^{-8}$\\
HR~8883$_\mathrm{r}$ & ~ & ~ & 191 & 210 & 0.24 & 0.38 & ~ & ~ & -0.10 & 0.05 & -0.48 & 0.074\\

\hline 
\end{tabular}
\end{table*}

\begin{table*}
\caption{\label{Tab:fwhm}Correlations of RV with FWHM for the HARPS data.
Listed are the mean, the rms, the slope, the correlation coefficient
$r$ and its false alarm probability. Significant correlation coefficients
with $\mathrm{FAP}<0.01$ are in bold font.}

\centering{}
\begin{tabular}{l|rrr@{$\pm$}rrl}
\hline 
\hline Star & $\langle\mathrm{FWHM}\rangle$ & $\mathrm{rms}_{\mathrm{FWHM}}$ & \multicolumn{2}{c}{slope} & $r_{\mathrm{RV-FWHM}}$ & FAP\\
 & {[}km/s{]} & {[}km/s{]} & \multicolumn{2}{c}{ {[}m/s/(km/s){]}} &  & \\
\hline 
$\zeta$~Tuc & 7.283 & 0.004 & 32 & 27 & 0.10 & 0.23\\
$\beta$~Hyi & 8.032 & 0.005 & 266 & 82 & \bf{0.41} & 0.0021\\
HR~209 & 7.503 & 0.013 & 447 & 125 & \bf{0.69} & 0.003\\
$\nu$~Phe & 8.540 & 0.005 & 102 & 128 & 0.19 & 0.44\\
HR~506 & 9.374 & 0.011 & 250 & 200 & 0.24 & 0.22\\
HR~506$_\mathrm{r}$ & ~ & ~ & 187 & 99 & 0.35 & 0.07\\
$\tau$~Cet & 6.279 & 0.005 & -62 & 13 & \bf{-0.28} & $1.9\cdot10^{-6}$\\
$\kappa$~For & 7.844 & 0.004 & -71209 & 26361 & -0.44 & 0.011\\
$\kappa$~For$_\mathrm{r}$ & ~ & ~ & -56 & 82 & -0.12 & 0.5\\
$\iota$~Hor & 9.905 & 0.020 & 353 & 198 & 0.23 & 0.08\\
$\iota$~Hor$_\mathrm{r}$ & ~ & ~ & 207 & 67 & \bf{0.38} & 0.003\\
$\alpha$~For & 8.992 & 0.005 & -113 & 465 & -0.05 & 0.81\\
$\alpha$~For$_\mathrm{r}$ & ~ & ~ & -40 & 93 & -0.09 & 0.68\\
$\zeta^1$~Ret & 7.084 & 0.026 & 383 & 51 & \bf{0.84} & $1.1\cdot10^{-7}$\\
$\zeta^2$~Ret & 7.026 & 0.005 & 315 & 58 & \bf{0.72} & $8.8\cdot10^{-6}$\\
$\epsilon$~Eri & 6.478 & 0.029 & 33 & 46 & 0.14 & 0.48\\
$\delta$~Eri & 6.489 & 0.003 & 138 & 127 & 0.19 & 0.29\\
$\alpha$~Men & 6.964 & 0.008 & 186 & 36 & \bf{0.61} & $6.1\cdot10^{-6}$\\
HR~2400 & 10.288 & 0.007 & -20630 & 5269 & \bf{-0.67} & 0.00093\\
HR~2400$_\mathrm{r}$ & ~ & ~ & -149 & 74 & -0.42 & 0.057\\
HR~2667 & 6.893 & 0.004 & 43 & 76 & 0.12 & 0.58\\
HR~3259 & 6.152 & 0.004 & -55 & 80 & -0.06 & 0.49\\
HR~3677 & 8.617 & 0.009 & 66798 & 17643 & \bf{0.65} & 0.0012\\
HR~3677$_\mathrm{r}$ & ~ & ~ & 188 & 177 & 0.23 & 0.3\\
HR~4523 & 6.649 & 0.004 & 128 & 88 & 0.19 & 0.15\\
HR~4979 & 7.516 & 0.005 & 71 & 67 & 0.11 & 0.29\\
$\alpha$~Cen~A & 7.465 & 0.006 & -6760 & 1860 & \bf{-0.48} & 0.00071\\
$\alpha$~Cen~A$_\mathrm{r}$ & ~ & ~ & -69 & 45 & -0.22 & 0.13\\
$\alpha$~Cen~B & 6.282 & 0.016 & 6323 & 1872 & \bf{0.61} & 0.0032\\
$\alpha$~Cen~B$_\mathrm{r}$ & ~ & ~ & 41 & 40 & 0.23 & 0.32\\
GJ~570~A & 6.202 & 0.021 & 136 & 38 & \bf{0.65} & 0.0024\\
HR~6416 & 6.413 & 0.004 & 968 & 393 & 0.48 & 0.023\\
HR~6416$_\mathrm{r}$ & ~ & ~ & 66 & 60 & 0.24 & 0.28\\
HR~6998 & 6.492 & 0.005 & -87 & 57 & -0.33 & 0.14\\
HR~7703 & 5.773 & 0.006 & 293 & 140 & 0.39 & 0.047\\
HR~7703$_\mathrm{r}$ & ~ & ~ & 41 & 33 & 0.25 & 0.22\\
$\phi^2$~Pav & 9.906 & 0.007 & -47 & 112 & -0.10 & 0.68\\
HR~8323 & 7.275 & 0.005 & 325 & 69 & \bf{0.47} & $1.2\cdot10^{-5}$\\
$\epsilon$~Ind~A & 6.167 & 0.016 & 177 & 25 & \bf{0.66} & $8.8\cdot10^{-10}$\\
$\epsilon$~Ind~A$_\mathrm{r}$ & ~ & ~ & 46 & 19 & 0.28 & 0.02\\
HR~8501 & 7.034 & 0.005 & 1969 & 778 & 0.52 & 0.021\\
HR~8501$_\mathrm{r}$ & ~ & ~ & 218 & 85 & 0.53 & 0.02\\
HR~8883 & 17.317 & 0.236 & -53 & 70 & -0.20 & 0.46\\
HR~8883$_\mathrm{r}$ & ~ & ~ & 23 & 32 & 0.19 & 0.49\\

\hline 
\end{tabular}
\end{table*}
\end{appendix}

\Online

\begin{appendix} 

\section{Online figures: HARPS time series and  Correlation plots for RVs,
$\log R'_{\mathrm{HK}}$ and BIS}

\newcommand{\figmacro}[2]{\begin{figure*}
\includegraphics[width=0.5\textwidth]{onlFig/#1.ps}

\includegraphics[width=0.5\textwidth]{onlFig/#1.rhk.ps}
\includegraphics[width=0.5\textwidth]{onlFig/#1.rv_rhk.ps}
\includegraphics[width=0.5\textwidth]{onlFig/#1.bis.ps}
\includegraphics[width=0.5\textwidth]{onlFig/#1.rv_bis.ps}
\includegraphics[width=0.5\textwidth]{onlFig/#1.fwhm.ps}
\includegraphics[width=0.5\textwidth]{onlFig/#1.rv_fwhm.ps}
\caption{\label{onlFig:#1}Activity indictors and correlations with HARPS RVs for #2 (2\,h binned data, secular acceleration substracted, see Sect.~\ref{Sub:Jitter}).}
\end{figure*}
\clearpage
}
\figmacro{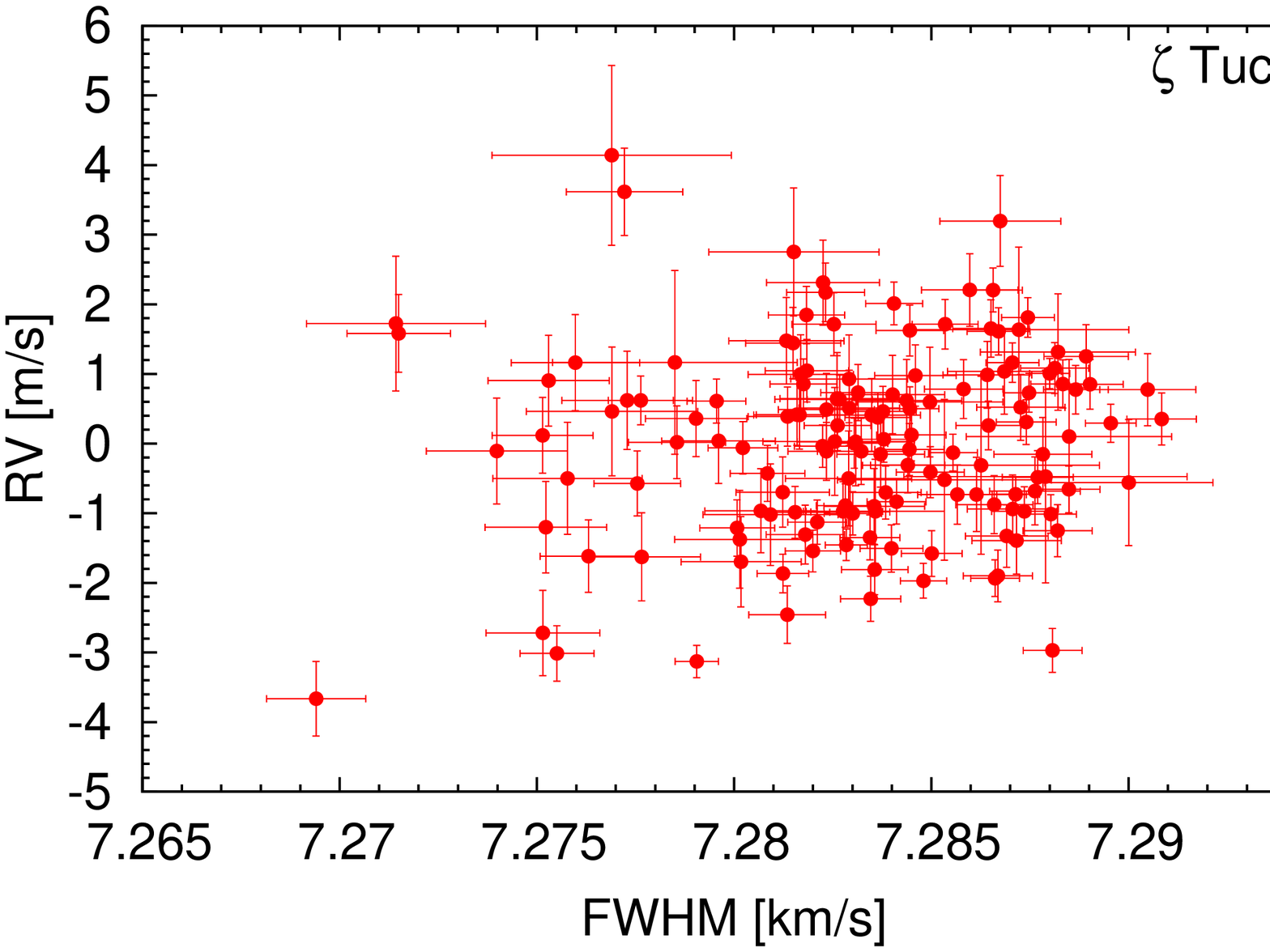}{$\zeta$~Tuc}
\figmacro{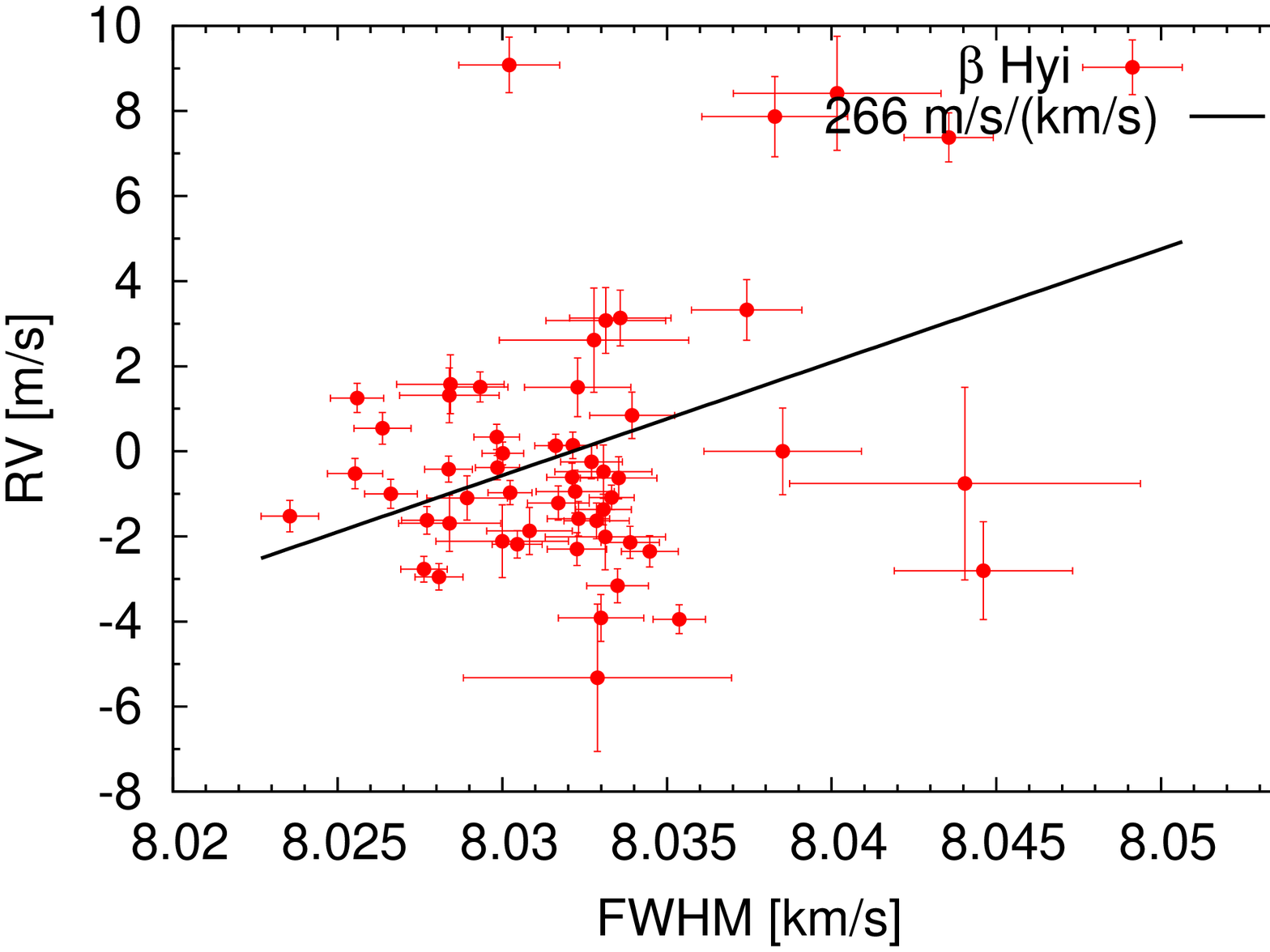}{$\beta$~Hyi}
\figmacro{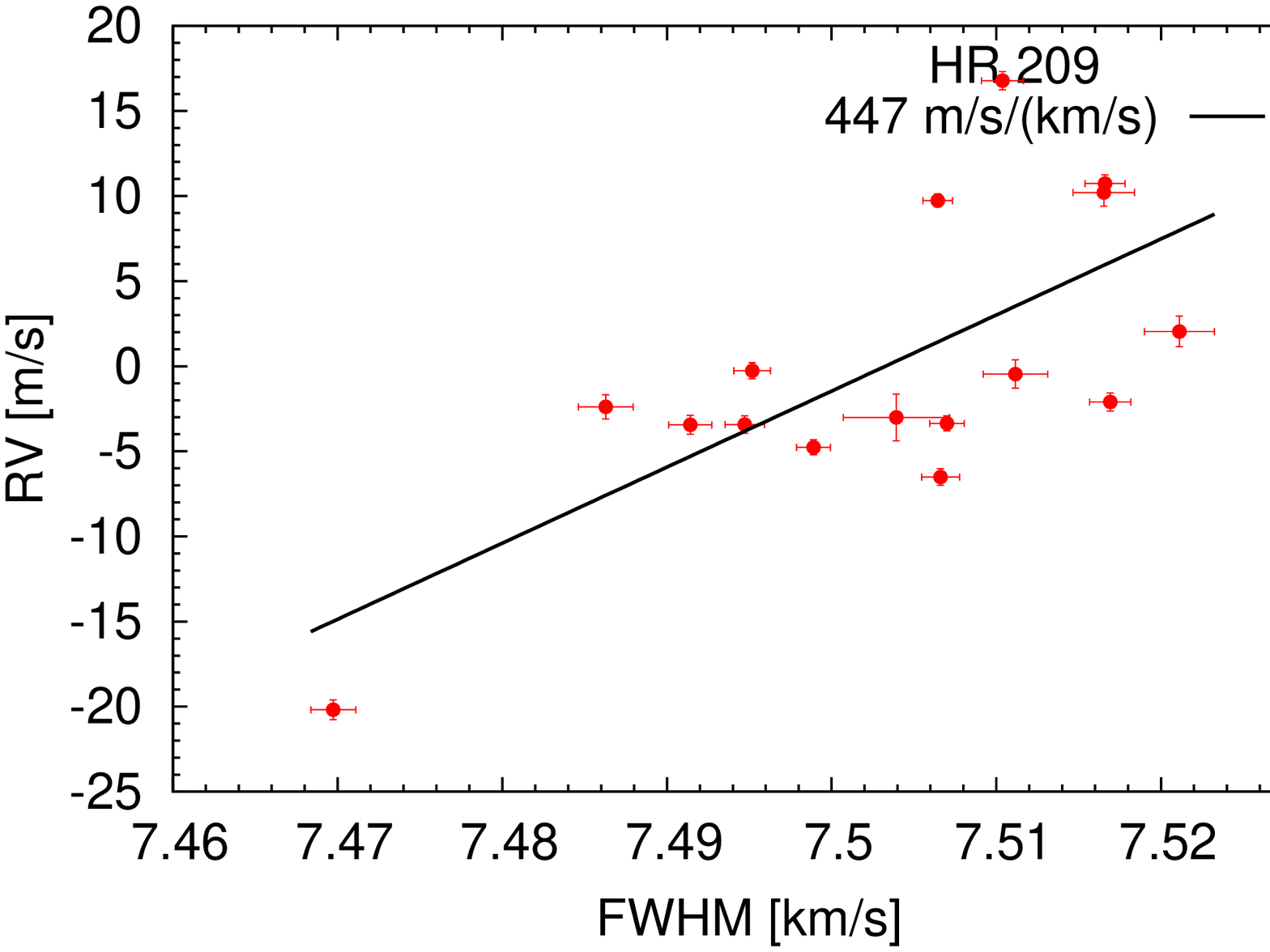}{HR 209}
\figmacro{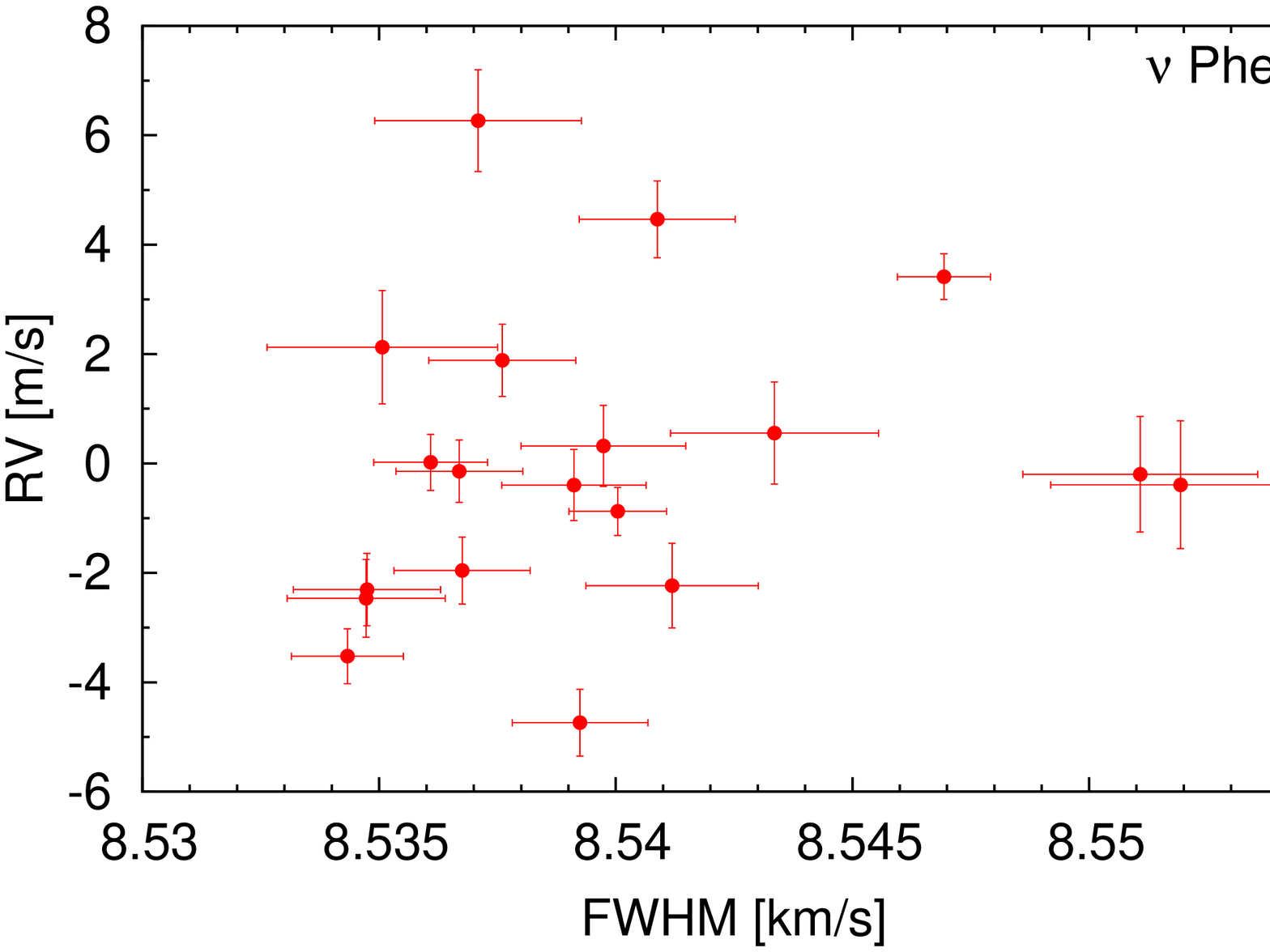}{$\nu$~Phe}
\figmacro{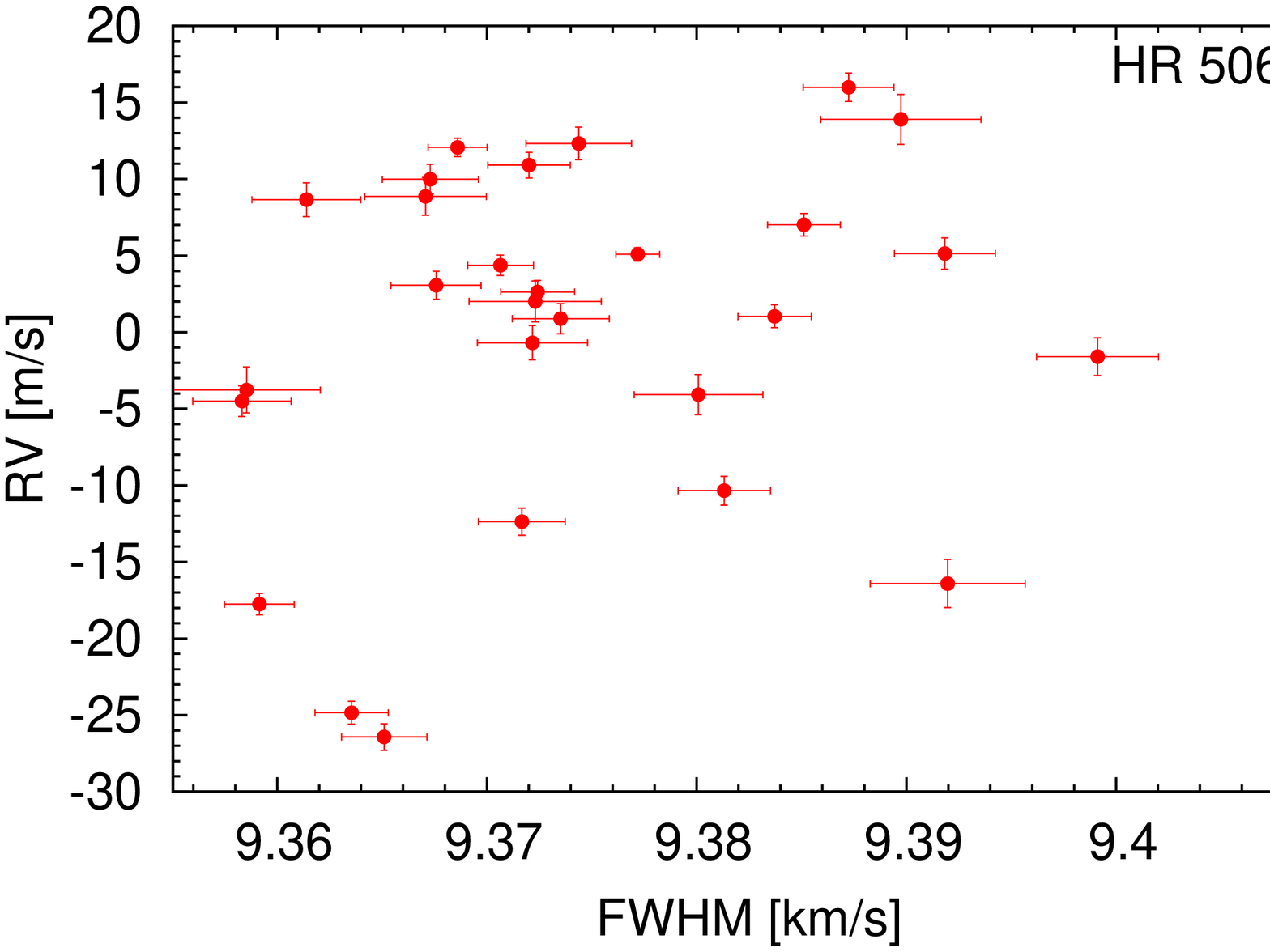}{HR 506}
\figmacro{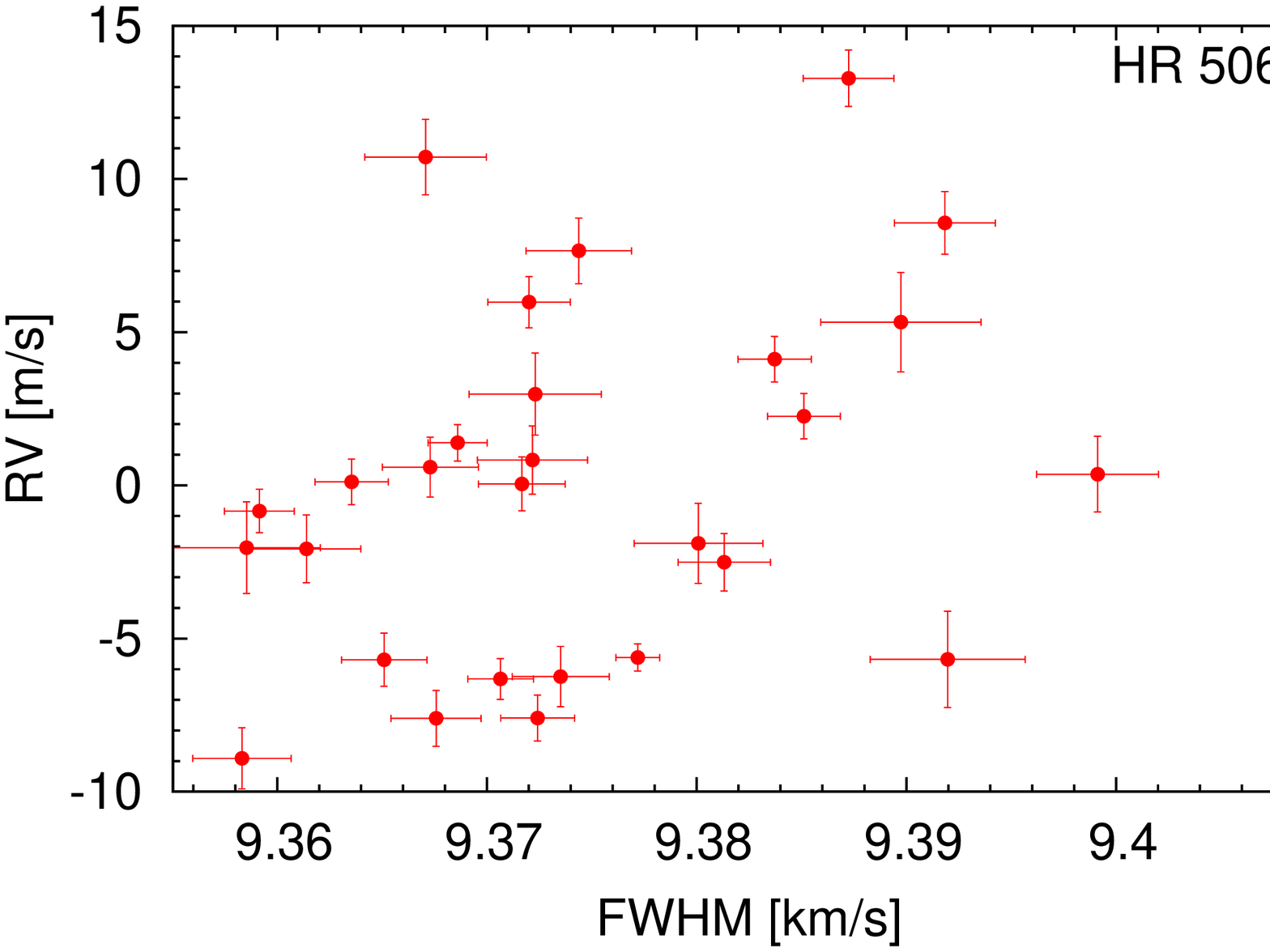}{the HR~506 residuals}
\figmacro{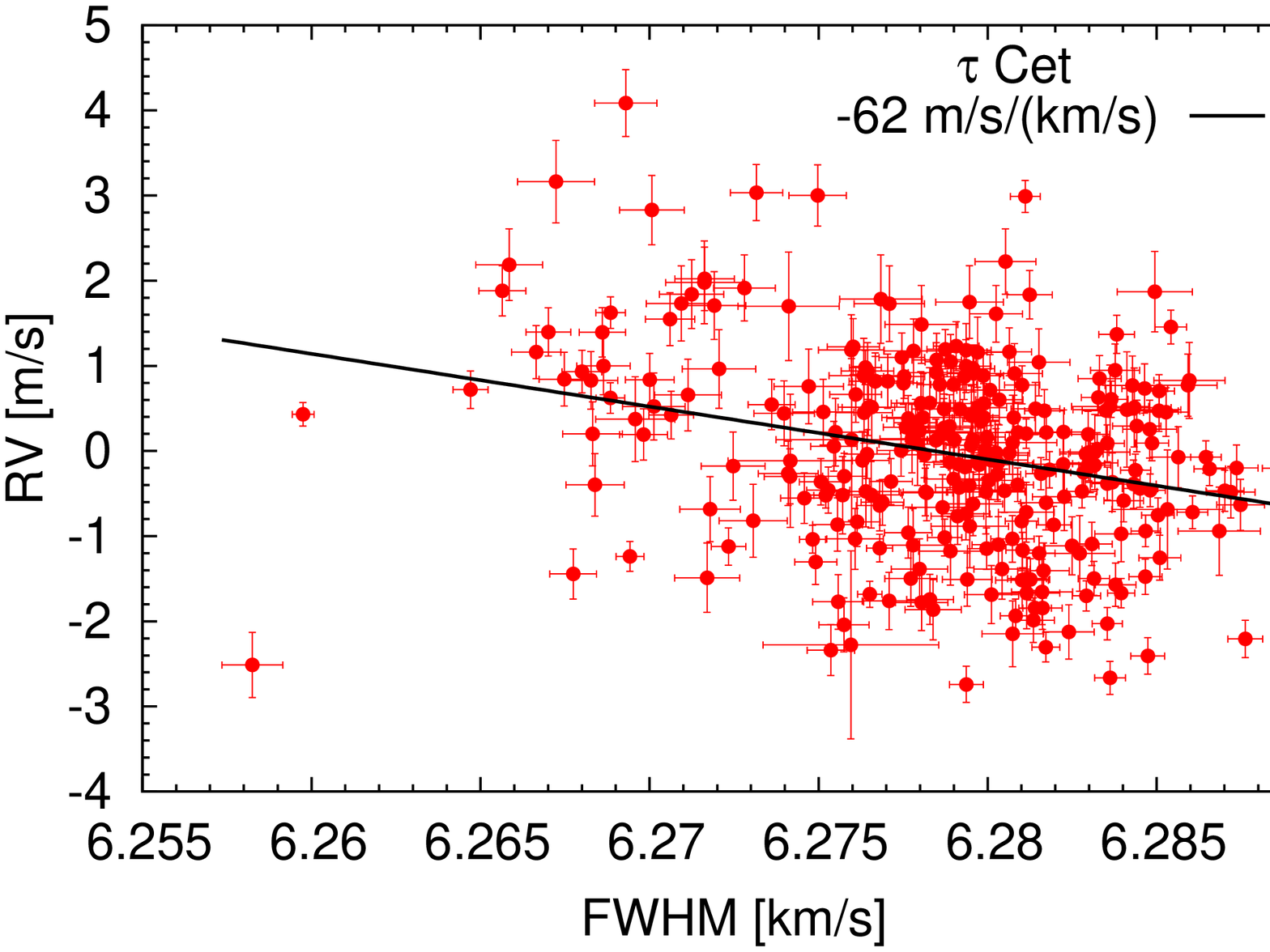}{$\tau$~Cet}
\figmacro{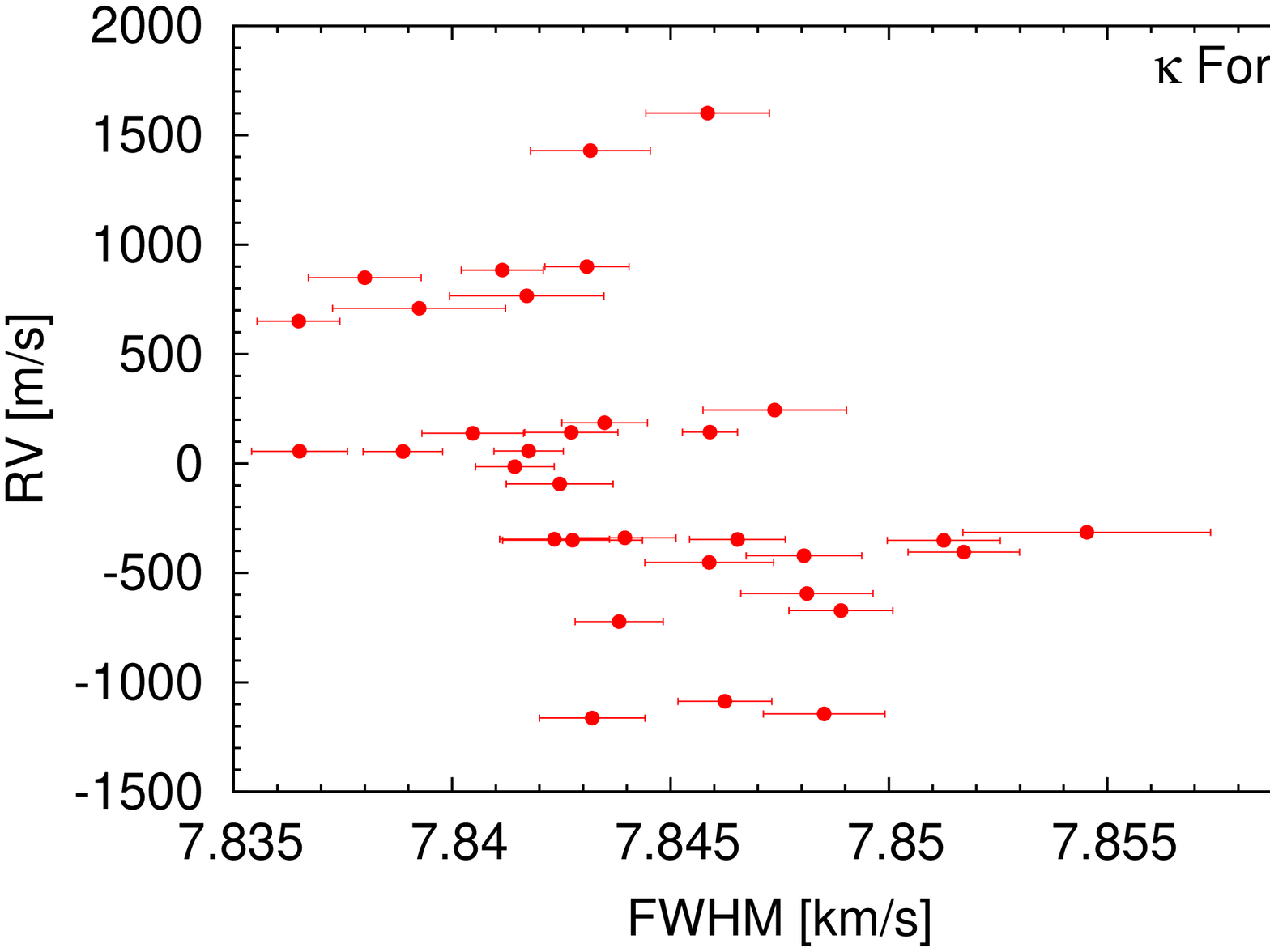}{$\kappa$~For}
\figmacro{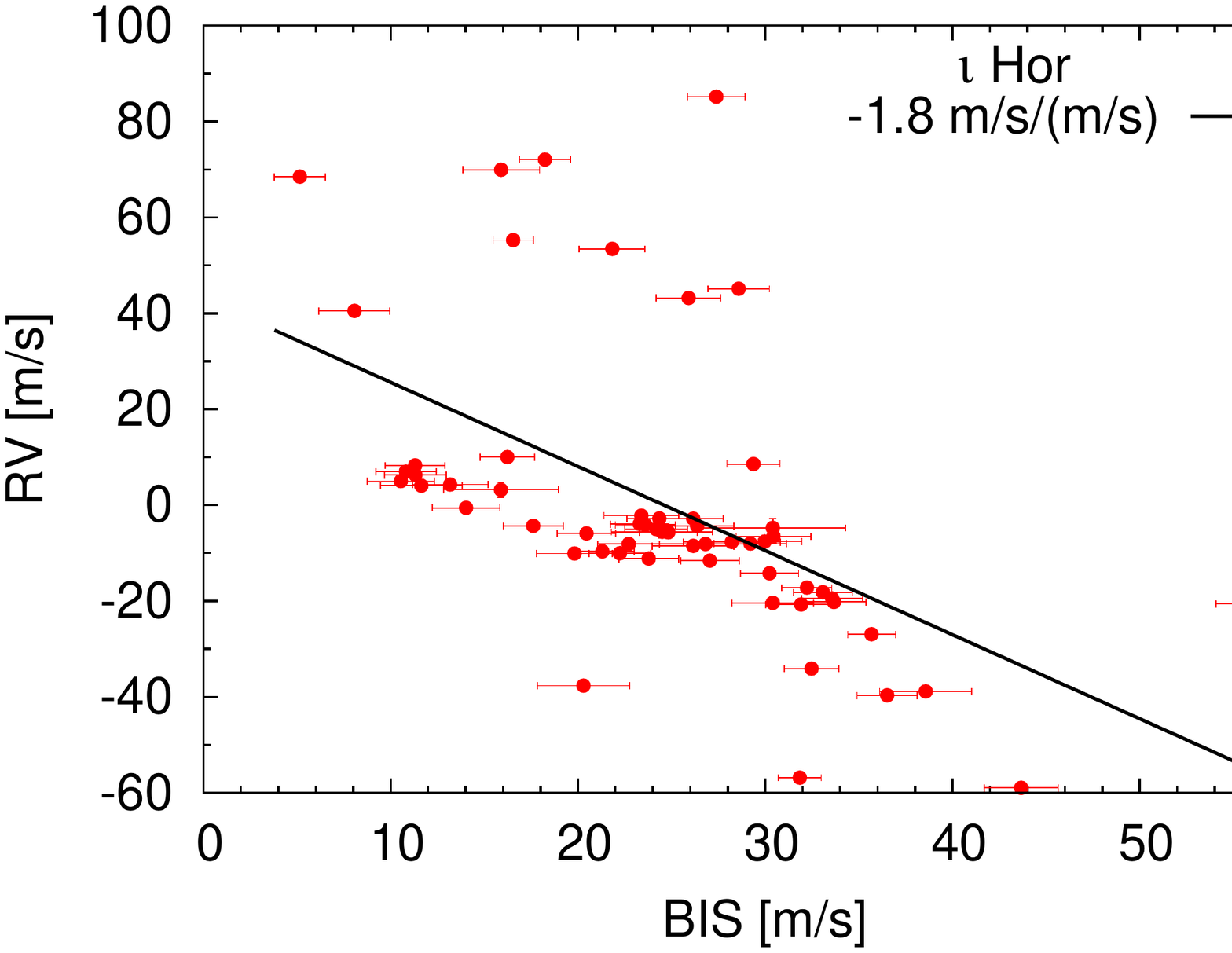}{$\iota$~Hor}
\figmacro{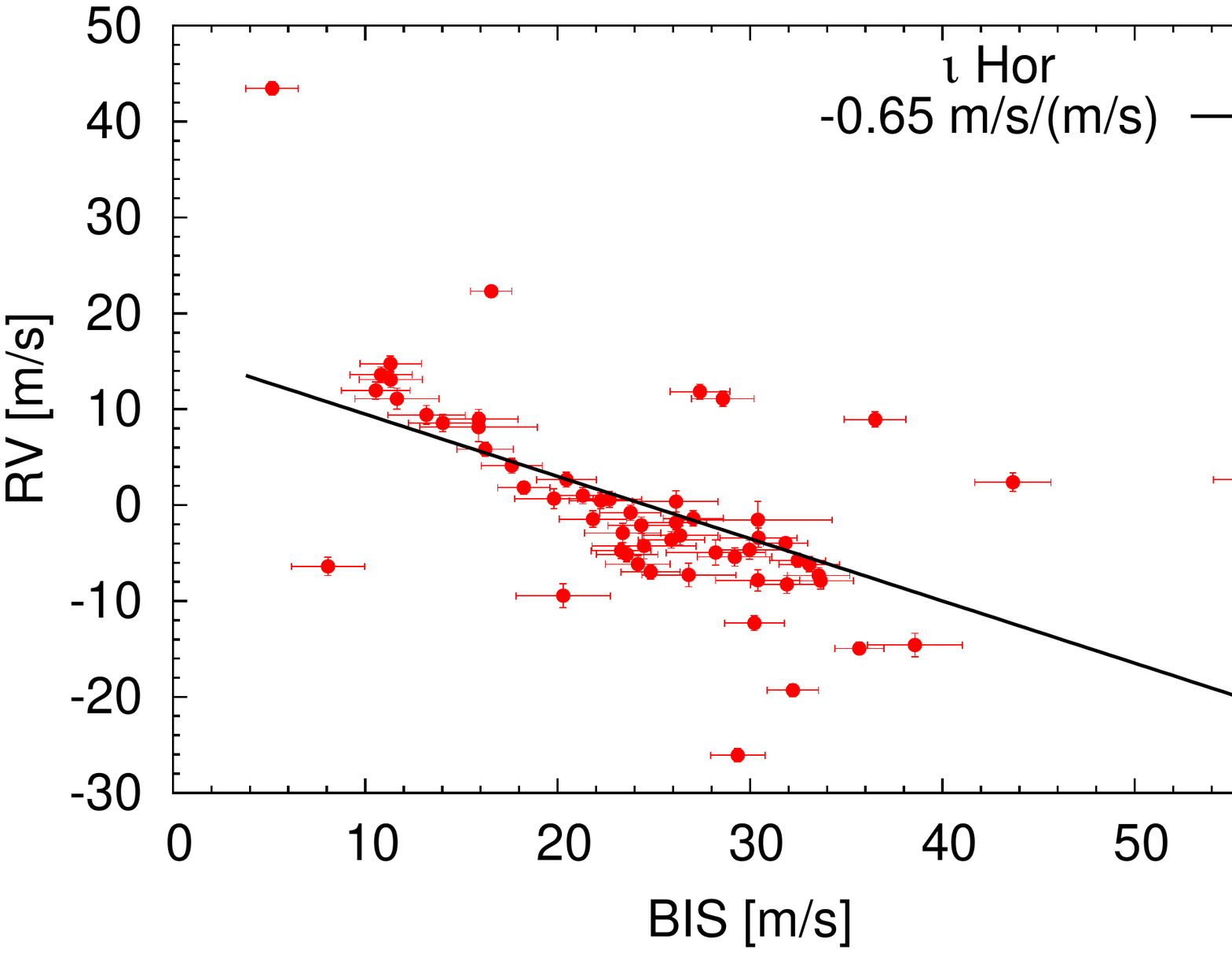}{the $\iota$~Hor residuals}
\figmacro{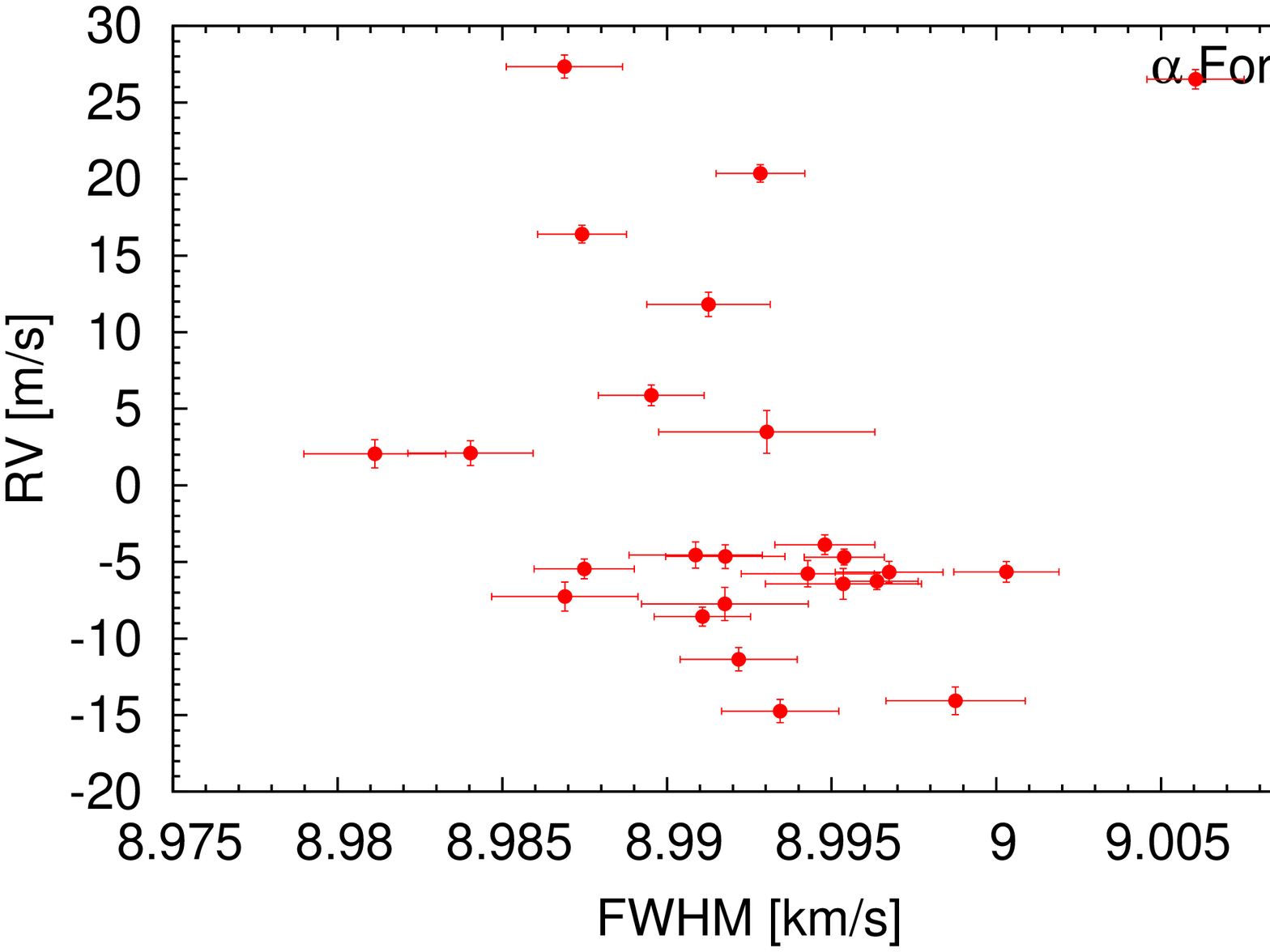}{$\alpha$~For}
\figmacro{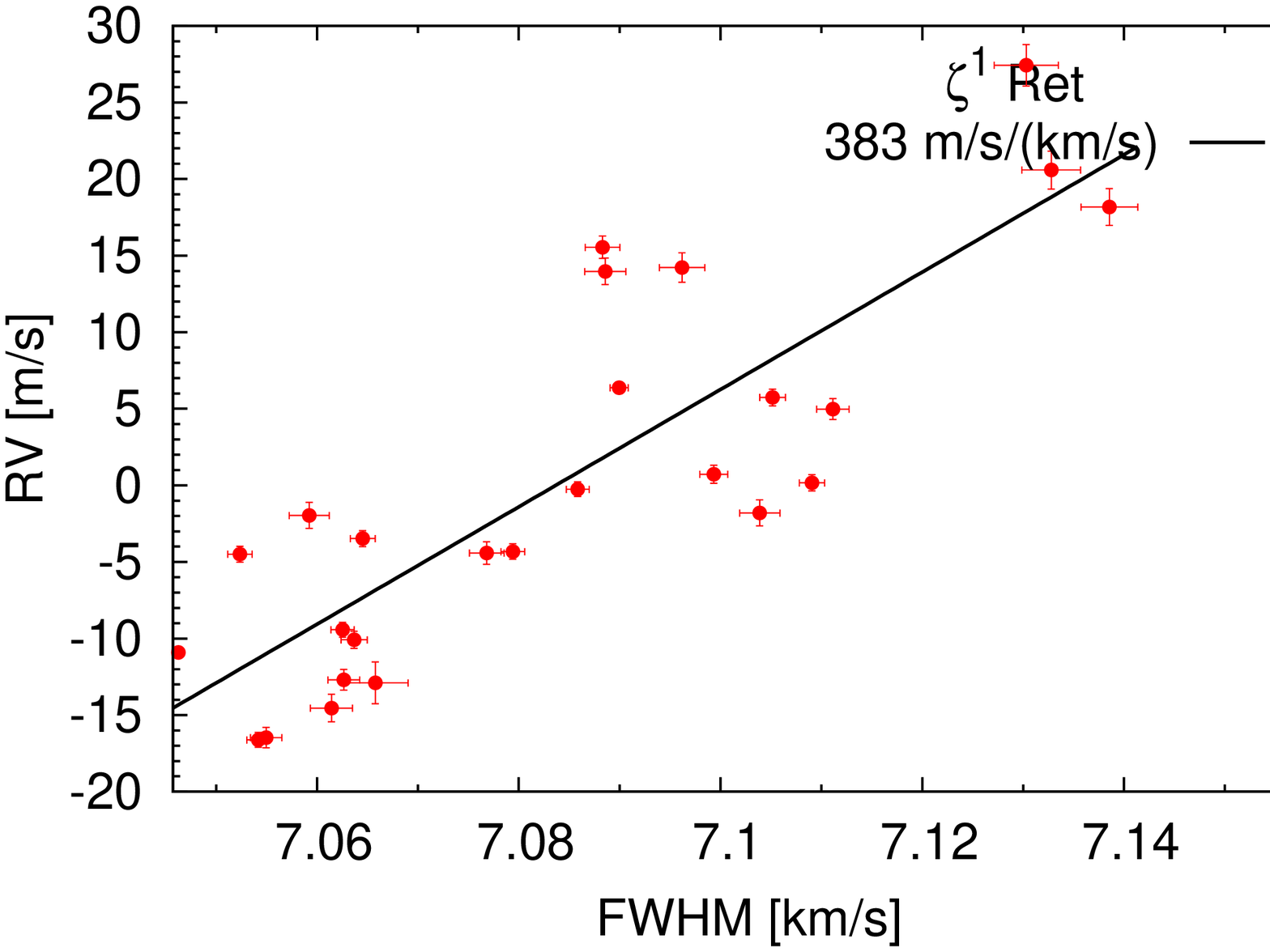}{$\zeta^1$~Ret}
\figmacro{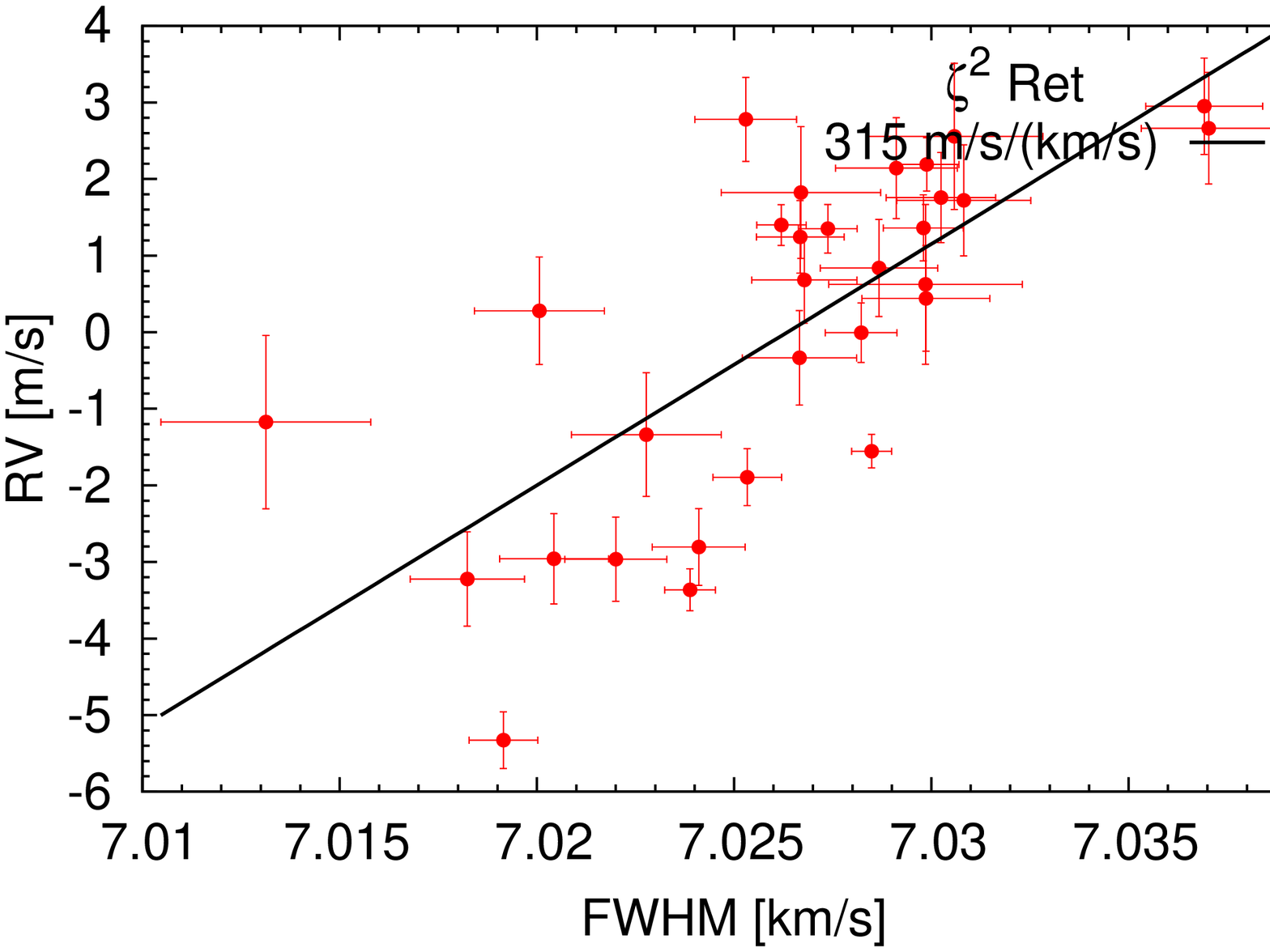}{$\zeta^2$~Ret}
\figmacro{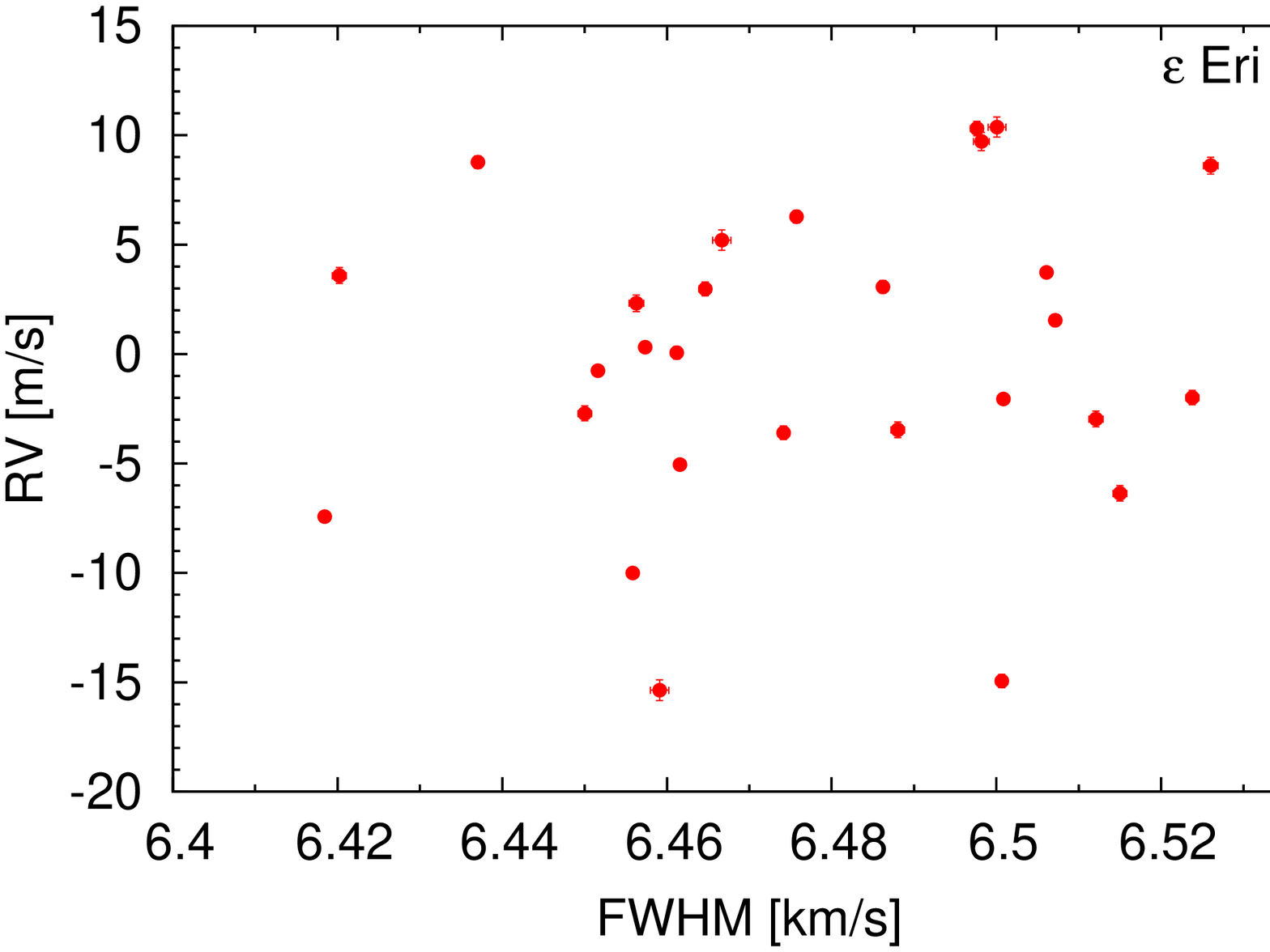}{$\epsilon$~Eri}
\figmacro{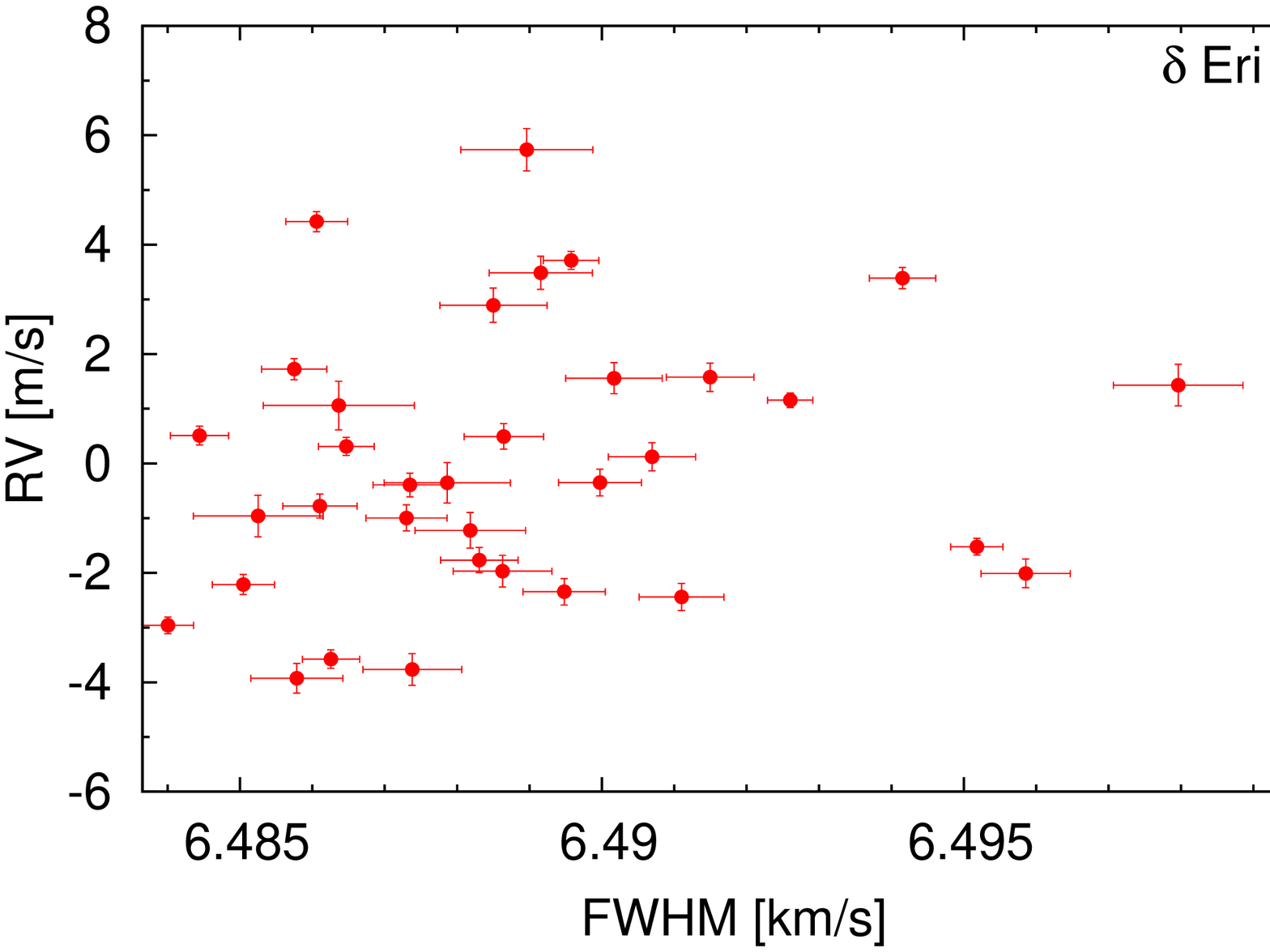}{$\delta$~Eri}
\figmacro{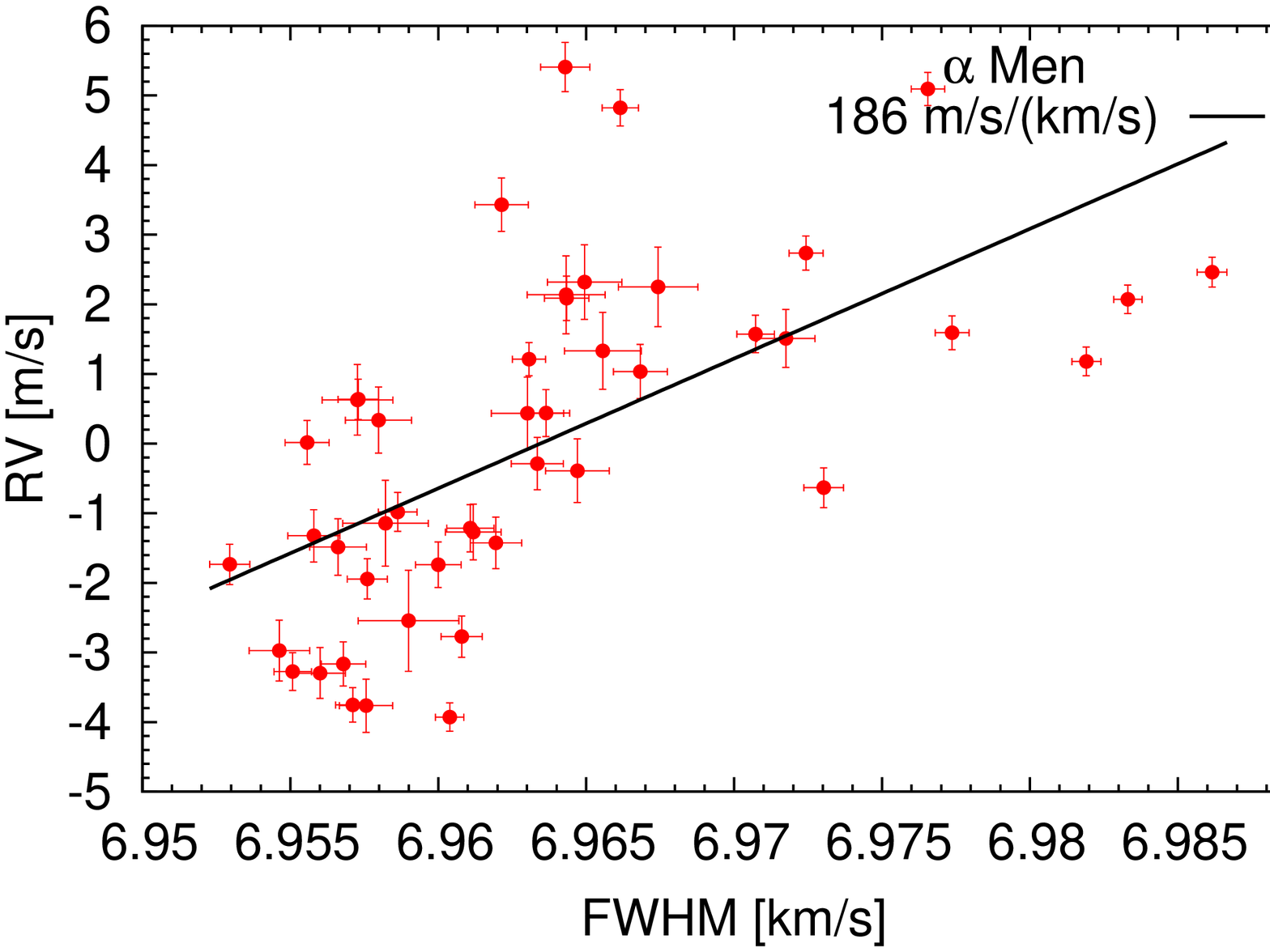}{$\alpha$~Men}
\figmacro{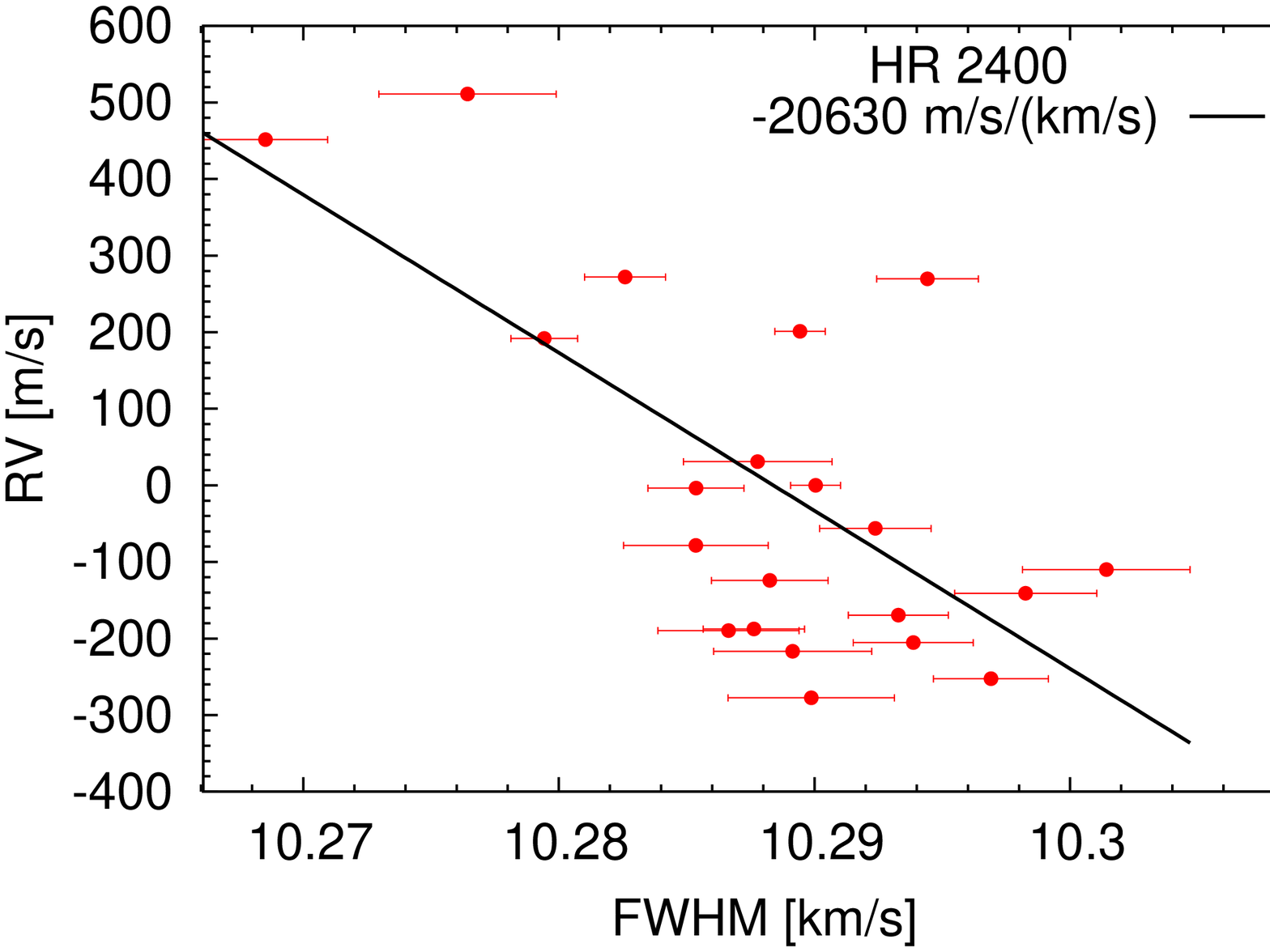}{HR 2400}
\figmacro{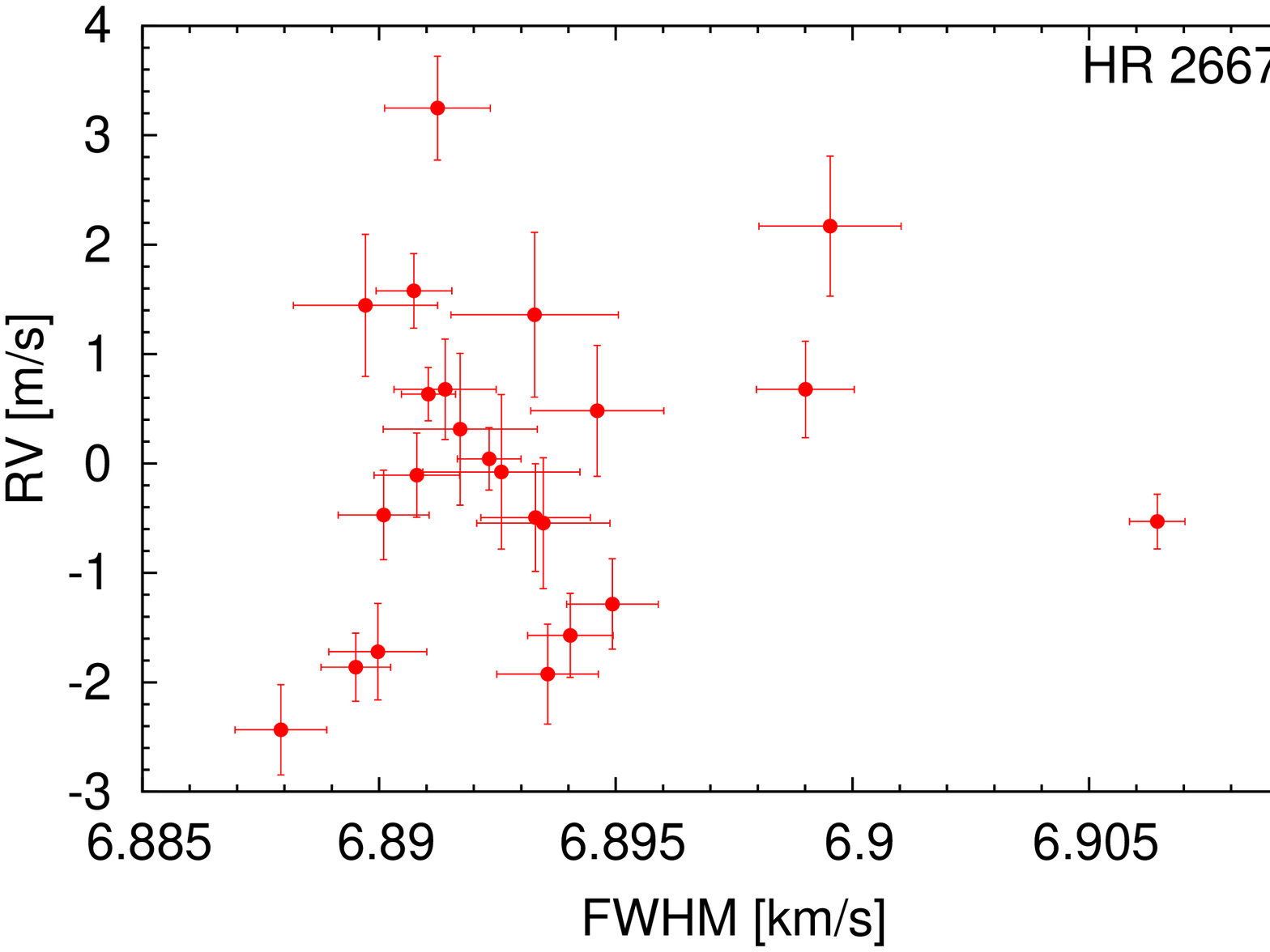}{HR 2667}
\figmacro{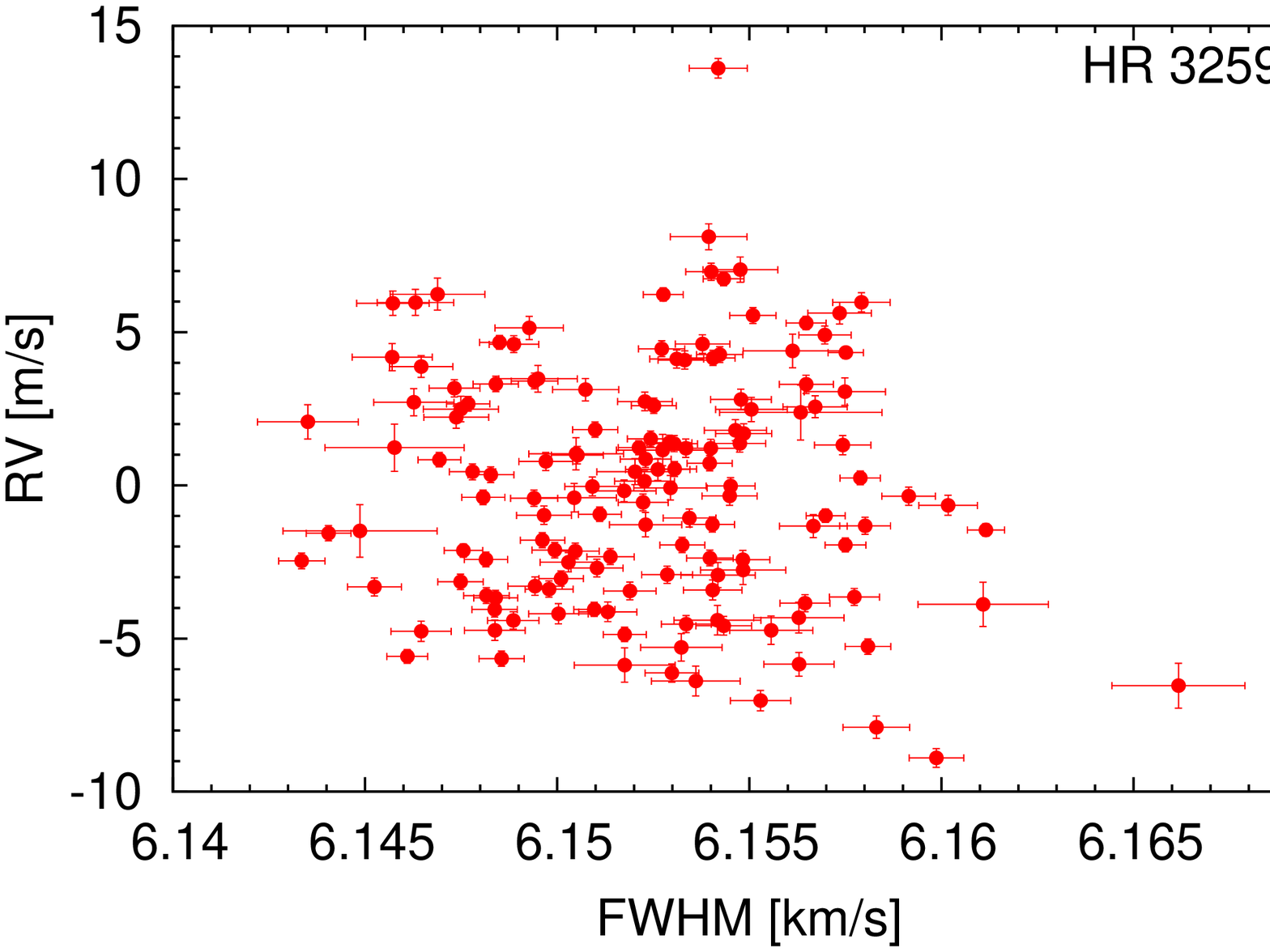}{HR 3259}
\figmacro{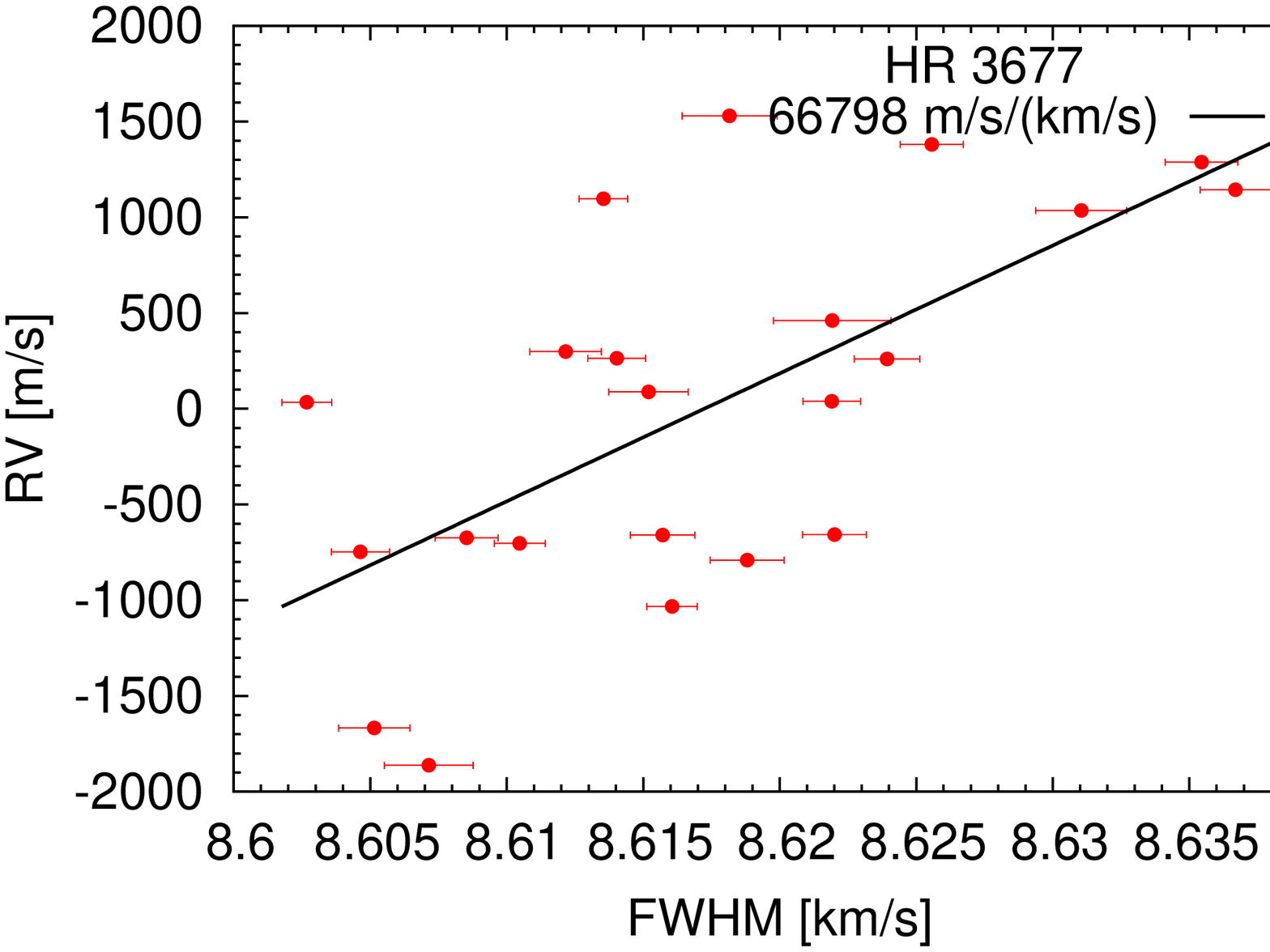}{HR 3677}
\figmacro{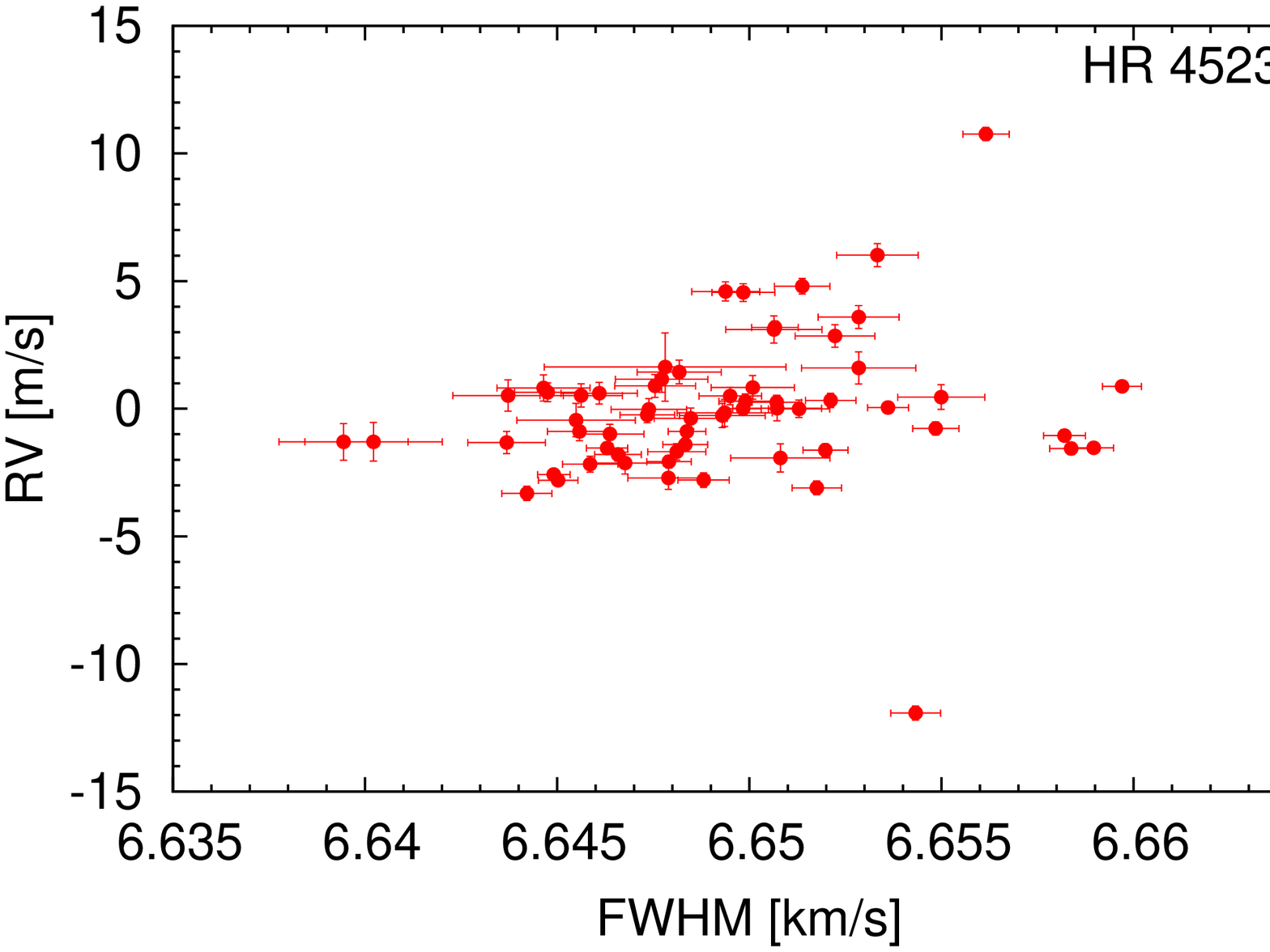}{HR 4523}
\figmacro{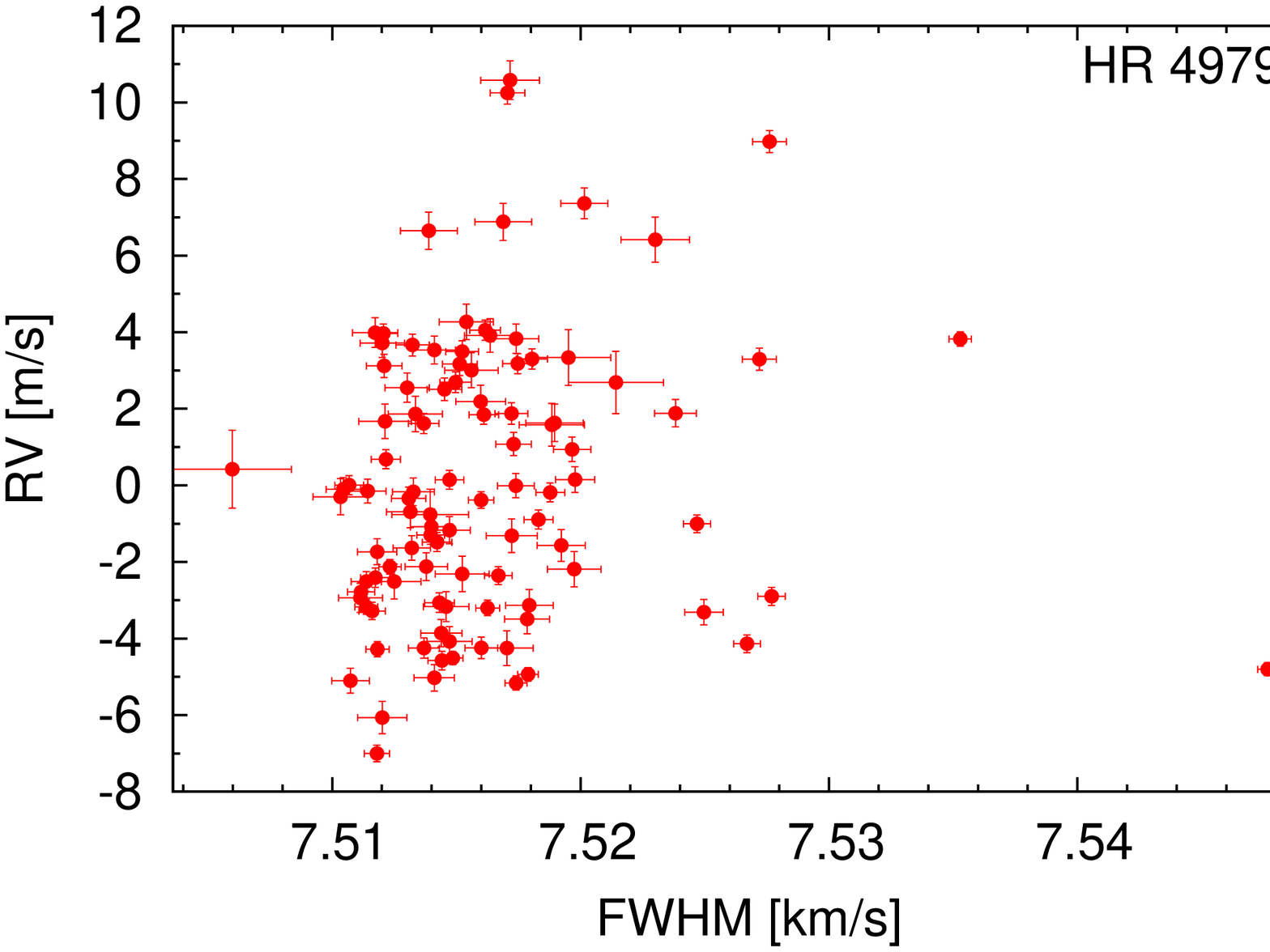}{HR 4979}
\figmacro{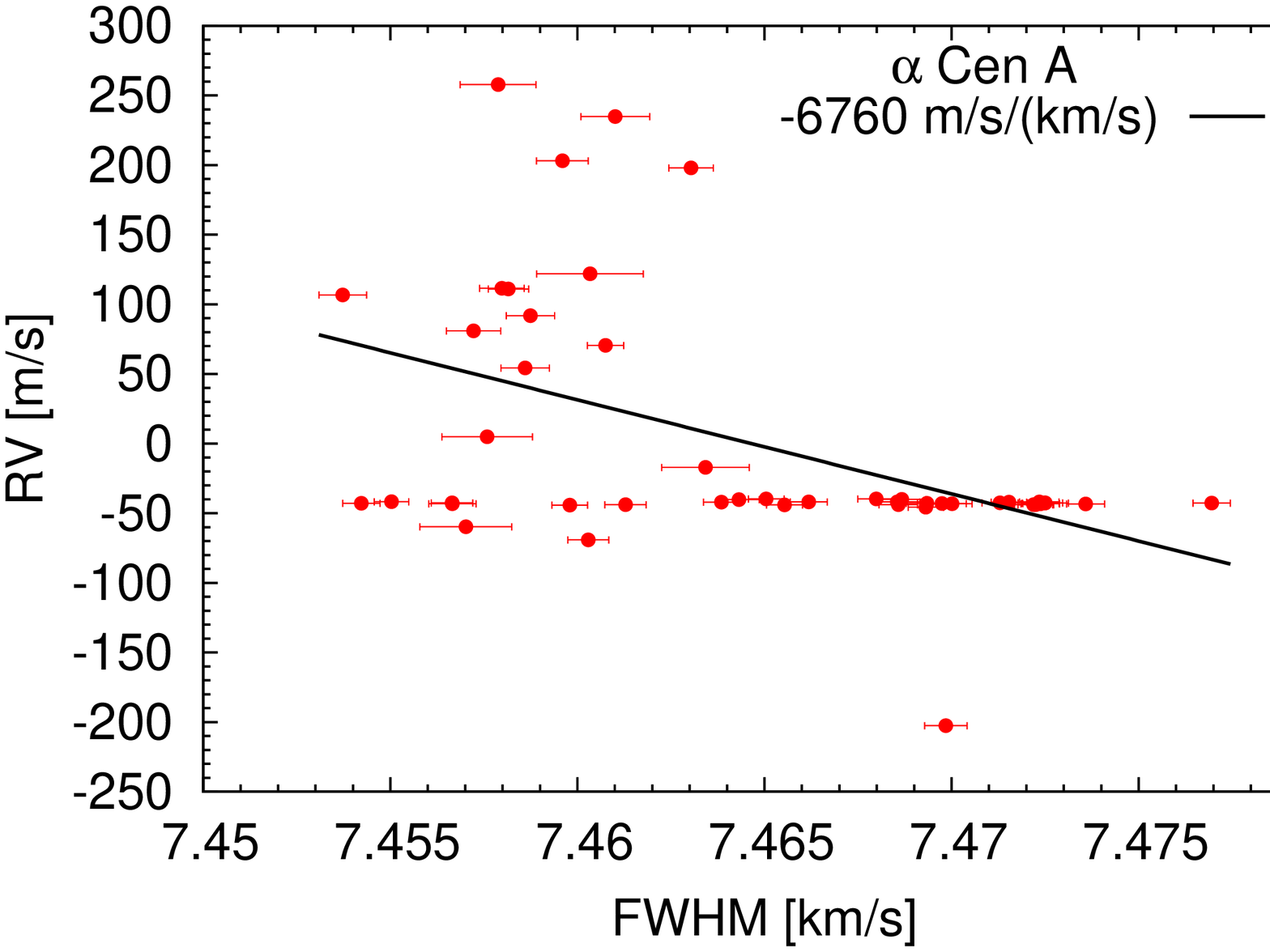}{$\alpha$~Cen~A}
\figmacro{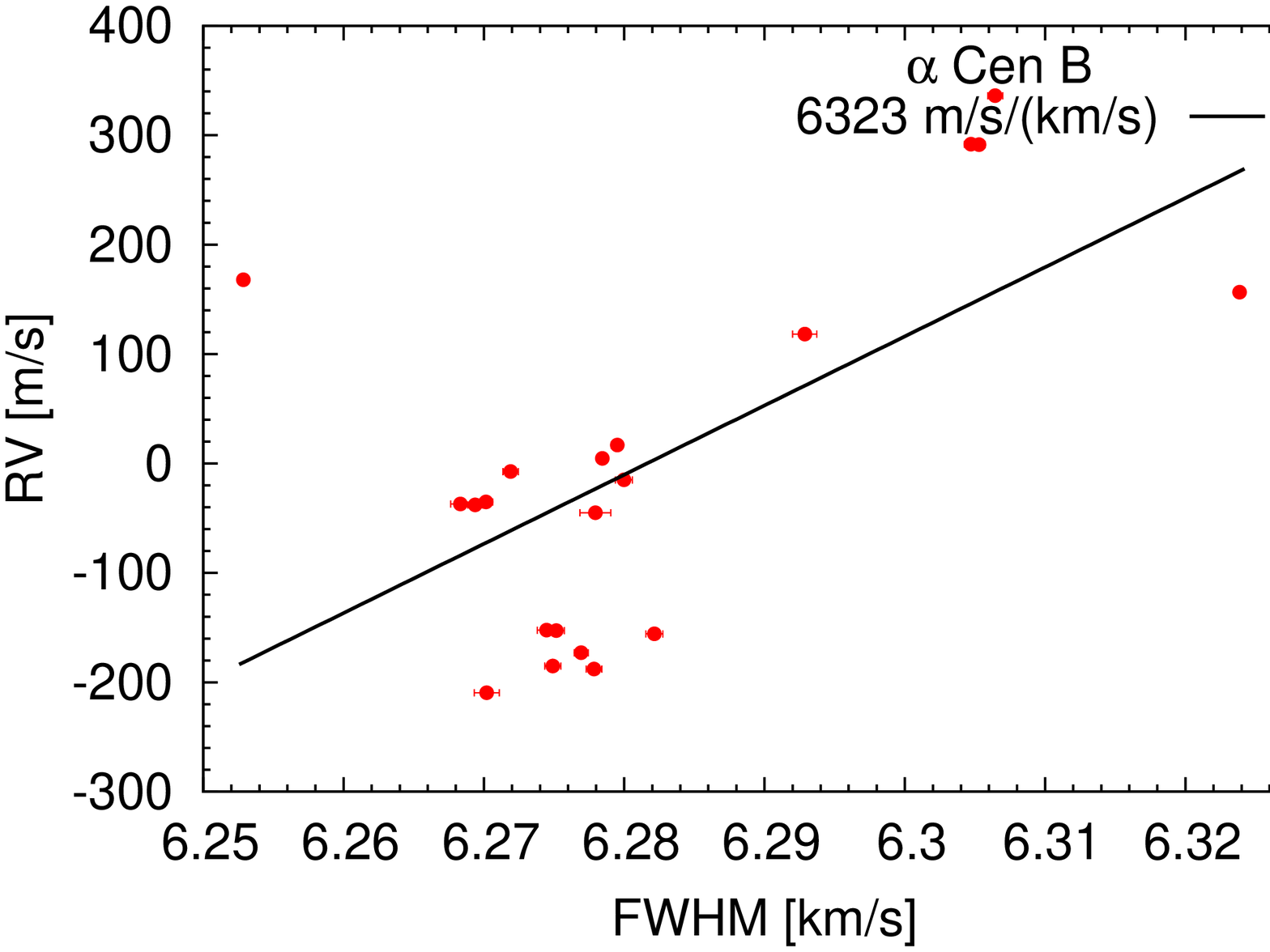}{$\alpha$~Cen~B}
\figmacro{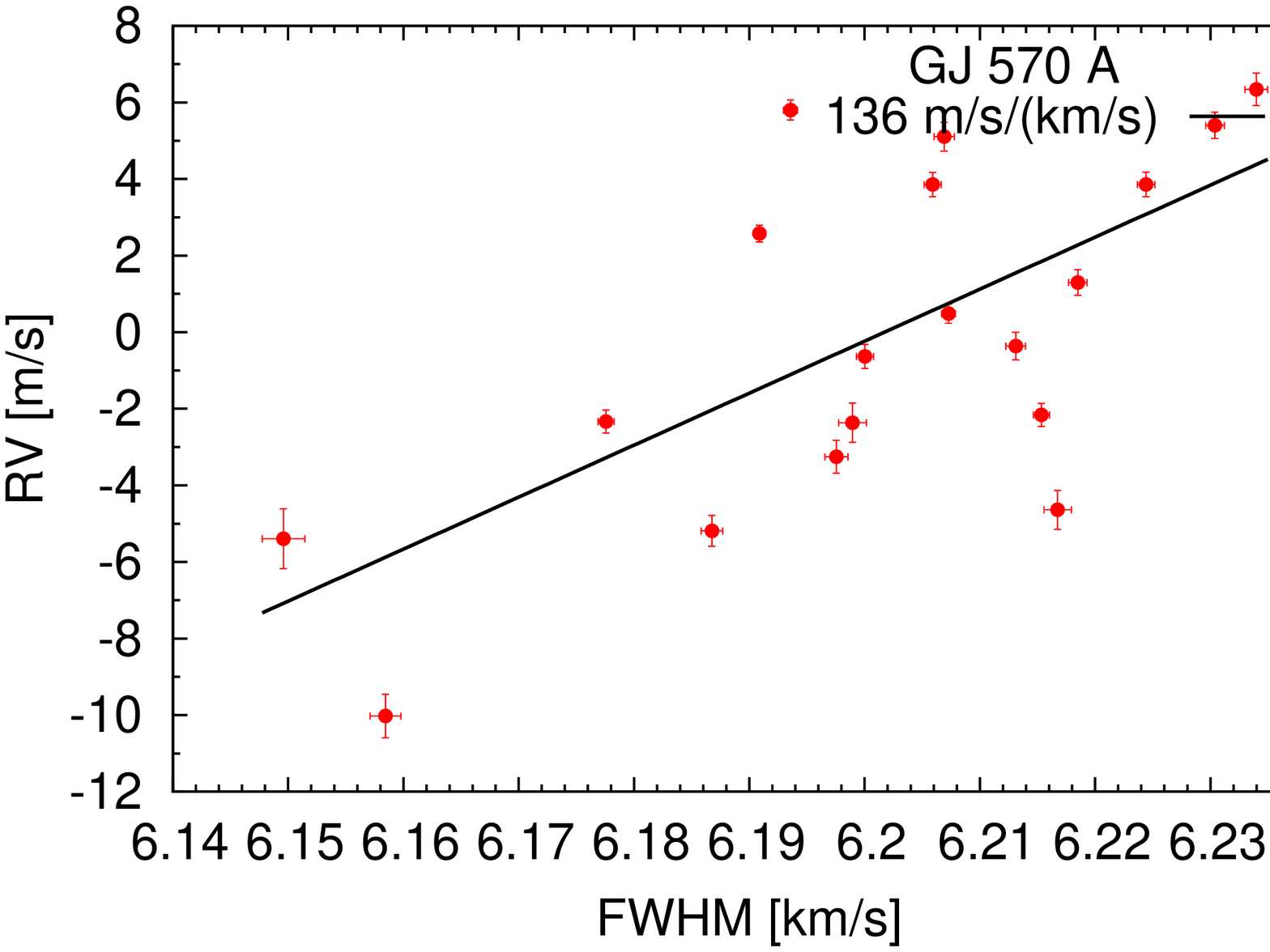}{GJ 570A}
\figmacro{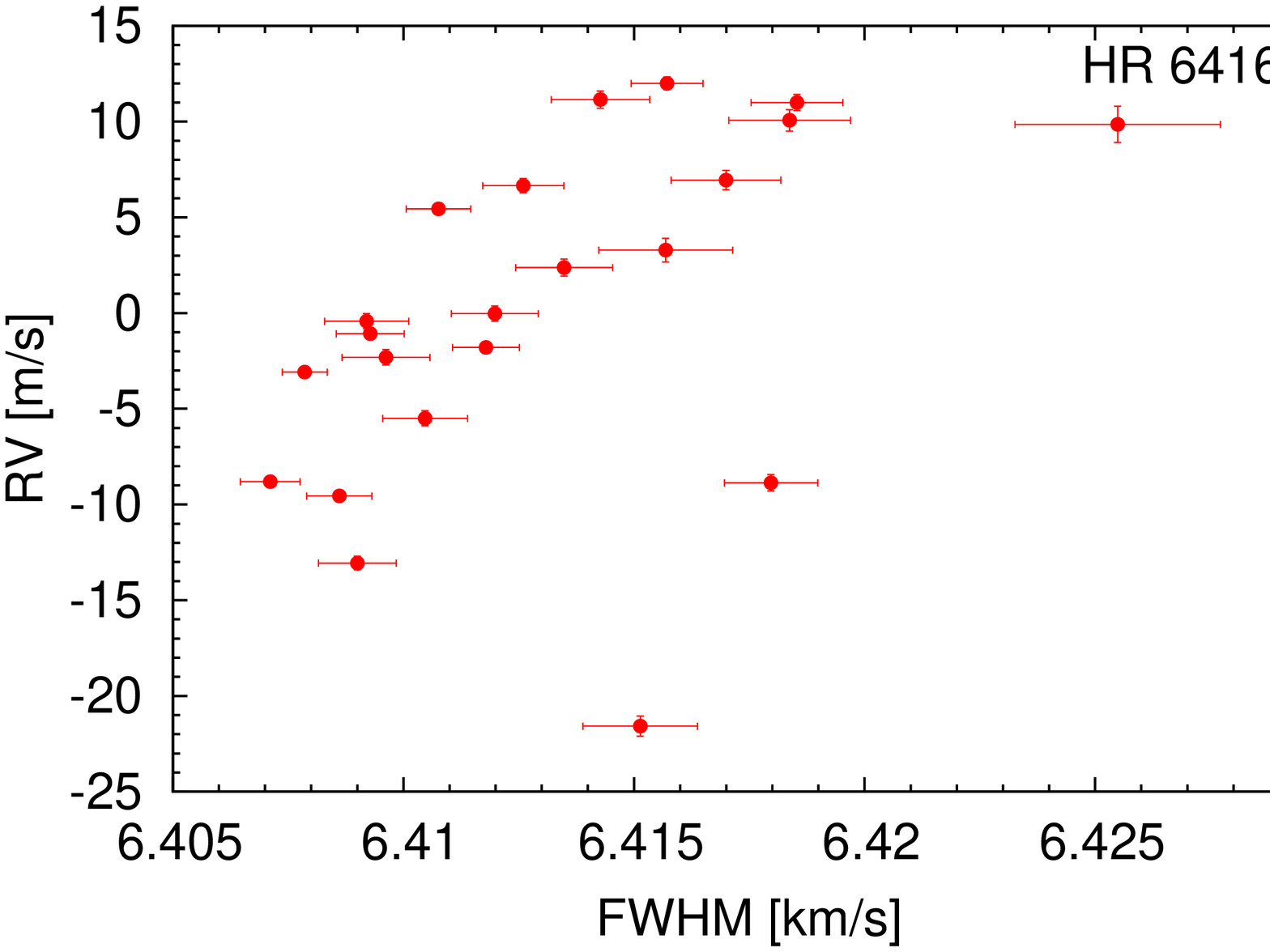}{HR 6416}
\figmacro{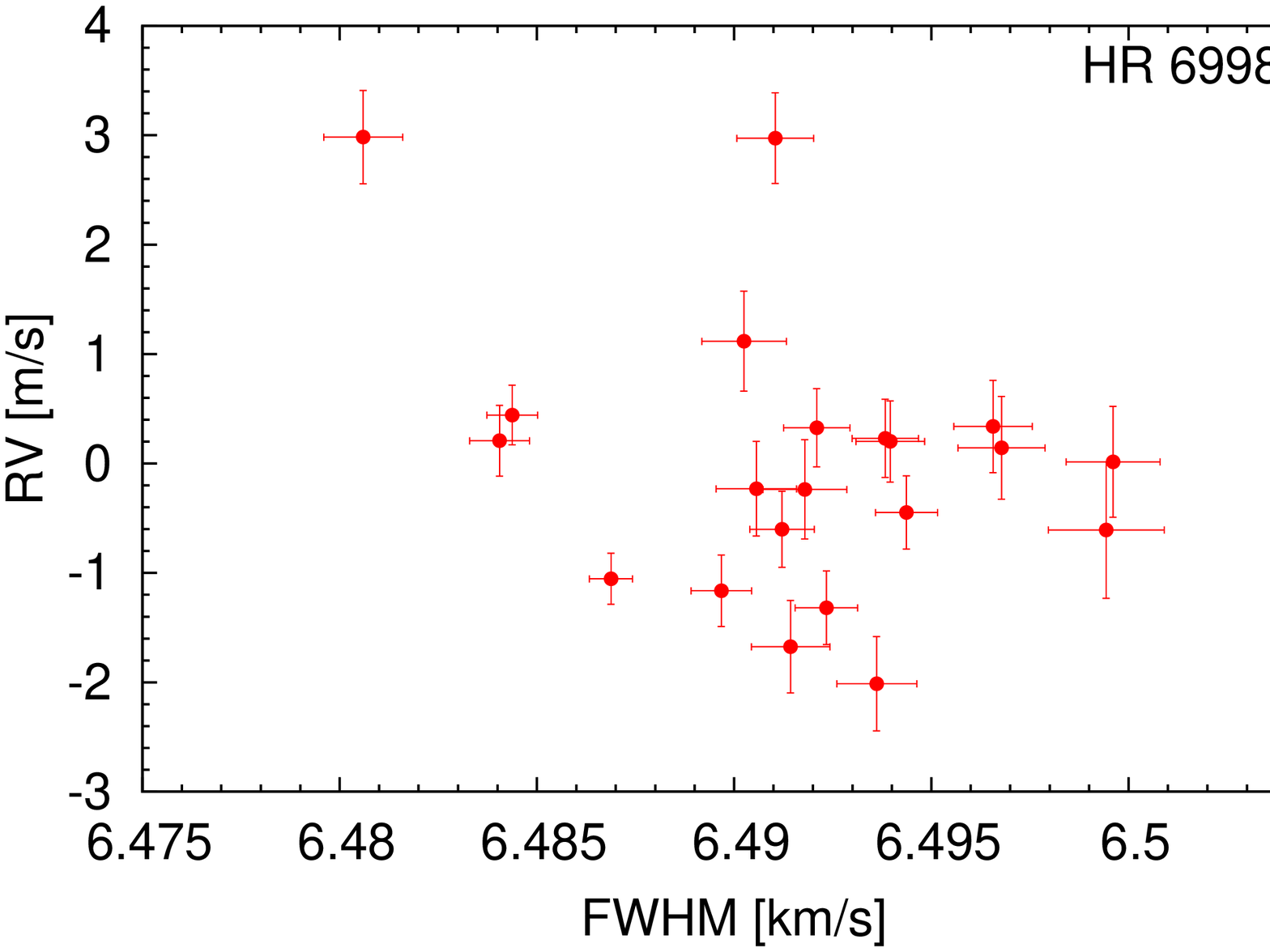}{HR 6998}
\figmacro{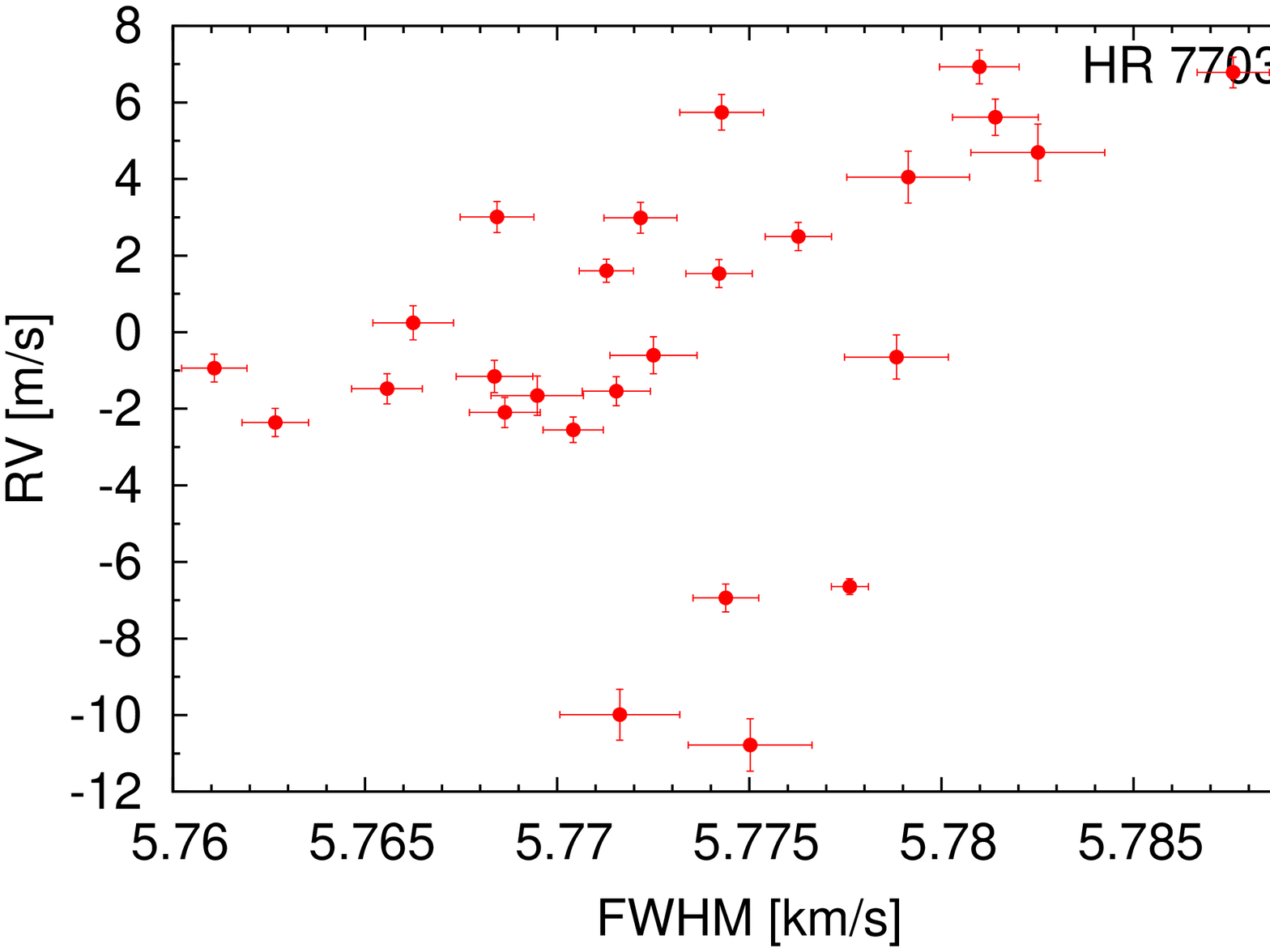}{HR 7703}
\figmacro{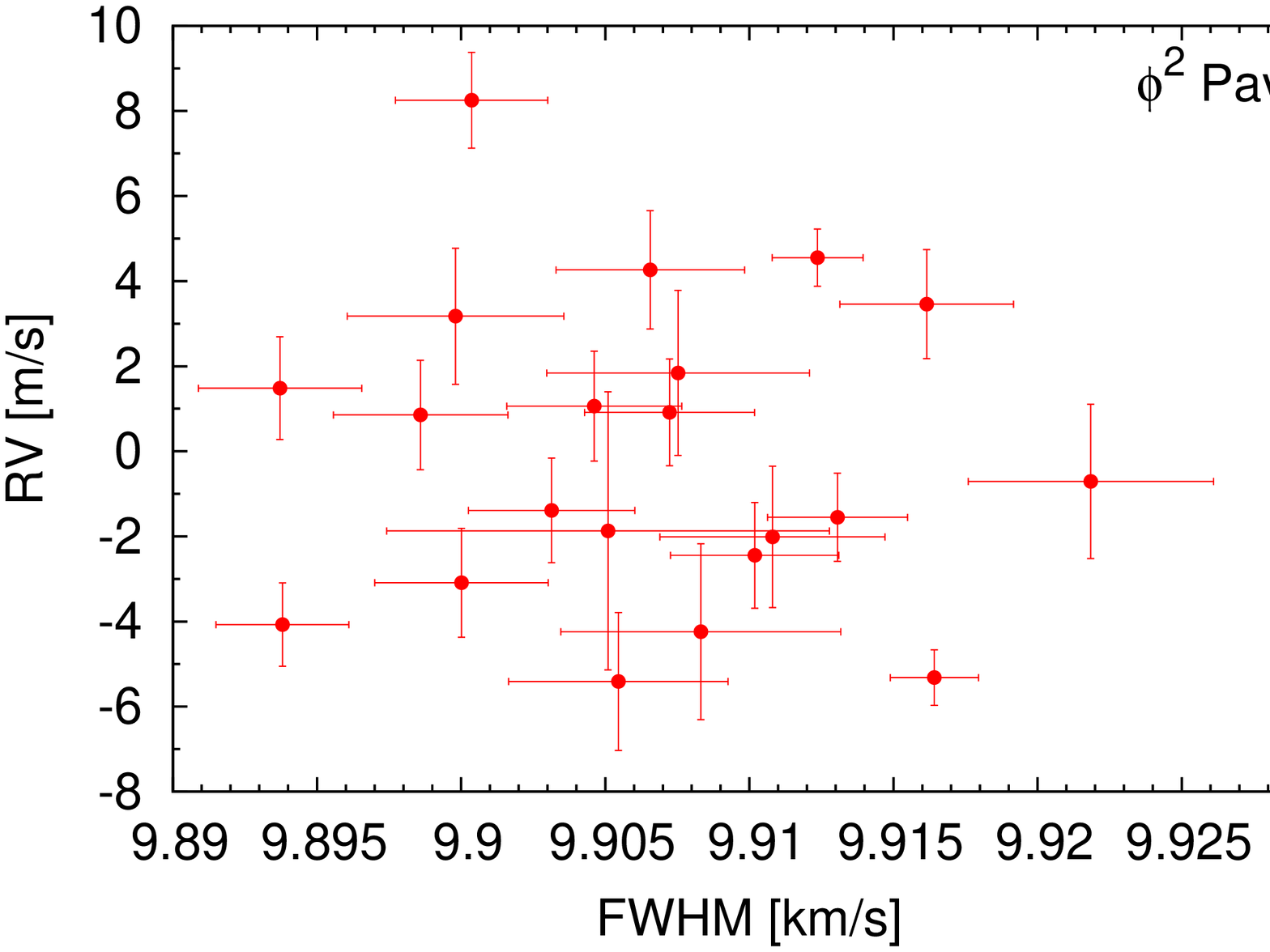}{$\phi^2$~Pav}
\figmacro{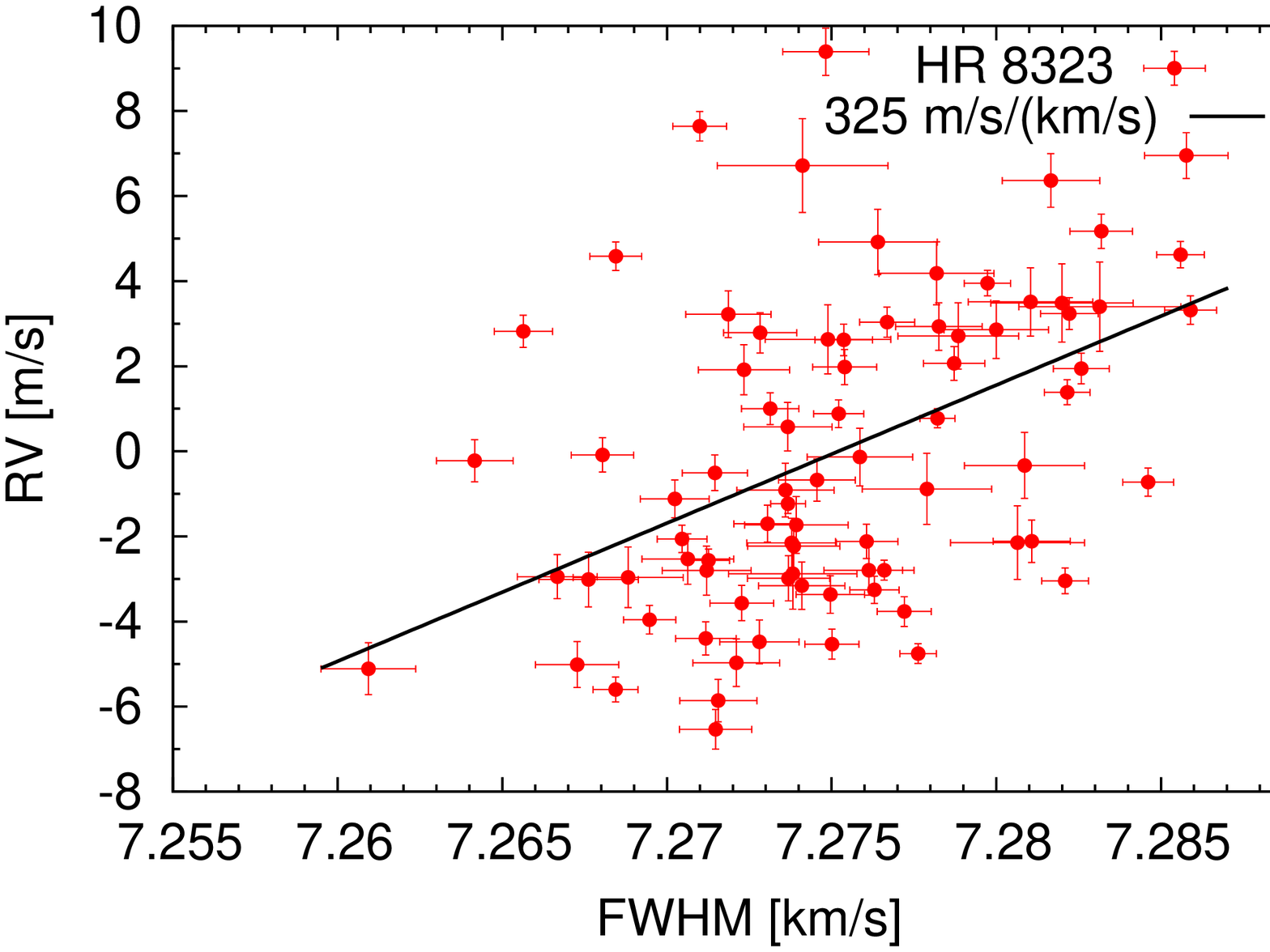}{HR 8323}
\figmacro{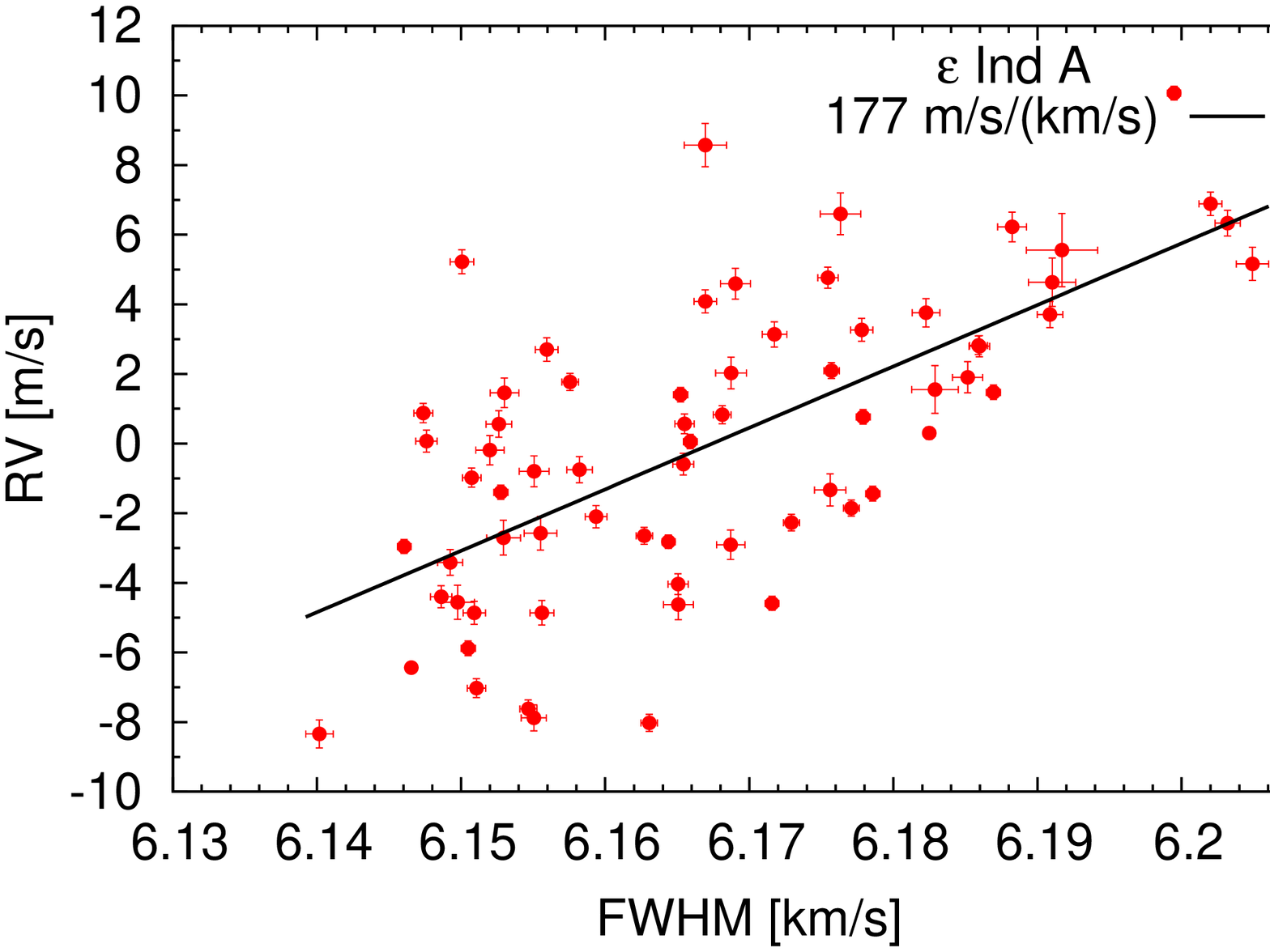}{$\epsilon$~Ind~A}
\figmacro{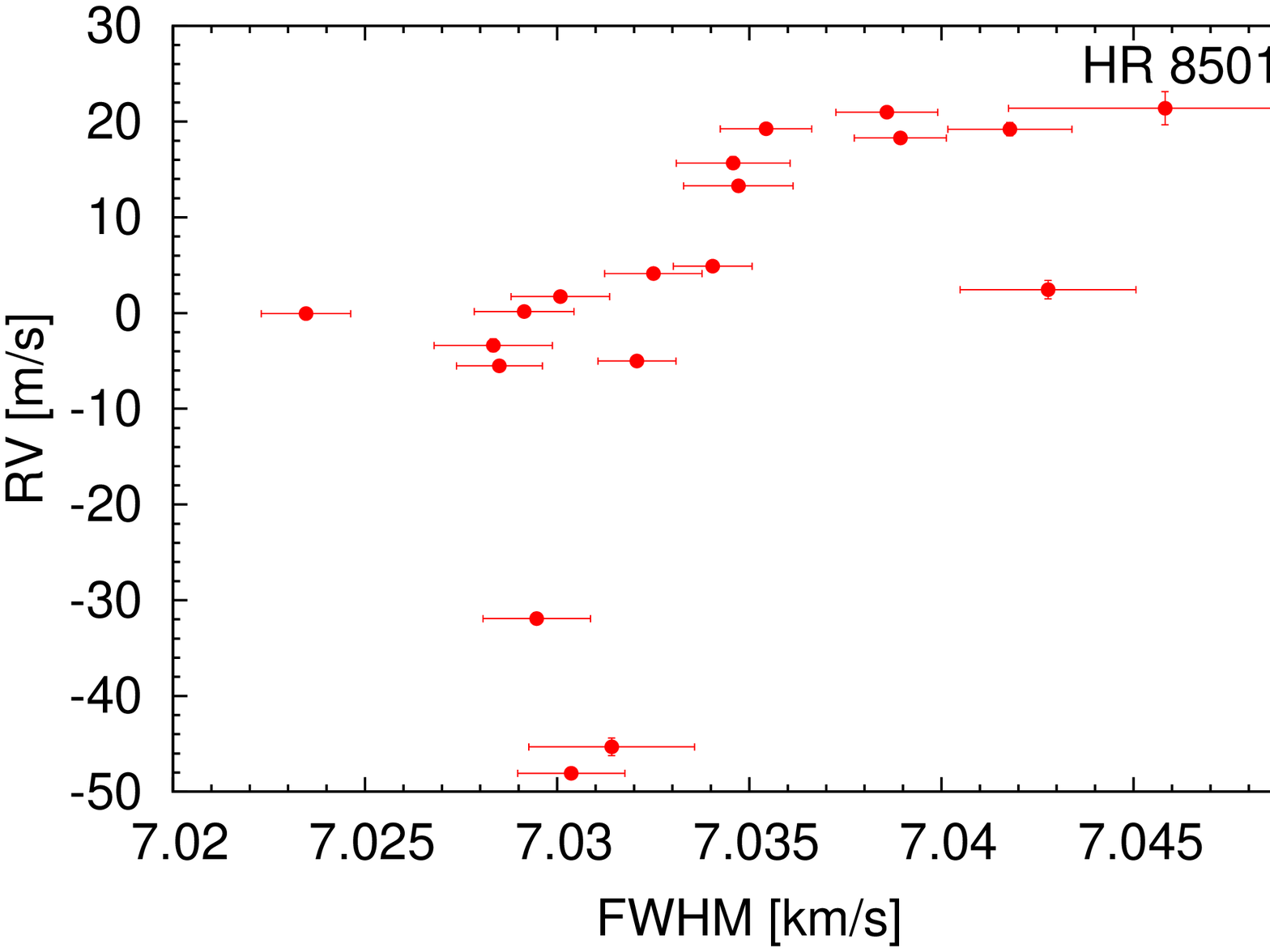}{HR 8501}
\figmacro{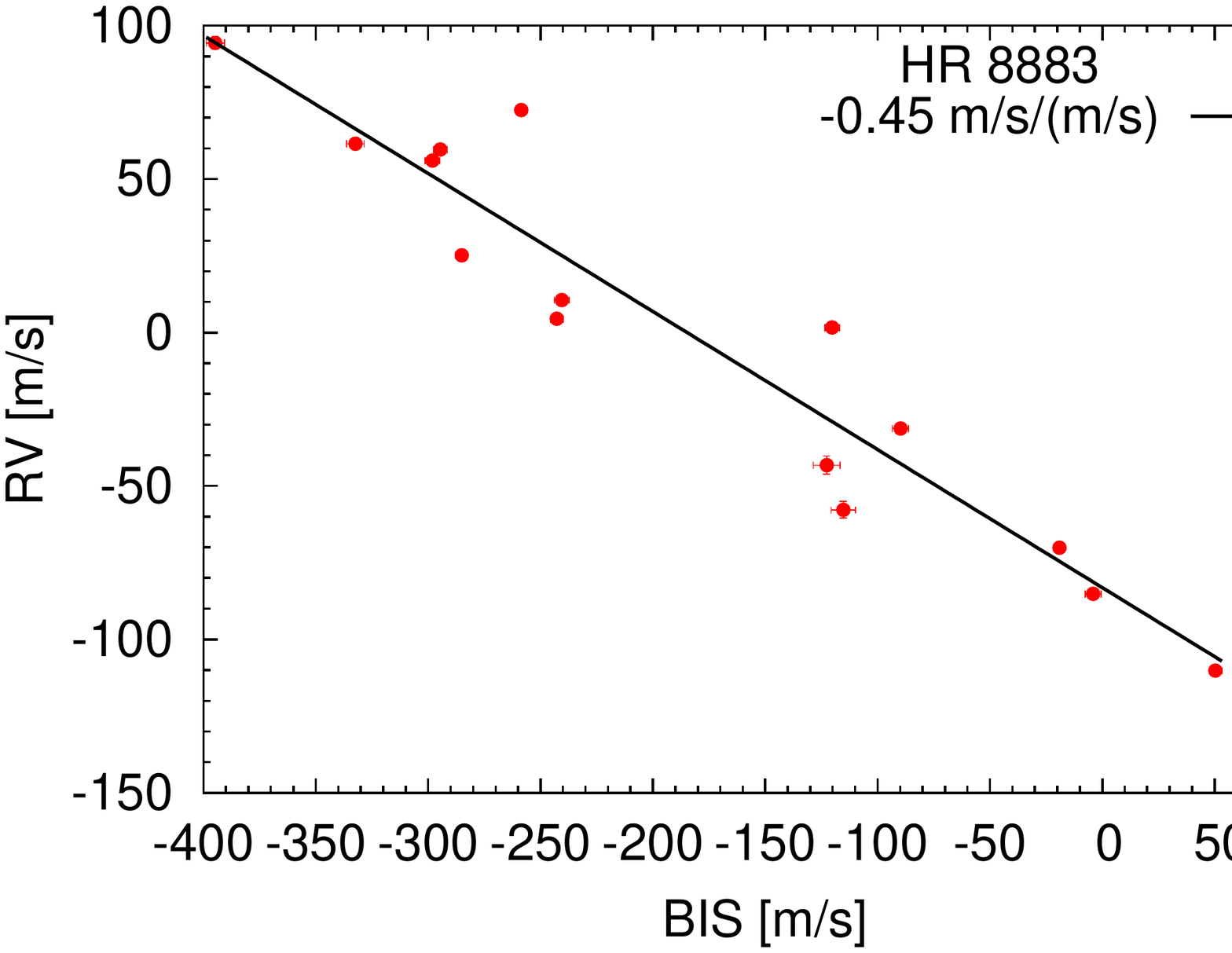}{HR 8883}
\end{appendix}
\end{document}